\documentclass[10pt]{amsart}

\usepackage{amsmath, latexsym, amsfonts, amssymb, amsthm,MnSymbol,stmaryrd}
\usepackage{graphicx, xcolor,hyperref,dsfont,epsfig,caption,wrapfig,subfig,pifont,tikz,float}
\usepackage{booktabs}
\usetikzlibrary{decorations.pathmorphing,positioning}
\usetikzlibrary{calc}
\usepackage{datetime}
\usepackage{movie15}
\hypersetup{
	linktoc=page,
	linkcolor=red,          
	citecolor=blue,        
	filecolor=blue,      
	urlcolor=cyan,
	colorlinks=true           
}

\newcommand{\nocontentsline}[3]{}
\newcommand{\tocless}[2]{\bgroup\let\addcontentsline=\nocontentsline#1{#2}\egroup}
\numberwithin{equation}{section}

\newtheorem{theorem}{Theorem}[section]
\newtheorem{lemma}[theorem]{Lemma}
\newtheorem{proposition}[theorem]{Proposition}
\newtheorem{corollary}[theorem]{Corollary}

\newtheorem{remark}[theorem]{Remark}
\newtheorem{definition}[theorem]{Definition}

\theoremstyle{definition}

\renewcommand{\tilde}{\widetilde}          
\DeclareMathSymbol{\leqslant}{\mathalpha}{AMSa}{"36} 
\DeclareMathSymbol{\geqslant}{\mathalpha}{AMSa}{"3E} 
\DeclareMathSymbol{\eset}{\mathalpha}{AMSb}{"3F}     
\renewcommand{\leq}{\;\leqslant\;}                   
\renewcommand{\geq}{\;\geqslant\;}                   
\newcommand{\dd}{\text{\rm d}}             

\newcommand{\C}{\mathbb{C}}
\renewcommand{\H}{\mathbb{H}}
\newcommand{\D}{\mathbb{D}}
\newcommand{\R}{\mathbb{R}}
\newcommand{\Z}{\mathbb{Z}}
\newcommand{\N}{\mathbb{N}}

\newcommand{\Q}{\mathbb{Q}}
\newcommand{\A}{\mathbb{A}}

\newcommand{\E}{\mathds{E}}
\renewcommand{\P}{\mathds{P}}

\newcommand{\cjd}{\rangle}
\newcommand{\cjg}{\langle}

\newcommand{\hf}{\frac{_1}{^2}}
\newcommand{\cC}{\mathcal{C}}

\newcommand{\bA}{\mathbf{A}}
\newcommand{\bL}{\mathbf{L}}

\newcommand{\bs}{\boldsymbol}

\def\k{\mathbf{k}}

\newcommand{\pl}{\partial}
\newcommand{\bbar}{\overline}
\newcommand{\mc}{\mathcal}
\newcommand{\la}{\lambda}
\newcommand{\til}{\widetilde}

\newcommand{\bzeta}{\boldsymbol{\zeta}}

\usepackage{xparse}

\def\eps{\epsilon}

\def\T{\mathbb{T}}

\def\bi{\begin{itemize}}
	\def\ei{\end{itemize}}

\def\bnum{\begin{enumerate}}
	\def\enum{\end{enumerate}}
\def\<#1{\langle #1 \rangle}

\newcommand{\caA}{{\mathcal A}}

\newcommand{\caP}{{\mathcal P}}

\newcommand{\bH}{\mathbf{H}}

\author{Colin Guillarmou}
\address{Universit\'e Paris-Saclay, CNRS,  Laboratoire de math\'ematiques d'Orsay, 91405, Orsay, France.}
\email{colin.guillarmou@universite-paris-saclay.fr}

\author{R\'emi Rhodes}
\address{Aix-Marseille University, CNRS, Institut de Math\'ematiques de Marseille (I2M), and Institut Universitaire de France (IUF)}
\email{remi.rhodes@univ-amu.fr}

\author{Baojun Wu}
\address{Beijing Institute of Technology}
\email{wubaojunmathe@outlook.com}

\title[Bootstrap for boundary LCFT II: spectral resolution and bootstrap]{Conformal Bootstrap for surfaces with boundary in Liouville CFT. 
	Part II: spectral resolution and bootstrap}    

\date{}
\begin{document}

\begin{abstract}This paper is the second part of the proof of the conformal bootstrap for
Liouville conformal field theory on compact surfaces with boundary. It is
devoted to the spectral theory of the half-annulus semigroup and to the
resulting bootstrap formula. Building on the Segal axioms and gluing
properties established in Part~1~\cite{GRW1}, we identify the generator of
the half-annulus semigroup with the boundary Hamiltonian and establish its
spectral decomposition by scattering methods. This yields a direct-integral
decomposition of the boundary state space into irreducible Virasoro
representations, governed by the boundary spectral measure. Combining this
decomposition with Ward identities for boundary deformations, we express
general correlation functions as integrals over spectral parameters attached
to cutting curves, with integrands given by products of bulk and boundary
structure constants and the corresponding conformal blocks. These results
also underpin the analyticity and symmetry properties of conformal
blocks~\cite{GRSSbloc}, and contribute to the construction of mapping
class group representations on spaces of conformal blocks~\cite{PapierBlocs}.
\end{abstract}
\maketitle
\tableofcontents

\section{Introduction}

 Conformal Field Theory (CFT), in its Euclidean formulation, was originally developed in physics to describe universality classes of statistical systems undergoing second-order phase transitions at criticality.  Its success owes much to the conformal bootstrap program \cite{Polyakov74,Ferrara_Grillo_Gatto73}. In the seminal work of Belavin-Polyakov-Zamolodchikov \cite{BPZ84}, this method, based on the  operator product expansion combined with conformal symmetry, led to the solution of important two-dimensional theories--the rational CFTs--by expressing their correlation functions on closed Riemann surfaces in terms of structure constants and representation-theoretic special functions known as conformal blocks.
 
The study of CFTs on Riemann surfaces with boundary, known as boundary CFT, began roughly forty years ago in physics and has since found wide applications ranging from condensed matter physics to cosmology and string theory. Foundational ideas were developed by Cardy \cite{CARDY1984}, who also applied them successfully to rational CFTs, notably to the minimal models \cite{CARDY1986200,CARDYaspm}. Boundary CFT exhibits a richer structure than its closed-surface counterpart, with deep connections to representation theory (e.g. it provides  a projective representation of the  mapping class group in the space of conformal blocks) and even to three-dimensional Reshetikin-Turaev type Topological Quantum Field Theory, as anticipated by Witten \cite{Witten89} in the context of 3D Chern-Simons theory.

Rational CFTs have a relatively simple algebraic structure: the Hamiltonian (generator of dilations) has a discrete spectrum, and the Hilbert space decomposes into a finite direct sum of irreducible Virasoro representations.  Attention later shifted to non-compact CFTs, where the Hamiltonian has a continuous spectrum and the Hilbert space decomposes over a continuum of irreducible representations. The prototype of such theories is Liouville CFT. Originally introduced by Polyakov \cite{Polyakov81} as a model for random metrics in two dimensions arising in string theory, Liouville theory was already the central motivation behind \cite{BPZ84}. At the physics level, on a closed Riemann surface $\Sigma$ with metric $g$, it is formally described by the path integral
 \begin{equation}\label{thePI}
F\mapsto \int F(\phi) e^{-S_{{\rm L}}(\phi ,g  )} \,D\phi 
\end{equation}
where $D\phi$ is the formal Lebesgue measure on some space of maps $\phi:\Sigma\to\R$,  $F$ is a  test function, and   $S_{{\rm L}} $ is the Liouville action
 \begin{equation}\label{AL}
S_{{\rm L}}(\phi ,g  )=       \frac{1}{4\pi} \int_{\Sigma}\big(|d\phi |^2_{{g }}+QK_{{g }} \phi  +4\pi \mu e^{ \gamma \phi  }\big)\, {\rm dv}_{{g }}  .
\end{equation}
Here $|\cdot|_g$ is the induced metric on $T^\ast\Sigma$,  $K_g$ is the scalar curvature, ${\rm dv}_g$ the Riemannian area element, and the parameters are $\gamma\in(0,2)$, $\mu>0$, and $Q=\tfrac{\gamma}{2}+\tfrac{2}{\gamma}$. Local (primary) fields are formally given by $V_\alpha(z,\phi)=e^{\alpha\phi(z)}$, with $\alpha\in\C$, and correlation functions are
 \begin{equation}\label{correl}
\langle\prod_{i=1}^mV_{\alpha_i}(z_i)\rangle=\int  \Big(\prod_{i=1}^me^{\alpha_i\phi(z_i)}\Big)e^{- S_{{\rm L}}(\phi ,g  )}D\phi 
\end{equation}
 where $z_1,\dots,z_m$ are distinct points on $\Sigma$ and $\alpha_1,\dots,\alpha_m\in\C$ are complex numbers.  A breakthrough was achieved by Dorn-Otto and Zamolodchikov-Zamolodchikov \cite{DornOtto94,Zamolodchikov96}, who conjectured the celebrated DOZZ formula for the structure constants, later refined by Teschner \cite{Tesc}, thereby providing a consistent solution to the bootstrap equations after the spectrum was determined by Curtright-Thorn \cite{CurtrightThorn82} and further analyzed in \cite{Teschner_revisited}.
 
 On the mathematical side, a probabilistic construction of the Liouville path integral was achieved in \cite{DKRV16,Guillarmou2019} using Gaussian multiplicative chaos (GMC). The structure constants were computed in \cite{KRV19_local,KRV_DOZZ} and shown to coincide with the DOZZ formula. The correlation functions were organized into amplitudes through Segal's functor \cite{GKRV21_Segal}, while the analysis of annulus amplitudes led to a representation of the Virasoro algebra on a Hilbert space \cite{BGKRV,BGKR1}. The spectral analysis of the Hamiltonian in \cite{GKRV20_bootstrap} provided a complete basis of generalised eigenstates, reducing the conformal bootstrap on closed surfaces to the computation of matrix elements of pair-of-pants amplitudes in this basis.

 A natural and essential extension is to consider Liouville theory on surfaces with boundary. Physically, boundary Liouville theory plays a central role in string theory, where it describes D-branes (the so-called FZZT and ZZ branes), and has deep connections with statistical mechanics models with boundary conditions.  On a Riemann surface $\Sigma$ with boundary $\partial \Sigma$ and metric $g$, the Liouville action acquires additional boundary terms:
\begin{equation}\label{ALboundary}
S_{{\rm L}}(\phi ,g  )=       \frac{1}{4\pi} \int_{\Sigma}\big(|d\phi |^2_{{g }}+QK_{{g }} \phi  +4\pi \mu e^{ \gamma \phi  }\big)\, {\rm dv}_{{g }} +  \frac{1}{2\pi} \int_{\partial\Sigma}\big(Qk_{{g }} \phi  +2\pi \mu^{\rm b} e^{\frac{ \gamma}{2} \phi  }\big)\, \dd \ell_g 
\end{equation}
where $k_g$ is the geodesic curvature, $\dd \ell_g$ the boundary length element, $\mu\geq 0$ and $\mu^{\rm b}$ is a nonnegative piecewise constant function on $\partial\Sigma$. Local fields are now of two types: bulk insertions $V_\alpha(z,\phi)=e^{\alpha\phi(z)}$ with $z\in\mathring{\Sigma}$, and boundary insertions $V_\beta(s,\phi)=e^{\frac{\beta}{2}\phi(s)}$ with $s\in \partial \Sigma$. The correlation functions then formally read
  \begin{equation}\label{correlboundary}
\langle\prod_{i=1}^mV_{\alpha_i}(z_i)\prod_{j=1}^{m^{\rm b}}V_{\beta_j}(s_j)\rangle=\int   \Big(\prod_{i=1}^me^{\alpha_i\phi(z_i)}\prod_{j=1}^{m^{\rm b}}e^{\frac{\beta_j}{2}\phi(s_j)}\Big)e^{- S_{{\rm L}}(\phi ,g  )}D\phi 
\end{equation}
 where $z_1,\dots,z_m$ are distinct points in $\mathring{\Sigma}$, $s_1,\dots, s_{m^{\rm b}}$ are distinct points on $\partial\Sigma$, and $\alpha_1,\dots,\alpha_m,\beta_1,\dots,\beta_{m^{\rm b}}$ are complex numbers.  A rigorous path integral construction using probabilistic methods was initiated in \cite{huang2018,MR3843631} and extended to arbitrary bordered surfaces in \cite{Wu1}. The construction, recalled in Section \ref{sec:open}, involves the Gaussian Free Field (GFF, see Section \ref{sec:back}) with Neumann condition on $\pl \Sigma$.

\begin{figure}[h] 
 \begin{center}
\begin{tikzpicture}[scale=0.8]
\draw[fill=lightgray,draw=black,very thick] (0,0) circle (1.5) ;
\draw (0,0) node{$\times$} ;
\draw (0,0) node[right]{$\alpha$} ;
\draw (-0.6,0.7) node[right]{$\mu$} ;
\draw (1.1,1.2) node[right]{$\mu^{\rm b}$} ;
\draw[fill=lightgray,draw=black,very thick] (4,0) circle (1.5) ;
\draw (4,0) node{$\times$} ;
\draw (4,0) node[right]{$\alpha$} ;
\draw (3.6,0.7) node[right]{$\mu$} ;
\draw (5.1,1.2) node[right]{$\mu^{\rm b}$} ;
\draw (5.5,0) node{$\times$} ;
\draw (5.5,0) node[left]{$\beta$} ;
\draw[fill=lightgray,draw=black,very thick] (8,0) circle (1.5) ;
\draw (6.5,0) node{$\times$} ;
\draw (6.5,0) node[right]{$\beta$} ;
\draw (7.6,0.7) node[right]{$\mu$} ;
\draw (9.1,1.2) node[right]{$\mu_1$} ;
\draw (9.1,-1.2) node[right]{$\mu_2$} ;
\draw (9.5,0) node{$\times$} ;
\draw (9.5,0) node[left]{$\beta$} ;
\draw[fill=lightgray,draw=black,very thick] (12,0) circle (1.5) ;
\draw (11,1.11) node[rotate=30]{$\times$} ;
\draw (11.2,1.11) node[right,below]{$\beta_2$} ;
\draw (11,-1.11) node[rotate=30]{$\times$} ;
\draw (11.2,-1.11) node[right,above]{$\beta_3$} ;
\draw (6.5,0) node[right]{$\beta_1$} ;
\draw (11.8,0.7) node[right]{$\mu$} ;
\draw (13.1,1.2) node[right]{$\mu_2$} ;
\draw (10.6,0) node[left]{$\mu_3$} ;
\draw (13.1,-1.2) node[right]{$\mu_1$} ;
\draw (13.5,0) node{$\times$} ;
\draw (13.5,0) node[left]{$\beta_1$} ;
\end{tikzpicture}
\end{center}
\caption{Boundary structure constants (SC). From left to right:  bulk SC, bulk-boundary SC, boundary 2pt SC, boundary 3pt SC.}\label{pic:boundarystructure}
\end{figure}
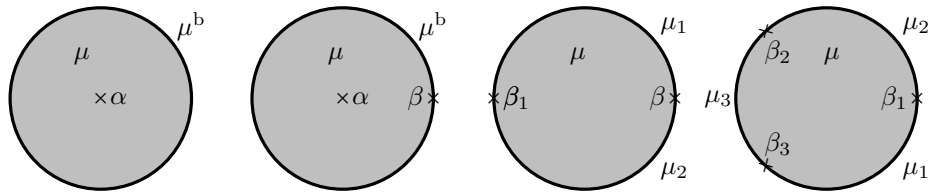

 The purpose of this manuscript is to establish the conformal bootstrap for boundary Liouville CFT. While the case of closed surfaces already provides a blueprint for the strategy, the presence of boundaries introduces new geometric, analytic, and algebraic features, which we address here  in detail. In particular,   boundary Liouville theory involves several structure constants in addition to the usual DOZZ structure constant, as illustrated in Figure \ref{pic:boundarystructure}. These constants correspond to disk correlations with either bulk or boundary vertex insertions.

The bulk structure constant, introduced by Fateev-Zamolodchikov-Zamolodchikov (FZZ) \cite{FZZ}, was recently established in full mathematical rigor by Ang, Remy, Sun \cite{AngRemySun21_FZZ}. The bulk-boundary, boundary two-point, and boundary three-point structure constants--originally computed in the physics literature by  Hosomichi \cite{Hosomichi}, FZZ \cite{FZZ},  and Ponsot-Teschner \cite{PT_boundary_3pt}, respectively--were derived rigorously in \cite{ang2024derivationstructureconstantsboundary}. These results build on earlier works of Remy and Zhu \cite{FBformula,RemyZhu20,MR4483018} in the special case $\mu=0$, which already led to a number of exact formulas for GMC measures on the circle and interval (e.g. the Fyodorov-Bouchaud formula). See Subsection \ref{sub:structure}.

From a probabilistic perspective, the boundary structure constants carry significant   information. For instance, the boundary two-point structure constant encodes the joint distribution of the area and boundary length of two-pointed quantum disks \cite{MatingOfTrees,CWquantumdisk}, while the boundary three-point structure constant does the same for a more general class of Liouville quantum gravity surfaces known as quantum triangles \cite{AHS21_welding,Ang:2022yoz}.

The Segal functor was constructed by the present authors in \cite{GRW1}. In this framework, amplitudes are assigned to Riemann surfaces with boundary. Boundaries come in two types of gluing components: boundary circles and boundary half-circles, the latter joining Neumann-type boundary components that cannot be glued.  Accordingly, two distinct Hilbert spaces are attached to the theory: one for the boundary circles and another for the boundary half-circles (see Section \ref{sec:amplitude}). Two semigroups are extracted from the amplitudes \cite{GRW1}: the semigroup of annuli and the semigroup of half-annuli, from which a great deal of structure can be inferred.

First, we obtain representations of the symmetry algebra acting on the theory. The annuli semigroup carries two commuting representations of the Virasoro algebra, while the half-annuli semigroup carries one   (see Section \ref{sec:semigroups}). The respective dilation generators are the bulk Hamiltonian and the boundary Hamiltonian (see Section \ref{s:Hamiltonians_bulk}). The spectral analysis of the bulk Hamiltonian was carried out in \cite{GKRV20_bootstrap}; here we perform this task for the boundary Hamiltonian (Section \ref{sec:spectral}), obtaining a spectral resolution for each Hilbert space, each organized as a direct sum of irreducible Virasoro representations. In physics, this spectrum was studied by Teschner \cite{Teschner:2000hD}. We emphasize a limitation of our methods: the boundary Hamiltonian is substantially more delicate than the bulk one and has a subtle domain, so our analysis is restricted to   $\gamma<\sqrt{2}$ in the case when $\mu>0$, or $\gamma\in (0,2)$ when $\mu=0$.

Second, we establish the differentiability of   amplitudes with respect to boundary deformations (Subsection \ref{sub:derivamp}), leading to   the Ward identities in Subsections   \ref{sub:ward1} and \ref{sec: variation_corner}. The Segal functor then allows us to decompose correlation functions into pairings of amplitudes of geometric building blocks (see Subsection \ref{geometric_decomp}) in the relevant Hilbert spaces. Inserting the spectral decomposition at each gluing curve (circle or half-circle) reduces the computation of correlation functions to the evaluation of    matrix elements of building block amplitudes (See Section \ref{sec:Ward}). Evaluating  these amplitudes at primary states (highest weight vectors of irreducible Virasoro representations) produces the structure constants, while the action of descendants on the primaries contributes a multiplicative factor, which will serve to constitute   the conformal blocks. At this stage, we implement an idea originating with Cardy in the  Rational CFTs case:  the conformal blocks of a bordered surface can be identified with the holomorphic part of the conformal block of its Schottky double. The final outcome is our main theorem, stated in Subsection \ref{CBB} and summarized below.

 \subsection{Informal statement of the main result}

Our main statement holds under the condition   $\gamma<\sqrt{2}$ in the case when $\mu>0$, and for $\gamma\in (0,2)$ when $\mu=0$. We fix a maximal system of cutting curves that decomposes the surface $\Sigma$ into geometric building blocks. Roughly speaking, this means that the Neumann double (including the marked points) of each geometric building block along its Neumann boundary is a pair of pants, with the important convention that each marked point on the double is counted as an end of a pair of pants. This decomposition is encoded by a graph $\mc{G}$ (see Subsection~\ref{sub:graph}). To each geometric building block we associate a structure constant, as illustrated in Figure~\ref{pic:boundarystructure}, a metric-dependent constant, and a matrix element. These matrix elements are then assembled to form the conformal block associated with the surface $\Sigma$ and the graph $\mc{G}$. A key result is that this conformal block coincides with the conformal block of the Neumann double of $\Sigma$, equipped with the graph $\mc{G}^\#$ obtained by doubling the system of cutting curves.

Our main theorem can then be informally stated as follows.
\begin{theorem}[\textbf{Conformal bootstrap}]\label{th:fullintro} 
The correlation function of the   surface  $(\Sigma,g)$ with $m$ 
interior insertions $({\bf x},\boldsymbol{\alpha})$ and $m^{\rm b}$ boundary insertions $({\bf s},\boldsymbol{\beta})$ is given by the expression
\begin{equation}\label{theoremmodifiedconfblock}
\cjg \prod_{i=1}^m V_{\alpha_i}(x_i)\prod_{i=1}^{m^{\rm b}}V_{\beta_i}(s_i)\cjd_{\Sigma,g} =\frac{C_{\rm gluing}}{(2\pi)^L(4\pi)^{L^+}}\int_{\R_+^L\times \R_+^{L^+}}C_{\Sigma,g}(\boldsymbol{\alpha},\boldsymbol{\beta},{\bf p}) \boldsymbol{\rho}(\boldsymbol{\alpha},\boldsymbol{\beta},{\bf p},{\bf p}^+)
\tilde{\mc{F}}^{\mc{G}^\#}_{\bf p,\bf p,\bf p^+}(\boldsymbol{\alpha}^\#)\dd{\bf p}\dd{\bf p}^+.
\end{equation}
Here $C_{\rm gluing}$ is given by \eqref{Cgluing} and depends only on the number of cutting curves: $L$ stands for the number of interior cutting curves and $L^+$ for the number of boundary-to-boundary cutting curves. The factors $\boldsymbol{\rho}(\boldsymbol{\alpha},\boldsymbol{\beta},{\bf p},{\bf p}^+)$ and $C_{\Sigma,g}(\boldsymbol{\alpha},\boldsymbol{\beta},{\bf p},{\bf p}^+)$ are respectively the products of the structure constants and of the metric-dependent constants associated with each geometric building block, as defined in Subsection~\ref{constantpj}. The conformal block is denoted by $\tilde{\mc{F}}^{\mc{G}^\#}_{{\bf p},{\bf p},{\bf p}^+}(\boldsymbol{\alpha}^\#)$.
The integration is performed over the spectral parameters ${\bf p}$ and ${\bf p}^+$: one spectral parameter is attached to each cutting curve, with the superscript $+$ indicating those associated with boundary-to-boundary cuts, while ${\bf p}$ collects the parameters corresponding to interior cutting curves.
\end{theorem}
\begin{remark}
In~\cite{GKRV21_Segal}, it was shown that the conformal block
		$\tilde{\mc{F}}^{\mc{G}^\#}_{\bf p,\bf p',\bf p^+}(\boldsymbol{\alpha}^\#)$
		exists for almost every $(\bf p,\bf p',\bf p^+)\in\R_+^{2L+L^+}$.
		However, this does not imply existence for almost every
		$(\bf p,\bf p,\bf p^+)\in\R_+^{2L+L^+}$, since the diagonal set
		$\{\bf p=\bf p'\}$ has Lebesgue measure zero in $\R_+^{2L}$.
		Moreover, in~\cite{GKRV21_Segal} the conformal block is shown to be an $L^2$
		function with respect to the DOZZ measure, whereas here we prove that the
		(diagonal) conformal block is $L^1$ with respect to the bulk and boundary
		structure constants.
\end{remark}

Conceptually, our approach shows how the blueprint developed for closed Liouville CFT adapts to bordered surfaces:
\begin{itemize}
\item Segal's functor incorporates both circle and half-circle gluings;
\item the analysis of the boundary Hamiltonian yields the relevant Hilbert space decomposition;
\item Ward identities control the action of descendants;
\item and Cardy's doubling identifies the boundary conformal blocks with the holomorphic parts of blocks on the double surface.
\end{itemize}
Altogether, these ingredients produce a rigorous bootstrap formalism for boundary Liouville theory, providing a model case for more general boundary CFTs.

\subsection{Applications and further directions}
 
The bootstrap for boundary Liouville CFT opens the way to several further developments. 
 
On the probabilistic side, Wu initiated in \cite{WuConfBoot} the study of the conformal bootstrap for surfaces with boundary in the case of the annulus with two boundary insertions, a setup where only bulk technology is needed for the cuts. Building on this work, Ang-R\'emy-Sun obtained in \cite{ARS25} the law of the random modulus for random planar maps with annulus topology. These preliminary results suggest a broader program, where the boundary bootstrap could serve as a key input to determine the distribution of the random modulus in the scaling limit of random planar maps. 
 
 The structure constants of Liouville CFT are also expected to encode information about many other critical statistical physics models, particularly those described by Schramm-Loewner Evolution (SLE) or Conformal Loop Ensembles (CLE). The bridge between these theories is provided by couplings between the Gaussian Free Field and SLE/CLE \cite{MatingOfTrees,AHS21_welding}. This connection has already led to striking results, such as the computation of three-point correlation functions of CLE \cite{ACSW24}, shown to match the imaginary DOZZ formula, and the determination of the backbone exponent in critical percolation \cite{nolin}. A major challenge is to extend these bridges to correlation functions with nontrivial moduli dependence, since the structure constants are trivial from the viewpoint of moduli. 
 
 Some specific conformal blocks have been shown to admit probabilistic representations \cite{GRSS24,GRSSbloc}. The key input is the bootstrap formula for the disk four-point function with boundary insertions  (see Subsection \ref{special}), which can be Fourier transformed to yield such a representation. Understanding whether this probabilistic formulation extends systematically is important, as it provides information--such as analyticity in certain parameters--that is otherwise difficult to access.

 \medskip
 On the geometric side, the structure of conformal blocks and their relation to the quantization of Teichm\"uller spaces remain to be explored. In a forthcoming work \cite{BGKRX}, it is proven that conformal blocks are holomorphic sections of a holomorphic line bundle over the Teichm\"uller space of Riemann surfaces with marked points, and a projective representation of the mapping class group on the space of conformal blocks is shown. A key input in that work comes from the boundary bootstrap in the present manuscript. Indeed, the existence of the fusion kernel for two elementary moves in the Moore-Seiberg groupoid comes as a consequence of the boundary bootstrap formula, as shown by \cite{GRSSbloc}, where the Ponsot-Teschner formulas (in \cite{ponsot1999,PT_boundary_3pt}) for fusion kernels has been proved, under the assumption that the boundary bootstrap holds.
 
 \medskip   
 On the physics side, boundary Liouville theory has close ties with two-dimensional gravity and minimal string theory. In particular, the semiclassical limit of minimal gravity connects with Jackiw-Teitelboim gravity \cite{JACKIW1985343,TEITELBOIM198341}, as reviewed in \cite{mertens}. On surfaces with boundary, JT gravity reduces to the Schwarzian theory, recently constructed in \cite{bauerschmidtST}. The semiclassical limit of boundary Liouville CFT should therefore provide a framework for computing the correlation functions of this model, with the bootstrap formulas serving as key inputs.\\

\textbf{Ackowledgements.}  C. Guillarmou and R. Rhodes are supported by the Simons Collaboration grant  "Probabilistic Paths to Quantum Field Theory". B. Wu was supported by the National Ket R\& D Program of China 
(No 2023YFA1010700) and Natural Science Foundation of China Grant (No. 12526204).
We thank B. Cercl\'e and N. Huguenin for sharing their result (in preparation) on the relection coefficients for boundary Liouville and Toda theories. B. Wu thanks X. Sun for helpful discussions.\\

\textbf{AI disclosure}. No AI was use for the mathematical content of the paper, which was finished more than a year ago. AI was only used for english corrections and a couple of proofread.\\

\textbf{Notation list.}
 \begin{itemize}
 	\item $\Sigma$ is the bordered Riemann surface; $\partial\Sigma_{\rm D}$ and $\partial\Sigma_{\rm N}$ denote Dirichlet and Neumann boundary components; $\Sigma^{\#2}$ is the Neumann double, with involution $\tau_\Sigma$.
 	\item $\mc C_j$ and $\mc C_j^+$ denote D-marked cutting circles and half-circles, with parametrisations $\zeta_j^\ell:\T\to\mc C_j$ and $\zeta_j^{\rm h}:\T^+\to\mc C_j^+$. The signs $\sigma_j^\ell,\sigma_j^{\rm h}\in\{-1,1\}$ encode outgoing/incoming orientation.
 	\item $g$ is a compatible metric; $K_g$ and $k_g$ are scalar and geodesic curvature; $\gamma\in(0,2)$, $Q=\gamma/2+2/\gamma$, $\mu$ is the bulk cosmological constant, and $\mu^{\rm b}$ is the boundary cosmological constant.
 	\item $X_{g,{\rm D}},X_{g,{\rm N}},X_{g,{\rm m}}$ are the Dirichlet, Neumann, and mixed-boundary GFFs. Boundary fields are $\tilde\varphi^\ell=c+\varphi^\ell$ on $\T$ and $\tilde\varphi^{\rm h}=c+\varphi^{\rm h}$ on $\T^+$; the Liouville field is $\phi_g=X_{g,{\rm m}}+P\tilde{\bs\varphi}$ when $\partial\Sigma_{\rm D}\ne\emptyset$, and $\phi_g=c+X_{g,{\rm N}}$ in the purely Neumann case.
 	\item $\mc H=L^2(H^s(\T),\mu_0)$ and $\mc H^+=L^2(H^s_{\rm even}(\T),\mu^+_0)$ are the bulk and boundary Hilbert spaces; $\mathbf H$ and $\mathbf H^+$ are the bulk and boundary Liouville Hamiltonians.
 	\item $\mathbf L_n,\widetilde{\mathbf L}_n$ are the two bulk Virasoro families; $\mathbf L_n^+$ is the boundary Virasoro family. Primary and descendant states are denoted $\Psi_{Q+ip,\nu,\tilde\nu}$ and $\Psi^+_{Q+ip,\nu}$. Their orthonormal version: $H_{Q+ip,\nu,\tilde\nu}$ and $H^+_{Q+ip,\nu}$.
 	\item $\mc A_{\Sigma,g,\ldots,\bs\zeta}$ denotes a Segal amplitude. The structure constants are $C^{\rm DOZZ}$, $\widetilde{\rm G}$, $\mathrm H$, and $\mathrm U$ for sphere, bulk-boundary, boundary three-point, and disk one-point amplitudes.
 	\item $\mc T_n$ denotes Young diagrams of size $n$; $F_\alpha$ is the Schapovalov form; $\tilde{\mc F}^{\mc G^\#}$ denotes the normalized conformal block attached to the doubled graph.
 	\item Internal spectral parameters attached to cutting circles are denoted $p$; those attached to half-circle cuts are denoted $p^+$.
 \end{itemize}

\section{Background and notations}\label{sec:back}

We will use all the material of the first part \cite{GRW1} of this work. For obvious reasons, 
we do not recall everything here and recommend the reader to have Part I as a companion paper while reading this work. 
We will however recall some notations. 

\subsection{Riemann surfaces with boundary/corners}\label{sec:surfaces}

This is a summary of the material of Sections 2.1 to 2.6 in \cite{GRW1}, the reader can consult that reference for details.

\subsubsection*{Model surfaces} For  $\delta<1$, we let 
\[ \D := \{z\in \C\,|\, |z|\leq 1\}, \quad \H=\{z\in \C\,|\, {\rm Im}(z)\geq 0\},\quad  \D^+:=\D\cap \H, \quad  \D^{++}:=\D^+\cap \{{\rm Re}(z)\geq 0\}\]
\[ \T :=\{z\in \C\,|\, |z|=1\}, \quad \T^+:=\T\cap \H=\{ z\in \C\,|\, |z|=1, {\rm Im}(z)\geq 0\}.\]
\[\mathbb{A}_{\delta}:=\{z\in \C\,|\, |z|\in [\delta,1]\}, \quad \mathbb{A}_{\delta_1,\delta_2}:=\{z\in \C\,|\, |z|\in [\delta_1,\delta_2]\} , \quad \A^+_\delta:=\H\cap \A_\delta, \quad  \A^+_{\delta_1,\delta_2}:=\H\cap \A_{\delta_1,\delta_2}.\]
We will use also the notation $\hat{\C}$ for the Riemann sphere, viewed as $\C$ with the point at infinity, equipped with its canonical complex structure, and $\hat{\H}$ for the Riemann half-sphere, which corresponds to the region  ${\rm Im}(z)\geq 0$ in $\hat{\C}$ (i.e. the compactification of $\H$), equipped with the complex structure induced by $\hat{\C}$.

\subsubsection*{Boundary parametrisations} We will consider smooth Riemann surfaces $(\Sigma,J)$ with boundary $\pl \Sigma=\sqcup_{j=1}^b\pl_j\Sigma$. Each boundary circle $\pl_j\Sigma$ has a marking D (for Dirichlet) or N (for Neumann), 
and we rename the D-marked boundary circles by $\mc{C}_j$ for $j=1,\dots,b^\ell$:
 \[ \partial \Sigma_{\rm D}=\sqcup_{j=1}^{b^\ell}\mc{C}_j.\]
We require that there are analytic parametrisations $\zeta^\ell_j:\T\to \mc{C}_j$ of the Dirichlet boundary circles. These parametrisations $\zeta^\ell_j$ are such that 
$(\zeta^\ell_j)^{-1}$ admit a holomorphic extension denoted  $\omega^\ell_j:V_j\to \A_{\delta}$ (in which case the parametrisation is called 
\textbf{outgoing} and we set $\sigma_j^\ell:=-1$) or $\omega^\ell_j:V_j\to \A_{1,\delta^{-1}}$ (parametrisation is called 
\textbf{incoming} and we set $\sigma_j^\ell:=1$)
for some $\delta<1$ and $V_j$ a neighborhood of $\mc{C}_j$ in $\Sigma$.

\subsubsection*{Riemann surfaces with corners} We also consider Riemann  surfaces $(\Sigma,J)$ with corners in the sense of  \cite[Definition 2.2]{GRW1}, whose topological boundary is piecewise real analytic and consists in real analytic circles as above, and real analytic half-circles: this definition says that there are holomorphic charts covering $\Sigma$, which are either biholomorphic to the local model $\D,\D^+$ or $\D^{++}$ and the transition maps are biholomorphic maps on a neighborhood of these local models. In particular the boundary has a piecewise real analytic structure, and $\Sigma$ can be extended to an open Riemann surface $\tilde{\Sigma}$ in a way that $\pl \Sigma$ is a union of piecewise analytic curves, in $\tilde{\Sigma}$, and the set $\mc{K}\in\pl\Sigma$ of corner points is such that the angle between two half-circles is always $\pi/2$.
As above, we mark each circle and half-circle  with a marking D or N, and we denote by $\mc{C}_j$ and $\mc{C}_j^+$ the D-marked circles and half-circles. We assume that the D-marked circles and half-circles admit respective analytic parametrisations $\zeta_j^\ell:\T \to \mc{C}_j$ and  
 $\zeta_j^{\rm h}:\T^+\to \mc{C}_j^+$. The parametrisations $(\zeta_j^{\ell})^{-1}$ extend holomorphically to some $\omega_j^\ell:V_j\to \A_{\delta}$ or $\omega_j^\ell:V_j\to \A_{1,\delta^{-1}}$ as above, and similarly $(\zeta_j^{\rm h})^{-1}$ extend holomorphically to some $\omega_j^{\rm h}:V_j\to \A^+_{\delta}$  with $\omega_j^{\rm h}(V_j\cap \pl \Sigma_{\rm N})=\A_{\delta}^+\cap \R$ (the parametrisation is called 
\textbf{outgoing} and we set $\sigma_j^{\rm h}:=-1$) or $\omega_j^{\rm h}:V_j\to \A^+_{1,\delta^{-1}}$  with $\omega_j^{\rm h}(V_j\cap \pl \Sigma_{\rm N})=\A_{1,\delta^{-1}}^+\cap \R$ (the parametrisation is called 
\textbf{incoming} and we set $\sigma_j^{\rm h}:=1$); moreover $\omega_j^{\rm h}$ extend holomorphically to a neighborhood of $V_j\cap \pl \Sigma$
in $\tilde{\Sigma}$. The holomorphic charts $\omega_j^\ell$ are called \textbf{annular holomorphic charts} while 
$\omega_j^{\rm h}$ are called \textbf{half-annular holomorphic charts}, the other charts mapping to neighborhoods of interior points will be denoted $\omega_j$.
 \begin{figure}[h] 
\begin{tikzpicture}[scale=0.6]
\draw[very thick] (1,0) arc (0:180:1);
\draw[very thick] (2,0) arc (0:180:2);
\draw[very thick] (-2,0) -- (-1,0);
\draw[very thick] (1,0) -- (2,0);
\coordinate (Z1) at (1.5,0) ;
\coordinate (Z2) at (-1,0) ;
\coordinate (Z3) at (-0.8,1.3) ;
\coordinate (Z4) at (-0.7,-0.9) ;
 \draw (0,2) node[above]{${\rm D}$} ;
 \draw (-1.5,-0.8) node[above]{${\rm N}$} ;
 \draw (1.5,-0.8) node[above]{${\rm N}$} ;
  \draw (-2.5,-0.2) node[above]{$-1$} ;  
  \draw (-0.5,-0.2) node[above]{$-\delta$} ;
   \draw (2.4,-0.2) node[above]{$1$} ;  
  \draw (0.6,-0.2) node[above]{$\delta$} ;
\end{tikzpicture}
\caption{Half-annular holomorphic chart $\mathbb{A}_\delta^+$ with its D/N boundary markings}\label{pic2}
\end{figure}
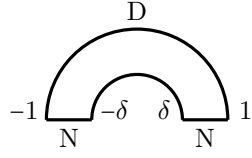
 Since the surface is oriented, there is a non-vanishing $2$-form $\omega_\Sigma$ which has the property that for each holomorphic chart $\omega_j$, $(\omega_j^{-1})^*\omega_\Sigma=e^{\rho_j}dx\wedge dy$ where $z=x+iy$ is the complex coordinate on $\C$ and $\rho_j$ is a smooth function. Each boundary circle and half-circle inherits an orientation by pulling back the $1$-form $-\iota_{\nu}\omega_\Sigma$, if $\nu$ is a non-vanishing inward pointing vector field.
 
 \subsubsection*{Metrics} A smooth Riemannian metric $g$ is \textbf{compatible} with the complex structure if in each holomorphic chart $\omega_i:V_i \to \C$, one has $g|_{V_i}=\omega_i^*(e^\rho |dz|^2)$ for some smooth $\rho\in C^\infty(\omega_i(V_i))$.
Such a metric induces a volume form ${\rm dv}_g$ and a boundary measure ${\rm d\ell}_g$ on $\partial \Sigma$. We will denote $K_g$ the scalar curvature and $k_g$ the geodesic curvature along the interior of the boundary hypersurfaces.  A fixed set of points $z_i\in\Sigma$, $i=1,\dots,m$ on $\Sigma$ are called \emph{marked points} on $\Sigma$; they could belong to $\partial \Sigma$. We will denote them 
by ${\bf z}=(z_1,\dots, z_m)$ and $\{{\bf z}\}:=\cup_{j=1}^m\{z_j\}\subset \Sigma$.

\subsubsection*{Neumann double} We denote the \textbf{Neumann double} $\Sigma^{\# 2}$ the Riemann surface obtained by identifying $\pl \Sigma_{\rm N}$ with $\pl \bbar{\Sigma}_N$ in $\Sigma\sqcup \bbar{\Sigma}$, where $\bbar{\Sigma}$ is a copy of $\Sigma$ equipped with the complex structure $-J$ (i.e. opposite orientation); the resulting surface $\Sigma^{\# 2}$ is a surface with Dirichlet boundary circles (or possibly a closed surface if $\pl \Sigma_{\rm D}=\emptyset$), equipped with an anti-holomorphic involution $\tau_\Sigma$ fixing 
$\pl \Sigma_{\rm N}$.  For a function, distribution in the Sobolev space $H^s(\Sigma)$ for $s>-1/2$ (see \eqref{SobolevHs}), or a 
Riemannian metric $f$ on $\Sigma$, we will denote by
\[f^\#:=f{\bf 1}_{\Sigma}+\tau_\Sigma^*f{\bf 1}_{\bbar{\Sigma}}\]
the symmetrised function, distribution or metric on $\Sigma^{\#2}$; notice that if $s>-1/2$, $f\in H^s(\Sigma)$ can be extended by $0$ on $\Sigma^{\#2}$ as an element in $H^s(\Sigma^{\#2})$ and $f{\bf 1}_{\Sigma}$ denotes this extension.
We say that a compatible metric $g$ is \textbf{Neumann extendible} if $g$ extends smoothly on the Neumann double $\Sigma^{\#2}$ in a way that it is symmetric under the involution $\tau_\Sigma:\Sigma^{\#2}\to \Sigma^{\#2}$, that is if $g^\#$ is smooth on $\Sigma^{\#2}$. 
We say that $g$ is \textbf{admissible} if in addition, each boundary circle $\mc{C}_j$ (resp. half-circle $\mc{C}_j^+$) with Dirichlet  marking has a neighborhood isometric to the flat annulus $(\A_\delta,|\dd z|^2/|z|^2)$  (resp. the flat half-annulus $(\A_\delta^+,|\dd z|^2/|z|^2)$) for some $\delta>0$.
If $\zeta_j^\ell,\zeta_j^{\rm h}$ denote real analytic parametrisations of the $j$-th D-marked circle and half-circle as above, 
the Neumann double has induced parametrisations of its D-marked boundary by setting $\tau_\Sigma\circ \zeta_j^\ell$ for the reflected boundary $j$-th circle in $\bbar{\Sigma}$, and 
\[  e^{i\theta}\in \T^+ \mapsto \zeta_j^{\rm h}(e^{i\theta}), \quad e^{i\theta}\in \T^- \mapsto \tau_\Sigma\circ \zeta_j^{\rm h}(e^{-i\theta})\]
for the circle obtained as the gluing $\mc{C}^+_j\cup \tau_\Sigma(\mc{C}^+_j)$ of two half-circles (here $\T^-=\T\setminus \T^+$).

\subsubsection*{Gluing of surfaces} On a Riemann surface $\Sigma$ with boundary (resp. with corners), not necessarily connected, we can glue two Dirichlet boundary circles (resp. two half-circles). This can be done via the parametrisations: for example if $\mc{C}_j$ (resp. $\mc{C}_j^+$) 
is outgoing and $\mc{C}_k$ (resp. $\mc{C}_k^+$) is incoming, we can identify 
$\mc{C}_j$ with $\mc{C}_k$ by identifying $\zeta_{j}^\ell(e^{i\theta})$ with $\zeta_{k}^\ell(e^{i\theta})$ for each $\theta\in [0,2\pi]$ (resp. $\theta\in [0,\pi]$). The resulting surface is a surface with corners (which may be a surface with boundary or a closed surface).

 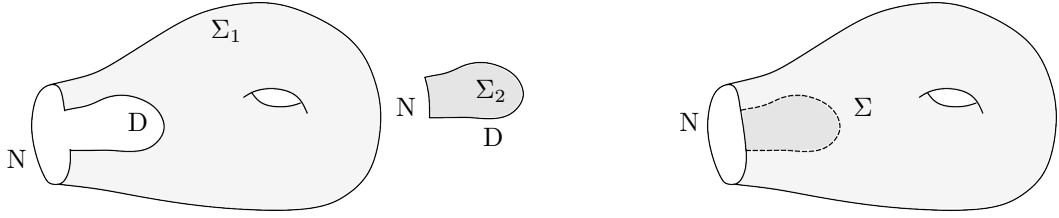
\begin{figure}[h]
  \begin{tikzpicture}
    \node[inner sep=0pt] (pant) at (0,0)
{ \includegraphics[scale=0.4]{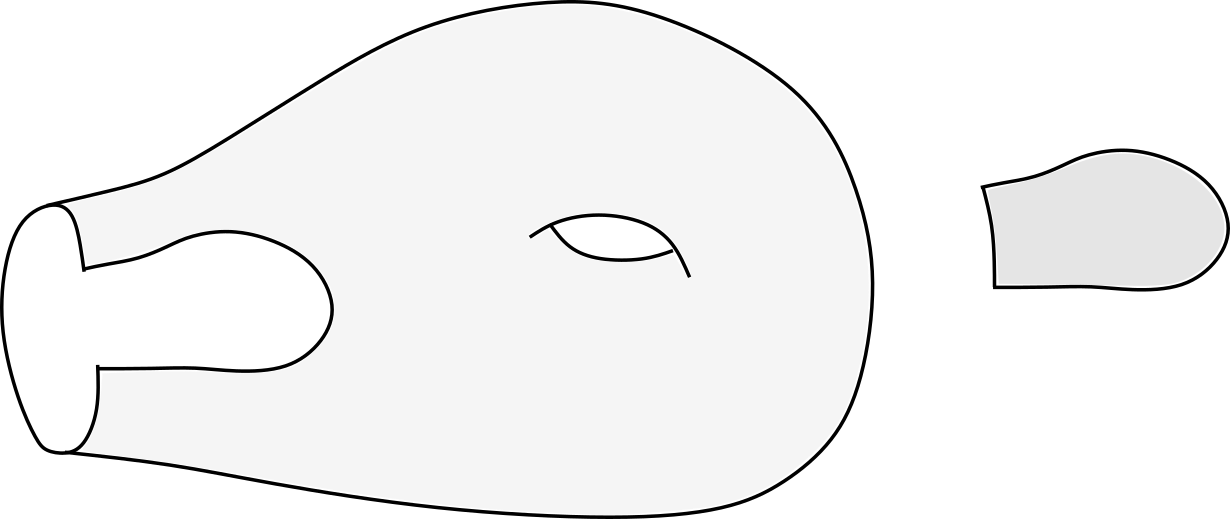}};
  \node[inner sep=0pt] (pant) at (8,0)
{ \includegraphics[scale=0.4]{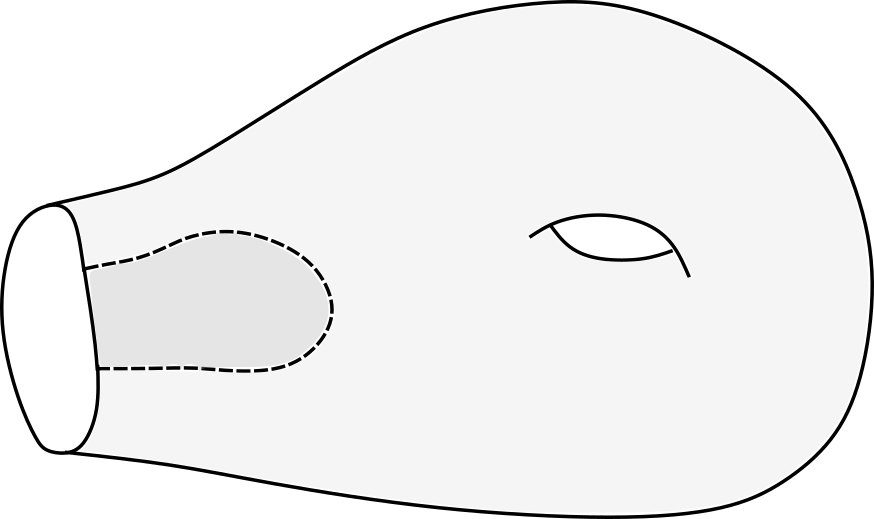}};
  \draw (7.5,0) node[right,black]{$\Sigma$} ;
    \draw (-1,1) node[right,black]{$\Sigma_1$} ; 
     \draw (2.5,0.2) node[right,black]{$\Sigma_2$} ; 
      \draw (2.6,-0.4) node[right,black]{${\rm D}$} ; 
       \draw (1.45,0) node[right,black]{${\rm N}$} ; 
        \draw (-3.7,-0.7) node[right,black]{${\rm N}$} ; 
         \draw (-2.1,-0.2) node[right,black]{${\rm D}$} ;
          \draw (5.2,-0.2) node[right,black]{${\rm N}$} ; 
\end{tikzpicture}
\caption{Two surfaces with corners $\Sigma_1,\Sigma_2$ glued along half-circles to produce the surface $\Sigma$ with N-marked boundary.}
\label{glue1}
\end{figure}


\subsection{Green function and GFF}

On a surface with corners, we consider the standard non-negative Laplacian $\Delta_{g}=\dd^*\dd$ where $\dd^*$ is the $L^2$ adjoint of the exterior derivative $\dd$. We will consider respectively the Dirichlet condition at the D-marked boundary and Neumann condition at the N-marked boundary, we denote this operator by $\Delta_{g,{\rm m}}$ (m stands for mixed). The case where there are only D or N marked boundary is also contained in our setting, and in that case $\Delta_{g,{\rm m}}$ just reduces to the Dirichlet Laplacian $\Delta_{g,{\rm D}}$ or Neumann Laplacian $\Delta_{g,{\rm N}}$.
We denote by $\partial_\nu$   the (unit) normal inward derivative along the interior of $\pl \Sigma_{\rm N}$ or  $\pl \Sigma_{\rm D}$ with respect to the metric $g$ (at a corner point, there is a normal to $\pl \Sigma_{\rm N}$ and a normal to $\pl \Sigma_{\rm D}$). The regularised determinants of these operators will be denoted 
\[   {\det}(\Delta_{g,{\rm D}}), \quad  {\det} '(\Delta_{g,{\rm N}}), 
 \quad {\det} (\Delta_{g,{\rm m}})\]
where the prime denotes the fact we have removed the kernel of the operator (i.e. the constants) in the definition of the spectral zeta function. The Green functions are defined to be the integral kernel, with respect to the volume measure ${\rm v}_g$, of $2\pi$ times the inverse of the operators  $\Delta_{g,{\rm D}}$, $\Delta_{g,{\rm N}}$ and $\Delta_{g,{\rm m}}$, on the orthogonal of their kernel. They are denoted by 
\[ G_{g,{\rm D}}(x,x'), \quad G_{g,{\rm N}}(x,x'), \quad G_{g,{\rm m}}(x,x')\]
for $x\not= x'$ in $\Sigma$. Near any $x'\in \Sigma^\circ$ in the interior, these functions have the form $-\log d_g(x,x')+F_g(x,x')$ with $F_g$ smooth near $x=x'$, and $d_g(x,x')$ is the Riemannian distance between $x,x'$.

On $\Sigma$ with compatible metric $g$, the Gaussian Free Fields (GFF in short) are defined as follows. 
Let $(a_j)_j$ be a sequence of i.i.d. real Gaussians   $\mc{N}(0,1)$ with mean $0$ and variance $1$, defined on some probability space   $(\Omega,\mc{F},\mathbb{P})$. Let $(e_j)_{j=1}^\infty$ be an orthonormal basis of $(\ker \Delta_{g,{\rm m}})^\perp$ of eigenfunctions of $\Delta_{g,{\rm m}}$ with eigenvalues $\la_j>0$. 
We define the Gaussian Free Field as the random distribution
\begin{equation}\label{GFFong}
X_{g,{\rm m}}:= \sqrt{2\pi}\sum_{j\geq 1}a_j\frac{e_j}{\sqrt{\la_j}}
\end{equation}
 where  the sum converges almost surely in the Sobolev space  $H^{s}(\Sigma)$ for each $s\in (-1/2,0)$ defined by
\begin{equation}\label{SobolevHs} 
H^{s}(\Sigma):=\{f=\sum_{j\geq 0}f_je_j\,|\, \|f\|_{H^s}^2:=|f_0|^2+\sum_{j\geq 1}\lambda_j^{s}|f_j|^2<+\infty\}.
\end{equation} 
We will sometimes write $X_{g,{\rm D}}$ or $X_{g,{\rm N}}$ instead of $X_{g,{\rm m}}$ when $\pl\Sigma=\emptyset$, resp. when  $\pl\Sigma_{\rm N}=\emptyset$, when $\pl \Sigma_{\rm D}=\emptyset$. 
The covariance of $X_{g,{\rm D}}, X_{g,{\rm N}}$ and $X_{g,{\rm m}}$ is the Green function when viewed as a distribution 
\[ \mathbb{E}[X_{g,*}(x)X_{g,*}(x')]= \,G_{g,*}(x,x')\]
for $*\in \{{\rm D},{\rm N},{\rm m}\}$.
We will sometimes add the surface $\Sigma$ as a subscript in the notation, for example by writing $ {\det}(\Delta_{\Sigma,g,{\rm D}})$, $G_{\Sigma,g,{\rm m}}$ or $X_{\Sigma,g,{\rm N}}$, to avoid confusion when several surfaces are considered.

\subsection{Boundary fields and Liouville field}
If $\pl \Sigma_{\rm D}=\emptyset$ , the \textbf{Liouville field}  is defined by 
\[\phi_g:=c+X_{g,{\rm N}} \] 
where $c\in\R$ is a constant that stands for the constant mode of the field. 

In the case  $\pl \Sigma_{\rm D}\not=\emptyset$ we need to introduce boundary fields on the Dirichlet boundary, which are GFF on the $\T$ and $\T^+$ via the parametrisations of $\mc{C}_j$ and $\mc{C}_j^+$.  A distribution on $\T$ can be written as   
\begin{equation}\label{GFFcircle0}
		\tilde\varphi^\ell=c+\varphi^\ell,\quad \text{and }\quad \forall \theta\in\R, \quad\varphi^\ell(e^{i\theta})=\sum_{n\not=0}\varphi^\ell_ne^{in\theta}
	\end{equation}
	with $(\varphi^\ell_n)_{n\not=0}$ the Fourier coefficients, and $\tilde{\varphi}^\ell\in H^s(\T)$ if 
		\begin{equation}\label{outline:ws}
		\|\tilde\varphi^\ell\|_{H^s(\T)}^2:=c^2+\sum_{n\not=0}|\varphi^\ell_n|^2(|n|+1)^{2s} <\infty.
	\end{equation}
If $x_n^\ell,y_n^\ell$ are i.i.d. real Gaussians in $\mc{N}(0,1)$ on a probabilistic space denoted $(\Omega_\T,\P_\T)$, we set for each $n>0$ $\varphi_n^\ell:=\frac{x^\ell_n+iy^\ell_n}{2\sqrt{n}}$  and $\varphi^\ell_{-n}:=\frac{x^\ell_n-iy^\ell_n}{2\sqrt{n}}$  for $n>0$ and $\varphi^\ell$ is 
a random Gaussian variable in $H^{s}(\T)$ for all $s<0$ with covariance kernel 
\[ \E[ \varphi^\ell(\theta)\varphi^\ell(\theta')]=-\log|e^{i\theta}-e^{i\theta'}|.\]
Similarly, we consider distributions on $\T^+$ written under the form   
\begin{equation}\label{GFFcircle+}
\tilde\varphi^{\rm h}=c+\varphi^{\rm h},\quad \text{and }\quad \forall \theta\in\R, \quad   \varphi^{\rm h}(e^{i\theta})= \sum_{n>0}\varphi^{\rm h}_n(e^{in\theta}+e^{-in\theta}). 
\end{equation}
If $x_n^{\rm h}$ are i.i.d. real Gaussians in $\mc{N}(0,1)$ on a probabilistic space, denoted $(\Omega_{\T^+},\P_{\T^+})$, we define $\varphi^{\rm h}_{n}:=\frac{\sqrt{2}x^{\rm h}_n}{2\sqrt{|n|}}$ and the random variable 
$\varphi^{\rm h}$ can be and will often be viewed as defined on $\T$, invariant by the involution $\theta\mapsto -\theta$, 
can be considered as an element in $H^s_{\rm even}(\T):=\{u\in H^s(\T) \, |\, u(\theta)=u(-\theta)\}$ for all $s<0$. Its covariance kernel is 
\[ \E[\varphi^{\rm h}(\theta)\varphi^{\rm h}(\theta')]=-\log|e^{i\theta}-e^{i\theta'}|-\log|e^{i\theta}-e^{-i\theta'}|.\]

Let $\boldsymbol{\varphi}=(\boldsymbol{\tilde \varphi}^\ell,\boldsymbol{\tilde \varphi}^{\rm h})=(\tilde{\varphi}_1^\ell,\dots,\tilde{\varphi}_{b^\ell}^\ell,\tilde{\varphi}_1^{\rm h},\dots,\tilde{\varphi}_{b^{\rm h}}^{\rm h})$ be a collection of independent random variables defined as above, one $\tilde{\varphi}^\ell_j$ for each boundary circle $\mc{C}_j$ and one $\tilde{\varphi}^{\rm h}_j$ for each boundary circle $\mc{C}_j^+$. We will 
denote   by $P\tilde{\boldsymbol{\varphi}}$ the harmonic function on $\Sigma$  with boundary value 
$\tilde{\varphi}_j^{\rm h}\circ (\zeta^{\rm h}_j)^{-1}$ on $\mc{C}_j^+$ and $\tilde{\varphi}_j^{\rm \ell}\circ (\zeta^{\ell}_j)^{-1}$ on $\mc{C}_{j}$, and with Neumann boundary condition at $\pl\Sigma_{\rm N}$. As before, when there are possible confusions about the surface we consider, we will sometimes write $P_\Sigma\tilde{\boldsymbol{\varphi}}$ instead of $P\tilde{\boldsymbol{\varphi}}$.
The Liouville field is the random variable 
\begin{equation}\label{liouville_field_general}
\phi_g= X_{g,{\rm m}}+P\tilde{\boldsymbol{\varphi}} \in H^{s}(\Sigma), \quad \forall s<0.
\end{equation}


\subsection{Gaussian Multiplicative Chaos and vertex operators} 
We call \textbf{$g$-regularisation} of the GFF  its regularisation by averaging it along geodesic circles of radius $\epsilon$ in the metric $g$ (we refer the reader to \cite[Section 3.2]{Guillarmou2019} and \cite{Wu1} for details). Let $x\in \Sigma$ and  $C_g(x,\epsilon):=\{z\in\Sigma\mid d_g(z,x)=\epsilon\}$ (for $x\in\partial\Sigma_{\rm N}$ and $\epsilon$ small enough, $C_g(x,\epsilon)$ is an arc, with its two ends lying on $\partial\Sigma_{\rm N}$). The regularisation of $X_{g,{\rm m}}$ at $x\in \Sigma$ is defined by $X_{g,{\rm m},\eps}(x):=\lim_{n\to \infty}\cjg X_{g,{\rm m}},f_{x,\eps}^n\cjd$ for some  
smooth approximation $f^{n}_{x,\eps}$ of the  uniform measure on $C_g(x,\epsilon)$. As above, we denote by $X_{g,{\rm D},\eps}$ and $X_{g,{\rm N},\eps}$ the $g$-regularisation of $X_{g,{\rm D}},X_{g,{\rm N}}$ respectively, when $\pl \Sigma=\emptyset$, $\pl \Sigma_{\rm N}=\emptyset$ and $\pl \Sigma_{\rm D}=\emptyset$. More generally, for a random distribution $h\in H^{s}(\Sigma)$ admitting a $g$-regularisation (i.e. the limit above exists), we write $h_{\epsilon}$ 
for its $g$-regularization. The Liouville field admits such a $g$-regularisation.
For $\gamma\in\R$ and $h$ a random distribution admitting a  $g$-regularization $(h_\epsilon)_\epsilon$, we define the measures 
\begin{equation}\label{GMCg}
M^g_{\gamma,\eps}(h,\dd x):= \eps^{\frac{\gamma^2}{2}}e^{\gamma h_{ \eps}(x)}{\rm dv}_g(x),\quad \text{and }\quad M^{{\rm b},g}_{\gamma,\eps}(h,\dd s):= \eps^{\frac{\gamma^2}{4}}e^{\frac{\gamma}{2} h_{ \eps}(s)}{\rm d}\ell_g(s).
\end{equation}
We will focus in particular on the case when $h=X_{g,{\rm D}},X_{g,{\rm N}}$ or $X_{g,{\rm m}}$. 
When $\pl\Sigma_{\rm D}\not=\emptyset$,  
$M^{g}_{\gamma,\eps}(h,\dd x)$ converges to some non-trivial random measure $M^g_{\gamma}(h,\dd x)$.
When $\pl\Sigma_{\rm D}=\emptyset$  and $g$ is Neumann extendible, 
on each set $\{ d_g(\cdot,\pl \Sigma)>\delta\}$ with $\delta>0$,  
the measure $M^{g}_{\gamma,\eps}(X_{g,{\rm N}},\dd x)$ converges in probability and weakly in the space of Radon measures towards a  non trivial random measure $M^g_{\gamma}(X_{g,{\rm N}},\dd x)$ and  $M^g_{\gamma}(X_{g,{\rm N}},\Sigma)<+\infty $ almost surely, with the expectation being finite iff $\gamma^2<2$. For the boundary measure, the measures $M^{{\rm b},g}_{\gamma,\eps}(X_{g,{\rm N}},\dd s)$ or $M^{{\rm b},g}_{\gamma,\eps}(X_{g,{\rm m}},\dd s)$ converge   as $\eps\to 0$ to some non-trivial random measure $M^{{\rm b},g}_{\gamma}(X_{g,{\rm N}},\dd s)$ or $M^{{\rm b},g}_{\gamma}(X_{g,{\rm m}},\dd s)$, which give finite mass to $\pl\Sigma_{\rm N}$.

We introduce the regularised vertex operators, for fixed $\alpha\in\R$ and $x\in \Sigma^\circ$ 
\begin{equation*}
V_{\alpha,g,\eps}(x)=\eps^{\alpha^2/2}  e^{\alpha  \phi_{g,\epsilon}(x) } .
\end{equation*}
while for   $s\in \partial\Sigma_{\rm N}$, we define the boundary vertex operators, for fixed $\beta\in\R$   
\begin{equation*}
V_{\beta,g,\eps}(s)=\eps^{\frac{\beta^2}{4}}  e^{\frac{\beta}{2}  \phi_{g,\epsilon}(s) } .
\end{equation*}
If $g'=e^{\omega}g$ with $g$ admissible and $\omega$ smooth, we have the relation (see \cite[Section 2.10]{GRW1})
\begin{align}\label{scalingmeasure1}
&  M^{g'}_\gamma(X_{g',{\rm m}},\dd x)= e^{\gamma(\frac{Q}{2} \omega(x)-A)} M^g_\gamma(X_{g,{\rm m}},\dd x)\\
\label{scalingmeasure2}
& M^{{\rm b},g'}_\gamma(X_{g',{\rm m}},\dd s)= e^{\frac{\gamma}{2}(\frac{Q}{2} \omega(s)-A)} M^{{\rm b},g}_{\gamma}(X_{g,{\rm m}},\dd s)\quad \text{ on }\partial\Sigma_{\rm N}.
\end{align}
with $A=0$ if $\partial\Sigma_D\neq\emptyset$ and $A=\frac{1}{{\rm v}_{g'}(\Sigma)} \int_{\Sigma} X_{g,{\rm m}}(x)\, \dd{\rm v}_{g'}(x)$ if $\partial\Sigma_D=\emptyset$. As $\eps\to 0$, for $x\in \Sigma^\circ$ and $s\in \pl \Sigma_{\rm N}$
\begin{equation}\label{scalingvertex}
V_{\alpha,g',\eps}(x)=(1+o(1)) e^{\frac{\alpha^2}{4} \omega(x)} V_{\alpha,g,\eps}(x)\quad \text{ and }\quad V_{\beta,g',\eps}(s)=(1+o(1)) e^{\frac{\beta^2}{8} \omega(s)} V_{\beta,g,\eps}(s).
\end{equation}

\section{Liouville Conformal Field Theory on compact Riemann surfaces with Neumann boundary}\label{sec:open}
In this section we recall the definition of correlation functions, and more generally the path integral, for Liouville conformal field theory on Riemann surfaces with Neumann boundary. We summarize here the relevant material, details can be found in 
\cite{huang2018,Wu1,GRW1}. 
\subsection{Correlation functions}
Let $\Sigma$ be a compact Riemann surface with boundary and $g$ be a compatible metric. Recall from last section that $g$ is  conformal to a Neumann extendible metric, but we do not need to assume that $g$ is Neumann extendible for this Section.
Correlation functions will always be considered in the case with $b$ boundary circles and Neumann boundary condition (i.e. $\partial\Sigma_{\rm D}=\emptyset$). We will denote by $\partial_j\Sigma$  ($j=1,\dots,b$) the boundary circles. We fix parameters $\gamma\in(0,2)$, $\mu\geq 0$ and $Q=\frac{\gamma}{2}+\frac{2}{\gamma}$ and consider a piecewise non-negative  constant function $\mu^{\rm b}:\partial\Sigma_{\rm N}\to[0,\infty)$, which we also consider as function on $\pl \Sigma$ by extending it by $0$ on $\pl \Sigma_{\rm D}$.
We assume 
\begin{equation}\label{hyp:cosmo}
\quad \mu+\mu^{\rm b}\not =0 
\end{equation}
For $F:  H^{s}(\Sigma)\to\R$ (with $s<0$) a bounded continuous functional, we set (recall the definition of the Liouville field   $\phi_g=c+X_{g,{\rm N}}$ in the case of Neumann b.c.)
 \begin{align}\label{def:pathintegralFb}
	 \cjg F(\phi)\cjd_{\Sigma, g}   :=&  \big(\frac{{\rm v}_{g}(\Sigma)}{{\det}'(\Delta_{g,{\rm N}})}\big)^\hf e^{-\frac{1}{8\pi}\int_{\partial\Sigma_N} k_g\dd \ell_g} \\
	 &\times \int_\R  \E\Big[ F( \phi_g) e^{ -\frac{Q}{4\pi}\int_{\Sigma}K_{g}\phi_g\,{\rm dv}_{g}  -\frac{Q}{2\pi}\int_{\partial\Sigma}k_{g}\phi_g\,{\rm d}\ell_{g} - \mu   M^g_\gamma(\phi_g,\Sigma)-     M^{{\rm b},g}_\gamma(\phi_g,\mu^{\rm b}) } \Big]\,\dd c. \nonumber
\end{align} 
Here $K_g$ and $k_g$ are respectively the scalar curvature of $g$ and geodesic curvature of $\pl \Sigma$. This quantity  defines a measure on $H^{s}(\Sigma)$. Then we consider the limit
\begin{equation}\label{defcorrelgbound}
	\big\cjg F(\phi_g)\prod_i V_{\alpha_i,g}(x_i)\prod_j V_{\beta_j,g }(s_j)  \big\cjd_ {\Sigma, g}:=\lim_{\eps \to 0} \: \big\cjg F(\phi_g)\prod_i V_{\alpha_i,g,\eps}(x_i)\prod_j V_{\beta_j,g,\eps}(s_j) \big\cjd_ {\Sigma, g}
\end{equation}
where we have fixed $m$   distinct bulk points $x_1,\dots,x_m$  on $\mathring{\Sigma}$ with respective associated weights $\alpha_1,\dots,\alpha_m\in\R$ and $m^{\rm b}$ distinct boundary points $s_1,\dots,s_{m^{\rm b}}$ on $\partial\Sigma$. We will always denote by ${\bf x}=(x_1,\dots,x_m)$ the collection of interior marked points, and $\{{\bf x}\}:=\cup_{j=1}^m \{x_j\}\subset \Sigma$. And similarly for the boundary marked points  with the notations  ${\bf s}$ and $\{{\bf s}\}$.

Correlation functions are defined by taking $F=1$ and are denoted by 
\[ \textbf{(Correlation functions)}\qquad \big\cjg \prod_i V_{\alpha_i,g}(x_i)\prod_j V_{\beta_j,g }(s_j)  \big\cjd_ {\Sigma, g}.\]
Existence and non-triviality of the limit \eqref{defcorrelgbound} defining the correlation functions hold under the following condition called Seiberg bounds (see \cite[Proposition 6.1]{Wu1}, \cite{huang2018}):
	\begin{align}\label{seiberg1bound}
		& \sum_{i}\alpha_i + \sum_j \frac{\beta_j}{2}>Q\chi(\Sigma) ,\\ 
		&\forall i,\quad \alpha_i<Q\quad \quad \forall j,\quad \beta_j<Q\label{seiberg2bound}
	\end{align}
	where  $\chi(\Sigma)=2-2g(\Sigma)-b$ ($b$ is the number of boundary components).

The correlation functions satisfy diffeomorphism invariance and conformal covariance with respect to the background Riemannian metric (\cite[Proposition 5.5 and 5.6]{Wu1}): assume that \eqref{seiberg1bound} hold, then  if $\psi:\Sigma\to \Sigma'$ is an orientation preserving diffeomorphism, we have  
\begin{equation}\label{diff_corel}
	\cjg \prod_i V_{\alpha_i,\psi^*g}(x_i) \prod_j V_{\beta_j,\psi^*g }(s_j) \cjd_ {\Sigma, \psi^*g}=\cjg \prod_i V_{\alpha_i,g}(\psi(x_i))\prod_j V_{\beta_j,g }(\psi(s_j))  \cjd_ {\Sigma', g},\end{equation}
and if $g,g'=e^{\omega}g$ are two compatible metrics on $\Sigma$ with Neumann boundary (for some $\omega\in C^\infty(\Sigma)$), and if \eqref{seiberg1bound} hold, then  
	\begin{align}\label{confan3} 
		\log\Big(\frac{\cjg \prod_i V_{\alpha_i,g'}(x_i)\prod_j V_{\beta_j,g' }(s_j)  \cjd_ {\Sigma, g'}}{\cjg \prod_i V_{\alpha_i,g}(x_i)\prod_j V_{\beta_j,g }(s_j)  \cjd_ {\Sigma, g}}\Big)=& 
		\frac{1+6Q^2}{96\pi}\Big(\int_{\Sigma}(|d\omega|_g^2+2K_g\omega) {\rm dv}_g+4\int_{\partial\Sigma}k_g\omega \dd\ell_g\Big)
		\\
		&-\sum_i\Delta_{\alpha_i}\omega(x_i)-\frac{1}{2}\sum_j\Delta_{\beta_j}\omega(s_j)\nonumber
	\end{align}
	where the conformal weights $\Delta_{\alpha_i},\Delta_{\beta_j}$ are given by $\Delta_\alpha=\frac{\alpha}{2}(Q-\frac{\alpha}{2})$ for any $\alpha\in\C$.
	
\section{Liouville amplitudes and Segal gluing}\label{sec:amplitude}

The bootstrap approach is a method to express the correlation functions of boundary Liouville theory in terms of structure constants and conformal blocks. It is based on decomposing the probabilistic path integral representing the correlation function on $(\Sigma,g)$ in terms of pairings (or compositions) of Hilbert-Schmidt operators on a Hilbert space, called \textbf{amplitudes}.  These amplitudes can be represented in terms of integral kernels and are associated to surfaces with corners (or boundaries), containing parametrised D-marked boundaries, and decomposing the Riemannian surface $(\Sigma,g)$ into geometric building blocks. Probabilistically, the  integral kernels of these amplitudes   are obtained by conditioning the value of our random field on the D-marked boundary.

In this section, we recall briefly the necessary material introduced in \cite[Section 5]{GRW1} (or \cite{GKRV21_Segal} for the case $\pl \Sigma_{\rm N}=\emptyset$)  in order to bridge the gap between the probabilistic expressions for the Liouville CFT and the Hilbert space picture, namely a representation of the symmetry algebra acting on some Hilbert space. We will more specifically focus on the generators of dilations, usually called bulk Hamiltonian and boundary Hamiltonian. These operators are self-adjoint and both provide a spectral resolution, based on which the conformal bootstrap will later be interpreted as a Plancherel type formula.

	\subsection{Hilbert spaces of LCFT}\label{sub:hilbert} 
	
The construction of the {\bf bulk Hilbert space} of LCFT relies on the real-valued  Fourier series defined on  the unit circle $\T=\{z\in \C\,|\, |z|=1\}$. Consider the random variable $\tilde\varphi^\ell=c+\varphi^\ell \in H^s(\T)$ defined in \eqref{GFFcircle0}, where the constant mode 
$c$ is distributed with the Lebesgue measure $\dd c$. Its law defines a measure $\mu_0$ on  $H^s(\T)$ where $s<0$ is any fixed number, 
and our bulk Hilbert space is  
\[\mc{H}:=L^2(H^s(\T),\mu_0),\] 
which is also isomorphic to $L^2(\R\times \Omega_\T,\dd c\otimes \P_\T)$ 
where  $\Omega_\T=(\R^{2})^{\N^*}$ is equipped with the cylinder algebra and the Gaussian  product measure 
	\begin{align}\label{Pdefin}
		\P_\T:=\bigotimes_{n\geq 1}\frac{1}{2\pi}e^{-\frac{1}{2}((x_n^\ell)^2+(y^\ell_n)^2)}\dd x^\ell_n\dd y^\ell_n.
	\end{align}
It is a separable Hilbert space, with Hermitian product denoted by $\langle\cdot,\cdot\rangle_{\mc{H}}$; 
see \cite[Section 4]{GKRV20_bootstrap}.  A function in $\mc{H}$ can also be considered as a function of $(c,(x_n^{\ell},y_n^\ell)_n)$.
	
The {\bf boundary Hilbert space} is defined  similarly by 
\[\mc{H}^+:=L^2(H^s_{\rm even}(\T),\mu_0^+)\] 
where $\mu_0^+$ is the law of the random variable $\tilde\varphi^{\rm h}=c+\varphi^{\rm h}\in H^s_{\rm even}(\T)$ defined in \eqref{GFFcircle+}, where the constant mode $c$ is distributed with the Lebesgue measure $\dd c$. The space $\mc{H}^+$ is also isomorphic to $L^2(\R\times \Omega_{\T^+},\dd c\otimes \P_{\T^+})$ where 
$\Omega_{\T^+}=\R^{\N^*}$  is equipped with the cylinder sigma-algebra   and the Gaussian  product measure 
\[\mathbb{P}_{\mathbb{T}^{+}}:=\bigotimes_{n\geq 1} \frac{1}{\sqrt{2\pi}}e^{-\frac{1}{2}(x^{\rm h}_n)^2}\dd x^{\rm h}_n.\]
It is a separable Hilbert space, with Hermitian product denoted by $\langle\cdot,\cdot\rangle_{\mc{H}^+}$. It means that a function in $\mc{H}^+$ can be considered as afunction of $(c,(x_n^{\rm h})_n)$.
We notice that there is a natural linear injective map 
\begin{equation}\label{embeddingH_+}
 \iota_{\mc{H}^+}:  \mc{H}^+\to \mc{H}, \quad (\iota_{\mc{H}^+}F)(c,(x_n^\ell,y_n^\ell)_n)=F(c,(x_n^{\ell})_n)
\end{equation}
so that $\mc{H}^+$ can be viewed as a subspace of $\mc{H}$ and 
\[  \|\iota_{\mc{H}^+}F\|_{\mc{H}}=\|F\|_{\mc{H}^+}.\]
The orthogonal projection $\pi_{\mc{H}^+}:\mc{H}\to \mc{H}^+$ is given by 
\[(\pi_{\mc{H}^+}F)(c,(x_n^{\rm h})_n)=\E [F(c,(x_n^{\rm h},\cdot)_n)]\]
 where the expectation is over the $y_n^\ell$ Gaussian variables, and $\pi_{\mc{H}^+}\iota_{\mc{H}^+}={\rm Id}$.

\subsection{Amplitudes}\label{sub:ampli} 
Let  $\Sigma$ be a compact  Riemann surface with corners, with parametrized boundary $\partial\Sigma=\partial\Sigma_{\rm N}\cup \partial\Sigma_{\rm D}$ as defined in Section \ref{sec:surfaces}, and let $g$ be a metric that is Neumann extendible. Our setting also contains the case where there are no corners, i.e. $b^{\rm h}=0$, and when there is no Neumann boundary $\pl \Sigma_{\rm N}=\emptyset$. 
We consider a family of random fields as in the previous subsection 
\[\boldsymbol{\tilde \varphi}=(\boldsymbol{\tilde \varphi}^{\ell},\boldsymbol{\tilde \varphi}^{\rm h}) =(\tilde\varphi^\ell_1,\dots,\tilde\varphi^\ell_{b^\ell},\tilde\varphi^{\rm h}_1,\dots,\tilde\varphi^{\rm h}_{b^{\rm h}})\in (H^{s}(\T))^{b^\ell}\times  (H^{s}_{\rm even}(\T))^{b^{\rm h}},\] 
in which case the Fourier coefficients of the $j$-th field will be augmented with an index $j$, namely $c_j$, $\varphi^\ell_{j,n}$ or $\varphi^{\rm h}_{j,n}$,  $x^\ell_{j,n}$ or $x^{\rm h}_{j,n}$, $y^\ell_{j,n}$.  We will also denote by $(\cdot,\cdot)_{2,\ell}$ the pairing between $(H^{s}(\T))^{b^\ell}$ and  $(H^{-s}(\T))^{b^\ell}$, and by $(\cdot,\cdot)_{2,{\rm h}}$ the pairing between $(H_{\rm even}^{s}(\T))^{b^{\rm h}}$ and $(H_{\rm even}^{-s}(\T))^{b^{\rm h}}$, which are extensions of the products 
\[
(\boldsymbol{\tilde \varphi}^\ell,\boldsymbol{\tilde \phi}^\ell)_{2,\ell}:=\frac{1}{2\pi}\sum_{j=1}^{b^\ell}\int_0^{2\pi}  \tilde\varphi^\ell_j(e^{i\theta})\tilde\phi^\ell_j (e^{i\theta}) \dd \theta,
\quad (\boldsymbol{\tilde \varphi}^{\rm h},\boldsymbol{\tilde \phi}^{\rm h})_{2,{\rm h}}:=\frac{1}{\pi}\sum_{j=1}^{b^{\rm h}}\int_0^\pi  \tilde\varphi^{\rm h}_j(e^{i\theta})\tilde\phi^{\rm h}_j (e^{i\theta}) \dd\theta.
\]
on smooth functions.
Finally,   we will denote by $(\cdot,\cdot)_{2}$ the total  pairing
\begin{equation}\label{unpairing}
(\boldsymbol{\tilde \varphi},\boldsymbol{\tilde \phi})_{2}=  (\boldsymbol{\tilde \varphi}^\ell,\boldsymbol{\tilde \phi}^\ell)_{2,\ell}+\tfrac{1}{2}(\boldsymbol{\tilde \varphi}^{\rm h},\boldsymbol{\tilde \phi}^{\rm h})_{2,{\rm h}}.
\end{equation}
Such families of fields   serve as boundary values in the   definition of the amplitudes. 
We assume \eqref{hyp:cosmo} and consider possible marked points $({\bf x},{\bf s})$ 
with weights $(\boldsymbol{\alpha},\boldsymbol{\beta})$ satisfying the Seiberg bounds.
 The amplitude of $(\Sigma,g,{\bf x},\boldsymbol{\alpha},{\bf s},\boldsymbol{\beta}, \boldsymbol{\zeta})$ is defined to be the non-negative function on $(H^{s}(\T))^{b^\ell}\times  (H^{s}_{\rm even}(\T))^{b^{\rm h}}$ defined by 
\begin{align}\label{amplitude}
&\caA_{\Sigma,g,{\bf x},\boldsymbol{\alpha},{\bf s},\boldsymbol{\beta},\boldsymbol{\zeta}}(\tilde{\boldsymbol{\varphi}})= \lim_{\eps\to 0}Z_{\Sigma,g,{\rm m}}\caA^0_{\Sigma,g}(\tilde{\boldsymbol{\varphi}})\nonumber\\
&\times \E \big[\prod_{i=1}^m V_{\alpha_i,g,\epsilon}(x_i)\prod_{i=1}^{m^{\rm b}} V_{\beta_i,g,\epsilon}(s_i)e^{-\frac{Q}{4\pi}\int_\Sigma K_g\phi_g\dd {\rm v}_g-\frac{Q}{2\pi}\int_{\partial \Sigma}k_g\phi_g\dd \ell_g -\mu M_\gamma^g (\phi_g,\Sigma)- M^{{\rm b},g}_\gamma(\phi_g,\mu^{\rm b}) }\big]
\end{align}
where $\phi_g= X_{g,{\rm m}}+P \boldsymbol{\tilde \varphi}$ (recall $P \boldsymbol{\tilde \varphi}$ is the harmonic function with boundary values $\tilde{\varphi}_j^{\rm h}\circ (\zeta^{\rm h}_j)^{-1}$ on $\mc{C}^+_j$, $\tilde{\varphi}_j^{\ell}\circ (\zeta^{\ell}_j)^{-1}$ on $\mc{C}_{j}$ and Neumann condition on $\pl \Sigma_{\rm N}$), the expectation $\E$ is over the  GFF $X_{g,{\rm m}}$, $M_\gamma^g $ and $M^{{\rm b},g}_\gamma$  are defined as limit of \eqref{GMCg} as $\eps\to 0$ and $Z_{\Sigma,g,{\rm m}}$ and $\caA^0_{\Sigma,g}(\tilde{\boldsymbol{\varphi}})$ are defined by~\footnote{In general, $Z_{\Sigma,g,{\rm m}}=\det (\Delta_{g,{\rm m}})^{-\hf}\exp\big(\frac{1}{8\pi}\int_{\partial\Sigma_D}k_g\,\dd\ell_g-\frac{1}{8\pi}\int_{\partial\Sigma_N}k_g\,\dd\ell_g\big)$, since the metric $g$ is Neumann extendible, we have $k_g=0$ on $\partial\Sigma_N$. And here $\det (\Delta_{g,{\rm m}})$ means $\frac{\det' (\Delta_{g,{\rm N}})}{\mathrm{v}_g(\Sigma)}$ if $\partial\Sigma_D=\emptyset$. }  
\begin{align}\label{znormal}
Z_{\Sigma,g,{\rm m}}=\det (\Delta_{g,{\rm m}})^{-\hf}\exp\big(\frac{1}{8\pi}\int_{\partial\Sigma_D}k_g\,\dd\ell_g\big), \quad \caA^0_{\Sigma,g}(\tilde{\boldsymbol{\varphi}}):=e^{-\frac{1}{2}( \tilde{\boldsymbol{\varphi}}, ({\bf D}_\Sigma-{\bf D})  \tilde{\boldsymbol{\varphi}})_2}
\end{align}
Here ${\bf D}_\Sigma: (H^{1}(\T))^{b^\ell}\times  (H^{1}_{\rm even}(\T))^{b^{\rm h}}\to (L^2(\T))^{b^\ell}\times  (L^2_{\rm even}(\T))^{b^{\rm h}}$	
is the Dirichlet-to-Neumann map for $\Sigma$ defined for smooth $\tilde{\boldsymbol{\varphi}}$ by 
\[\int_{\Sigma} |\dd P\tilde{\boldsymbol{\varphi}}|_g^2{\rm dv}_g = 2 \pi (  \boldsymbol{\tilde \varphi} ,{\bf D}_\Sigma \boldsymbol{\tilde \varphi} )_2  \]
and ${\bf D}$ is the Fourier multiplier by $|n|$ on each Fourier mode, acting diagonally on 	$(H^{1}(\T))^{b^\ell}\times  (H^{1}_{\rm even}(\T))^{b^{\rm h}}$ (see \cite[Section 5.2]{GRW1} for details where it is shown that ${\bf D}_\Sigma-{\bf D}$ is smoothing, so that $\caA^0_{\Sigma,g}(\tilde{\boldsymbol{\varphi}})$ is well-defined and continuous in $\tilde{\boldsymbol{\varphi}}$). When there are no marked points, we simply write $\caA_{\Sigma,g,\boldsymbol{\zeta}}(\tilde{\boldsymbol{\varphi}})$ for the amplitude.

By \cite[Proposition 5.12]{GRW1}, the amplitudes are \textbf{Hilbert-Schmidt} if $s:=\sum_i\alpha_i+\sum_j\frac{\beta_j}{2}-\frac{Q}{2}\chi(\Sigma^{\#2})>0$, in the following sense: 
\begin{equation}\label{HSamplitudes}
\caA_{\Sigma,g,{\bf x},\boldsymbol{\alpha},{\bf s},\boldsymbol{\beta},\boldsymbol{\zeta}}\in L^2(H^{s}(\T)^{b^\ell}\times  H^{s}_{\rm even}(\T)^{b^{\rm h}}, \mu_0^{\otimes b^{\ell}}\otimes (\mu_0^+)^{\otimes b^{\rm h}})=\mc{H}^{\otimes b^{\ell}}\otimes (\mc{H}^+)^{\otimes b^{\rm h}}.
\end{equation}
When all the parametrisations $\zeta_j^\ell,\zeta^{\rm h}_j$ are outgoing, we  consider $\caA_{\Sigma,g,{\bf x},\boldsymbol{\alpha},{\bf s},\boldsymbol{\beta},\boldsymbol{\zeta}}$ as a state in $\mc{H}^{\otimes b^{\ell}}\otimes (\mc{H}^+)^{\otimes b^{\rm h}}$, while if they are all
incoming, we will freely consider $\caA_{\Sigma,g,{\bf x},\boldsymbol{\alpha},{\bf s},\boldsymbol{\beta},\boldsymbol{\zeta}}$ as a Hilbert-Schmidt operator $\mc{H}^{\otimes b^{\ell}}\otimes (\mc{H}^+)^{\otimes b^{\rm h}}\to \C$, since $\mc{H}^+,\mc{H}$ are Hilbert spaces, thus canonically isomorphic to their dual. 
More generally, up to reordering the boundary components (and the associated fields $\tilde{\varphi}_j^{\ell},\tilde{\varphi}_j^{\rm h}$) so that the $(b^{\ell,+},b^{{\rm h},+})$ outgoing parametrised circles/half-circles come first and   the $(b^{\ell,-},b^{{\rm h},-})$ incoming parametrised circles/half-circles come last, we will consider $\caA_{\Sigma,g,{\bf x},\boldsymbol{\alpha},{\bf s},\boldsymbol{\beta},\boldsymbol{\zeta}}$ as a Hilbert-Schmidt operator 
\begin{equation}\label{amplitude_HS_op}
\caA_{\Sigma,g,{\bf x},\boldsymbol{\alpha},{\bf s},\boldsymbol{\beta},\boldsymbol{\zeta}}: \mc{H}^{\otimes b^{\ell,-}}\otimes (\mc{H}^+)^{\otimes b^{{\rm h},-}}\to \mc{H}^{\otimes b^{\ell,+}}\otimes (\mc{H}^+)^{\otimes b^{{\rm h},+}}
\end{equation}
by setting 
\[ (\caA_{\Sigma,g,{\bf x},\boldsymbol{\alpha},{\bf s},\boldsymbol{\beta},\boldsymbol{\zeta}}F)(\tilde{\bs{\varphi}})=\int  \caA_{\Sigma,g,{\bf x},\boldsymbol{\alpha},{\bf s},\boldsymbol{\beta},\boldsymbol{\zeta}}(\tilde{\bs{\varphi}},\tilde{\bs{\varphi}}')F(\tilde{\bs{\varphi}}')\dd (\mu_{0}^{\otimes b^{\ell,-}}\otimes (\mu_0^+)^{\otimes b^{{\rm h},-}})(\tilde{\bs{\varphi}}')
\]
for all $F\in \mc{H}^{\otimes b^{\ell,-}}\otimes (\mc{H}^+)^{\otimes b^{{\rm h},-}}$.

The amplitudes also admit \textbf{conformal covariance} and \textbf{diffeomorphism invariance} similar to \eqref{diff_corel} and \eqref{confan3} (see \cite[Prop. 5.9]{GRW1}): if $g'=e^{\omega}g$ is Neumann extendible with $\omega|_{\pl \Sigma_{\rm D}}=0$ and $\psi:\Sigma\to \Sigma'$ is a diffeomorphism, we have 
\begin{equation}\label{Weyl_amp}
\caA_{\Sigma, g',{\bf x},\boldsymbol{\alpha},{\bf s},\boldsymbol{\beta},\boldsymbol{\zeta}}(\tilde{\boldsymbol{\varphi}})= \caA_{\Sigma,g,{\bf x},\boldsymbol{\alpha},{\bf s},\boldsymbol{\beta},\boldsymbol{\zeta}}(\tilde{\boldsymbol{\varphi}}\big)e^{(1+6Q^2)S_{\rm L}^0(\Sigma,g,g')-\sum_{j=1}^m\Delta_{\alpha_j}\omega(x_j) -\sum_{j=1}^{m^{\rm b}}\frac{1}{2}\Delta_{\beta_j}\omega(s_j)}
\end{equation}
\begin{equation}\label{diffeo_amp} 
\caA_{\Sigma,  \psi^*g,{\bf x},\boldsymbol{\alpha},{\bf s},\boldsymbol{\beta},\boldsymbol{\zeta}}(\tilde{\boldsymbol{\varphi}})= \caA_{\Sigma', g,\psi({\bf x}),\boldsymbol{\alpha},\psi({\bf s}),\boldsymbol{\beta},\psi\circ\boldsymbol{\zeta}}\big(\tilde{\boldsymbol{\varphi}}\big)
\end{equation}
where we have used the notation 
\begin{equation}\label{def_SL0}
S_{\rm L}^0(\Sigma,g,g')=\frac{1}{96\pi}\int_\Sigma (|d\omega|_g^2+2K_g\omega){\rm dv}_g
\end{equation}
and $\psi\circ \boldsymbol{\zeta}$ is the parametrisation obtained by left-composing each $\zeta_j^{\ell}$ and $\zeta_j^{\rm h}$ by $\psi$ and $\psi({\bf x})$ and $\psi({\bf s})$ are the collection of marked points $\psi(x_j)$ and $\psi(s_j)$.

We will need the following version of Weyl anomaly, which generalizes \eqref{Weyl_amp} to metrics that are Neumann extendible but not admissible:
\begin{proposition}\label{Weyl2}
Assume $\partial\Sigma_{\rm D}\not=\emptyset$. Let $g_0$ be admissible on $\Sigma$ and $g$ be a Neumann extendible metric of the form $g=e^{\omega}g_0$ for some smooth map $\omega: \Sigma\to\R$, with restriction to the $\partial\Sigma_D$ denoted $ \omega_\partial:=\omega\mid_{\partial\Sigma_{\rm D}}$. Then   
\[\begin{split}
\caA_{\Sigma, g,{\bf x},\boldsymbol{\alpha},{\bf s},\boldsymbol{\beta},\boldsymbol{\zeta}}(\tilde{\boldsymbol{\varphi}})= &\caA_{\Sigma,g_0,{\bf x},\boldsymbol{\alpha},{\bf s},\boldsymbol{\beta},\boldsymbol{\zeta}}\big(\tilde{\boldsymbol{\varphi}}+\frac{Q}{2}\omega_\partial\big)\\
& \times e^{(1+6Q^2)S_{\rm L}^0(\Sigma,g_0,g)-\sum_{j=1}^m\Delta_{\alpha_j}\omega(x_j) -\sum_{j=1}^{m^{\rm b}}\frac{1}{2}\Delta_{\beta_j}\omega(s_j)-\frac{1}{32}\sum_{j=1}^{n_{\mc{K}}}\omega(\kappa_j)} e^{ -\frac{Q}{2}({\bf D}\omega_\pl,  \tilde{\boldsymbol{\varphi}})_{2}-\frac{Q^2}{8}({\bf D}\omega_\pl,\omega_\pl )_{2}}
\end{split}\]
where $(\kappa_j)_{j=1,\dots, n_{\mc{K}}}$ is the set of corner points.
\end{proposition}

\begin{proof} The proof is   similar to \cite[Proposition 2.12]{BGKR1} so we only write the main steps, explaining the differences. First we decompose $\omega=\omega_0+P\omega_\partial$ with $\omega_0|_{\partial\Sigma_{\rm D}}=0$ and $P\omega_\partial$ the harmonic extension of $\omega_\partial$ obeying $\partial_\nu P\omega_\partial=0$ on $\partial\Sigma_{\rm N}$. Let us introduce the metric $g_0':=e^{\omega_0}g_0=e^{-P\omega_\pl}g$. 

First, we make a few computational observations. Since $g_0'=e^{\omega_0}g_0$, $\omega_0=0$ on
$\partial\Sigma_{\rm D}$, and $k_{g_0}=0$ there, the conformal
transformation rule for the geodesic curvature, with our inward normal
convention, gives
\[
k_{g_0'}=-\frac12\partial_\nu\omega_0
=-\frac12\partial_\nu\omega
+\frac12\partial_\nu P\omega_\partial
\qquad\text{on }\partial\Sigma_{\rm D}.
\]
Using Green's formula (with $\nu$ the inward pointing normal),  
\[
\int_{\partial\Sigma_{\rm D}}\omega_\partial\,\partial_\nu
P\omega_\partial\,d\ell_{g_0}
=
-2\pi({\bf D}_\Sigma\omega_\partial,\omega_\partial)_2,
\]
we obtain
\begin{equation}\label{IPPgeod}
\int_{\partial\Sigma_{\rm D}}k_{g_0'}\omega\,d\ell_{g_0'}
=
-\frac12\int_{\partial\Sigma_{\rm D}}
\omega\,\partial_\nu\omega\,d\ell_{g_0}
-\pi({\bf D}_\Sigma\omega_\partial,\omega_\partial)_2.
\end{equation}

Next, we claim that $g_0'$ is Neumann extendible, since the harmonic extension $P\omega_{\pl}$ on $\Sigma$ with Neumann condition at $\pl \Sigma_{\rm N}$ is the restriction on $\Sigma$ of the harmonic extension of $\omega_\pl^\#$ on the Neumann double $\Sigma^{\#2}$. Then the formula relating the amplitude with metric $g_0'$ and the amplitude with metric $g_0$ follows from \eqref{Weyl_amp}. Thus it suffices to compare the amplitudes with metric $g_0'$ and with metric $g$.
	We notice that $k_g=e^{-\frac{P\omega_\partial}{2}}(k_{g_0'}-\frac{1}{2}\partial_\nu P\omega_\partial)$, so $k_g=k_{g_0'}=k_{g_0}=0$ on $\partial\Sigma_{\rm N}$, with $\nu$ the inward pointing normal. Furthermore $K_{g}=e^{-P\omega_\pl} K_{g_0'}$ using $\Delta_{g_0}P\omega_\pl=0$. Then
\begin{equation}\label{boundgeod}
\begin{gathered}
\frac{Q}{2\pi}\int_{\partial\Sigma}k_{g}P\tilde{\bs{\varphi}}\,\dd \ell_{g}=\frac{Q}{2\pi}\int_{\partial\Sigma}k_{g_0'}P\tilde{\bs{\varphi}}\,\dd \ell_{g_0'}+\frac{Q}{2} ( {\bf D}_\Sigma\omega_\pl,\tilde{\bs{\varphi}})_2,\\
\frac{Q}{4\pi}\int_{\Sigma} K_{g} (X_{g,D} +P\tilde{\bs{\varphi}})\dd {\rm v}_{g}= \frac{Q}{4\pi}\int_{\Sigma}  K_{g_0'} (X_{g_0',D}+P\tilde{\bs{\varphi}})\dd {\rm v}_{g_0'}.
\end{gathered}
\end{equation}
Using the identity 
$\caA^0_{\Sigma,g_0'}(\tilde{\bs{\varphi}})=\caA^0_{\Sigma,g_0'}(\tilde{\bs{\varphi}}+\frac{Q}{2}\omega_\pl)e^{ \frac{Q}{2}(({\bf D}_\Sigma-{\bf D})\omega_\pl,  \tilde{\bs{\varphi}})_2  +\frac{Q^2}{8}(\omega_\pl, ({\bf D}_\Sigma -{\bf D})\omega_\pl)_2}$, the scaling expressions \eqref{scalingmeasure1}, \eqref{scalingmeasure2} and \eqref{scalingvertex},  we end up with the formula
\[\begin{split}
\caA_{\Sigma,g,{\bf x},\bs{\alpha},{\bf s},\boldsymbol{\beta},\bzeta}(\tilde{\bs{\varphi}})=&\frac{Z_{\Sigma,g,{\rm m}}}{Z_{\Sigma,g_0',{\rm m}}} \caA_{\Sigma,g_0',{\bf x},\bs{\alpha},{\bf s},\boldsymbol{\beta},\bzeta}\big(\tilde{\bs{\varphi}}+\frac{Q}{2}\omega_\pl\big)  \\
& \times   
    e^{ -\frac{Q}{2}({\bf D}\omega_\pl,   \tilde{\bs{\varphi}})_2-\frac{Q^2}{8}({\bf D}\omega_\pl,\omega_\pl )_2+\frac{Q^2}{4\pi}\int_{\pl \Sigma}k_{g_0'}\omega \dd\ell_{g_0'} }
e^{6Q^2S_{\rm L}^0(\Sigma,g_0', g)-\sum_i\Delta_{\alpha_i}P\omega_\pl(x_i)-\sum_{j=1}^{m^{\rm b}}\frac{1}{2}\Delta_{\beta_j}\omega(s_j)} .
\end{split}\]
The Polyakov anomaly for the Dirichlet Laplacian on curvilinear polygonal domains, proved in \cite[Theorem 1.6]{Aldana-Kirsten-Rowlett}, reads
\[ 
\frac{Z_{\Sigma,g,{\rm D}}}{Z_{\Sigma,g'_0,{\rm D}}}=\frac{\det(\Delta_{\Sigma,g_0',{\rm D}})^{1/2}}{\det(\Delta_{\Sigma,g,{\rm D}})^{1/2}}=\exp\Big(S_{\rm L}^0(\Sigma,g'_0,g)+\frac{1}{24\pi}\int_{\pl \Sigma_{\rm D}}k_{g_0'}\omega \dd \ell_{g_0'}+\frac{1}{32}\sum_{j=1}^{n_{\mc{K}}}\omega(\kappa_j)\Big).
\]
Here, we have also used that $\int_{\pl \Sigma_{\rm D}}k_{g}\dd \ell_g=\int_{\pl \Sigma_{\rm D}}k_{g_0'}\dd \ell_{g'_0}-\frac{1}{2}\int_{\pl \Sigma_{\rm D}}\pl_\nu P\omega_\pl \dd \ell_{g_0'}=\int_{\pl \Sigma_{\rm D}}k_{g_0'}\dd \ell_{g'_0}$ and that the Neumann boundary has zero geodesic curvature for both $g_0'$ and $g$. 
Combining with the identity $Z_{\Sigma,g,{\rm D}}Z_{\Sigma,g,{\rm m}}=Z_{\Sigma^{\# 2},g^{\# 2},{\rm D}}$ (\cite[eq. (2.5)]{GRW1}) 
and the Polyakov anomaly formula for the partition function $Z_{\Sigma^{\# 2},g^{\# 2},{\rm D}}$ of the Neumann double (with no corners), 
we obtain 
\begin{align}\label{formula: anomaly corner}
\frac{Z_{\Sigma,g,{\rm m}}}{Z_{\Sigma,g'_0,{\rm m}}}= \frac{\det(\Delta_{\Sigma,g'_0,{\rm m}})^{1/2}}{\det(\Delta_{\Sigma,g,{\rm m}})^{1/2}}=\exp\Big(S_{\rm L}^0(\Sigma,g'_0,g)+\frac{1}{24\pi}\int_{\pl \Sigma_{\rm D}}k_{g_0'}\omega \dd \ell_{g_0'}
-\frac{1}{32}\sum^ {n_{\mc{K}}}_{j=1}\omega(\kappa_j)\Big).\end{align}
We may now use \eqref{IPPgeod}: this gives (for $c_{\rm L}=1+6Q^2$)
\[\begin{split}
\caA_{\Sigma,g,{\bf x},\bs{\alpha},{\bf s},\boldsymbol{\beta},\bzeta}(\tilde{\bs{\varphi}})=&e^{ -\frac{Q}{2}({\bf D}\omega_\pl,   \tilde{\bs{\varphi}})_2-\frac{Q^2}{8}({\bf D}\omega_\pl,\omega_\pl )_2+c_{\rm L}(S_{\rm L}^0(\Sigma,g_0', g)-\frac{1}{48\pi}\int_{\pl \Sigma}\pl_\nu \omega.\omega \dd\ell_{g_0} -\frac{1}{24}({\bf D}_{\Sigma}\omega_\pl,\omega_\pl)_2)} \\
& \times   
e^{-\frac{1}{32}\sum^ {n_{\mc{K}}}_{j=1}\omega(\kappa_j)-\sum_{j=1}^m\Delta_{\alpha_j}P\omega_\pl(x_j) -\sum_{j=1}^{m^{\rm b}}\frac{1}{2}\Delta_{\beta_j}\omega(s_j)}\caA_{\Sigma,g_0',{\bf x},\bs{\alpha},{\bf s},\boldsymbol{\beta},\bzeta}\big(\tilde{\bs{\varphi}}+\frac{Q}{2}\omega_\pl\big) .
\end{split}\]
Applying now \eqref{Weyl_amp} to compare  $\caA_{\Sigma,g_0',{\bf x},\bs{\alpha},{\bf s},\boldsymbol{\beta},\bzeta}$ and $\caA_{\Sigma,g_0,{\bf x},\bs{\alpha},{\bf s},\boldsymbol{\beta},\bzeta}$, and the identity 
\[S_{\rm L}^0(\Sigma,g_0',g)+S_{\rm L}^0(\Sigma,g_0,g'_0)=
S_{\rm L}^0(\Sigma,g_0,g)+\frac{1}{48\pi}\int_{\pl \Sigma_{\rm D}}\omega \pl_\nu\omega. \dd\ell_{g_0}+\frac{1}{24}({\bf D}_\Sigma \omega_\pl,\omega_\pl)_2,\]
we complete the proof.
\end{proof}

\subsection{Gluing of amplitudes} If $(\Sigma_1,g_1,\bs{\zeta}_1)$ and $(\Sigma_2,g_2,\bs{\zeta}_2)$ are two Riemann surfaces with corners, each equipped with an admissible Riemannian metric and analytic parametrisations of their D-marked boundaries, they can be glued along some boundary circles and/or half-circles to produce a new Riemann surface with corners  $(\Sigma,g,\bs{\zeta})$  endowed with an admissible metric.

Let $\Sigma_j$ have  $b_j^\ell$ D-marked boundary circles and  $b_j^{\rm h}$ D-marked boundary half-circles. If the gluing is performed along 
$k^\ell$ circles and $k^{\rm h}$ half-circles, we may, up to relabelling the D-marked boundary components, assume that 
$\pl_i\Sigma_1$ is glued to $\pl_i\Sigma_2$ for each $i$. The identification is then given by
$$\zeta^\ell_{1i}(e^{i\theta})\simeq \zeta^\ell_{2i}(e^{i\theta}),\quad i\leq k^\ell,\qquad   \zeta^{\rm h}_{1i}(e^{i\theta})\simeq \zeta^{\rm h}_{2i}(e^{i\theta}), \quad i\leq k^{\rm h},$$
where $(\zeta_{j1}^{\ell},\dots,\zeta_{jb^\ell_j}^{\ell})$ (resp. $(\zeta_{j1}^{\rm h},\dots,\zeta_{jb^{\rm h}_j}^{\rm h})$)  are the boundary parametrisations of the D-marked boundary circles (resp. half-circles) of $\Sigma_j$.   By \cite[Proposition 6.1]{GRW1}, if $({\bf x}_j,\bs{\alpha}_j)$ (for $j=1,2$) denote the collections of interior marked points with weights on $\Sigma_j^\circ$, and $({\bf s}_j,\bs{\beta}_j)$ ($j=1,2$) denote the boundary marked points on  $(\pl \Sigma_j)_{\rm N}$ with weights satisfying the Seiberg bounds, then for almost all
$$\tilde{\bs{\varphi}}_j\in H^{-s}(\T)^{b_j^\ell-k^\ell}\times H^{-s}_{\rm even}(\T)^{b_j^{\rm h}-k^{\rm h}},\quad j=1,2$$ one has
\begin{equation}\label{glue_ampl}
\mc{A}_{\Sigma,g,{\bf x},\bs{\alpha},{\bf s},\bs{\beta},\bs{\zeta}}(\tilde{\bs{\varphi}}_1,\tilde{\bs{\varphi}}_2)=C\int \caA_{\Sigma_1,g_1,{\bf x}_1,\boldsymbol{\alpha}_1,{\bf s}_1,\boldsymbol{\beta}_1,\boldsymbol{\zeta}_1}(\tilde{\boldsymbol{\varphi}} ,\tilde{\boldsymbol{\varphi}}_1)\caA_{\Sigma_2,g_2,{\bf x}_2,\boldsymbol{\alpha}_2,{\bf s}_2,\boldsymbol{\beta}_2,\boldsymbol{\zeta}_2}(\tilde{\boldsymbol{\varphi}},\tilde{\boldsymbol{\varphi}}_2) \dd\mu_0^{\otimes {k^\ell}} (\tilde{\boldsymbol{\varphi}}^\ell){\dd\mu_0^+}^{\otimes{k^{\rm h}}} (\tilde{\boldsymbol{\varphi}}^{\rm h})
\end{equation}
where $({\bf x},\bs{\alpha})$ and $({\bf s},\bs{\beta})$ are obtained by taking the unions of the corresponding data from $\Sigma_1$ and $\Sigma_2$.
The constant $C$ is given by
\begin{align*}
 C= \frac{1}{2^{\frac{k^\ell}{2}+\frac{3 k^{\rm h}}{4}} \pi^{k^\ell+\frac{3k^{\rm h}}{4}}} \textrm{ if }\partial\Sigma_{\rm D} \not =\emptyset,\qquad C= \frac{1}{2^{\frac{k^\ell}{2}+\frac{3 k^{\rm h}}{4}-1} \pi^{k^\ell+\frac{3k^{\rm h}}{4}-1}} \textrm{ if }\partial\Sigma_{\rm D} =\emptyset.
 \end{align*}
By \eqref{HSamplitudes}, if 
\begin{equation}\label{12seiberg}
\sum_i\alpha_{ji}+\sum_i\frac{\beta_{ji}}{2}-\frac{Q}{2}\chi(\Sigma_j^{\#2})>0, \quad j=1,2,
\end{equation}
then the amplitudes associated with $\Sigma_1$ and $\Sigma_2$ are Hilbert-Schmidt. Applying \cite[Lemma 7.2]{GKRV21_Segal}, 
we deduce that for 
$$b^\ell=b_1^{\ell}+b^{\ell}_2-2k^\ell, \quad b^{\rm h}=b_1^{\rm h}+b^{\rm h}_2-2k^{\rm h},$$
 the glued amplitude is also Hilbert-Schmidt:
\begin{equation}\label{glued_amp}
\mc{A}_{\Sigma,g,{\bf x},\bs{\alpha},{\bf s},\bs{\beta},\bs{\zeta}}\in  L^2(H^{s}(\T)^{b^\ell}\times  H^{s}_{\rm even}(\T)^{b^{\rm h}}, \mu_0^{\otimes b^{\ell}}\otimes (\mu_0^+)^{\otimes b^{\rm h}})=\mc{H}^{\otimes b^{\ell}}\otimes (\mc{H}^+)^{\otimes b^{\rm h}}.
\end{equation}

Next, we turn to the case of self-gluing.   Let  $(\Sigma,g,\boldsymbol{\zeta})$  be a Riemann surface with corners, equipped with an admissible metric $g$, and having   $b^\ell+b^{\rm h}$ parametrised D-marked boundary components. We assume that the boundary contains either an  outgoing boundary component  $\mc{C}_j \subset\partial\Sigma_{\rm D}$ and an incoming boundary component $\mc{C}_k\subset\partial\Sigma_{\rm D}$, or an  outgoing boundary half-circle  $\mc{C}^+_j \subset\partial\Sigma_{\rm D}$ and an incoming boundary half-circle $\mc{C}^+_k\subset\partial\Sigma_{\rm D}$.  In both cases,  we can glue the corresponding boundary components   to produce a new Riemann  surface $(\Sigma_{j\# k},g,\boldsymbol{\zeta}_\#)$. 

In the first case, we  write the boundary fields for $\Sigma$ as  
$$(\tilde\varphi^\ell_j,\tilde\varphi^\ell_k  ,\tilde{\boldsymbol{\varphi}}') \in   H^{s}(\T) \times H^{s}(\T) \times (H^{s}(\T))^{b^\ell-2}\times (H^{s}_{\rm even}(\T))^{b^{\rm h}},$$ where $\tilde\varphi_j^\ell$ and  $\tilde\varphi_k^\ell$  correspond respectively  to  $\mc{C}_j$ and   $\mc{C}_k$, while    $\tilde{\boldsymbol{\varphi}}'$ corresponds to the remaining  boundary fields of $(\Sigma_{j\# k}, g,\boldsymbol{\beta}_\# )$. Then
 \[
\caA_{\Sigma_{j\# k},g,{\bf x}, \boldsymbol{\alpha},{\bf s},\boldsymbol{\beta},\boldsymbol{\zeta}_\#}
(\tilde{\boldsymbol{\varphi}}')=C \int \caA_{\Sigma,g,{\bf x},\boldsymbol{\alpha},{\bf s},\boldsymbol{\beta},\boldsymbol{\zeta}}(\tilde\varphi^\ell,\tilde\varphi^\ell, \tilde{\boldsymbol{\varphi}}') \dd\mu_0(\tilde{\varphi}^\ell).
\]
where $C= \frac{1}{\sqrt{2} \pi }$ if $(\partial{\Sigma_{j\# k}})_{\rm D} \not =\emptyset$ and $C= \sqrt{2}  $ if $(\partial{\Sigma_{j\# k}})_{\rm D} =\emptyset$.

In the second case, we  write the boundary fields of $\Sigma$ as  
$$(\tilde\varphi^{\rm h}_j,\tilde\varphi^{\rm h}_k  ,\tilde{\boldsymbol{\varphi}}') \in   (H^{s}_{\rm even}(\T)) \times (H^{s}_{\rm even}(\T)) \times(H^{s}(\T))^{b^\ell} \times(H^{s}_{\rm even}(\T))^{b^{\rm h}-2}   ,$$ where $\tilde\varphi^{\rm h}_j$ and $\tilde\varphi^{\rm h}_k$  correspond respectively  to  $\mc{C}^+_j$ and $\mc{C}^+_k$, and   $\tilde{\boldsymbol{\varphi}}'$   to the remaining ones  of $\Sigma_{j\# k}$. We then have 
 \begin{equation}\label{self_gluing_formula}
\caA_{\Sigma_{j\#k},g, \boldsymbol{\alpha},{\bf s},\boldsymbol{\beta},\boldsymbol{\zeta}_\#}
(\tilde{\boldsymbol{\varphi}}')=C \int \caA_{\Sigma,g,{\bf x},\boldsymbol{\alpha},{\bf s},\boldsymbol{\beta},\boldsymbol{\zeta}}(\tilde\varphi^{\rm h} , \tilde\varphi^{\rm h}, \tilde{\boldsymbol{\varphi}}') \dd\mu_0^+(\tilde{\varphi}^{\rm h}).
\end{equation}
where $C= \frac{1}{(2 \pi)^{3/4}}$ if $\partial\Sigma_{\rm D} \not =\emptyset$ and $C=(2\pi)^{1/4}$ if $\partial\Sigma_{\rm D} =\emptyset$.

In both situations, if the Seiberg bound condition~\eqref{12seiberg} is satisfied for $\Sigma$, then the amplitude associated with the glued surface $\Sigma_{j\#k}$  is also Hilbert-Schmidt.

 \section{Geometric decomposition of surfaces into building blocks and structure constants}\label{sec:geom}

\subsection{Geometric/complex building blocks}\label{geometric_decomp}

In this section, we discuss the decomposition of a compact Riemann surface with N-marked boundary components and marked points into elementary Riemann surfaces with boundary, having both N-marked and D-marked components, that correspond to the structure constants of the boundary CFT.

First, we introduce for $\eps\geq 0$ the spaces
\begin{equation}\label{Holeps}
\begin{gathered}
{\rm Hol}_\eps(\D):=\{\psi:(1+\eps)\D\to \C \textrm{ holomorphic}\,|\, \psi\in C^\infty((1+\eps)\D)\} ,\quad
{\rm Hol}^+_\eps(\D):=\{\psi\in {\rm Hol}_\eps(\D)\,|\, \bar{\psi}(z)=\psi(\bar{z})\,\, \forall z\},\\
{\rm BiHol}_\eps(\D):=\{\psi\in {\rm Hol}_\eps(\D)\,|\, \psi \textrm{ diffeomorphism}\}, \quad
{\rm BiHol}^+_\eps(\D):={\rm BiHol}_\eps(\D)\cap {\rm Hol}^+_\eps(\D).
\end{gathered}
\end{equation} 
We will  specify  a topology for these spaces later  in Section \ref{sec:semigroups}.

Consider a compact Riemann surface $(\Sigma,J)$, of genus $g(\Sigma)$, with N-marked boundary, together with  a collection of bulk marked points ${\bf x}=(x_1,\dots,x_m)$, i.e. $x_j\in \Sigma^\circ$, and boundary marked points ${\bf s}=(s_1,\dots,s_{m^{\rm b}})$ on the Neumann boundary, i.e. $s_i\in \pl\Sigma$. We will assume that the Seiberg bounds \eqref{seiberg1bound} and \eqref{seiberg2bound} are satisfied.

We will decompose  the surface into building blocks $(\mc{P}_j)_j$, which are Riemann surfaces with corners and marked points, each associated with a Liouville amplitude in 
$\mc{H}^{\otimes b^\ell_j}\otimes (\mc{H}^+)^{\otimes b^{\rm h}_j}$, where $b^\ell_j$ (resp. $b^{\rm h}_j$) denotes 
the number of D-marked (resp. N-marked) boundary components  of $\mc{P}_j$.  The first step is to cut the surface along a family of curves.\\

\begin{figure}[H] 
 \centering
\begin{tikzpicture}[scale=0.6]
\draw[fill=lightgray,draw=black,very thick] (0,0) circle (1.5) ;
\draw[fill=white,draw=black,very thick] (-0.7,0.3) circle (0.4) ;
\draw[fill=white,draw=black,very thick] (0.7,-0.3) circle (0.3) ;
 \draw (0,1.5) node[above]{(A)} ; 
\draw[fill=lightgray,draw=black,very thick] (4,0) circle (1.5) ;
\draw (4.5,0) node{$\times$} ;
 \draw (4,1.5) node[above]{(B)} ; 
\draw[fill=white,draw=black,very thick] (3.3,0.3) circle (0.4) ;
\draw[fill=lightgray,draw=black,very thick] (8,0) circle (1.5) ;
\draw (7,0.3) node{$\times$} ;
 \draw (8,1.5) node[above]{(C)} ; 
\draw (8,-0.5) node{$\times$} ;
\shade[ball color=lightgray,opacity=0.9] (12,0) circle (1.5);
\draw[draw=black,very thick] (12,0) circle (1.5);
\draw (12,1.0) node{$\times$};
\draw (11.2,-0.4) node{$\times$};
\draw (12.9,-0.6) node{$\times$};
\draw[dashed,thin] (10.5,0) arc[start angle=180,end angle=360,x radius=1.5,y radius=0.4];
\draw[thin] (10.5,0) arc[start angle=180,end angle=0,x radius=1.5,y radius=0.4];

  \path[fill=lightgray,draw=black,very thick]
    (1,-4) arc[start angle=0,end angle=180,radius=1.5] -- (-2,-4) -- cycle;
    \draw[fill=white,draw=black,very thick] (-1,-4) circle (0.5) ;
    \draw[fill=white,draw=black,very thick] (0.2,-4) circle (0.3) ;
 \draw [fill=white,draw=white] (-2,-4.05) rectangle (2,-5);
 \draw (0.8,-4.8) node[above]{${\rm N}$} ; 
  \draw (-0.3,-4.8) node[above]{${\rm N}$} ; 
   \draw (-1.7,-4.8) node[above]{${\rm N}$} ; 
   \draw (-0.5,-2.6) node[above]{(A2)} ; 
   \path[fill=lightgray,draw=black,very thick]
    (5,-4) arc[start angle=0,end angle=180,radius=1.5] -- (2,-4) -- cycle;
    \draw[fill=white,draw=black,very thick] (3,-4) circle (0.5) ;
 \draw [fill=white,draw=white] (2,-4.05) rectangle (6,-5);
  \draw (4.1,-4.8) node[above]{${\rm N}$} ; 
   \draw (2.3,-4.8) node[above]{${\rm N}$} ;  
   \draw (4.5,-4) node{$\times$};  
      \draw (3.5,-2.6) node[above]{(B2)} ; 
   \path[fill=lightgray,draw=black,very thick]
    (9,-4) arc[start angle=0,end angle=180,radius=1.5] -- (6,-4) -- cycle;
     \draw (7.7,-4.8) node[above]{${\rm N}$} ; 
   \draw (8.5,-4) node{$\times$};  
      \draw (6.7,-4) node{$\times$};  
       \draw (7.5,-2.6) node[above]{(C2)} ; 
   \path[fill=lightgray,draw=black,very thick]
    (13,-4) arc[start angle=0,end angle=180,radius=1.5] -- (10,-4) -- cycle;
     \draw (11.5,-4.8) node[above]{${\rm N}$} ;
      \draw (11.5,-2.6) node[above]{(A1)} ; 
 \draw[fill=white,draw=black,very thick] (11.5,-3.5) circle (0.3) ;
   \path[fill=lightgray,draw=black,very thick]
    (17,-4) arc[start angle=0,end angle=180,radius=1.5] -- (14,-4) -- cycle;
     \draw (15.5,-4.8) node[above]{${\rm N}$} ; 
   \draw (15.5,-3.4) node{$\times$};  
    \draw (15.5,-2.6) node[above]{(C1)} ; 
\draw[fill=lightgray,draw=black,very thick] (19.5,-3) circle (1.5) ;
\draw (21,-3) node{$\times$} ;
\draw[fill=white,draw=black,very thick] (19.5,-3) circle (0.3) ;
     \draw (17.7,-3.8) node[above]{${\rm N}$} ; 
      \draw (19.5,-1.6) node[above]{(B1)} ; 
    \end{tikzpicture}
\caption{Building blocks satisfying \eqref{nondegpant}. Only the N-marked boundaries are shown in the figures (the remaining boundary components are therefore D-marked).
The first row corresponds to the case of an empty Neumann boundary. The first three surfaces are disks with marked points or holes, and the last one is the Riemann sphere with three marked points.
The second row corresponds to the case of a non-empty Neumann boundary, which should be supplemented with the structure constants of Figure~\ref{pic:boundarystructure}.
Note that the surfaces depicted in the second row, together with those of Figure~\ref{pic:boundarystructure},  exhaust all possible Riemann surfaces whose Neumann doubles reproduce the figures in the first row.}
\label{pic:geomblock}
\end{figure}
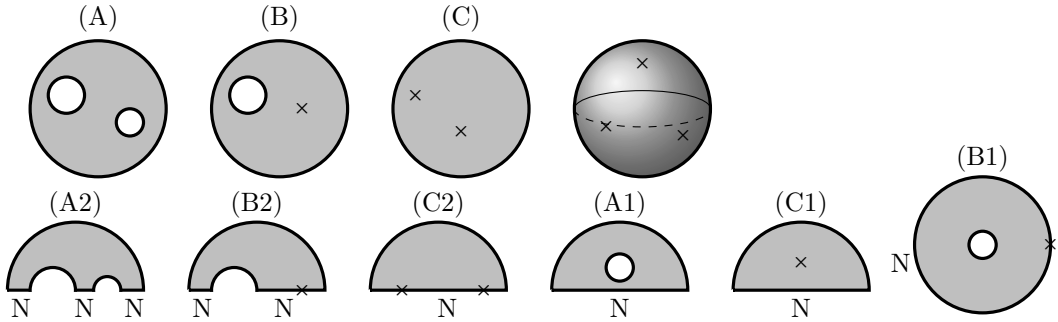

\noindent \textbf{Cutting curves.}  We choose a finite family of analytic parametrized curves in $\Sigma$ (see \cite[Section 2.2]{GRW1} for a detailed definition\footnote{The parametrisations $\zeta_j^{\rm h}$ of the half-circles are required to be real-analytic, which makes their existence not completely trivial. The existence of such maps  arbitrarily close to smooth parametrisations is proven in \cite[Lemma 2.6]{GRW1}.
}),  called cutting curves, satisfying the following properties:\\
$\bullet$ each curve is 
\begin{itemize}
\item[-] either an analytic simple loop $\xi^\ell_j:\T\to \Sigma^\circ\setminus \{{\bf x}\}$ with image 
 $\xi^\ell_j(\T)=\mc{C}_j$,
 \item[-] or  an analytic simple half-circle $\xi_j^{\rm h}:\T^+\to \Sigma$ with image $\xi_j^{\rm h}(\T^+)=\mc{C}^+_j$ and endpoints 
 $\xi_j^{\rm h}(-1),\xi_j^{\rm h}(1)\in \pl \Sigma \setminus \{{\bf s}\}$. At the endpoints, the curves make    a $\pi/2$ angle with $\pl \Sigma$.
\end{itemize}
$\bullet$  the curves are disjoint in the sense that  $\mc{C}_j\cap \mc{C}_{j'}=\emptyset$ if $j\not=j'$, $\mc{C}^+_j\cap \mc{C}^+_{j'}=\emptyset$ for all $j\not=j'$, and $\mc{C}^+_j\cap \mc{C}_i=\emptyset$ for all $i,j$.\\
$\bullet$ Let $(\mc{P}_j)_j$ be the connected components of $\Sigma\setminus (\cup_{j}\mc{C}_j\cup \cup_{i}\mc{C}^+_i)$, and denote $\mc{P}_j^{\#2}$ their Neumann double    along their N-marked boundary (if non trivial). Let $\{{\bf x}_j\}:=\mc{P}_j\cap \{{\bf x}\}$ and $\{{\bf s}_j\}:=\mc{P}_j\cap \{{\bf s}\}$. We require that 
\begin{equation}\label{nondegpant}
 \chi(\mc{P}_j)-\# \{{\bf x}_j\} <0 \textrm{ if }(\pl \mc{P}_j)_{\rm N}=\emptyset, \quad  \chi(\mc{P}_j^{\#2})-2\#\{{\bf x}_j\} -\# \{{\bf s}_j\} <0 \textrm{ if }(\pl \mc{P}_j)_{\rm N}\not=\emptyset
\end{equation}
where $\#$ denotes the cardinal and $\chi(\cdot)$ the Euler characteristic as before. See Figure \ref{pic:geomblock}.

Under these conditions, we say that the cutting is \textbf{non-degenerate}. As mentioned in Section \ref{sec:surfaces}, the Neumann double $\Sigma^{\#2}$ of $\Sigma$ (with involution $\tau_\Sigma$) admits a family of parametrised cutting circles obtained from the circles $\mc{C}_j$, their reflected images $\tau_\Sigma(\mc{C}_j)$ and the circles formed by gluing the half-circles $\mc{C}^+_j$ to their reflections $\tau_\Sigma(\mc{C}_j^+)$ at the endpoints. This procedure yields a decomposition of $\Sigma^{\#2}\setminus (\cup_{j}(\mc{C}_j\cup\tau_\Sigma(\mc{C}_j))\cup (\cup_{j}(\mc{C}^+_j\cup \tau_\Sigma(\mc{C}_j^+)))$ into 
connected components $\mc{P}'_j$ satisfying
\[ \chi(\mc{P}_j')-\# \{\textrm{marked points in }\mc{P}_j\}<0.\] 

Notice that the boundary hypersurfaces (loops or half-circles) of $\mc{P}_j$ inherit an orientation from the orientation induced by the parametrisations of $\mc{C}_j$ (resp. $\mc{C}_j^+$). We say that they are \textbf{outgoing} if $(\dot{\xi}^\ell_j,\nu)$ (resp. $(\dot{\xi}_j^{\rm h},\nu)$) is positively oriented in $\mc{P}_j$ for $\nu$ any normal  vector field pointing inside $\mc{P}_j$, and \textbf{incoming} otherwise. It is possible to reverse the orientation of a $\mc{C}_j$  (resp. $\mc{C}_j^+$) by replacing $\xi_j^\ell$ (resp. $\xi_j^{\rm h}$) by 
\[ \xi_j^\ell \circ o  \qquad (\textrm{resp. }  \, \xi_j^{\rm h}\circ (-o)) ,\quad \textrm{ with } o(z):=1/z.\]

The last condition on the Euler characteristic implies that there is a maximal number of such cutting curves. We denote by $L$ the number of cutting loops and $L^+$ the number of cutting half-circles on $\Sigma$.
Two families of cutting curves that are related by a diffeomorphism homotopic to the identity, relative to the marked points ${\bf x},{\bf s}$ (i.e. the homotopy fixes the marked points for all times), are called equivalent families. The set of equivalence classes of such families is a topological data, which can be encoded in a graph as in the closed case (see \cite[Section 3.2.5]{GKRV21_Segal}). We investigate now  all the possible situations.\\

\textbf{Case 1}: First, it may happen that no non-degenerate family of cutting curves exists.
In that case, the surface with marked points $(\Sigma, {\bf x}, {\bf s})$ is diffeomorphic to one of the following:\\
1) A sphere with $3$ marked points (in case $\pl \Sigma=\emptyset$),\\
2) A disk with $1$ boundary point and $1$ interior point,\\  
3) A disk with $3$ boundary points.\\
We shall refer to these as the \textbf{irreducible surfaces}, since they cannot be decomposed into subsurfaces with D-marked boundaries whose associated amplitudes are Hilbert-Schmidt operators.
The correlation functions corresponding to these irreducible surfaces define the   \textbf{structure constants} of the theory. 
Notice that in both cases 2) and 3), the Neumann double of the surface  is a sphere with $3$ marked points.\\

\textbf{Case 2}:  If a family of non-degenerate cutting curves does exist, we consider a maximal family of such cutting curves and assign a D-marking  to each $\mc{C}_j$ and $\mc{C}^+_i$. Then we consider the closures of  the connected components of $\Sigma\setminus (\cup_{j}\mc{C}_j\cup \cup_i \mc{C}^+_i)$, which are oriented surfaces with corners and marked points,  with boundary hypersurfaces assigned with marking D or N. The D-marked boundary curves are oriented by the orientations of $\mc{C}_i$ and $\mc{C}_i^+$.We denote these connected components by $\mc{P}_1,\dots,\mc{P}_{N+N^+}$ and we order them so that $\mc{P}_j$ have only D-marked boundary for $j\leq N$ and $\mc{P}_j$ has non-empty N-marked boundary for $N+1\leq j\leq N+N^+$. Notice that $\mc{P}_j$ has no corners for $j\leq N$ while $\mc{P}_j$  may have corners for $j>N$. 
Denote $\{{\bf x}_j\}:=\{{\bf x}\}\cap \mc{P}_j$ the set of marked points in $\mc{P}_j$ and 
  $\{{\bf s}_j\}=\{{\bf s}\}\cap (\pl \mc{P}_j)_{\rm N}$ those in the N-marked boundary of  $\mc{P}_j$.  
As they are marked, we will label them by (notice $m^{\rm b}_j=0$ if $j\leq N$)
\[{\bf x}_j=(x_{j1},\dots x_{jm_j}), \quad \quad {\bf s}_j=(s_{j1},\dots, s_{jm^{\rm b}_j}).\]
Denote by $\zeta^{\ell}_{jk}$ (if $j\leq N$) and $\zeta^{\rm h}_{jk}$ (if $j>N$) the parametrisations of the boundary components of $\mc{P}_j$ induced by the parametrisations $\xi^{\ell}_i,\xi^{\rm h}_i$ of the loops $\mc{C}_i$ and half-circles $\mc{C}_i^+$ bounding $\mc{P}_j$, and write 
\[\boldsymbol{\zeta}_j=(\zeta_{j1}^\ell,\dots \zeta^\ell_{jb_j^{\ell}},\zeta_{j1}^{\rm h},\dots,\zeta_{jb^{{\rm h}}_j}^{\rm h}) .\] 
We write as before $\sigma_{jk}^{\ell}=-1$ (resp. $\sigma_{jk}^{{\rm h}}=-1$) if $\zeta^{\ell}_{jk}$ is outgoing in $\mc{P}_j$ and $\sigma_{jk}^{\ell}=1$ (resp. $\sigma_{jk}^{{\rm h}}=1$) if it is incoming. On the Neumann double $\Sigma^{\#2}$, the surfaces 
$(\mc{P}_j)_{j\leq N}$, $(\tau_\Sigma(\mc{P}_j))_{j\leq N}$ and $(\mc{P}_j^{\#2}=\mc{P}_j\# \tau_\Sigma( \mc{P}_j))_{j>N}$ provide a decomposition into building blocks of the closed surface $\Sigma^{\#2}$ (see \cite[Section 3.2.3]{GKRV21_Segal}). The set $\bs{\zeta}_j^{\#}$ of parametrisations 
of the double $\mc{P}_j^{\#2}$ is given by $\zeta_{jk}^\ell$, $\tau_\Sigma\circ \zeta_{jk}^{\ell}\circ o$ (for the loops and their reflection by the anti-holomorphic involution $\tau_\Sigma$ on $\Sigma^{\#2}$), and by 
$e^{i\theta}\mapsto \zeta_{jk}^{\rm h}(e^{i\theta})$ for $\theta\in [0,\pi]$ and $e^{i\theta}\mapsto \tau_\Sigma(\zeta_{jk}^{\rm h}(e^{i(2\pi-\theta)}))$ for $\theta\in [\pi,2\pi]$ (for the doubled half-circles $\mc{C}_i^+\cup\tau_\Sigma(\mc{C}_i^+)$ if $\zeta_{jk}^{\rm h}=\xi_i^{\rm h}$).\\

In what follows, we fix the triplet of points in the complex plane as
\[{\bf z}^0=(z^0_1,z^0_2,z^0_3) \in \{ (-1,0,1), (i,0,-i)\}.\]
This choice serves to fix the M\"obius invariance in the uniformisation of the building blocks onto the Riemann sphere. 
Later on, however, we shall need to relate the uniformisation of pairs of pants (counting marked points as boundary ends) to that of half-pants.
Depending on the situation, it will be more convenient to use one triplet of coordinates or the other.

By \cite[Section 8.2 \& 8.3]{GKRV21_Segal}   each component $\mc{P}_j$, for $j\leq N$,  is biholomorphic to one of the following surfaces with marked points, which we refer to as   \textbf{model forms}:

\medskip

\noindent(A) A \textbf{pair of pants} with $3$ boundary circles corresponding to loops in $\mc{P}_j\cap \cup_i \mc{C}_i$, hence D-marked. More precisely, for    $\eps>0$ small enough,  there exist  unique maps $\psi^\ell_{jk}\in {\rm BiHol}_\eps(\D)$ for   $k=1,2,3$, and a unique 
biholomorphism $\Phi:\mc{P}_j\to \hat{\C}\setminus \cup_{k=1}^3\psi_{jk}^\ell(\D)$, which extends holomorphically over a neighborhood of 
$\mc{P}_j$, such that  
\[(\psi_{j1}^\ell(0),\psi_{j2}^\ell(0),\psi_{j3}^{\ell}(0))={\bf z}^0,\hspace{1.5cm} \psi^{\ell}_{jk}|_\T=\Phi \circ \zeta_{jk}^{\ell} \circ o^{(1-\sigma_{jk}^\ell)/2},\quad k=1,2,3.\]
The surface $\hat{\C}\setminus \cup_{k=1}^3\psi_{jk}^\ell(\D)$, endowed with its canonical complex structure and boundary parametrisations 
\[\psi_{j1}^\ell\circ o^{(1-\sigma_{j1}^\ell)/2}|_{\T}, \quad \psi_{j2}^\ell\circ o^{(1-\sigma_{j2}^\ell)/2}|_{\T}, \quad \psi_{j3}^\ell\circ o^{(1-\sigma_{j3}^\ell)/2}|_{\T},\]
is called the model form of $\mc{P}_j$.

\medskip
 (B) An \textbf{annulus} with $1$ interior marked point  ${\bf x}_j:=x_{j1}$
and  $2$ boundary circles corresponding to $2$ loops of $\cup_j \mc{C}_j$, hence D-marked. 
More precisely,  for    $\eps>0$ small enough,  there exist  unique maps  $\psi^\ell_{jk}\in {\rm BiHol}_\eps(\D)$ for $k=1,2$, and a unique 
biholomorphism $\Phi:\mc{P}_j\to \hat{\C}\setminus \cup_{k=1}^2\psi_{jk}^\ell(\D)$, which extends holomorphically over a neighborhood of 
$\mc{P}_j$, such that  
\[(\psi_{j1}^\ell(0),\psi_{j2}^\ell(0),\Phi(x_{j1}))={\bf z}^0,\hspace{1.5cm} \psi^{\ell}_{jk}|_\T=\Phi \circ \zeta_{jk}^{\ell} \circ o^{(1-\sigma_{jk}^\ell)/2},\quad k=1,2.\]
The surface $\hat{\C}\setminus \cup_{k=1}^2\psi_{jk}^\ell(\D)$, endowed with its canonical complex structure and boundary parametrisations
\[\psi_{j1}^\ell\circ o^{(1-\sigma_{j1}^\ell)/2}|_{\T}, \quad \psi_{j2}^\ell\circ o^{(1-\sigma_{j2}^\ell)/2}|_{\T} \]
is called the model form of $\mc{P}_j$.  

\medskip
(C) A \textbf{disk} with $2$ interior marked points   ${\bf x}_j:=(x_{j1},  x_{j2})$ and $1$ boundary circle corresponding to a loop of $\cup_j \mc{C}_j$, hence D-marked. More precisely,  for    $\eps>0$ small enough,  there exists a unique map  $\psi^\ell_{j1}\in {\rm BiHol}_\eps(\D)$, and a unique 
biholomorphism $\Phi:\mc{P}_j\to \hat{\C}\setminus  \psi_{j1}^\ell(\D)$, which extends holomorphically over a neighborhood of 
$\mc{P}_j$, such that  
\[(\psi_{j1}^\ell(0),\Phi(x_{j1}),\Phi(x_{j2}))={\bf z}^0,\hspace{1.5cm} \psi^{\ell}_{j1}|_\T=\Phi \circ \zeta_{j1}^{\ell} \circ o^{(1-\sigma_{j1}^\ell)/2} .\]
The surface $\hat{\C}\setminus  \psi_{j1}^\ell(\D)$, endowed with its canonical complex structure and boundary parametrisation 
\[\psi_{j1}^\ell\circ o^{(1-\sigma_{j1}^\ell)/2}|_{\T} \]
is called the model form of $\mc{P}_j$. \\

Now, for the components $\mc{P}_j$ with $j>N$, the doubled surfaces $\mc{P}_j^{\#2}$ are building blocks biholomorphic to one of the forms (A), (B), or (C) described above, where the points $({\bf x}_j, \tau_{\Sigma}({\bf x}_j), {\bf s}_j)$ play the role of marked points in $\mc{P}_j^{\#2}$. The corresponding uniformising map $\Phi$ for the building block $\mc{P}_j^{\#2}$ satisfies that $\overline{\Phi} \circ \tau_\Sigma$ is holomorphic and has the same properties as $\Phi$. By uniqueness of the uniformising map, it follows that 
$$\Phi\circ \tau_\Sigma=\bbar{\Phi},\qquad \Phi((\pl\mc{P}_j)_{\rm N})\subset \R$$
 in the Riemann sphere $\hat{\C}$. Choosing ${\bf z}^0 = (-1, 0, 1)$ in the case ${\bf x}_j = \emptyset$, and ${\bf z}^0 = ({\rm i}, 0, -{\rm i})$ when ${\bf x}_j \neq \emptyset$, we obtain that each $\mc{P}_j$ with $j > N$ is biholomorphic to one of the following surfaces with marked points:
 
 \medskip
   
\noindent (A1) A \textbf{half pair of pants} with one boundary circle corresponding to a loop in $\cup_i\mc{C}_i$, hence marked D, 
 and two boundary half-circles, one of them corresponding to a half-circle in $\cup_i\mc{C}^+_i$, thus marked D, and the other one corresponding to an interval contained in $\pl \Sigma_{\rm N}$,  hence   marked N. More precisely, for $\epsilon$ small enough, there exist 
 $\psi^{\ell}_{j1}\in {\rm BiHol}_\eps(\D)$, $\psi^{\rm h}_{j2}\in {\rm BiHol}^+_\eps(\D)$, and  a 
 biholomorphism $\Phi:\mc{P}_j\to \hat{\H}\setminus (\psi^{\ell}_{j1}(\D)\cup \psi^{\rm h}_{j2}(\D^+))$, which extends holomorphically to   a neighborhood of $\mc{P}_j$, such that
 $$(\psi^{\ell}_{j1}(0)  , \psi^{\rm h}_{j2}(0))=( {\rm i} , 0) ,\qquad  \psi^{\ell}_{j1}|_{\T^+}=\Phi \circ \zeta_{j1}^{\ell} \circ o^{(1-\sigma_{j1}^\ell)/2} ,\qquad \psi^{\rm h}_{j2}|_{\T^+}=\Phi \circ \zeta_{j2}^{\rm h} \circ (-o)^{(1-\sigma_{j2}^{\rm h})/2} .
$$
 The surface $\hat{\H}\setminus (\psi^{\ell}_{j1}(\D)\cup \psi^{\rm h}_{j2}(\D^+))$, endowed with its canonical complex structure and boundary parametrisation  
\[\psi^{\ell}_{j1}\circ o^{(1-\sigma_{j1}^{\ell})/2}|_{\T}, \quad \psi_{j2}^{\rm h}\circ (-o)^{(1-\sigma_{j2}^{\rm h})/2}|_{\T^+}\]
is called the model form of $\mc{P}_j$. The model form of $\mc{P}_j^{\#2}$ is given by 
$\hat{\C}\setminus (\psi^{\ell}_{j1}(\D)\cup \psi^{\rm h}_{j2}(\D)\cup \tau(\psi^{\ell}_{j1}(\D)))$  with parametrisations
\[\psi_{j1}^{\ell}\circ o^{(1-\sigma_{j1}^{\ell})/2}|_{\T}, \quad \psi_{j2}^{\rm h}\circ (-o)^{(1-\sigma_{j2}^{\rm h})/2}|_{\T}, \quad 
\tau\circ \psi_{j1}^{\ell}\circ o^{(1+\sigma_{j1}^{\ell})/2}|_{\T}.\]

 \medskip
(A2) A \textbf{half pair of pants} with 3 boundary half-circles corresponding to half-circles in $\cup_i\mc{C}^+_i$, hence marked D, 
 and three disjoint boundary intervals   of $\pl \Sigma_{\rm N}$, hence marked N. More precisely,  for $\epsilon$ small enough, there exist
 maps $\psi^{\rm h}_{jk}\in {\rm BiHol}^+_\eps(\D)$ for $k=1,2,3$ and a biholomorphism
 $\Phi:\mc{P}_j\to \hat{\H}\setminus \cup_{k=1}^3\psi^{\rm h}_{jk}(\D^+)$, which extends holomorphically to a neighborhood of $\mc{P}_j$, such that
 \[\psi_{j1}^{\rm h}(0)=-1, \quad \{\psi_{j2}^{\rm h}(0),\psi_{j3}^{\rm h}(0)\}= \{0,1\}, \qquad \psi^{\rm h}_{jk}|_{\T^+}=\Phi \circ \zeta_{jk}^{\rm h} \circ (-o)^{(1-\sigma_{jk}^{\rm h})/2},\quad k=1,2,3.\]
 Here, the order in which the half-circles $\mc{C}^+_i$ are positioned determines if $\psi_{j2}^{\rm h}(0)=0$ or $\psi_{j2}^{\rm h}(0)=1$.
 The surface $\hat{\H}\setminus \cup_{k=1}^3\psi_{jk}^{\rm h}(\D^+)$, endowed with its canonical complex structure and boundary parametrisations 
\[\psi_{j1}^{\rm h}\circ (-o)^{(1-\sigma_{j1}^{\rm h})/2}|_{\T^+}, \quad \psi_{j2}^{\rm h}\circ (-o)^{(1-\sigma_{j2}^{\rm h})/2}|_{\T^+}, \quad 
\psi_{j3}^{\rm h}\circ (-o)^{(1-\sigma_{j3}^{\rm h})/2}|_{\T^+}\]
is called the model form of $\mc{P}_j$. The model form of $\mc{P}_j^{\#2}$ is given by 
$\hat{\C}\setminus \cup_{k=1}^3\psi^{\rm h}_{jk}(\D)$  with parametrisations
\[\psi_{j1}^{\rm h}\circ (-o)^{(1-\sigma_{j1}^{\rm h})/2}|_{\T}, \quad \psi_{j2}^{\rm h}\circ (-o)^{(1-\sigma_{j2}^{\rm h})/2}|_{\T}, \quad 
\psi_{j3}^{\rm h}\circ (-o)^{(1-\sigma_{j3}^{\rm h})/2}|_{\T}.\]
 
 \medskip
 (B1) A \textbf{half-annulus} with $1$ boundary marked point ${\bf s}_j=s_{j1}$ and a boundary circle corresponding to a circle in $\cup_i\mc{C}_i$, hence marked   D, and a circle  corresponding to a circle of $\pl \Sigma_{\rm N}$,  hence marked N, and containing $s_{j1}$. More precisely, for $\epsilon$ small enough, there exist
 a map  $\psi^{\ell}_{j1}\in {\rm BiHol}_\eps(\D)$ and a  biholomorphism $\Phi:\mc{P}_j\to \hat{\H}\setminus  \psi^{\ell}_{j1}(\D)$, which extends holomorphically to a neighborhood of   $\mc{P}_j$,      such that 
 $$(\Phi(s_{j1}), \psi^{\ell}_{j1}(0)  )=(0,{\rm i}) ,\qquad  \psi^{\ell}_{j1}|_{\T}=\Phi \circ \zeta_{j1}^{\ell} \circ o^{(1-\sigma_{j1}^{\ell})/2}.$$
 The surface $\hat{\H}\setminus\psi_{j1}^\ell(\D)$, endowed  with its canonical complex structure and boundary  parametrisation  
 $\psi_{j1}^\ell \circ o^{(1-\sigma_{j1}^{\ell})/2}|_{\T}$
is called the model form of $\mc{P}_j$. 

 \medskip
 (B2) A \textbf{half-annulus} with $1$ boundary marked point ${\bf s}_j=s_{j1}$, 
and two boundary half-circles  belonging to $\cup_i\mc{C}_i^+$, hence marked D, and two disjoint boundary intervals of $\pl \Sigma_{\rm N}$, hence marked N, one of which containing $s_{j1}$. More precisely, for $\epsilon>0$ small enough, there exist maps $\psi^{\rm h}_{jk}\in {\rm BiHol}^+_\eps(\D)$ for $k=1,2$ and   a  biholomorphism  $\Phi:\mc{P}_j\to \hat{\H}\setminus \cup_{k=1}^2\psi^{\rm h}_{jk}(\D^+)$, which extends holomorphically to a neighborhood of   $\mc{P}_j$,      such that 
 \[ \Phi(s_{j1})=1, 
 \{\psi^{\rm h}_{j1}(0),\psi^{\rm h}_{j2}(0)\} = \{-1,0\} ,\qquad \psi^{\rm h}_{jk}|_{\T^+}=\Phi \circ \zeta_{jk}^{\rm h} \circ (-o)^{(1-\sigma_{jk}^{\rm h})/2},\quad k=1,2.\]
 The surface $\hat{\H}\setminus \cup_{k=1}^2\psi_{jk}^{\rm h}(\D^+)$ with the canonical complex structure and with parametrisations 
 $$\psi_{j1}^{\rm h}\circ (-o)^{(1-\sigma_{j1}^{\rm h})/2}|_{\T^+},\qquad \psi_{j2}^{\rm h}\circ (-o)^{(1-\sigma_{j2}^{\rm h})/2}|_{\T^+}$$   is called the model form of $\mc{P}_j$.
 
 \medskip
(C1) A \textbf{half-disk} with $1$ interior marked point $x_{j1}$,  one boundary half-circle belonging to  $\cup_i\mc{C}^+_i$ (hence D-marked), and a single boundary interval  of $\pl \Sigma_{\rm N}$ (hence N-marked).
More precisely, for  $\epsilon$ small enough, there exist
 a map $\psi^{\rm h}_{j1}\in {\rm BiHol}^+_\eps(\D)$ and   a biholomorphism  $\Phi:\mc{P}_j\to \hat{\H}\setminus  \psi^{\rm h}_{j1}(\D^+)$, which extends holomorphically to a neighborhood of $\mc{P}_j$, such that
 $$(\Phi(x_{j1}),\psi^{\rm h}_{j1}(0))=({\rm i},0 ),\qquad  \psi^{{\rm h}}_{j1}|_{\T^+}=\Phi \circ \zeta_{j1}^{\rm h} \circ (-o)^{(1-\sigma_{j1}^{\rm h})/2}.$$
 The surface $\hat{\H}\setminus \psi_{j1}^{\rm h}(\D^+)$, endowed  with its canonical complex structure and boundary  parametrisation   $\psi_{j1}^{\rm h}\circ (-o)^{(1-\sigma_{j1}^{\rm h})/2}|_{\T^+}$ is called the model form of $\mc{P}_j$.
 
  \medskip
(C2) A \textbf{half-disk} with $2$ marked points on the boundary ${\bf s}_j=(s_{j1},s_{j2})$, one boundary half-circle belonging to  $\cup_i\mc{C}^+_i$ (hence D-marked), and a single boundary interval  of $\pl \Sigma_{\rm N}$ (hence N-marked). More precisely, 
for  $\epsilon$ small enough, there exist
 a map $\psi^{\rm h}_{j1}\in {\rm BiHol}^+_\eps(\D)$ and   a biholomorphism  $\Phi:\mc{P}_j\to \hat{\H}\setminus  \psi^{\rm h}_{j1}(\D^+)$, which extends holomorphically to a neighborhood of $\mc{P}_j$, such that
\[ \psi^{\rm h}_{j1}(0)=-1, \quad \{\Phi(s_{j1}),\Phi(s_{j2})\}=\{0,1\},\qquad   \psi^{{\rm h}}_{j1}|_{\T^+}=\Phi \circ \zeta_{j1}^{\rm h} \circ o^{(1-\sigma_{j1}^{\rm h})/2}.\]
 The surface $\hat{\H}\setminus \psi_{j1}^{\rm h}(\D^+)$ with the canonical complex structure and with parametrisations $\psi_{j1}^{\rm h}|_{\T^+}$ is called the model form of $\mc{P}_j$.\\

\begin{figure}
	\begin{tikzpicture}
		\node[inner sep=0pt] (pant) at (0,0)
		{\includegraphics[width=0.4\textwidth]{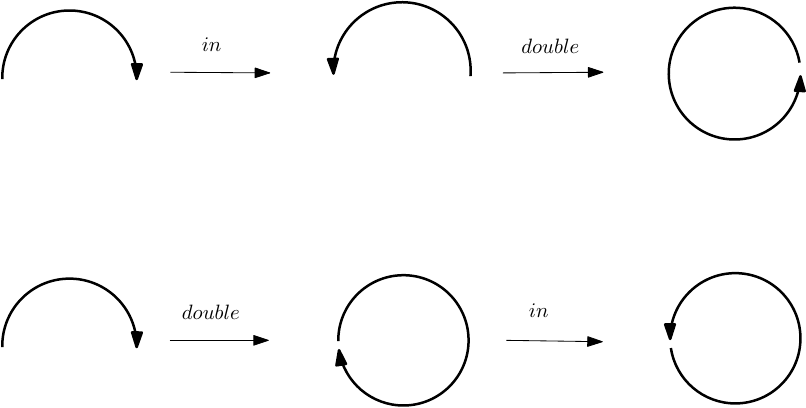}};
	\end{tikzpicture}
	\caption{Starting from an outgoing arc, the first line corresponds to first changing the orientation to incoming and then doubling the building blocks; the second line corresponds to first doubling the building blocks and then changing the orientation to incoming. The resulting orientations are both incoming, but they differ by a rotation of $e^{i\pi}$.}\label{fig: orientation and double}
\end{figure}

The components $\mc{P}_j$ are called the \textbf{complex building blocks} of the decomposition. Let us note a useful fact for later use.
If $\mc{P}_j$ is a complex building block with outgoing boundary components, it can be transformed into a building block  $(\mc{P}_j^{\rm in},
\bs{\zeta}_j^{\rm in})$ having only incoming boundaries by composing each outgoing boundary parametrisation as follows: for each outgoing half-circle $\zeta_{jk}^{\mathrm{h}}$, we set
$$\zeta_{jk}^{{\rm h},{\rm in}}:=\zeta_{jk}^{\rm h}\circ (-o) ,$$
and for each outgoing circle $\zeta_{jk}^{\ell}$, we set
$$\zeta_{jk}^{\ell,{\rm in}}:=\zeta_{jk}^{\ell}\circ o.$$
In the same way, one defines $((\mc{P}_j^{\#2})^{\rm in},(\bs{\zeta}_j^\#)^{\rm in})$ for the double $\mc{P}_j^{\#2}$, see Figure~\ref{fig: orientation and double}.
Then, the model form of $(\mc{P}_j^{\rm in},\bs{\zeta}_j^{\rm in})$ is related to the model form of $((\mc{P}_j^{\#2})^{\rm in},(\bs{\zeta}_j^\#)^{\rm in})$ as follows (recall that $\tau(z)=\bar{z}$): 
\begin{align*} 
\textrm{(A2)} \,\,\quad  & (\mc{P}_j^{\rm in},\bs{\zeta}_j^{\rm in}) \longrightarrow  (\hat{\H} \setminus \cup_{k=1}^3\psi_{jk}^{\rm h}(\D^+), (\psi_{j1}^{\rm h}|_{\T^+},\psi_{j2}^{\rm h}|_{\T^+},\psi_{j3}^{\rm h}|_{\T^+}))\\
&  (\mc{P}_j^{\#2})^{\rm in},(\bs{\zeta}_j^\#)^{\rm in}) \longrightarrow    (\hat{\C} \setminus \cup_{k=1}^3\psi_{jk}^{\rm h}(\D), (\psi_{j1}^{\rm h}\circ e^{{\rm i}\pi(1-\sigma_{j1}^{\rm h})/2}|_{\T},\psi_{j2}^{\rm h}\circ e^{{\rm i}\pi(1-\sigma_{j2}^{\rm h})/2}|_{\T},\psi_{j3}^{\rm h}\circ e^{{\rm i}\pi(1-\sigma_{j3}^{\rm h})/2}|_{\T})\\
\textrm{(A1)} \,\,\quad  &  (\mc{P}_j^{\rm in},\bs{\zeta}_j^{\rm in}) \longrightarrow (\hat{\H} \setminus ( \psi_{j1}^{\ell}(\D)\cup \psi_{j2}^{\rm h}(\D^+)), (\psi_{j1}^{\ell}|_{\T}, \psi_{j2}^{\rm h}|_{\T^+}))\\
&  (\mc{P}_j^{\#2})^{\rm in},(\bs{\zeta}_j^\#)^{\rm in}) \longrightarrow  (\hat{\C} \setminus (\psi_{j1}^{\ell}(\D)\cup  \tau(\psi_{j1}(\D))\cup \psi_{j2}^{\rm h}(\D)), (\psi_{j1}^{\ell}|_{\T},  \tau\circ \psi_{j1}^\ell\circ o|_{\T}, \psi_{j2}^{\rm h}\circ e^{{\rm i}\pi(1-\sigma_{j2}^{\rm h})/2}|_{\T} )).
\end{align*}
Analogous relations hold for the models (B1), (B2), (C1), (C2) and for their corresponding Neumann doubles.

\subsection{Structure constants} \label{sub:structure}
In this section, we introduce the structure constants of the boundary Liouville CFT, which consist of the three-point function on the sphere together with the disk correlation functions depicted in Figure \ref{pic:boundarystructure}.
As discussed above, these quantities correspond to the correlation functions associated with the irreducible surfaces.

For convenience, we shall express them on the surface $\Sigma$ given by the extended upper half-plane $\hat\H$, endowed with the specific metric $g_{\rm dozz}=|z|^{-4}_+|\dd z|^2,$ which we refer to as the DOZZ metric. By the Weyl anomaly relation \eqref{confan3}, their expressions on any domain conformally equivalent to the unit disk, equipped with an arbitrary conformal metric, can then be recovered. \\

\noindent (1) {\bf Sphere with $3$ points.}
Assume the cosmological constant $\mu$ is positive and let $\alpha_1,\alpha_2,\alpha_3$ satisfy $\alpha_1+\alpha_2+\alpha_3>2Q$ and $\alpha_i<Q$. Then one has 
\begin{align}\label{dozz}
	& \left\langle V_{\alpha_1}\left(x_1\right) V_{\alpha_2}\left(x_2\right) V_{\alpha_3}\left(x_3\right)\right\rangle_{\hat{\mathbb{C}}, g} \nonumber\\
	& =\left|x_1-x_3\right|^{2\left(\Delta_{\alpha_2}-\Delta_{\alpha_1}-\Delta_{\alpha_3}\right)}\left|x_2-x_3\right|^{2\left(\Delta_{\alpha_1}-\Delta_{\alpha_2}-\Delta_{\alpha_3}\right)}\left|x_1-x_2\right|^{2\left(\Delta_{\alpha_3}-\Delta_{\alpha_1}-\Delta_{\alpha_2}\right)}\left(\prod_{i=1}^3 g\left(x_i\right)^{-\Delta_{\alpha_i}}\right) \nonumber\\
	&\quad  \times\frac{1}{2} {\rm C}_{\gamma, \mu}^{\mathrm{DOZZ}}\left(\alpha_1, \alpha_2, \alpha_3\right)\left(\frac{\mathrm{v}_g(\hat{\mathbb{C}})}{\operatorname{det}^{\prime}\left(\Delta_g\right)}\right)^{\frac{1}{2}} e^{-6 Q^2 S_{\mathrm{L}}^0(\hat{\mathbb{C}}, g, \omega)}   
\end{align}
with $\omega: \hat{\mathbb{C}} \rightarrow \mathbb{R}$ defined by $ e^\omega g=g_{\rm dozz}$, with $g$ smooth metric on $\hat{\mathbb{C}}$, and ${\rm C}_{\gamma, \mu}^{\mathrm{DOZZ}}\left(\alpha_1, \alpha_2, \alpha_3\right)$ is the DOZZ function given in Appendix \ref{app:DOZZ}. This result was proven in \cite{KRV_DOZZ}.

\noindent (2) {\bf Disk with $1$ boundary point and $1$ bulk point.}
Assume that the boundary cosmological constant $\mu^{\rm b}$ is constant and let $\alpha,\beta\in\R$ satisfy the Seiberg bounds   $\alpha,\beta<Q$ and $\alpha+\beta/2>Q$. Let $\omega:\hat\H\to \R$ be piecewise smooth and such that $g= e^{-\omega} g_{\rm dozz}=g(z)|\dd z|^2$ is smooth and $k_g=0$ on $\R$. For $x\in\H$ and $s\in\R$, we have
\begin{align}\label{G2}
	\langle V_{\alpha,g}(x)  V_{\beta,g}(s)  \rangle_{\hat\H,g}=
	g(x)^{-\Delta_{\alpha}}g(s)^{-\frac{\Delta_{\beta}}{2}}\big(\frac{{\rm v}_{g}(\hat\H)}{{\det}'(\Delta_{g,{\rm N}})}\big)^{\frac{1}{2}} e^{-6Q^2 S_{\rm L}^0(\H,g,\omega)}  \frac{\tilde {\rm G}_{\mu,\mu^{\rm b}}(\alpha,\beta)}{|x-\bar{x}|^{2\Delta_\alpha-\Delta_\beta}|x-s|^{2\Delta_\beta}}.
\end{align} 
Here $\tilde {\rm G}_{\mu,\mu^{\rm b}}(\alpha,\beta):=\mu^{\frac{2Q -2 \alpha - \beta}{2\gamma}}{\rm G}_{\mu,\mu^{\rm b}}(\alpha,\beta)$, and ${\rm G}_{\mu,\mu^{\rm b}}(\alpha,\beta)$ is independent of $g,x,s$ and admits a meromorphic extension to $\C^2$. Its expression in terms of special functions was derived in the physics literature~\cite{Hosomichi}, and is recalled    in Appendix \ref{app:defG}; a mathematical derivation appears    in \cite{ang2024derivationstructureconstantsboundary}.\\

\noindent (3) {\bf Disk with $3$ boundary points.}
The cosmological constant is piecewise constant  and depends on the insertions $s_1,s_2,s_3$ on $\R$.
Here we assume $s_1<s_2<s_3$ and the boundary cosmological constant 
\[\mu^{\rm b}=1_{(-\infty,s_1)\cup(s_3,\infty)}\mu_1+1_{(s_1,s_2)}\mu_2+1_{(s_2,s_3)}\mu_3.\] 
Consider $\beta_1,\beta_2,\beta_3\in\R$ satisfying the Seiberg bounds: $\beta_1,\beta_2,\beta_3<Q$ and $\beta_1+\beta_2+\beta_3>2Q$. Then 
\begin{align}\label{G3}
	\cjg  V_{\beta_1,g}(s_1)  V_{\beta_2,g }(s_2)   V_{\beta_3,g}(s_3)\cjd_ {\H, g}
	=&g(s_1)^{-\frac{\Delta_{\beta_1}}{2}}g(s_2)^{-\frac{\Delta_{\beta_2}}{2}}g(s_3)^{-\frac{\Delta_{\beta_3}}{2}}\big(\frac{{\rm v}_{g}(\hat{\H})}{{\det}'(\Delta_{\hat{\H},g,{\rm N}})}\big)^{\frac{1}{2}} e^{-6Q^2 S_{\rm L}^0(\H,g,\omega)}\nonumber \\
	&\frac{\widetilde{\rm H}_{\mu_1,\mu_2,\mu_3}^{(\beta_1,\beta_2,\beta_3)}}{|s_1-s_2|^{\Delta_{\beta_1}+\Delta_{\beta_2}-\Delta_{\beta_3}}|s_1-s_3|^{\Delta_{\beta_1}+\Delta_{\beta_3}-\Delta_{\beta_2}}|s_2-s_3|^{\Delta_{\beta_2}+\Delta_{\beta_3}-\Delta_{\beta_1}}}.
\end{align}
Here $\widetilde{\rm H}_{\mu_1,\mu_2,\mu_3}^{(\beta_1,\beta_2,\beta_3)}$ is independent of  $g,s_i$ ($i=1,2,3$) and admits a meromorphic extension to $\C^3$, whose exact expression is expanded in Appendix \ref{app:defH}. Its expression    was  derived in physics  by Ponsot-Teschner \cite{PT_boundary_3pt}, and this was proved in \cite{ang2024derivationstructureconstantsboundary}.

Finally, we include one last case of structure constant, which does not fall within the setting above, as its weight violates the Seiberg bounds. Consequently, the corresponding probabilistic expression does not converge and must be renormalised.\\
(4) {\bf Bulk 1 point structure constant and the FZZ formula.}
 The path integral \eqref{def:pathintegralFb} is divergent for the bulk 1-point function with weight $\alpha<Q$. It can however be defined via an analytic continuation argument, as proved in \cite{AngRemySun21_FZZ}:
\begin{align*}
\langle V_{\alpha,g}(x) \rangle_{\H,g}:= \big(\frac{{\rm v}_{g}(\H)}{{\det}'(\Delta_{g,{\rm N}})}\big)^{\frac{1}{2}}
\int_\R  \E\Big[  V_{\alpha,g}(x)e^{ -\frac{Q}{4\pi}\int_{\Sigma}K_{g}\phi \,{\rm dv}_{g} -\frac{Q}{2\pi}\int_{\partial \Sigma}k_{g}\phi{\rm d\ell}_g}\Big(e^{- \mu  M_{\gamma}^g(\phi,\Sigma)-M^{{\rm b},g}_{\gamma}(\phi,\mu^{\rm b})}-1\Big) \Big]\,\dd c .
\end{align*}
Then, the symmetries reduce this expression to
\begin{align}\label{probfzz}
\langle V_{\alpha}(x)   \rangle_{g,\H}=
g(x)^{-\Delta_{\alpha}}  \big(\frac{{\rm v}_{g}(\H)}{{\det}'(\Delta_{g,{\rm N}})}\big)^{\frac{1}{2}} e^{-6Q^2 S_{\rm L}^0(\H,g,\omega)}  \frac{\widetilde{{\rm U}}_{\mu,\mu^{\rm b}}(\alpha)}{|x-\bar{x}|^{2 \Delta_{\alpha}}}
\end{align} 
 where $\widetilde{{\rm U}}_{\mu,\mu^{\rm b}}(\alpha)$ is the bulk (or FZZ) structure constant. Its expression was derived in physics \cite{FZZ}, and proved in \cite{AngRemySun21_FZZ}, see Appendix \ref{app:FZZ}. 
For cases (2), (3), and (4), one can also take the bulk cosmological constant $\mu=0$. The only difference is that we must replace the structure constants $\mathrm{G}$, $\mathrm{H}$, and $\mathrm{U}$ by their $\mu=0$ counterparts; see~\cite{RemyZhu20} for explicit expressions.


\section{The Hamiltonians}\label{s:Hamiltonians_bulk} 
As explained in Segal \cite{Segal87}, the amplitudes of analytic annuli  encode the local conformal symmetries of the model, namely the Virasoro algebra.	By restricting to the annuli $\A_{e^{-t}}$ for $t>0$, a semigroup of operators is constructed in  \cite{GKRV20_bootstrap}. The generator of this semigroup is a selfadjoint operator on $\mathcal{H}$, called the Hamiltonian of Liouville CFT.  
A spectral resolution for this operator is proved in \cite{GKRV20_bootstrap},  providing a convenient orthogonal basis for the Hilbert space $\mc{H}$. This basis can be described in terms of the action of the Virasoro algebra on $\mc{H}$. The strategy developed in \cite{GKRV20_bootstrap,GKRV21_Segal} allows one to interpret the conformal bootstrap as a Plancherel formula with respect to this basis, by
expanding the pairing of Liouville amplitudes 
along boundary circles. Since this approach will also be used for Liouville theory with Neumann boundary, we recall below the main facts and results needed in this work. This will also serve as a guiding framework and as a tool for the analogous analysis of the Hamiltonian and the Virasoro representation in the boundary theory, which involves gluing Liouville amplitudes of surfaces with corners along half-circles.

\subsection{Free theory, no boundary: Hamiltonian and associated representation of Virasoro algebra in $\mc{H}$}\label{s:Hamiltonian_free_bulk} 

We now recall the main facts concerning the free Hamiltonian and its associated representation of the Virasoro algebra on $\mc{H}$.
These structures were originally introduced to study Liouville correlation functions on closed surfaces, by relating the free-field and Liouville CFTs through a scattering theory argument developed in \cite{GKRV20_bootstrap}.
For further details on the material summarized below, we refer the reader to Sections 4.1-4.4 of \cite{GKRV20_bootstrap}.
 
Let us denote by $\mathcal{S}(\Omega_\T)\subset L^2(  \Omega_{\T},\P_{\T})$ the dense subspace 
of smooth functions depending on finitely many coordinates, i.e. of the form
	$F(x^\ell_1,y^\ell_1, \dots,x_n^\ell ,y^\ell_n)$  for some $n\geq 1$ and $F\in C^\infty((\R^2)^n)$, with at most polynomial growth at infinity for $F$ and its derivatives.
Then consider the space of smooth functions on $\mc{H}$ of the form
\begin{equation}\label{smoothexpgrowth}
\begin{gathered}
\mathcal{C}_c(\R\times \Omega_\T):=\mathrm{Span}\{ \psi(c)F\,|\,\psi\in C_c^\infty(\R)\text{ and }F\in\mathcal{S}(\Omega_\T) \}, \\
\mathcal{C}_\infty(\R\times \Omega_\T):=\mathrm{Span}\{ \psi(c)F\,|\,\psi\in C^\infty(\R)\text{ and }F\in\mathcal{S}(\Omega_\T)\}.
\end{gathered}
\end{equation}
On $\mathcal{S}(\Omega_\T), \mc{C}_c(\R\times \Omega_\T), \mc{C}_\infty(\R\times \Omega_\T)$ we consider  the  operator ${\bf P}$ defined by 
\begin{align}\label{hdefi}
{\bf P}=\sum_{n=1}^\infty n(\partial_{x^{\ell}_n}^\ast \partial_{x^{\ell}_n}+\pl_{y^{\ell}_n}^*\pl_{y^{\ell}_n}).
\end{align}
	where $\pl_{x^{\ell}}^*,\pl_{y^{\ell}_n}^*$ are the $L^2(\Omega_{\T},\P_{\T})$ (formal) adjoint of $\pl_{x^{\ell}_n},\pl_{y^{\ell}_n}$ (only finite number of terms in the sum contributes when acting on $\mathcal{S}(\Omega_\T)$). This operator extends uniquely to an unbounded 
	self-adjoint positive operator on $L^2(\Omega_{\T},\P_{\T})$, with discrete spectrum given by 
	$\sigma({\bf P})=\{|{\bf k}|+|{\bf l}|\,|\, {\bf k},{\bf l}\in \mc{N}\}$ with multiplicity, where $\mc{N}$ is the set  of sequences ${\bf k}=(k_n)_{n\in \N}$ with values in $\N$ with only finitely many non-zero $k_n$ and 
	$|{\bf k}|=\sum_{n=1}^\infty nk_n$.  The free (bulk) Hamiltonian is defined by the expression, on $\mc{C}_c(\R\times \Omega_\T)$,
\begin{equation}\label{free_bulk_Hamiltonian}
\qquad {\bf H}^0:=-\frac{1}{2}\pl_c^2+\frac{Q^2}{2}+{\bf P}  \qquad \textbf{(free Hamiltonian)}.
\end{equation}
It is an unbounded operator on $\mc{H}$ which admits a self-adjoint extension with only continuous spectrum given by $[Q^2/2,\infty)$. By \cite[Section 4.4]{GKRV20_bootstrap},  there are two families of operators acting on $\mc{C}_c(\R\times \Omega_\T)$, denoted by $({\bf L}_n^0)_{n\in \Z}$ and $(\widetilde{{\bf L}}_n^0)_{n\in \Z}$, satisfying the commutation relations $ [{\bf L}^{0}_n, \widetilde{{\bf L}}^{0}_m]=0$ and 
\begin{equation}\label{Virasoro_free} 
[{\bf L}^{0}_n,{\bf L}^{0}_m]= (n-m){\bf L}^{0}_{n+m}+\frac{c_{\rm L}}{12}\delta_{n,-m}(n^3-n), \quad 
[\widetilde{{\bf L}}^{0}_n,\widetilde{{\bf L}}^{0}_m]= (n-m)\widetilde{{\bf L}}^{0}_{n+m}+\frac{(1+6Q^2)}{12}\delta_{n,-m}(n^3-n).
\end{equation}
Here the central charge is $c_{\rm L}:=1+6Q^2$, and   $({\bf L}^{0}_n)^*={\bf L}^{0}_{-n}$ and 
$(\widetilde{{\bf L}}^{0}_n)^*=\widetilde{{\bf L}}^{0}_{-n}$ where the adjoint denotes the formal adjoint for the scalar product on 
$\mc{H}$. These operators have explicit expressions, which are not needed here, see \cite[Section 4.4.]{GKRV20_bootstrap} for details.
Moreover, the free (bulk) Hamiltonian is given by 
\[{\bf H}^0={\bf L}_0^0+\tilde{{\bf L}}_0^0\] 
and its associated quadratic form is denoted
\begin{equation}\label{defQ^0}
\mc{Q}^0(u)=\frac{1}{2}\|\pl_cu\|^2_{\mc{H}}+\frac{Q^2}{2}\|u\|^2_{\mc{H}}+\sum_{n\in \N}n(\|\pl_{x^{\ell}_n}u\|_{\mc{H}}^2+\|\pl_{y^{\ell}_n}u\|_{\mc{H}}^2)
\end{equation}
with domain $\mc{D}(\mc{Q}^0)=\{u\in \mc{H}\,|\, \mc{Q}^0(u)<\infty\}$.   It is closable and we write $(\mc{Q}^0,\mc{D}(\mc{Q}^0))$ for its closure. Let  $\mc{D}'(\mc{Q}^0)$ be the dual space.
Note that $\bH^0:\mc{D}(\mc{Q}^0)\to \mc{D}'(\mc{Q}^0)$ is bounded.
By \cite[Lemma 2.1]{BGKRV}, the operators ${\bf L}^{0}_n, \widetilde{{\bf L}}^{0}_n$ are bounded as maps $\mc{D}(\mc{Q}^0)\to \mc{D}'(\mc{Q}^0)$ with norm 
\begin{equation}\label{normL_n0}
\|{\bf L}_n^0\|_{\mc{D}(\mc{Q}^0)\to \mc{D}'(\mc{Q}^0)}\leq C(1+n)^{3/2},\quad \|\tilde{{\bf L}}_n^0\|_{\mc{D}(\mc{Q}^0)\to \mc{D}'(\mc{Q}^0)}\leq C(1+n)^{3/2}
\end{equation}
for some $C$ independent of $n$. They also act  on $\mc{C}_\infty(\R\times \Omega_\T)$.
For $\alpha\in \C$, we define (recall \eqref{GFFcircle0})
\begin{equation}\label{psialphadef}
\Psi^0_\alpha(\tilde\varphi^\ell):=e^{(\alpha-Q)c}\in \mathcal{C}_\infty(\R\times \Omega_\T), \qquad \textbf{(free primary states)}.
\end{equation}
For $\alpha\in \C$, these maps are eigenstates of ${\bf H}^0$, although     they do not belong to $\mc{H}$ itself but rather to certain weighted spaces of the form $e^{\beta |c|}\mc{H}$ for $\beta>|{\rm Re}(\alpha)-Q|$. Moreover, they are simultaneous eigenstates of both ${\bf L}_0^0$ and $\widetilde{{\bf L}}_0^0$
\begin{equation}\label{L0psialpha}
{\bf L}_0^0\Psi^0_\alpha=\widetilde{{\bf L}}_0^0\Psi^0_\alpha=\Delta_{\alpha}
\Psi^0_\alpha,  \qquad {\bf L}_n^0\Psi^0_\alpha=\widetilde{{\bf L}}_n^0\Psi^0_\alpha=0,\ \ \ n>0,
\end{equation}
where  $\Delta_\alpha=\frac{\alpha}{2}(Q-\frac{\alpha}{2})$ is the conformal weight. 
	The generators ${\bf L}_n^0,\widetilde{{\bf L}}_n^0$ generate two commuting representations of the Virasoro algebra as follows. Let $\mc{T}$ be the set of Young diagrams, that is, the set of  integer sequences $\nu$  such that $\nu(k)\geq \nu(k+1)$ for all $k$ and $\nu(k)=0$ for $k$ large enough. For a Young diagram $\nu$ we denote its length by $|\nu|=\sum_k\nu(k)$ and its size by $s(\nu)=\max\{k\,|\,\nu(k)\not=0\}$. Given two Young diagrams $\nu= (\nu(i))_{i \in [1,k]}$ and $\tilde{\nu}= (\tilde{\nu}(i))_{i \in [1,j]}$, we define the operators 
	$$ {\bf L}_{-\nu}^0={\bf L}_{-\nu(k)}^0 \dots \, {\bf L}_{-\nu(1)}^0,\qquad \tilde{{\bf L}}_{-\tilde \nu}^0=\tilde{{\bf L}}_{-\tilde\nu(j)}^0 \dots\, \tilde{{\bf L}}_{-\tilde\nu(1)}^0,$$ and the free (bulk) descendant states
\begin{align}\label{psibasis}
\qquad  \Psi^0_{\alpha,\nu, \tilde\nu}={\bf L}_{-\nu}^0\tilde{{\bf L}}_{-\tilde\nu}^0  \: \Psi^0_\alpha  \qquad \textbf{(free descendant states)}\end{align}
with the convention that $\Psi^0_{\alpha,\emptyset, \emptyset}=\Psi^0_{\alpha}$. We recall now their main properties (see \cite[Proposition 4.9]{GKRV20_bootstrap}).
	
\begin{proposition}\label{prop:mainvir0}  
{\bf (1) } For each pair of Young diagrams $\nu,\tilde{\nu}\in \mathcal{T}$,  $\Psi^0_{\alpha,\nu, \tilde\nu}$ can be written as   	
$\Psi^0_{\alpha,\nu, \tilde\nu}=\mathcal{Q}_{\alpha,\nu,\tilde \nu}\Psi^0_\alpha$ where $\mathcal{Q}_{\alpha,\nu,\tilde\nu}\in \ker ({\bf P}- |\nu|-|\tilde\nu|)$ satisfies
\begin{equation}\label{scapo}  
\langle \mathcal{Q}_{2Q-\bar\alpha,\nu,\tilde\nu}, \mathcal{Q}_{\alpha,\nu',\tilde\nu'} \rangle_{L^2(\Omega_\T)}=\delta_{|\nu| ,|\nu'|}\delta_{|\tilde\nu| ,|\tilde\nu'|}F_{\alpha}(\nu,\nu')F_{\alpha}(\tilde\nu,\tilde\nu')
\end{equation} 
for some coefficients $F_{\alpha}(\nu,\nu')$, polynomial in  ${\alpha}$, called the {\it Schapovalov form}.  The functions $(\mathcal{Q}_{\alpha,\nu,\tilde \nu})_{\nu,\tilde\nu\in\mathcal{T}}$ are linearly independent for $\alpha\notin \{{\alpha_{r,s}} \mid \, r,s\in \N^\ast ,rs \leq \max(|\nu|, |\tilde \nu |)\}$ with ${\alpha_{r,s}}=Q-r\frac{\gamma}{2}-s\frac{2}{\gamma}$.
\\
{\bf (2) }  For all $\alpha \in \C$		
\[
{\bf L}_0^0\Psi^0_{\alpha,\nu, \tilde\nu} = (\Delta_{\alpha}
+|\nu|)\Psi^0_{\alpha,\nu ,\tilde\nu},\ \ \  \tilde{{\bf L}}_0^0\Psi^0_{\alpha,\nu, \tilde\nu} = (\Delta_{\alpha}+|\tilde\nu|)\Psi^0_{\alpha,\nu ,\tilde\nu},\quad
{\bf H}^0\Psi^0_{\alpha,\nu,\tilde{\nu}} = (2 \Delta_\alpha+|\nu|+|\tilde{\nu}| )\Psi^0_{\alpha,\nu,\tilde{\nu}}  .
\]
{\bf (3) } If $u_1,u_2\in \mc{H}$ then
\begin{align}\label{fcomplete}
\langle u_1, u_2\rangle_{\mc{H}}=\frac{1}{2 \pi}\sum_{\nu,\tilde\nu,\nu',\tilde\nu'\in \mc{T}}\int_\R \langle u_1,
\Psi^0_{Q+iP,\nu',\tilde{\nu}'} \rangle_{\mc{H}} \langle \Psi^0_{Q+iP,\nu,\tilde{\nu}}, u_2\rangle _{\mc{H}}F^{-1}_{Q+iP}(\nu,\nu')F^{-1}_{Q+iP}(\tilde\nu,\tilde\nu')\, \dd P.
\end{align}
where $F^{-1}_\alpha$, on the set $\mc{T}_n$ of Young diagrams of size $n$, 
denotes the inverse of the matrix $(F_{\alpha}(\nu,\nu'))_{\nu,\nu'\in \mc{T}_n}$.
\end{proposition}

\subsection{Free theory with boundary: Hamiltonian and representation of Virasoro algebra in $\mc{H}^+$}\label{s:semigroup_half}

The boundary Liouville Hamiltonian and the boundary Virasoro algebra are defined in a way analogous to their bulk counterparts. They are associated with cutting a surface with Neumann boundary components along half-circles whose endpoints lie on the Neumann boundary, or equivalently, with gluing surfaces with corners along D-marked half-circles. We begin by discussing the case of the free field.

As we will see, the free boundary theory can be compared with the free
bulk theory through an isometric embedding $\mathbf U:\mc{H}^+\to\mc{H}$
which rescales the zero mode. This embedding is defined below.
Let  $\mc{S}(\Omega_{\T^+})\subset L^2(\Omega_{\T^+})$ be the space of  functions of the form $F(x^{\rm h}_1,\dots, x^{\rm h}_N)$ for some $N \geq 1$ where $F\in C^\infty(\R^N)$ with at most polynomial growth at infinity for $F$ and its derivatives. Obviously $\mc{S}(\Omega_{\T^+})$ is  dense in $L^2(\Omega_{\T^+},\P_{\T^+})$. Next consider
\begin{equation}\label{core}
\begin{gathered}
\mc{C}_c(\R\times \Omega_{\T^+})=\mathrm{Span}\{ \psi(c)F\,|\,\psi\in C_c^\infty(\R)\text{ and }F\in \mc{S}(\Omega_{\T^+})\}\subset \mc{H}^+,\\
\mc{C}_{\infty}(\R\times \Omega_{\T^+})=\mathrm{Span}\{ \psi(c)F\,|\,\psi\in C^\infty(\R)\text{ and }F\in \mc{S}(\Omega_{\T^+})\}	
\end{gathered}
\end{equation}
On $\mc{S}(\Omega_{\T^+})$ and $\mc{C}_c(\R\times \Omega_{\T^+})$ we consider  the  operator ${\bf P}^+$ defined by 
	\begin{align}\label{hdefi+}
		{\bf P}^+:=\sum_{n=1}^\infty n\partial_{x^{\rm h}_n}^\ast \partial_{x^{\rm h}_n}
	\end{align}
	where $\pl_{x^{\rm h}_n}^*$ is the $L^2(\Omega_{\T^+},\P_{\T^+})$ (formal) adjoint of $\pl_{x^{\rm h}_n}$ (only finite number of terms in the sum contributes when acting on $ \mc{S}(\Omega_{\T^+})$). This operator extends uniquely to an unbounded 
	self-adjoint positive operator on $L^2(\Omega_{\T^+},\P_{\T^+})$, with discrete spectrum given by $\{|{\bf k}|=\sum_{n=1}^\infty nk_n\,|\, {\bf k}\in \mc{N}\}$ with multiplicity.  We define the operators  on $\mc{C}_c(\R\times \Omega_{\T^+})$ (with $\partial_{\varphi^{\rm h}_n}:=\sqrt{2n}\pl_{x_n^{\rm h}}$)
\[{\bf A}^+_0=i(\partial_c+\frac{Q}{2}),\quad  \forall n\geq 1, \quad {\bf A}^+_n=\frac{i}{2}\partial_{\varphi^{\rm h}_n}, \quad  {\bf A}^+_{-n}=\frac{i}{2}(\partial_{\varphi^{\rm h}_n}-2n\varphi^{\rm h}_n).\]
The normal ordered product on $\mc{C}_c(\R\times \Omega_{\T^+})$ of such operators is defined as usual by
	$:\!{\bf A}^+_n{\bf A}^+_m\!\!:\,={\bf A}^+_n{\bf A}^+_m$ if $m>0$ and ${\bf A}^+_m{\bf A}^+_n$ if $n>0$ (i.e. annihilation operators are on the right).
We next define the free boundary Virasoro generators, acting on $\mc{C}_c(\R\times \Omega_{\T^+})$ and on $\mc{C}_\infty(\R\times \Omega_{\T^+})$,  for any $n\in \Z$ by 
	\begin{equation}\label{Ln^+0}
		{\bf L}^{+,0}_n:=-i(n+1)Q{\bf A}^+_n+\sum_{m\in\mathbb{Z}}:{\bf A}^+_{n-m}{\bf A}^+_{m}: .
	\end{equation}
They obey the commutation relations of the Virasoro algebra
\begin{equation}\label{Virasoro _commutation} 
[{\bf L}^{+,0}_n,{\bf L}^{+,0}_m]= (n-m){\bf L}^{+,0}_{n+m}+\frac{(1+6Q^2)}{12}\delta_{n,-m}(n^3-n).
\end{equation}
The free  boundary Hamiltonian  studied in \cite[Proposition 7.2]{GRW1} is the operator acting on $\mc{C}_c(\R\times \Omega_{\T^+})$
\[\qquad {\bf H}^{+,0}:={\bf L}^{+,0}_0=-\pl_c^2+\frac{Q^2}{4}+{\bf P}^+\qquad \textbf{(Boundary free Hamiltonian)}.\]
Define the quadratic form 
\[ \mc{Q}^{+,0}(u)= \|\pl_cu\|^2_{\mc{H}^+}+\frac{Q^2}{4}\|u\|^2_{\mc{H}^+}+\sum_{n=0}^\infty n\| \pl_{x^{\rm h}_n}u\|^2_{\mc{H}^+}\]
with domain $\mc{D}(\mc{Q}^{+,0})=\{u \in \mc{H}^+\,|\, \mc{Q}^{+,0}(u)<\infty\}$. It is closable and we call $\mc{Q}^{+,0}$ its closure. Note that $\mc{Q}^{+,0}$ are all comparable for different values of $Q>0$. We denote by $\mc{D}'(\mc{Q}^{+,0})$ the topological dual space of $\mc{D}(\mc{Q}^{+,0})$. 

The operator ${\bf H}^{+,0}:\mc{D}(\mc{Q}^{+,0})\to \mc{D}'(\mc{Q}^{+,0})$ is bounded.
 We define the isometric embedding $\mathbf U:\mc{H}^+\to\mc{H}$ by
$$
(\mathbf UF)(t,(x_n^\ell,y_n^\ell)_{n\geq1})
:=2^{1/4}F(\sqrt2\,t,(x_n^\ell)_{n\geq1}).
$$
Some elementary algebra leads to the relation, for $F\in \mc{D}(\mc{Q}^{+,0})$
\begin{equation}\label{relationVir0}
\mathbf U^*
 \bigl({\bf L}_n^{0,Q/\sqrt2}
       +\widetilde{\bf L}_n^{0,Q/\sqrt2}\bigr)
 \mathbf U
={\bf L}_n^{+,0,Q}.
\end{equation}
where we have added the $Q$-dependence in the Virasoro generators. 
Using that ${\bf L}_n^{0,Q}:\mc{D}(\mc{Q}^0)\to \mc{D}'(\mc{Q}^0)$ are bounded  with norm $\mc{O}(1+|n|)^{3/2}$ for each $Q$ 
(from \eqref{normL_n0}), we deduce that ${\bf L}^{+,0}_n: \mc{D}(\mc{Q}^{+,0})\to \mc{D}'(\mc{Q}^{+,0})$ is bounded with norm 
\begin{equation}\label{normLn^+0} 
\|{\bf L}^{+,0}_n\|_{\mc{D}(\mc{Q}^{+,0})\to \mc{D}'(\mc{Q}^{+,0})}\leq C(1+|n|)^{3/2}
\end{equation}
for some uniform constant $C>0$. 

As before, we can reconstruct the whole spectrum of the free boundary Hamiltonian by means of the Virasoro generators. For this, and given a Young diagram $\nu= (\nu(i))_{i \in [1,k]}$,   we define the operators ${\bf L}_{-\nu}^{+,0}={\bf L}_{-\nu(k)}^{+,0} \dots \, {\bf L}_{-\nu(1)}^{+,0}$ on $\mc{C}_c(\R\times \Omega_{\T^+})$. For $\alpha\in \C$, we consider next the  states (recall \eqref{GFFcircle+})
	\begin{align}\label{psialphadef+}
		\Psi^{+,0}_\alpha(\tilde\varphi^{\rm h}):=e^{(\alpha-Q)\frac{c}{2}}\in \mc{C}_\infty(\R\times \Omega_{\T^+}) \qquad \textbf{(boundary free primary states)}.
	\end{align}
For $\alpha\in \C$, these are eigenstates of ${\bf H}^{+,0}$ but they are not in $\mc{H}^+$. They obey ${\bf L}_0^{+,0}\Psi^{+,0}_\alpha  =\Delta_{\alpha}\Psi^{+,0}_\alpha$ and ${\bf L}_n^{+,0}\Psi^{+,0}_\alpha =0 $ for $n>0$ with  $\Delta_\alpha=\frac{\alpha}{2}(Q-\frac{\alpha}{2})$  the conformal weight. As such, $\Psi^{+,0}_\alpha$ is a  {\it highest weight state}, with highest weight $\Delta_{\alpha}$, for the representation $({\bf L}_{n}^{+,0})_n$ of the Virasoro algebra.
Given a Young diagram $\nu$, we define the descendant state  
\begin{equation}\label{Psi^+0alphanu}
\Psi^{+,0}_{\alpha,\nu}={\bf L}_{-\nu}^{+,0} \: \Psi^{+,0}_\alpha, \qquad \textbf{(boundary free descendant states)}
\end{equation}
with the convention that $\Psi^{+,0}_{\alpha,\emptyset}=\Psi^{+,0}_{\alpha}$.

\medskip
  They obey basically the same properties as  Proposition \ref{prop:mainvir0}  (and the proof is the same, see \cite[Proposition 4.9]{GKRV20_bootstrap})
	\begin{proposition}\label{prop:mainvir+}  
		1) For each Young diagrams $\nu\in \mathcal{T}$,  the  descendant state $\Psi^{+,0}_{\alpha,\nu}$ can be written as  
		 $\Psi^{+,0}_{\alpha,\nu}=\mathcal{Q}^+_{\alpha,\nu}\Psi^{+,0}_\alpha$ for some polynomial  $\mathcal{Q}^+_{\alpha,\nu}  $  in the coefficients $(\varphi^{\rm h}_n)_n$.\\
		2)  for all $\alpha \in \C$, ${\bf H}^{+,0}\Psi^{+,0}_{\alpha,\nu} = ( \Delta_\alpha+|\nu| )\Psi^{+,0}_{\alpha,\nu}  $.\\
		3)   the inner products of the descendant states obey
		\begin{equation}\label{scapo+}  
			\langle \mathcal{Q}^+_{2Q-\bar\alpha,\nu}, \mathcal{Q}^+_{\alpha,\nu'} \rangle_{L^2(\Omega_{\T^+})}=\delta_{|\nu| ,|\nu'|}F_{\alpha}(\nu,\nu')
		\end{equation} 
		where     $F_{\alpha}(\nu,\nu')$  is   the {\it Schapovalov form}.  The functions $(\mathcal{Q}^+_{\alpha,\nu})_{\nu\in\mathcal{T}}$ are linearly independent for $$\alpha\notin\{{\alpha_{r,s}} \mid \,\,r,s\in \N^\ast ,rs \leq |\nu| \}\quad \quad\text{with }\quad {\alpha_{r,s}}=Q-r\frac{\gamma}{2}-s\frac{2}{\gamma}.$$
	\end{proposition}
	We could provide as well a spectral resolution for the free boundary Hamiltonian but this will not be used in this paper.
	
	\subsection{Liouville theory, no boundary: the Hamiltonian}\label{s:Hamiltonian_LCFT_bulk} 

The Liouville bulk Hamiltonian is defined and analysed in \cite[Section 6]{GKRV20_bootstrap}.  We recall here its main properties and spectral resolution, restricting ourselves to the aspects that will be needed for the boundary bootstrap.
The reader is referred to \cite{GKRV20_bootstrap} for a detailed exposition.

Consider the annulus $\A_{e^{-t}}$ equipped with the metric $g_\A=|dz|^2/|z|^2$ and the boundary parametrisation $\boldsymbol{\zeta}_t=(\zeta_1,\zeta_2)$ given by $\zeta_2(e^{i\theta})=e^{i\theta}$ and $\zeta_1(e^{i\theta})=e^{-t+i\theta}$ for $e^{i\theta}\in \T$.
It is shown in \cite[Section 6.1]{GKRV21_Segal} that there is an unbounded non-negative self-adjoint operator ${\bf H}$ on $\mc{H}$ 
such that, for ${\rm c}_{\rm L}=1+6Q^2$, 
\[\caA_{\A_{e^{-t}},g_\A,\boldsymbol{\zeta}_t} = \sqrt{2}\pi e^{\frac{{\bf c}_{\rm L}}{24}t} e^{-t{\bf H}}\]
where the amplitude $\caA_{\A_{e^{-t}},g_\A,\boldsymbol{\zeta}_t}$ is defined by \eqref{amplitude} (with, in this case, the Liouville field being $\phi_{g_\A}=X_{g_{\A},{\rm D}}+P\boldsymbol{\varphi}^\ell$), and is viewed as the integral kernel of a bounded operator on $\mc{H}$. The generator ${\bf H}$ is called Liouville (bulk) Hamiltonian and takes the form 
\begin{equation}\label{Hamiltonian_bulk} 
\qquad {\bf H}={\bf H}^0+\mu e^{\gamma c}V \qquad \textbf{(Liouville Hamiltonian)}.
\end{equation}
Here $V$  is an unbounded non-negative symmetric operator on $L^2(\Omega_\T,\P_{\T})$. Although this will not be needed in the sequel, we recall from \cite[Section 6]{GKRV20_bootstrap} that when $\gamma<\sqrt{2}$, $V$ acts as multiplication by a positive function in $L^p(\Omega_{\T})$   for all $p<2/\gamma^2$, given by the GMC mass of $\T$, whereas $V$ ceases to be a multiplication operator when $\gamma\geq \sqrt{2}$.

The quadratic form $\mc{Q}$ associated to ${\bf H}$ is defined on the Hilbert subspace $\mc{D}(\mc{Q})=\{u\in \mc{H}\,|\, \mc{Q}(u)<\infty\}$ by 
\[ \mc{Q}(u)= \int_\R \Big( \frac{1}{2}\|\pl_cu\|^2_{L^2(\Omega_\T)}+\frac{Q^2}{2}\|u\|^2_{L^2(\Omega_\T)}+\sum_{n=0}^\infty n(\| \pl_{x^{\ell}_n}u\|^2_{L^2(\Omega_\T)} + \|\pl_{y^{\ell}_n}u\|^2_{L^2(\Omega_\T)})+\mu e^{\gamma c}\mc{Q}_V(u)\Big)\, \dd c\]
where $\mc{Q}_V$ is the quadratic form of $V$ on some Hilbert subspace $\mc{D}(\mc{Q}_{V})\subset L^2(\Omega_\T)$. The dual space   is denoted  $\mc{D}'(\mc{Q})$, and ${\bf H}:\mc{D}(\mc{Q})\to \mc{D}'(\mc{Q})$ is bounded.

\subsection{Liouville theory with boundary: the Hamiltonian}\label{s:Hamiltonian_LCFT_bdry}

The Liouville Hamiltonian for the theory with Neumann boundary (hereafter referred to as the boundary Hamiltonian) is defined and analysed in Part~I, \cite[Section~7.2]{GRW1}, of this work.
To keep the presentation concise, we define it directly in terms of the half-annulus amplitude (see \cite[Proposition~7.4]{GRW1}).

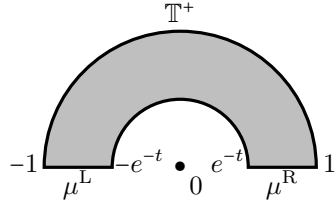
\begin{figure}[h!] 
		\begin{tikzpicture}[scale=0.9]
		  \path[fill=lightgray,draw=black,very thick] (1,0) arc[start angle=0,end angle=180,radius=1] -- (-1, 0) -- (-2, 0) arc[start angle=-180,end angle=-360,radius=2] -- (2,0) -- cycle;
			\draw (-1.5,-0.6) node[above]{$\mu^{\rm L}$} ;
			\draw (1.5,-0.6) node[above]{$\mu^{\rm R}$} ;
			\draw (-2.3,-0.2) node[above]{$-1$} ;  
			\draw (-0.6,-0.2) node[above]{$-e^{-t}$} ;
			\draw (2.2,-0.2) node[above]{$1$} ;  
			\draw (0.7,-0.2) node[above]{$e^{-t}$} ;
			\draw (0,0) node[below right] {$0$} node{$\bullet$};
			\draw (0,2) node[above]{$\T^+$} ;
				\end{tikzpicture}
		\caption{Half-annulus $\A^+_{e^{-t}}$.}\label{demianneau}
		\end{figure}

 Consider the half-annulus $\A_{e^{-t}}^+$ equipped with the metric $g_\A=|\dd z|^2/|z|^2$.  Its Dirichlet boundary is given by
$(\pl \A_{e^{-t}}^+)_{\rm D}=\T^+ \cup e^{-t}\T^+$, with  boundary parametrisation $\boldsymbol{\zeta}^+_t=(\zeta^+_1,\zeta^+_2)$ given by $\zeta^+_2(e^{i\theta})=e^{i\theta}$ and $\zeta^+_1(e^{i\theta})=e^{-t+i\theta}$ for $e^{i\theta}\in \T^+$, and its Neumann boundary by $(\pl \A_{e^{-t}}^+)_{\rm N}= \A_{e^{-t}}^+\cap \R$ (see Figure \ref{demianneau}). 

Consider $\boldsymbol{\mu}=(\mu,\mu_{\rm B})$ with $\mu,\mu_{\rm B}\in [0,\infty)$ satisfying $\mu+\mu_{\rm B}>0$, 
and $\gamma\in (0,2)$ such that 
\begin{equation}\label{assumptions_gamma_mu} 
\gamma\in (0,\sqrt{2})  \quad \textrm{ or } \quad \mu=0 \textrm{ and }\gamma\in (0,2)
\end{equation}
and the boundary cosmological constant is given by $\mu_{\rm B}=\mu^{\rm L} {\bf 1}_{[-1,-e^{-t}]}+\mu^{\rm R} {\bf 1}_{[e^{-t},1]}$ for some $\mu^{\rm L},\mu^{\rm R}\geq 0$.
Then under the conditions \eqref{assumptions_gamma_mu}, it is shown in \cite[Proposition 7.4]{GRW1} that there is an unbounded non-negative self-adjoint operator ${\bf H}^+$ on $\mc{H}^+$ 
such that 
\[\caA_{\A^+_{e^{-t}},g_\A,\boldsymbol{\zeta}_t^+} = (8\pi^3)^{\frac{1}{4}} e^{\frac{{\bf c}_{\rm L}}{24}t} e^{-t{\bf H}^{+}}\]
where the amplitude $\caA_{\A^+_{e^{-t}},g_\A,\boldsymbol{\zeta}_t^+}$ is defined by \eqref{amplitude} and is viewed as the integral kernel of a bounded operator on $\mc{H}^+$. By \cite[Section 7.2]{GRW1}, the boundary Hamiltonian has the form 
\begin{equation}\label{boundary_Hamiltonian} 
\begin{gathered}
{\bf H}^{+}={\bf H}^{+,0}+{\bf V}^{+,\bs{\mu}}, \quad \textbf{(boundary Liouville Hamiltonian)}.
\end{gathered}
\end{equation}
Here 
\[{\bf V}^{+,\bs{\mu}}:=\mu e^{\gamma c}V^{+,\cap} +\mu^{\rm L}e^{\frac{\gamma c}{2}}V^{+,{\rm L}}+\mu^{\rm R}e^{\frac{\gamma c}{2}}V^{+,{\rm R}}\]
for some unbounded symmetric non-negative operators $V^{+,\cap} ,V^{+, {\rm L}}, V^{+,{\rm R}}$ on $L^2(\Omega_{\T^+})$, defined as follows: 
\begin{itemize}
\item $V^{+,\cap}$ is the multiplication operator by the positive function in  $L^p(\Omega_{\T^+})$   (for $p<\frac{1}{4}+\sqrt{\frac{1}{16}+\frac{1}{\gamma^2}}$)
\begin{equation}\label{defV^+cap}
V^{+,\cap}(\varphi^{\rm h}):=  \int_{\T^+} e^{\gamma \varphi^{\rm h}(\theta)-\frac{\gamma^2}{2}\E[ \varphi^{\rm h}(\theta)^2]}\Big(\frac{1}{2|\sin(\theta)|}\Big)^{\frac{\gamma^2}{2}}\dd\theta
\end{equation}
\item  $V^{+, {\rm L}}$ and $V^{+, {\rm R}}$ are defined   as (unbounded) operators   on $L^2(\Omega_{\T^+})$ by their action on the dense subspace 
$\mc{S}(\Omega_{\T^+})$: if $u,v$ depend only on the   variables  $x_1,\dots,x_n$ for some $n>0$
\begin{equation}\label{girsuk}
\langle V^{+,{\rm R}}u,v \rangle_{\mc{H}^+}= \langle u_{\rm right},v_{\rm right} \rangle_{L^2(\Omega_{\T^+})} ,\qquad \langle V^{+,{\rm L}}u,v \rangle_{\mc{H}^+}= \langle u_{\rm left},v_{\rm left} \rangle_{L^2(\Omega_{\T^+})} 
\end{equation}
where the functions $u_{\rm right}$, $u_{\rm left}$ (and similarly for $v$) are defined by
\begin{align*}
u_{\rm right}(x_1, \dots,x_n ):=u \Big(x_1+\gamma  \sqrt{2}, \dots,x_n+\gamma  \sqrt{2/n}  \Big), \quad u_{\rm left}(x_1, \dots,x_n):=u \Big(x_1-\gamma  \sqrt{2}, \dots,x_n+\gamma (-1)^n \sqrt{2/n}  \Big).
\end{align*}
They extend to operators on $\mathcal{H}^+$ with a trivial action on the zero-mode variable $c$. We will sometimes use the following abuse of notation 
\begin{equation}\label{V^+Rdef/V^+Ldef}
V^{+, {\rm R}}(\varphi^{\rm h})=:e^{\frac{\gamma}{2} \varphi^{{\rm h}}(0)}\!:, \qquad V^{+, {\rm L}}(\varphi^{\rm h})=:e^{\frac{\gamma}{2} \varphi^{{\rm h}}(\pi)}\!: 
\end{equation}
since the action described in \eqref{girsuk} corresponds to applying the Girsanov transform to 
\[ \E[e^{\frac{\gamma}{2} \varphi^{{\rm h}}_\eps(0)-\frac{\gamma^2}{4}\E[ \varphi^{{\rm h}}_\eps(0)^2]} F( \varphi^{{\rm h}})G( \varphi^{{\rm h}})] \]
where $ \varphi^{{\rm h}}_\eps(\theta)$ is  the $\eps-$regularisation of $\varphi^{{\rm h}}_\eps$,  and then taking the limit $\eps \to 0$. 
 \end{itemize}
 
The operator ${\bf H}^+$ is associated to the quadratic form $\mc{Q}^+$  defined on a Hilbert subspace $\mc{D}(\mc{Q}^+)\subset \mc{H}^+$ by 
\[ \mc{Q}^+(u):= \int_\R \Big( \|\pl_cu\|^2_{L^2(\Omega_{\T^+})}+\frac{Q^2}{4}\|u\|^2_{L^2(\Omega_{\T^+})}+\sum_{n=0}^\infty n\| \pl_{x^{\rm h}_n}u\|^2_{L^2(\Omega_\T)}+\mu e^{\gamma c}\mc{Q}^{+,\cap}(u)+e^{\frac{\gamma}{2}c}(\mu^{\rm R}\mc{Q}^{+,{\rm R}}(u)+\mu^{\rm L}\mc{Q}^{+,{\rm L}}(u)\Big)\, \dd c.\]
Here $\mc{Q}^{+,\cap}(u)=\cjg V^{+,\cap}u,u\cjd_{L^2(\Omega_{\T^+})}$, and $\mc{Q}^{+,{\rm L}},\mc{Q}^{+,{\rm R}}$ are the quadratic forms associated with $V^{+,{\rm L}},V^{+,{\rm R}}$ (defined in \eqref{girsuk}) on some Hilbert subspaces 
$\mc{D}(\mc{Q}^{+,\cap})$, $\mc{D}(\mc{Q}^{+,{\rm L}})$ and $\mc{D}(\mc{Q}^{+,{\rm R}})$ of $L^2(\Omega_{\T^+})$. The space 
$\mc{D}(\mc{Q}^+)$ is the closure of $\mc{C}_c(\R\times \Omega_{\T^+})$ for the norm $\mc{Q}^+(u)^{1/2}$.
The dual space of $\mc{D}(\mc{Q}^+)$ is denoted by $\mc{D}'(\mc{Q}^+)$, and ${\bf H}^{+}:\mc{D}(\mc{Q}^+)\to \mc{D}'(\mc{Q}^+)$ is bounded. 

\section{Semi-group of annuli and half-annuli, Virasoro algebra}\label{sec:semigroups}
\subsection{Spaces of holomorphic maps}\label{sub:hol}
First, recall the spaces ${\rm Hol}_\eps (\D)$, ${\rm Hol}^+_\eps (\D)$, ${\rm BiHol}_\eps (\D)$ and ${\rm BiHol}^+_\eps (\D)$ introduced in \eqref{Holeps}. When $\eps=0$, we  shall simply omit the index $\eps$ in these notations, and write ${\rm Hol}(\D),{\rm Hol}^+ (\D)$, etc.
For $f\in{\rm Hol}_\eps (\D)$, the power series expansion 
\[f(z)=\sum_{n=-1}^\infty f_nz^{n+1}\]
allows us to define the countable family of seminorms (for $k\geq 0$) 
\begin{equation}\label{seminorm}
\|f\|^2_{\epsilon,k}=\sum_{n=-\infty}^\infty (1+\epsilon)^{2|n|+2}(1+|n|)^{2k}|f_n|^2. 
\end{equation}
We endow ${\rm Hol}_\eps(\D)$ with the family of seminorms $( \|\cdot\|^2_{\epsilon,k})_{k\geq 0}$ \footnote{Only coefficients $f_n$ with  $n\geq -1$ actually contribute, but we define the seminorms  in this way since this will later be used   for general holomorphic functions defined near $\T$.}, which turns it  into a Fr\'echet space. The subsets ${\rm Hol}^+_\eps (\D)$, ${\rm BiHol}_\eps (\D)$ and ${\rm BiHol}^+_\eps (\D)$ inherit the Fr\'echet topology from ${\rm Hol}_\eps(\D)$. 

Similarly, let ${\rm Hol}^{\bullet}_\epsilon(\D)$ (resp. ${\rm Hol}^{\bullet,+}_\epsilon(\D)$) denote the  closed subspace of ${\rm Hol}_\epsilon(\D)$ (resp. ${\rm Hol}_\epsilon^+(\D)$) consisting of functions $f$ such that $f(0)=0$. These spaces  are also  Fr\'echet when endowed with     the seminorms $( \|\cdot\|^2_{\epsilon,k})_{k\geq 0}$ (resp.  with the induced topology).
 
The Segal semigroup of annuli and its subgroup of half-annuli are respectively defined by 
\[ {\mc S}:=\{ f\in {\rm BiHol}^\bullet (\D)\,|\, f(\D)\subset \D^\circ\}, \quad \mc{S}^+:=\mc{S}\cap {\rm Hol}^+(\D) .\]	
Each $f\in \mc{S}$ determines an annulus $\A_f$ with boundary circles $\T$ and $f(\T)$, parametrised by $\bs{\zeta}_f=({\rm Id},f)$.

A representation $f\mapsto {\bf T}_f\in \mc{L}(\mc{H})$ of Segal's semigroup of annuli on the Hilbert space $\mc{H}$ was constructed in \cite{BGKR1}.  The "Lie algebra" of $\mc{S}$ is generated by the elements $({\bf L}_n)_{n\geq -1},(\tilde{{\bf L}}_n)_{n\geq -1}$ of two commuting Virasoro algebras with central charge $c_{\rm L}=1+6Q^2$,  obtained by differentiating $f\mapsto {\bf T}_f$. 

The operator ${\bf T}_f$ is closely related to the Liouville amplitude of $\A_f$ equipped with an admissible metric $g_f$, and the representation of $\mc{S}$ defined in terms of    amplitudes  is therefore projective. 
The negative generators $({\bf L}_n)_{n\leq -2},(\widetilde{{\bf L}}_n)_{n\leq -2}$ were obtained in \cite{BGKR1}  by differentiating the Liouville amplitude $\mc{A}_{\A_f,g_f,\bs{\zeta}_f}$ 
when $f$ is a   biholomorphic map  defined in a small annulus $\A_{\delta,\delta^{-1}}$  for $\delta<1$ close to $1$, satisfying  
 $f(\T)\subset \D^\circ$, and where $\A_f$  denotes the (compact) annulus in $\C$ bounded by $\T$ and $f(\T)$. 
 
 In the boundary Liouville theory, we shall define  a representation $f\in \mc{S}^+\mapsto {\bf T}^+_f\in \mc{L}(\mc{H}^+)$ using the boundary GMC measure, and apply the method of  \cite{BGKR1}  to generate one copy of the positive part of Virasoro algebra $(\bL^+_n)_{n\geq -1}$. The boundary Liouville amplitudes of half-annuli $\A^+_f=\A_f\cap \H$, with boundary parametrisation $\bs{\zeta}^+_f:=({\rm Id},f)$,  where $f$ is biholomorphic in a small annulus  $\A_{\delta,\delta^{-1}}$, will then be used  to generate the negative modes
$(\bL^+_n)_{n\leq -2}$.

Finally, we introduce the spaces
\begin{equation}\label{S_eps_S_+}
 \mc{S}_\epsilon^+:={\rm Hol}_\epsilon^\bullet (\D^+)\cap \mc{S}^+, \quad  
 \mc{S}^+_>:=\cup_{\eps\in (0,1)}\mc{S}_\epsilon^+=\{f\in \mc{S}^+\, |\, f_{|_\T}\text{ is analytic}\}.
\end{equation}
The space $  \mc{S}_\epsilon^+$, which is an open subset of ${\rm Hol}_\epsilon^{+,\bullet} (\D)$ (see \cite[Lemma 4.1]{BGKR1}), is equipped with the  induced  topology.

\subsection{Quantisation of the semi-group of annuli}
We call the semi-group of annuli the space $\mc{S}$ equipped with the composition.
For $f\in \mc{S}$, the (bulk) Liouville quantisation operator 
$${\bf T}_{f}:C^0_b(H^s(\T))\to L^\infty(H^s(\T),\mu_0)$$
  is defined in \cite{BGKR1}, for $\mu\geq 0$, by 
\begin{equation}\label{defTf} 
{\bf T}_f F(\widetilde{\varphi}^\ell ):=|f'(0)|^{\frac{Q^2}{2}} \mathbb{E}_{\varphi^\ell}\Big[F\big(\big(X_{\D}+P\tilde{\varphi}^\ell)\circ f+Q \log |\frac{f'}{f}|\big)|_{\mathbb{T}}\big)e^{-\mu \int_{\A_f}|x|^{-\gamma Q}M_\gamma(X_{\D}+P\tilde{\varphi}^\ell,\dd x)}\Big].
\end{equation}
where $M_\gamma=M_\gamma^{g_0}$ denotes the GMC measure defined with respect to the flat metric $g_0=|\dd z|^2$, and where $\tilde{\varphi}^\ell=c+\varphi^{\ell}$ as above.

When $\mu=0$, we   denote this operator by ${\bf T}_f^{0}$. It is shown in \cite[Proposition 3.6]{BGKR1} that  the restriction of ${\bf T}_f$   to $C^0_b(H^{s}(\T))\cap \mc{H}$ extends to a bounded operator  ${\bf T}_f: \mc{H}\to \mc{H}$ (including when $\mu=0$, i.e., for ${\bf T}_f^0$). Furthermore,  this operator satisfies the following composition property \cite[Proposition 3.3]{BGKRV}
\begin{equation}\label{composition}
\forall f_1,f_2\in \mc{S},\quad {\bf T}^0_{f_1\circ f_2}={\bf T}^0_{f_1}\circ {\bf T}^0_{f_2},\quad {\bf T}_{f_1\circ f_2}={\bf T}_{f_1}\circ {\bf T}_{f_2}.
\end{equation}

 There are particular $1$-parameter families  $t\mapsto f_t\in \mc{S}$ generated by the flow 
 of certain  holomorphic vector fields ${\rm v}=v(z)\pl_z$. 
A holomorphic vector field ${\rm v}=v(z)\pl_z$, defined in a neighborhood of the unit disk $\D$, is called 
\textbf{Markovian} if it satisfies  
 $$v(0)=0,\qquad {\rm Re}(\bar{z}v(z))<0.$$
 This condition guarantees that the flow $f_t:\D\to f_t(\D)$, defined by the differential equation 
 $$\pl_t f_t(z)=v(f_t(z)),\qquad f_0(z)=z$$ satisfies $f_t(\D)\subset f_{t'}(\D)$ for all $t>t'$ (see \cite[Lemma 2.2]{BGKRV}) with $f_t(0)=0$.
 
By \cite[Section 2.5 and 3.1]{BGKRV} (see also \cite{BGKR1}), there exist two families  $({\bf L}_n)_{n\in \Z}$ and $(\tilde{{\bf L}}_n)_{n\in \Z}$   of unbounded operators  on $\mc{H}$ satisfying the following properties:
\begin{description}
\item[(i) Boundedness] let $\rho\in C^\infty(\R)$ be a cutoff function such that   $\rho(c)=c$ for $c\leq -1$ and $\rho(c)=0$ for $c\geq 0$. For any $\beta\in \R$,  the operators 
${\bf L}_n,\tilde{\bf L}_n$ and ${\bf L}^0_n,\tilde{\bf L}^0_n$ are  bounded as maps 
\begin{equation}\label{LntildeLn} 
{\bf L}_n,\tilde{{\bf L}}_n: e^{\beta \rho}\mc{D}(\mc{Q})\to e^{\beta \rho}\mc{D}'(\mc{Q}) , \quad  {\bf L}_n^0, \widetilde{\bf L}^0_n: e^{\beta \rho}\mc{D}(\mc{Q})\to e^{\beta \rho}\mc{D}'(\mc{Q}),
\end{equation}
\item[(ii) Adjoint relations] for each $n\geq 0$, the following adjoint relations hold: 
$${\bf L}_n^*={\bf L}_{-n},\quad \tilde{{\bf L}}_n^*=\tilde{{\bf L}}_{-n},\quad ({\bf L}_n^0)^*={\bf L}^0_{-n},\quad (\tilde{{\bf L}}_n^0)*=\tilde{{\bf L}}^0_{-n}$$
in the sense that    for all $F,F'\in \mc{D}(\mc{Q})$ and $G,G'\in \mc{D}(\mc{Q}^0)$
 \begin{align*}
& \cjg {\bf L}_nF,F'\cjd_{\mc{H}}= \cjg F,{\bf L}_{-n}F'\cjd_{\mc{H}}, \quad \cjg \tilde{{\bf L}}_nF,F'\cjd_{\mc{H}}= \cjg F,\tilde{{\bf L}}_{-n}F'\cjd_{\mc{H}},\\
& \cjg {\bf L}^0_nG,G'\cjd_{\mc{H}}= \cjg G,{\bf L}^0_{-n}G'\cjd_{\mc{H}}, \quad \cjg \tilde{{\bf L}}^0_nG,G'\cjd_{\mc{H}}= \cjg G,\tilde{{\bf L}}^0_{-n}G'\cjd_{\mc{H}}.
\end{align*}
\item[(iii) Boundedness of $\bH_{\rm v}$]  if ${\rm v}=v(z)\pl_z$ with $v(z)=\sum_{n=0}^{\infty} v_n z^{n+1}\in {\rm Hol}_\epsilon^\bullet(\D)$, the following operators are bounded
\begin{equation}\label{eq:def_L}
\bH_{\rm v} :=-\sum_{n=0}^{\infty} (v_n {\bf L}_n+\bar{v}_n\tilde{{\bf L}}_n):\mc{D}(\mc{Q})\to \mc{D}'(\mc{Q}) ,\qquad \bH_{\rm v}^0 :=-\sum_{n=0}^{\infty} (v_n {\bf L}^0_n+\bar{v}_n\tilde{{\bf L}}^0_n):\mc{D}(\mc{Q}^0)\to \mc{D}'(\mc{Q}^0). 
\end{equation}
\item[(iv) Semigroup generation]  By \cite[Theorems 2.8 and 3.1]{BGKRV}, there is a constant $K>0$, independent of $v$, such that if 
$$v(z)=\sum_{n\geq 0} v_nz^{n+1}, \qquad \text{ with } {\rm Re}(v_0)<- K \sum_{n \geq 1}|v_n|n^2,$$
then  the quadratic forms of ${\bf H}_{\rm v}^0$ and ${\bf H}_{\rm v}$ are closable (with domains $\mc{D}(\mc{Q}^0)$ and $\mc{D}(\mc{Q})$)  and coercive. Consequently,  they generate two contraction semigroups on $\mc{H}$, with generators  ${\bf H}_{\rm v}^0$ and ${\bf H}_{\rm v}$,  satisfying
\begin{equation}\label{Tfsemigroup}
{\bf T}_{f_t}^0=e^{-t{\bf H}_{\rm v}^0} ,\quad {\bf T}_{f_t}=e^{-t{\bf H}_{\rm v}}.
\end{equation}
\end{description}

 \begin{remark}
The operators ${\bf L}_n$ and $\tilde{{\bf L}}_n$ provide a representation of the Virasoro algebra, but their unboundedness complicates the interpretation of their compositions and commutators. This will be addressed in subsequent discussion.
\end{remark}

\subsection{Semi-group of half-annuli for boundary Liouville theory and Virasoro generators}
We define a semigroup of operators associated with functions in the Segal semigroup $\mc{S}^+$, which we refer to as the semigroup of half-annuli.
To simplify the notation, we write $X_{\D^+}$ for the GFF on $\D^+$ with mixed boundary conditions (Dirichlet on $\T^+$ and Neumann on $[-1,1]$) and $X_\D$ for the GFF on $\D$ 
with Dirichlet boundary condition on $\T$, both defined with respect to the flat metric $g_0=|\dd z|^2$. 

The involution $\tau(z)=\bar{z}$ preserves $g_0$, which is Neumann extendible on $\D^+$. By abuse of notation, we also denote by $\tau$ the induced symmetry on the boundary    $\pl\D=\T$.  

 For $u \in H^s(\D^+)$ with $s\in (-1/2,0)$, we denote by $u^\#\in H^s(\D)$ the  reflected distribution  defined by 
 $$u^\#=u\text{ on }\D^+,\qquad   u^\# =u\circ \tau \text{ on } \tau(\D^+).$$
  Similarly, for $u\in H^s(\T^+)$, we define  $u^\#\in H^s(\T)$ by reflection across the real axis
  $$ u^\#=u\text{ on }\T^+,\qquad   u^\# =u\circ \tau \text{ on } \tau(\T^+).$$
 For instance $X_{\D^+}^\#$ has the same law as $2^{-1/2}(X_{\D}+X_{\D}\circ \tau)$ and its covariance is given by
 \begin{equation}\label{cov_X_Dsharp}
\E[X_{\D^+}^\#(z)X_{\D^+}^\#(z')]=\log\frac{|1-z\bar{z}'|\cdot |1-zz'|}{|z-z'|\cdot|z-\bar{z}'|}.
 \end{equation} 
Moreover, for $\tilde{\varphi}^{\rm h}\in H^s_{\rm even}(\T)$, the restriction of $P\tilde{\varphi}^{\rm h}$ to $\D^+$ coincides with the harmonic extension of  $\tilde{\varphi}^{\rm h}|_{\T^+}$ to $\D^+$ satisfying a Neumann boundary condition on $\D^+\cap \R$.   
  
 In what follows, we fix a Sobolev order $s\in (-1/2,0)$, so that our random fields take values  in $H^s(\T)$ (resp. $H^s_{\rm even}(\T)$), 
and we denote  by $C^0_b(H^s(\T))$ (resp. $C^0_b(H^s_{\rm even}(\T))$)   the space of bounded continuous functionals on  these spaces.

We are now ready to define the boundary Liouville quantisation of elements in 
$\mc{S}^+$, in a way entirely analogous to the bulk case.

\begin{definition}\label{quantTf}
Fix $\boldsymbol{\mu}:=(\mu, \mu^{\rm L},\mu^{\rm R})$ non-negative parameters. 
We define the boundary Liouville quantization of $f\in \mathcal{S}^+$ as the operator 
$${\bf T}^{+}_f:C_b^0\left(H^s_{\rm even}(\mathbb{T})\right) \rightarrow L^{\infty}\left(H^s_{\rm even}(\mathbb{T})\right) ,\qquad  s\in (-1/2,0)$$  
given by
\begin{equation}\label{FKv}
 {\bf T}^{+}_f F(\widetilde{\varphi}^{\rm h}):=|f^{\prime}(0)|^{\frac{Q^2}{4}} \mathbb{E}_{\varphi^{\rm h}}\Big[F\big(\big(X^\#\circ f+Q \log |\frac{f^{\prime}}{f}|\big)|_{\mathbb{T}}\big)e^{-\int_{\A^+_f}  M_\gamma^{\boldsymbol{\mu}}(X^\#,\dd x)}\Big]
 \end{equation}
where $\A^+_f=\D^+\setminus f(\D^+)$, $X^\#=X_{\D^+}^\#  + P\tilde{\varphi}^{\rm h}$ and  $ M^{\boldsymbol{\mu}}_\gamma(X^\#,\dd x)$ is the random measure on $\D^+$ defined by 
\[  M^{g,\boldsymbol{\mu}}_\gamma(X^\#,\dd x):= \mu |x|^{-\gamma Q}  M^{g}_\gamma (X^\#,\dd x) +
\mu^{\rm L}{\bf 1}_{x<0}  |x|^{-\frac{\gamma Q}{2}}M^{{\rm b},g}_\gamma (X^\#,\dd x) +\mu^{\rm R}{\bf 1}_{x>0}  |x|^{-\frac{\gamma Q}{2}}M^{{\rm b},g}_\gamma (X^\#,\dd x).
\]
When the metric is the flat one $g=g_0$,  we simply  write $M^{\boldsymbol{\mu}}_\gamma(X^\#,\dd x)$. Here, $\E_{\varphi^{\rm h}}$ denotes  expectation conditional on $\varphi^{\rm h}$. 

When  $\boldsymbol{\mu}:=(0,0,0)$,  we will write ${\bf T}^{+,0}_f$ instead of ${\bf T}^{+}_f$.
\end{definition}

A particular case of such an operator is the propagator $S^{+,\bs{\mu}}(t)$ studied in \cite[Section 7.2.3]{GRW1}, which  corresponds  to the contraction 
$f_t:z\mapsto e^{-t}z$,    the flow at time $t$ of the vector field $-z\pl_z$. It is given by 
\begin{equation}\label{propagator_half}
S^{+,\bs{\mu}}(t)F(\tilde{\varphi}^{\rm h})=e^{-\frac{Q^2}{4}t}\mathbb{E}_{\varphi^{\rm h}}\Big[F\big(X^\#(e^{-t}\cdot)|_{\mathbb{T}}\big)e^{-\int_{\A^+_{e^{-t}}}  M^{\boldsymbol{\mu}}_\gamma(X^\#,\dd x)}\Big]
\end{equation}
where $X^\#=X_{\D^+}^\#  + P\tilde{\varphi}^{\rm h}$ as above.

\begin{lemma}\label{contbyold}
For  $f\in \mathcal{S}^+$, the operator $ {\bf T}^{+}_f$ extends to a continuous operator on $\mc{H}^+$.
\end{lemma}

\begin{proof}
First we compare   ${\bf T}^{+,0}_{f}$ and ${\bf T}^{0}_{f}$. We  write temporarily ${\bf T}^{0}_{Q,f}$ and  ${\bf T}^{+,0}_{Q,f}$ for the operators ${\bf T}^{0}_{f}$ and ${\bf T}^{+,0}_{f}$, in order to keep track of the   parameter $Q$. For $\tilde{\varphi}^\ell \in H^s(\T)$ of the form
\eqref{GFFcircle0}, we will write $\tilde{\varphi}^{\rm h}$ and $\tilde{\varphi}^{\rm o}$ its even and odd parts. Consider \[\iota_{\mc{H}^+}:\mc{D}(\mc{Q}^{+,0})\to \mc{D}(\mc{Q}^0), \quad  (\iota_{\mc{H}^+}F)(c,(x_n^\ell,y_n^\ell)_n)=F(c,(x_n^\ell)_n)\]

 Then for   $F\in \mc{H}_+\cap L^\infty(H^s_{\rm even}(\T))$
\begin{align}\label{zizou}
({\bf T}_{Q,f}^{0}\iota_{\mc{H}_+}F)(\tilde{\varphi}^\ell)=& |f^{\prime}(0)|^{\frac{Q^2}{2}} \mathbb{E}_{\varphi^{\rm h}}\Big[F\big(\big((\tfrac{1}{2}(X_{\D}+X_{\D}\circ \tau)+P\tilde{\varphi}^{\rm h})\circ f+Q \log \Big|\frac{f^{\prime}}{f}\Big|\big)|_{\mathbb{T}}\big)\Big]\nonumber\\
=&  |f^{\prime}(0)|^{\frac{Q^2}{2}} \mathbb{E}_{\varphi^{\rm h}}\Big[F\big(\tfrac{1}{\sqrt{2}}((X_{\D^+}^\#+P(\sqrt{2}\tilde{\varphi}^{\rm h}))\circ f+\sqrt{2}Q \log \Big|\frac{f^{\prime}}{f}\Big|\big)|_{\mathbb{T}}\Big)\Big]\nonumber\\\
=&  ({\bf T}_{\sqrt{2}Q,f}^{+,0}(F(\tfrac{1}{\sqrt{2}}\cdot))(\sqrt{2}\tilde{\varphi}^{\rm h})
\end{align}
using the identity in law  $X_{\D^+}^\#\circ f= 2^{-1/2}(X_{\D}\circ f+X_{\D}\circ f\circ \tau)$. 

In particular, applying \eqref{zizou} with bulk background charge
$Q/\sqrt2$ and using the isometry $\mathbf U$ defined above, we obtain
\begin{equation}\label{zizouU}
{\bf T}_{Q/\sqrt2,f}^{0}\mathbf U
=\mathbf U{\bf T}_{Q,f}^{+,0}.
\end{equation}
Therefore, 
 we get (after a change of variables $c\to c/\sqrt{2}$ in the zero mode integration)
\begin{align*}
\int \big|({\bf T}_{Q,f}^{0}\iota_{\mc{H}_+}F(\sqrt{2}\cdot))(\tilde{\varphi}^\ell)\big|^2 \mu_0(\dd \tilde{\varphi}^\ell) = 2^{-1/2}\int | ({\bf T}_{\sqrt{2}Q,f}^{+,0}F)(\tilde{\varphi}^{\rm h})|^2\mu_0^+(\dd \tilde{\varphi}^{\rm h}) .
\end{align*}
Using the continuity of ${\bf T}_{Q,f}^{0}$ on $\mc{H}$, we deduce
\begin{align*}
\int \big|({\bf T}_{Q,f}^{0}\iota_{\mc{H}_+}F(\sqrt{2}\cdot))(\tilde{\varphi}^\ell)\big|^2 \mu_0(\dd \tilde{\varphi}^\ell) \leq C\int \big|(\iota_{\mc{H}_+}F(\sqrt{2}\cdot))(\tilde{\varphi}^\ell)\big|^2 \mu_0(\dd \tilde{\varphi}^\ell) =2^{-1/2}C\int |F(\tilde{\varphi}^{\rm h} )|^2 \mu_0^+(\dd \tilde{\varphi}^{\rm h}) .
\end{align*}
Combining, we complete the proof of the argument for ${\bf T}^{+,0}_f$. The case of  $ {\bf T}^{+}_f$ then follows by bounding the exponential of the potential by $1$.
\end{proof}

Next, we check that $f\mapsto {\bf T}_f^+$ forms a semi-group. 
\begin{proposition}\label{prop:semigroup}
The map $f\in \mathcal{S}^+\mapsto {\bf T}^{+}_f$ defines a semigroup of operators, in the sense that  
$$\forall f_1,f_2\in  \mathcal{S}^+,\qquad {\bf T}^{+}_{f_1}\circ {\bf T}^{+}_{f_2}={\bf T}^{+}_{f_1\circ f_2}.$$
\end{proposition}

\begin{proof} 
First we focus on ${\bf T}^{+,0}_f={\bf T}^{+,0}_{Q,f}$. By \cite[Proposition 3.3]{BGKRV}, we have 
${\bf T}^0_{Q',f_1}\circ {\bf T}^0_{Q',f_2}={\bf T}^0_{Q',f_1\circ f_2}$ for all $Q'>0$. Equation   \eqref{zizouU} implies  that ${\bf T}^0_{Q',f} $ preserves ${\rm Ran}(\mathbf U)$. Thus, because $\mathbf U\mathbf U^*$ is the orthogonal projector on ${\rm Ran}(\mathbf U)$,
$$
{\bf T}^{+,0}_{Q,f_1}{\bf T}^{+,0}_{Q,f_2}=\mathbf U^*{\bf T}^0_{Q/\sqrt2,f_1}\mathbf U\mathbf U^* {\bf T}^0_{Q/\sqrt2,f_2}\mathbf U
=\mathbf U^*{\bf T}^0_{Q/\sqrt2,f_1}{\bf T}^0_{Q/\sqrt2,f_2}\mathbf U=\mathbf U^*{\bf T}^0_{Q/\sqrt2,f_1\circ f_2}\mathbf U
 ={\bf T}^{+,0}_{Q,f_1\circ f_2}.
$$

Next we deal with the case $\boldsymbol{\mu}\not=0$. The Markov property of the GFF \cite[Prop A.1]{GRW1} implies that in $f_1(\D)$ we have ($P_{f_1(\D)}u$ denotes the harmonic extension of $u$ in $f_1(\D)$)
\[
X_{\D^+}^\#=  \tilde{X}^\#_{f_1(\D)}+ P_{ f_1(\D)  }(  X^\#_{\D^+}|_{\partial f_1(\D)})
\]
 where $\tilde X_{f_1(\D^+)}$ is a  GFF on $f_1(\D^+)$ with  mixed boundary conditions, namely Neumann on $f_1(\D^+)\cap \R$ and  Dirichlet on $f_1(\T^+)$, independent from $X_{\D^+}\circ f_1|_{\T^+}$. By conformal invariance of the GFF and since $f_1\in \mc{S}^+$ we have in $\D$
\begin{equation*}
X^\#_{\D^+} \circ f_1 =  \tilde{X}^\#_{\D^+}+ P_\D(  X_{\D^+}^\# \circ f_1|_{\T}).
\end{equation*}
for $\tilde{X}_{\D^+}$  a mixed condition GFF on $\D^+$, independent of $X_{\D^+}\circ f_1|_{\T^+}$.
Using the identities
\[\begin{gathered}
\log \Big|\frac{f_1'(z)}{f_1(z)}\Big|=    P_\D( \log \Big|\frac{f_1'}{f_1}\Big|{|_{\T}}   ) (z)+    \log \frac{1}{|z|},\\
\int_{\A^+_{f_1\circ f_2}}  M^{\boldsymbol{\mu}}_\gamma(X^\#,\dd x)= \int_{\A^+_{f_1}} M^{\boldsymbol{\mu}}_\gamma(X^\#,\dd x)+\int_{f_1(\A^+_{f_2}) } M^{\boldsymbol{\mu}}_\gamma(X^\#,\dd x)
\end{gathered}\]
and making the change of variable $x=f_1(y)$ in the second integral, we get that 
\[\begin{split}
\int_{f_1(\A^+_{f_2})} M^{\boldsymbol{\mu}}_\gamma(X^\#,\dd x)=&\mu \int_{\A^+_{f_2}}\Big(\frac{|f_1'(x)|}{|f_1(x)|}\Big)^{\gamma Q}  M^{g_0}_\gamma(X^\#\circ f_1,\dd x)+\mu^{\rm R}  \int_{\A^+_{f_2}\cap \R^+} \Big(\frac{|f_1'(x)|}{|f_1(x)|}\Big)^{\frac{\gamma Q}{2}}M^{{\rm b},g_0}_\gamma (X^\#\circ f_1,\dd x)\\
& +\mu^{\rm L}  \int_{\A^+_{f_2}\cap \R^-}\Big(\frac{|f_1'(x)|}{|f_1(x)|}\Big)^{\frac{\gamma Q}{2}}M^{{\rm b},g_0}_\gamma(X^\#\circ f_1,\dd x)\\
=&\int_{\A^+_{f_2}}M^{\boldsymbol{\mu}}_\gamma (X^\#\circ f_1+ QP_\D( \log \Big|\frac{f'_1}{f_1}\Big|{|_{\T}}   ) , \dd x)
\\
=&\int_{\A^+_{f_2}}M^{\boldsymbol{\mu}}_\gamma ( \tilde{X}^\#_{\D^+}+  P_\D\widetilde{\varphi}^{\rm h}_1    , \dd x)
\end{split}\]
where we have set
 \[
\widetilde{\varphi}^{\rm h}_1
:=
\left(
(X_{\D^+}^{\#}+P_{\D}\widetilde{\varphi}^{\rm h})\circ f_1
+
Q\log\left|\frac{f'_1}{f_1}\right|
\right)|_{\T}.
\]
Here we have used the conformal change of the GMC measures $M_\gamma$ and $M^{\rm b}_{\gamma}$, see for instance \cite{Berestycki_lqggff}.   Moreover, using
\[
\log\left|\frac{f'_1(w)}{f_1(w)}\right|
=
P_{\D}\left(
\log\left|\frac{f'_1}{f_1}\right|
\big|_{\T}
\right)(w)
+
\log\frac1{|w|},
\]
we get, for $z\in\T$,
\[
P_{\D}\left(
\log\left|\frac{f'_1}{f_1}\right|
\big|_{\T}
\right)(f_2(z))
+
\log\left|\frac{f'_2(z)}{f_2(z)}\right|
=
\log\left|
\frac{(f_1\circ f_2)'(z)}{(f_1\circ f_2)(z)}
\right|.
\]
Therefore
\begin{align*}
{\bf T}^{+}_{f_1\circ f_2}F(\widetilde{\varphi}^{\rm h})
&=
|f_1'(0)f_2'(0)|^{\frac{Q^2}{4}}
\mathbb E_{\varphi^{\rm h}}
\Bigg[
F\left(
\left(
\widetilde X_{\D^+}^{\#}\circ f_2
+
(P_{\D}\widetilde{\varphi}^{\rm h}_1)\circ f_2
+
Q\log\left|\frac{f'_2}{f_2}\right|
\right)\Big|_{\T}
\right)
\\
&\hspace{4em}\times
\exp\left(
-\int_{\A^+_{f_2}}
M^{\boldsymbol{\mu}}_\gamma
\left(
\widetilde X_{\D^+}^{\#}
+
P_{\D}\widetilde{\varphi}^{\rm h}_1,
\dd x
\right)
-\int_{\A^+_{f_1}}
M^{\boldsymbol{\mu}}_\gamma(X^\#,\dd x)
\right)
\Bigg]
\\
&=
|f_1'(0)|^{\frac{Q^2}{4}}
\mathbb E_{\varphi^{\rm h}}
\left[
{\bf T}^{+}_{f_2}F(\widetilde{\varphi}^{\rm h}_1)
\exp\left(
-\int_{\A^+_{f_1}}
M^{\boldsymbol{\mu}}_\gamma(X^\#,\dd x)
\right)
\right]
\\
&=
{\bf T}^{+}_{f_1}{\bf T}^{+}_{f_2}F(\widetilde{\varphi}^{\rm h}).
\end{align*}
 This completes the proof.
\end{proof}
Later we will show that the map $f\mapsto {\bf T}_f^+$ is continuous.  This, however, requires an explicit expression of the integral kernel of the operator ${\bf T}_f^+$ in terms of  the half-annulus amplitude on $\A_f^+$. For the moment, we will instead extract from this map certain one-parameter semigroups associated with the flows of holomorphic vector fields.
The ultimate goal is to prepare for the differentiability analysis of the map  $f\mapsto {\bf T}_f^+$.

 Let us consider a $1$-parameter family  of elements $f_t\in \mc{S}^+$ given by the flows of certain  holomorphic 
vector fields ${\rm v}=v(z)\pl_z$ defined in a neighbourhood of  $\D$. We say that ${\rm v}=v(z)\pl_z$ is \textbf{symmetric} if it satisfies $v(\bar{z})=\bbar{v(z)}$. If ${\rm v}$ is Markovian and symmetric, the flow $f_t:\D\to f_t(\D)$ of $v(z)\pl_z$, defined by 
$$\pl_t f_t(z)=v(f_t(z)),\quad t\geq 0,\qquad f_0(z)=z,$$
satisfies $f_t(\D)\subset f_{t'}(\D)$ for all $t>t'$, and $f_t(\D^+)=f_t(\D)\cap \H$, with $f_t(0)=0$. In particular, $f_t\in \mc{S}^+$ for all $t> 0$.

If ${\rm v}=v(z)\pl_z$ with $$v(z)=\sum_{n=0}^{\infty} v_n z^{n+1}\in {\rm Hol}^{+,\bullet}_\epsilon(\D),$$ we define the operator  $\bH_{\rm v}^{+,0}$ via the following formula: 
\begin{equation}\label{eq:def_L+} 
 \forall F\in\cC_c(\R\times \Omega_{\T^+}),\quad   \bH_{\rm v}^{+,0} F:=-\sum_{n=0}^{\infty} v_n {\bf L}^{+,0}_n F, 
\end{equation}
with ${\bf L}_n^{+,0}$ defined in \eqref{Ln^+0}. Since $F \in \cC_c(\R\times \Omega_{\T^+})$,  only finitely many terms in the above sum are non-zero, so that the expression is well defined. Moreover, using \eqref{normLn^+0} and the fast decay of $|v_n|$ as $n\to \infty$  since $\mathrm{v}\in{\rm Hol}^{+,\bullet}_\epsilon(\D)$, we see that $\bH_{\rm v}^{+,0}:\mc{D}(\mc{Q}^{+,0})\to \mc{D}'(\mc{Q}^{+,0})$ is a bounded operator. 

\begin{proposition}\label{prop_coercive}
Let $v\in   {\rm Hol}^{+,\bullet}_\epsilon(\D)$ for some $\eps>0$, such that ${\rm v}=v(z) \partial_z$ is  a Markovian vector field, and let $f_t$  be its flow at time $t$. 
There exists a  constant $K>0$ independent of $v$ such that if $ v_0<- K \sum_{n \geq 1}|v_n|n^2$ and $v(z)=\sum_{n\geq 0} v_nz^{n+1}$, then the operator $ \bH_{\rm v}^{+,0}$ defined in \eqref{eq:def_L+} admits a closed extension such that $e^{-t \bH_{\rm v}^{+,0} }$ is a continuous contraction semigroup on $\mc{H}^+$, and we have the relation
$${\bf T}^{+,0}_{f_t}= e^{-t \bH_{\rm v}^{+,0} }.$$
\end{proposition}

\begin{proof}
Consider  the  vector space $\mc{C}_{\rm exp}^+$ 
of smooth functions $F:\R\times \Omega_{\T^+}\to \C$ so that there is $N>0$ such that $F(c,\varphi^{\rm h})=F(c,x_1^{\rm h},  \dots, x_N^{\rm h})$ for all $c\in\R$ (recall \eqref{GFFcircle+} with $\varphi_n^{\rm h}=x_n^{\rm h}/\sqrt{2|n|}$), and 
\[ \forall k\in \N, \forall \alpha  \in \N^{N}, \exists L\geq 0, \forall M\geq 0,  \exists C>0, 
\quad  |\pl_c^k \pl_{x^{\rm h}}^\alpha  F(c,\varphi^{\rm h})|\leq Ce^{-M|c|}\cjg \varphi^{\rm h}\cjd_N^{L}
\]
if $\pl_{x^{\rm h}}^\alpha=\pl_{x_1^{\rm h}}^{\alpha_1}\dots \pl_{x_N^{\rm h}}^{\alpha_N}$ and $\cjg \varphi^{\rm h}\cjd_N=(1+\sum_{|n|\leq N}|\varphi^{\rm h}_n|^2)^{1/2}$ is the Japanese bracket. Using \eqref{zizou} and  \cite[Theorem 2.8]{BGKRV}, we observe that ${\bf T}^{+,0}_{f_t}$ preserves $\mc{C}_{\rm exp}^+$. Moreover, the proof of \cite[Theorem 2.8]{BGKRV} shows that there is $K>0$ such that if $v_0<- K \sum_{n \geq 1}|v_n|n^2 $, there is $C>0$ such that for all $F\in\mc{C}_{\rm exp}^+$
\begin{equation}\label{ineqQ0}
\frac{1}{C}\mc{Q}^{+,0}(F)\leq  {\rm Re}\cjg {\bf H}_{\rm v}^{+,0}F,F\cjd_{\mc{H}_+}\leq |\cjg {\bf H}_{\rm v}^{+,0}F,F\cjd_{\mc{H}_+}| \leq C\mc{Q}^{+,0}(F),
\end{equation}
that ${\bf H}_{\rm v}^{+,0}$ has a closed extension to a domain $\mc{D}({\bf H}_{\rm v}^{+,0})\subset \mc{D}(\mc{Q}^{+,0})$ and generates a contraction semi-group $e^{-t \bH^{+,0}_{\rm v}}$. By differentiating in $t$ the relation $({\bf T}_{Q/\sqrt{2},f_t}^{0} (F(\sqrt{2}\cdot)))(2^{-1/2}\tilde{\varphi}^{\rm h}) =  {\bf T}_{Q,f_t}^{+,0}F( \tilde{\varphi}^{\rm h})$ for $F\in \mc{C}_{\rm exp}^+$ and by using \eqref{relationVir0}, we deduce (using \cite[Th. 2.8]{BGKRV} to differentiate ${\bf T}_{\frac{Q}{\sqrt{2}},f_t}^{0} $)
\begin{align*}
\pl_t {\bf T}_{Q,f_t}^{+,0}F(\tilde{\varphi}^{\rm h})=&\pl_t({\bf T}_{\frac{Q}{\sqrt{2}},f_t}^{0} (F(\sqrt{2}\cdot)))(2^{-1/2}\tilde{\varphi}^{\rm h}) 
\\
=&\sum_{n=0}^\infty(v_n{\bf L}^{0,2^{-1/2}Q}_n+\bar{v}_n\widetilde{{\bf L}}^{0,2^{-1/2}Q}_n) ({\bf T}_{\frac{Q}{\sqrt{2}},f_t}^{0} (F(\sqrt{2}\cdot)))(2^{-1/2}\tilde{\varphi}^{\rm h}) 
\\
=&-\bH_{\rm v}^{+,0}{\bf T}_{Q,f_t}^{+,0}F(\tilde{\varphi}^{\rm h}).
\end{align*}
By uniqueness for this equation (with initial data ${\bf T}_{Q,f_0}^{+,0}F=F$), both semigroups coincide on $\mc{C}_{\rm exp}^+$, hence on $\mc{H}^+$ by density.
\end{proof}

We now investigate the semigroup ${\bf T}^+_{f_t}$ when $f_t$ is the flow of a Markovian vector field ${\rm v}=v(z)\pl_z$ with $v\in  {\rm Hol}^{+,\bullet}_\epsilon(\D)$. As before, we assume that the restriction  \eqref{assumptions_gamma_mu} is in force.
We shall comment on this condition when relevant in the sequel. 

We associate to ${\rm v}$ the positive measure on $\T^+$
\[\dd\varrho_{\rm v}(\theta)=-{\rm Re}(e^{-i\theta}v(e^{i\theta}))\dd\theta,\] 
and define the following real-valued potential in $L^p(\Omega_{\T^+})$    (using the Wick notation $:e^{\alpha h(x)}:$  for $e^{\alpha h(x)-\frac{\alpha^2}{2}\E[h^2(x)]}$)
\begin{align}\label{defVk}
V^{+,\cap}_{\rm v}(\varphi^{\rm h})=& \int_{\T^+} :e^{\gamma  \varphi^{\rm h}(\theta)}\!:\big( 2|\sin(\theta)|\big)^{-\frac{\gamma^2}{2}}\dd\varrho_{\rm v}(\theta) 
\end{align} 
for $p<\frac{1}{4}+\sqrt{\frac{1}{16}+\frac{1}{\gamma^2}}$. Note that the $L^p$-integrability follows from   that  of the potential 
$V^{+,\cap}$ defined in  \eqref{defV^+cap}, since $|V^{+,\cap}_{\rm v}(\varphi^{\rm h})|\leq \|v\|_{L^\infty}V^{+,\cap}(\varphi^{\rm h})$, and observe that $V^{+,\cap}=V^{+,\cap}_{{\rm v}_0}$ with ${\rm v}_0=-z\pl_z$.

\begin{equation}\label{defVk2}
 V^{+,{\rm R}}_{\rm v}:=-v(1)V^{+,{\rm R}},
 \qquad
 V^{+,{\rm L}}_{\rm v}:=v(-1)V^{+,{\rm L}},
\end{equation}
where $V^{+,{\rm R}},V^{+,{\rm L}}$ are defined in \eqref{girsuk}
(see also \eqref{V^+Rdef/V^+Ldef}).

This allows us to define an operator 
\begin{equation}\label{defHv^+}
\bH^+_{\rm v}:=\bH^{+,0}_{\rm v}+{\bf V}_{\rm v}^{+,\bs{\mu}}:\mc{D}(\mc{Q}^+)\to \mc{D}'(\mc{Q}^+)
\end{equation}
where
\begin{equation}\label{bfV+mu}
{\bf V}_{\rm v}^{+,\bs{\mu}}:=\mu e^{\gamma c}V_{\rm v}^{+,\cap} +\mu^{\rm L}e^{\frac{\gamma c}{2}}V_{\rm v}^{+,{\rm L}}+\mu^{\rm R}e^{\frac{\gamma c}{2}}V_{\rm v}^{+,{\rm R}}.
\end{equation}
One can then define the (non symmetric) bilinear form on $\mc{C}_c(\R\times \Omega_{\T^+})$ via the formula
\begin{equation} \label{defQ+v}
\mathcal{Q}^+_{\rm v}(F,G): =\cjg \bH_{\rm v}^{+,0} F , G \cjd_{\mc{H}^+} +   \cjg    {\bf V}_{\rm v}^{+,\bs \mu}  F , G \cjd_{\mc{H}^+} .
 \end{equation}  
 and remark that it extends continuously to $\mc{D}(\mc{Q}^+)$.
 Note that $\dd\varrho_v$ has a positive smooth density with respect to $\dd\theta$, and $v(1)<0<v(-1)$. This leads to the conclusion that there is a constant $C>1$ such that 
 \begin{equation}\label{comparisonVvtoV}
 C^{-1}{\bf V}^{+,\bs{\mu}} \leq {\bf V}_{\rm v}^{+,\bs{\mu}} \leq C{\bf V}^{+,\bs{\mu}}
 \end{equation} 
 as quadratic forms on $\mc{D}(\mc{Q}^+)$. Assuming furthermore  
 $v_0<- K \sum_{n \geq 1}|v_n|n^2$ (the condition of Proposition \ref{prop_coercive}),  
 we can use \eqref{ineqQ0}  to get that for all $F \in \mc{D}(\mc{Q}^+)$,
 \begin{equation}\label{fundinequality}
 C^{-1} \mathcal{Q}^+(F,F) \leq {\rm Re}(\mathcal{Q}^+_{\rm v}(F,F))\leq |\mathcal{Q}^+_{\rm v}(F,F)| \leq C \mathcal{Q}^+(F,F).
 \end{equation} 
 Since $\mathcal{Q}^+$ is closable, this implies that $ \mathcal{Q}_{{\rm v}}^+$ is strictly m-accretive with domain  $\mc{D}(\mathcal{Q}^+)$.  By \cite[Th VIII.16]{Reed-Simon}, the operator ${\bf H}^+_{{\rm v}}$ can be considered as an (unbounded) closed operator on $\mc{H}^+$ with domain $\mc{D}({\bf H}^+_{\rm v})\subset \mc{D}(\mathcal{Q}^+)$ such that  
 $$
  \mathcal{Q}_{{\rm v}}^+(F,G)=\langle {\bf H}^+_{\rm v}F,G \rangle_{\mc{H}^+},\qquad F\in \mc{D}({\bf H}^+_{\rm v}),G\in \mc{D}(\mathcal{Q}^+).
  $$
   By the Hille-Yosida theorem, ${\bf H}^+_{\rm v}$ is the generator of a contraction semi-group denoted $e^{-t{\bf H}^+_{\rm v}}$.
We can define the boundary Liouville Virasoro generators for $n\geq 0$ as the bounded operators ${\bf L}_n^+: \mc{D}(\mc{Q}^+)\to \mc{D}'(\mc{Q}^+)$ given by
 \begin{equation}\label{formulaLn}  
{\bf L}_n^+:=  {\bf L}_n^{+,0}+  \mu e^{\gamma c}V^{+,\cap}_n + \mu^{\rm R}e^{\frac{\gamma}{2}c} V_n^{+,{\rm R}} +\mu^{\rm L}(-1)^n e^{\frac{\gamma}{2}c}V_n^{+,{\rm L}}\qquad \textbf{(boundary Liouville Virasoro)}
\end{equation}
where ${\bf L}_n^{+,0}$ is defined by \eqref{Ln^+0} and $V^{+,\cap}_n$ is the multiplication operator by the $L^p(\Omega_{\T^+})$ potential 
\[V^{+,\cap}_n(\varphi^{\rm h}) =\int_{\T^+} :e^{\gamma  \varphi^{\rm h}(\theta)}\!:\big( 2|\sin(\theta)|\big)^{-\frac{\gamma^2}{2}}\cos(n\theta)\dd \theta.\]
For $n>0$, we define ${\bf L}_{-n}^+:\mc{D}(\mc{Q}^+)\to \mc{D}'(\mc{Q}^+)$ by ${\bf L}_{-n}^+:=({\bf L}_n^+)^*$ where the adjoint is defined for 
$F,G\in \mc{D}(\mc{Q}^+)$ by $\cjg ({\bf L}_n^+)^*F,G\cjd=\cjg F,{\bf L}_n^+G\cjd$, where $\cjg\cdot,\cdot \cjd$ 
denotes the duality pairing between $\mc{D}(\mc{Q}^+)$ and $\mc{D}'(\mc{Q}^+)$. This means
\[ \cjg ({\bf L}_n^+)^*F,G\cjd_{\mc{H}^+}=\cjg F,{\bf L}_n^+G\cjd_{\mc{H}^+},\qquad F,G\in \mc{C}_c(\R\times \Omega_{\T^+}).\]
In particular, we see that \eqref{formulaLn} also holds with $n\geq 0$ replaced by $-n$ in that formula and \eqref{normLn^+0} with \eqref{comparisonVvtoV} implies that there is $C>0$ such that for all $n\in \Z$, all $F,G\in \mc{D}(\mc{Q}^+)$:
\begin{equation}\label{boundLn}
\cjg \bL^+_n F,G\cjd_{\mc{H}^+}\leq C(1+|n|)^{3/2}\mc{Q}^+(F)^{1/2}\mc{Q}^+(G)^{1/2}.
\end{equation}
We also observe that, for $v(z)=\sum_{n=0}^{\infty} v_n z^{n+1}\in {\rm Hol}_\eps(\D)^+$, we have as bounded operators $\mc{D}(\mc{Q}^+)\to \mc{D}'(\mc{Q}^+)$,
\begin{equation}\label{Hv_Virasoro}  
\bH^+_{\rm v}= -\big(v_0 {\bf H}^++ \sum_{n\geq 1}v_n\, {\bf L}^+_n\big).  
\end{equation}   

The main result for the operators $\bH^+_{\rm v}$ is the following
\begin{proposition}\label{th:Tft} Assume the condition \eqref{assumptions_gamma_mu}. Let $v\in   {\rm Hol}^{+,\bullet}_\epsilon(\D)$ be  a Markovian vector field, and let $f_t\in \mc{S}^+$  be its flow. Assume that $v$ obeys the condition of Proposition \ref{prop_coercive}, namely $v_0<- K \sum_{n \geq 1}|v_n|n^2$. Then  ${\bf T}^+_{f_t}=e^{-t \bH^+_{\rm v}}$ on $\mc{H}^+$.
\end{proposition}

\begin{proof} First we rewrite the bulk potential in terms of an integral over time of the process $s\mapsto X^\#\circ f_s+Q\log |\frac{f_s'}{f_s}|$ defined on the circle.
The map $(s,\theta)\mapsto f_s(e^{i\theta})$ is a diffeomorphism from $(0,t)\times \T^+$ to $\A_{f_t}^+$, and its Jacobian is given by 
 $-{\rm Re} (e^{- i \theta} v(f_s (e^{i \theta}))\bbar{f'_s(e^{i \theta})})$. Therefore,  we get that (see the proof of \cite[Th 3.1]{BGKRV} for more details)
 \begin{equation*}
  \int_{\A_{f_t}^+}      |x| ^{-\gamma Q} M_\gamma^{g_0} (X^\#,\dd x)= -\int_{0}^t  \int_{0}^{  \pi}  
 e^{\gamma  ( X^\#\circ f_s+Q\log |\frac{f_s'}{f_s}|)(e^{i \theta}) }  {\rm Re} \Big(e^{- i \theta}   \frac{v(f_s (e^{i \theta}))}{f'_s(e^{i \theta})}  \Big)  \dd \theta \dd s
 \end{equation*}
From \cite[Lemma 2.4]{BGKRV}, the ratio $\frac{v(f_s (e^{i \theta}))}{f'_s(e^{i \theta})}$ does not depend on $s$, leading to the relation
\begin{equation}\label{polardecompositionctd}
 \int_{\A_{f_t}^+}      |x| ^{-\gamma Q} M_\gamma^{g_0} (X^\#,\dd x)=-\int_{0}^t  \int_{0}^{  \pi}  
 e^{\gamma  ( X^\#\circ f_s+Q\log |\frac{|f_s'}{|f_s}|)(e^{i \theta}) }  {\rm Re} (e^{- i \theta} v(e^{i \theta})) \dd \theta \dd s  .
\end{equation}
Now we proceed similarly for the boundary potential. Here we use that the map $s\mapsto f_s(1)$ (resp. $r\mapsto f_s(-1)$) is a diffeomorphism from $[0,t]$ to $[f_t(1),1]$ (resp.  $[0,t]$ to $[-1,f_t(-1)]$) with Jacobian $-v(f_s(1))$ (resp. $v(f_s(-1))$). Therefore
\begin{equation*}
  \int_{[f_t(1),1]}      |x| ^{-\frac{\gamma Q}{2}} M_{\partial,\gamma}^{g_0} (X^\#,\dd x)= -\int_{0}^t   
 e^{\frac{\gamma}{2}  ( X^\#\circ f_s(1)+Q\log |\frac{f_s'(1)}{f_s(1)}|)  }    \frac{v(f_s (1))}{f'_s(1)}    \dd s.
 \end{equation*}
Again, the fact that $ \frac{v(f_s (1))}{f'_s(1)}$ does not depend on $s$ reduces the above expression to 
\begin{equation}\label{polarboundaryR}
  \int_{[f_t(1),1]}      |x| ^{-\frac{\gamma Q}{2}} M^{{\rm b},g_0}_\gamma (X^\#,\dd x)= - v(1) \int_{0}^t   
 e^{\frac{\gamma}{2}  ( X^\#\circ f_s(1)+Q\log |\frac{f_s'(1)}{f_s(1)}|)  }      \dd s.
\end{equation}
The left boundary potential leads in the same way to
\begin{equation}\label{polarboundaryL}
  \int_{[-1,f_t(-1)]}      |x| ^{-\frac{\gamma Q}{2}} M^{{\rm b},g_0}_\gamma (X^\#,\dd x)=  v(-1) \int_{0}^t   
 e^{\frac{\gamma}{2}( X^\#\circ f_s(-1)+Q\log |\frac{f_s'(-1)}{f_s(-1)}|)  }      \dd s.
\end{equation}
The above formulas show that ${\bf T}^+_{f_t}$ is a semigroup (which was already proven in Prop.~\ref{prop:semigroup}) since it is obtained from ${\bf T}^{+,0}_{f_t}$ via a Feynman-Kac type formula.  

The next step is to connect ${\bf T}^+_{f_t}$ and $e^{-t \bH^+_{\rm v}}$ when the potential  ${\bf V}_{\rm v}^{+,\bs \mu} $ is replaced by a  regularized approximation. A similar argument is carried out in \cite[Prop. 5.3]{GKRV20_bootstrap} but the Trotter product formula used there is slightly different because it involves a self-adjoint operator ${\bf H}^0$ whereas, here, the operator $  \bH_{\rm v}^{+,0} $ is \textbf{not self-adjoint.}  So we use instead \cite[Exercise 5.11 in Chapter III]{EngelNagel}, which claims that, if $(T_t)_{t\geq 0}$ is a strongly continuous semigroup with generator ${\bf A}$ on a Banach space $X$ and if ${\bf B}\in \mc{L}(X)$, the semigroup $(U_t)_{t\geq 0}$ generated by ${\bf A}+{\bf B}$ obeys the Trotter product formula
$$\forall x\in X,\quad U_tx=\lim_{n\to\infty} \big[T_{t/n}e^{(t/n){\bf B}}\big]^nx.$$
Here the Banach space is $\mc{H}^+$ and we have to regularize the potential ${\bf V}_{\rm v}^{+,\bs \mu} $ in a way that it  becomes a bounded operator on $\mc{H}^+$. For this, we  consider the projector $\pi_N$ that truncates a trigonometric series on $\T^+$ of the type  \eqref{GFFcircle+} to its $N$ first components (i.e. the sum in \eqref{GFFcircle+} restricts to $|n|\leq N$). Now we define the regularized potential ${\bf V}_{{\rm v},N,M}^{+,\bs \mu} $ (resp. ${\bf V}_{{\rm v},N}^{+,\bs \mu} $) as follows. Recall that  ${\bf V}_{\rm v}^{+,\bs \mu} $ is defined in \eqref{bfV+mu}. 
We replace each term in the sum, call it $X(\tilde \varphi)$, by $X(\pi_N\tilde \varphi){\bf 1}_{|X(\pi_N\tilde \varphi)|\leq M}$ (resp. $X(\pi_N\tilde \varphi)$). The corresponding potential acting by multiplication is called ${\bf V}_{{\rm v},N,M}^{+,\bs \mu} $ (resp. ${\bf V}_{{\rm v},N}^{+,\bs \mu} $). We  define the semigroups ${\bf T}^{+}_{f_t,N,M}$ and $e^{-t \bH^+_{{\rm v},N,M}}$ (resp. ${\bf T}^{+}_{f_t,N}$ and $e^{-t \bH^+_{{\rm v},N}}$)  similarly to ${\bf T}^{+}_{f_t}$ and $e^{-t \bH^+_{\rm v}}$ with the potential ${\bf V}_{\rm v}^{+,\bs \mu} $ replaced by ${\bf V}_{{\rm v},N,M}^{+,\bs \mu} $ (resp. ${\bf V}_{{\rm v},N}^{+,\bs \mu} $). We  define the semigroups ${\bf T}^{+}_{f_t,N,M}$ and $e^{-t \bH^+_{{\rm v},N,M}}$ (resp. ${\bf T}^{+}_{f_t,N}$ and $e^{-t \bH^+_{{\rm v},N}}$)  similarly to ${\bf T}^{+}_{f_t}$ and $e^{-t \bH^+_{\rm v}}$ with the potential ${\bf V}_{\rm v}^{+,\bs \mu} $ replaced by ${\bf V}_{{\rm v},N,M}^{+,\bs \mu} $ (resp. ${\bf V}_{{\rm v},N}^{+,\bs \mu} $). For fixed $N,M$, the potential
${\bf V}_{{\rm v},N,M}^{+,\bs\mu}$ is a bounded multiplication operator on
$\mc H^+$. Hence the bounded perturbation version of the Trotter product
formula  \cite[Prop. 5.3]{GKRV20_bootstrap}  applies to
$\bH_{\rm v}^{+,0}+{\bf V}_{{\rm v},N,M}^{+,\bs\mu}$, to establish ${\bf T}^{+}_{f_t,N,M}=e^{-t \bH^+_{{\rm v},N,M}}$. Using dominated convergence, it is clear that, for $F\in \mc{H}^+$,  $\lim_{M\to \infty}{\bf T}^{+}_{f_t,N,M}F={\bf T}^{+}_{f_t,N}F$ and $\lim_{M\to \infty}{\bf R}^{+}_{\lambda,N,M}F={\bf R}^{+}_{\lambda,N}F$, with ${\bf R}^{+}_{\lambda,N,M}$ and ${\bf R}^{+}_{\lambda,N}$ the associated resolvents.  We further introduce the strictly m-accretive form associated to ${\bf R}^{+}_{\lambda,N}$ and $e^{-t \bH^+_{{\rm v},N}}$:
\begin{equation} 
\mathcal{Q}^+_{{\rm v},N}(F,G): =\cjg \bH_{\rm v}^{+,0} F , G \cjd_{\mc{H}^+} +   \cjg    {\bf V}_{{\rm v},N}^{+,\bs \mu}  F , G \cjd_{\mc{H}^+} 
 \end{equation}  
for $F,G$ in  the domain of the form $\mathcal{D}(\mathcal{Q}^+_{{\rm v},N})=\mathcal{D}(\mathcal{Q}^{+,0})\cap \mathcal{D}(\mathcal{Q}_{  {\bf V}_{{\rm v},N}^{+,\bs \mu}})$, where  $ \mathcal{D}(\mathcal{Q}_{  {\bf V}_{{\rm v},N}^{+,\bs \mu}})=\{F \mid \cjg   {\bf V}_{{\rm v},N}^{+,\bs \mu}   F , F \cjd_{\mc{H}^+}<+\infty\}$, which contains $\mc{C}_+$ as a dense subset.
 
Our goal is now to show that the semigroup ${\bf T}^{+}_{f_t}$ is associated to the  bilinear form $\mathcal{Q}^+_{\rm v}$.   This will be done by passing to the limit $N\to\infty$ in relations associated to the objects ${\bf R}^{+}_{\lambda,N}$ and $\mathcal{Q}^+_{{\rm v},N}$.  
In the sequel, we will consider the semigroups and the quadratic forms on the space of real functions (this is no restriction by a linearity argument). 
Let $\lambda>0$ and $F \in \mc{C}_c(\R\times \Omega_{\T^+})$ be real valued. 
We have for all  real valued
 $G \in \mathcal C_c(\R\times \Omega_{\T^+})$
\begin{equation*}
\lambda \cjg  {\bf R}^{+}_{\lambda,N}F  , G \cjd_{\mc{H}^+}+ \mathcal{Q}^+_{{\rm v},N}({\bf R}^{+}_{\lambda,N} F,G)=  \cjg F  ,G \cjd_{\mc{H}^+}.
\end{equation*} 
Since  $ {\bf R}^{+}_{\lambda,N} F$ and $G$ depends on the first $N$ harmonics of the field for $N$ large, the following holds by definition of $\mathcal{Q}^+_{\rm v}$, if $G$ depends also only on the first $N$ harmonics,
\begin{equation}\label{quadrelation}
\lambda \cjg  {\bf R}^{+}_{\lambda,N} F  , G \cjd_{\mc{H}^+}+ \mathcal{Q}^+_{\rm v}({\bf R}^{+}_{\lambda,N} F, G)=  \cjg F  , G \cjd_{\mc{H}^+}.
\end{equation} 
The above identity for $G= {\bf R}^{+}_{\lambda,N} F$ shows that 
\begin{align*}
\lambda \|{\bf R}^{+}_{\lambda,N} F\|^2_{\mc{H}^+}+ \mathcal{Q}^+_{\rm v}({\bf R}^{+}_{\lambda,N} F, {\bf R}^{+}_{\lambda,N} F) = \cjg F  , {\bf R}^{+}_{\lambda,N} F \cjd_{\mc{H}^+}
\end{align*}
Note in particular that this relation implies that ${\rm Im}(\mathcal{Q}^+_{\rm v}({\bf R}^{+}_{\lambda,N} F, {\bf R}^{+}_{\lambda,N} F)) =0$ since $ {\bf R}^{+}_{\lambda,N} F$ is real valued (because ${\bf T}^{+}_{f_t,N}F$ is).
Using next \eqref{fundinequality}, we get
\begin{align*}
\lambda \|{\bf R}^{+}_{\lambda,N} F\|^2_{\mc{H}^+}+ C^{-1}\mathcal{Q}^+({\bf R}^{+}_{\lambda,N} F,{\bf R}^{+}_{\lambda,N} F) \leq 
&
\lambda \|{\bf R}^{+}_{\lambda,N} F\|^2_{\mc{H}^+}+
\mathcal{Q}^+_{\rm v}({\bf R}^{+}_{\lambda,N} F, {\bf R}^{+}_{\lambda,N} F) 
\\ 
= & \cjg F  , {\bf R}^{+}_{\lambda,N} F \cjd_{\mc{H}^+}
\\
\leq &  \frac{1}{2\lambda}\|F  \|^2_{\mc{H}^+}+ \frac{\lambda}{2}\| {\bf R}^{+}_{\lambda,N} F\|^2_{\mc{H}^+}
\end{align*}
and this shows that 
\begin{equation}\label{aprioriestimate}
\sup_{N \geq 1} \mathcal{Q}^+({\bf R}^{+}_{\lambda,N} F, {\bf R}^{+}_{\lambda,N} F) < \infty.
\end{equation} 

Next, the resolvent ${\bf R}^{+}_{\lambda,N} F=\int_0^{\infty}e^{-\lambda r}{\bf T}^{+}_{f_r,N}F\,\dd r$ converges in $\mc{H}^+$ towards ${\bf R}^{+}_{\lambda} F:=\int_0^{\infty}e^{-\lambda r}{\bf T}^{+}_{f_r}F\,\dd r$ as can be seen from the Feynman-Kac representation of the type \eqref{FKv} with a $N$-regularized potential and the dominated convergence theorem. 

We claim now that ${\bf R}^{+}_{\lambda} F$ belongs to $\mc{D}(\mc{Q}^+)$. For this, we observe first that, because ${\bf R}^{+}_{\lambda,N} F\in \mc{D}( \mathcal{Q}^+_{{\rm v},N})$, there is a sequence $(F_k)_k$ such that $F_k\in \mc{C}^+$ and $\mathcal{Q}^+_{{\rm v}=-z,N}(F_k-{\bf R}^{+}_{\lambda,N} F,F_k-{\bf R}^{+}_{\lambda,N} F)\to 0$ as $k\to\infty$ (note that $\mathcal{Q}^+_{{\rm v}=-z,N}$ is just the quadratic form $\mathcal{Q}^+$ with the potential regularized to its first $N$ harmonics). Now we make the following series of observations: first, $\pi_N\circ \partial_{x_n^{\rm h}}= \partial_{x_n^{\rm h}}\circ \pi_N$ for $N\geq 1$ and $n\geq 1$, and therefore $\pi_N\circ \partial_{x_n^{\rm h}}^*= \partial_{x_n^{\rm h}}^*\circ \pi_N$. So $\pi_N\circ {\bf L}_n^{+,0} ={\bf L}_n^{+,0}\circ \pi_N$. Second, $ \cjg    {\bf V}_{{\rm v},N}^{+,\bs \mu}  \pi_NF , \pi_NF \cjd_{\mc{H}^+} \leq \cjg    {\bf V}_{{\rm v},N}^{+,\bs \mu}  F , F \cjd_{\mc{H}^+}  $ by Jensen for $F\in \mc{C}^+$. From this, we deduce that, up to taking $\pi_NF_k$ instead of $F_k$, the sequence $(F_k)_k$ can be chosen to depend only on the first $N$ harmonics. This implies that we can find a sequence $(F_N)_N$ in $ \mc{C}^+$ such that $\mathcal{Q}^+_{{\rm v}=-z,N}(F_N-{\bf R}^{+}_{\lambda,N} F,F_N-{\bf R}^{+}_{\lambda,N} F)\to 0$ as $N\to\infty$. Yet, as ${\bf R}^{+}_{\lambda,N} F$ depends also only on the first $N$ harmonics, we have 
$$
\mathcal{Q}^+(F_N-{\bf R}^{+}_{\lambda,N} F,F_N-{\bf R}^{+}_{\lambda,N} F)=\mathcal{Q}^+_{{\rm v}=-z,N}(F_N-{\bf R}^{+}_{\lambda,N} F,F_N-{\bf R}^{+}_{\lambda,N} F)\to 0,\quad \text{ as } N\to\infty
$$
 in such a way that ${\bf R}^{+}_{\lambda} F$ belongs to $\mc{D}(\mc{Q}^+)$ since this quadratic form is closed.

The a priori estimate \eqref{aprioriestimate} implies that the sequence $({\bf R}^{+}_{\lambda,N} F)_N$ weakly converges with respect to the $\mc{Q}^+$ norm towards ${\bf R}^{+}_{\lambda} F$, and the convergence also holds strongly in $\mc{H}^+$ as seen before. Because of the continuity of the bilinear form $ \mathcal{Q}^+_{{\rm v}}$ in $\mc{Q}^+$ norm, we can pass to the limit as $N\to\infty$ in \eqref{quadrelation} to get
\begin{equation}
\lambda \cjg  {\bf R}^{+}_{\lambda} F  , G \cjd_{\mc{H}^+}+ \mathcal{Q}^+_{{\rm v}}({\bf R}^{+}_{\lambda} F, G)=  \cjg F  , G \cjd_{\mc{H}^+}.
\end{equation} 
This shows that ${\bf R}^{+}_{\lambda}$ obeys the resolvent equation for the bilinear form $\mathcal{Q}^+_{v}$, and in particular we have thus  the equality of resolvents $\int_0^\infty e^{-\lambda r }{\bf T}^{+}_{f_r}F\,\dd r=\int_0^\infty e^{-\lambda r }e^{-r \bH^+_{\rm v} }F\,\dd r$ for $F\in \mc{C}_+$. By density, the resolvents coincide on  $\mc{H}^+$. So, the semigroups coincide as well.
\end{proof}

\subsection{Quantised Segal semigroup as Segal amplitudes of half-annuli}\label{subsec: relation T+}

The quantisation ${\bf T}^+_f$ of $f\in \mc{S}^+$	does not depend  on a choice of metric on the annulus $\A^+_f$. There is, however,  another natural operator associated to $\A_f^+$, namely  the Liouville amplitude, defined in Section \ref{sub:ampli}.

This amplitude is constructed for the half-annulus $\A_f^+$ equipped with the Dirichlet boundary parametrisation
   $\boldsymbol{\zeta}_f=(\zeta_1,\zeta_2)$ given by $\zeta_2(e^{i\theta})=e^{i\theta}$ and $\zeta_1(e^{i\theta})=f(e^{i\theta})$ for $e^{i\theta}\in \T^+$, with Neumann boundary $(\pl \A_{f}^+)_{\rm N}= \A_{f}^+\cap \R$. Unlike ${\bf T}^+_f$,  the Liouville amplitude depends on the choice of a  Neumann extendible metric $g$. We shall first observe that ${\bf T}_f^+$ can be related to the operator whose integral kernel is given by the amplitude $\mc{A}_{\A_f^+,g_\A,\boldsymbol{\zeta}_f}$, where $g_\A=|\dd z|^2/|z|^2$ is Neumann extendible but not admissible.
   
Let $f\in \mc{S}^+\cap {\rm Hol}_\eps(\D^+)$ and denote  by $\D_f:=f(\D)$ and $\D^+_f:=f(\D^+)$.  
We equip the annulus $\mathbb{A}^+_f$ with the metric  $g_{\mathbb{A}}$, which 
has vanishing scalar curvature and is admissible near the outer boundary $\T^+$, though not admissible near the inner boundary  $f(\T^+)$. 
		
\begin{figure}
\begin{tikzpicture}
\node[inner sep=0pt] (pant) at (0,0)
{\includegraphics[width=0.2\textwidth]{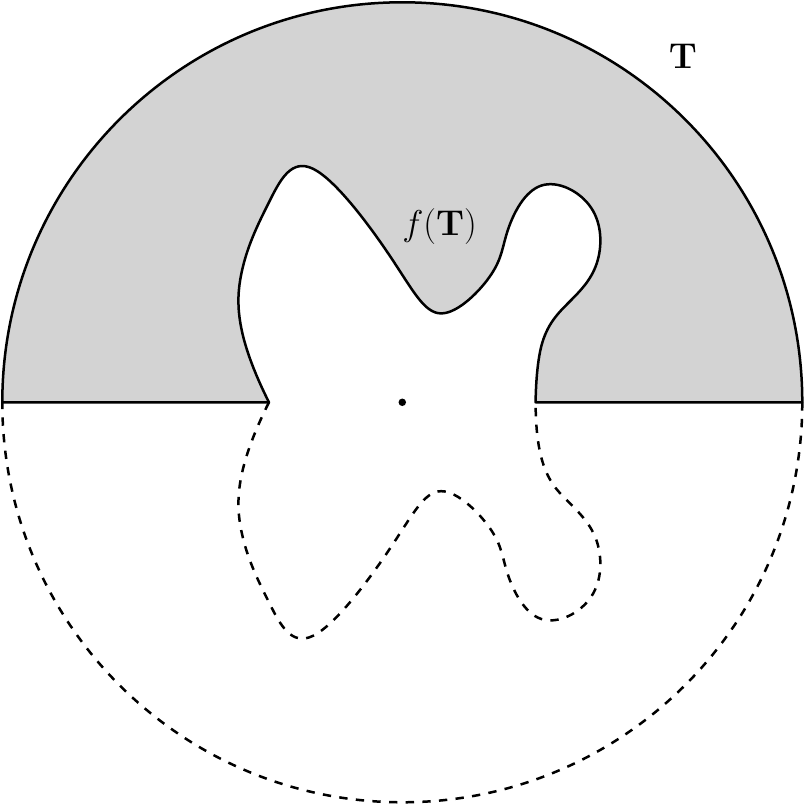}};
\end{tikzpicture}
\caption{The half annulus $\bA_f^+$.}\label{analytic half annulus}
\end{figure}
Let us consider the real valued harmonic function $h$ defined on a neighborhood of $ \D_f$ by 
\begin{equation}\label{defh}
\forall |z|\leq 1+\eps , \quad h(f(z))=\log\frac{|zf'(z)|}{|f(z)|}.
\end{equation} 
which satisfies $h\circ f(z)=h\circ f(\bar{z})$ and,
near $f(\T)$, $f^*g_\mathbb{A}= e^{2h\circ f} g_{\mathbb{A}}$.
Any admissible metric $g$ on $\mathbb{A}_f$ with parametrization $\bzeta_f=(\zeta_1,\zeta_2)$ of the boundary as above can be written under the form
\begin{equation}\label{defg_tadmissible}
g_f=e^{-2 \chi_f}g_{\mathbb{A}}
\end{equation}
where $\chi_f \in C^\infty(\mathbb{A}^+_f)$ is equal to $h$ near $f(\T^+)$ and $0$ near $\T^+$. We will  take $\chi_f=h\chi$ where $\chi \in C^\infty(\mathbb{A}_f)$ equals $1$ near $f(\T)$, $0$ near $\T$, with $\chi\circ \tau =\chi$ and  not depending on $f$ near a given $f_0$. 

We define
\begin{equation}\label{defomega}
\omega:=h\circ f|_{\T}=\log \frac{|f'|}{|f|}|_{\T},
\end{equation}	 
which satisfies the symmetry property $\omega\circ \tau=\omega$. We also set  $\bs{\omega}=(0,\omega)\in C^\infty(\T)^2$.
  		
Let $\mc{C}^+:=f(\T^+)$ and denote by 
$$ {\bf D}_{\D^+,\mc{C}^+}: H^{1}_{\rm even}(\T)\to L^2_{\rm even}(\T)$$   the Dirichlet-to-Neumann 
operator associated to the curve $\mc{C}^+$ parametrised by $f:\T^+\to \mc{C}^+$. It is  defined as the symmetric operator satisfying
\[ \int_{0}^\pi ({\bf D}_{\D^+,\mc{C}^+}u)(e^{i\theta})u(e^{i\theta})\dd \theta:= \int_{\D^+}|dP_{\D^+,\mc{C}^+}u(x)|^2 \dd x \]
for each  real-valued $u\in H^{1}_{\rm even}(\T)$, where $P_{\D^+,\mc{C}^+}u$ denotes the harmonic function in $\D^+$ with Neumann condition at $[-1,1]$, Dirichlet condition at $\T^+$ and  boundary value $P_{\D^+,\mc{C}^+}u(f(e^{i\theta}))=u(e^{i\theta})$. By \cite[Lemma 5.6]{GRW1},  this operator extends continuously to $H^s_{\rm even}(\T)$ for $s\in \R$, and satisfies, for all $u\in H^s_{\rm even}(\T)$,
\begin{equation}\label{invDNmap}
({\bf D}_{\D^+,\mc{C}^+}u,G_{\D^+,{\rm m}}(f(\cdot),f(e^{i\theta}))_2=u(e^{i\theta}).
\end{equation}

Similarly to~\cite[Proposition 3.4]{BGKR1} we have the following:
\begin{proposition}\label{prop:Tf}
If $f\in \mc{S}_>^+$ then 
\begin{equation}\label{prop_amplitudeA_t^v}
{\bf T}^{+}_fF(\tilde{\varphi}^{\rm h}_1)= C^+_f
\int F(\tilde{\varphi}^{\rm h}_2)e^{Q({\bf D} \omega ,  \tilde{ \varphi}^{\rm h}_2 )_{2}} \mc{A}_{\mathbb{A}^+_f,g_{\mathbb{A}},\bzeta^+_f}(\tilde{\bs{\varphi}}^{\rm h}-Q\bs{\omega})\dd\mu^+_0(\tilde{\varphi}^{\rm h}_2)
\end{equation}
where $\tilde{\bs{\varphi}}^{\rm h}-Q\bs{\omega}=(\tilde{\varphi}^{\rm h}_1,\tilde{\varphi}^{\rm h}_2-Q\omega)$ and, with $c_{\rm L}=1+6Q^2$,  (observe here that $(\mathbf{D}\omega,\omega)_2=\frac{1}{2}(\mathbf{D}\omega,\omega)_{2,\mathrm{h}}$, and this  will be used in the whole proof)
		\begin{align}\label{formula: Cf+}
			C^+_f=(2\pi)^{-\frac{3}{4}}e^{\frac{1}{16}(\omega(1)+\omega(-1))+\frac{c_{\rm L}}{24}\log|f'(0)|+\frac{1}{12}({\bf D}\omega,\omega)_2}	.	
		\end{align}
	\end{proposition}

\begin{proof} The proof follows closely the argument of \cite[Proposition 3.4]{BGKR1}. For the reader's convenience, and since the constants differ from those appearing in \cite[Proposition 3.4]{BGKR1}, we reproduce the details below.

By \cite[Equation 6.6]{GRW1}, 
for any   bounded measurable functional $F$ on $H_{\rm even}^{-s}(\T)$ with $s>0$, one has
\begin{equation}\label{densityDN}
\E[F(X_{\D^+}^\# \circ f |_{\T})]=\frac{1}{ (2\pi)^{1/2}\det({\bf D}_{\D^+,{\mc{C}^+}}(2{\bf D}_0)^{-1})^{-1/2} }\int F( \tilde{ \varphi}^{\rm h})e^{-\frac{1}{2}(\tilde{ \varphi}^{\rm h}, ({\bf D} _{\D^+,\mc{C}^+} -2{\bf D})\tilde{ \varphi}^{\rm h})_2}\dd\mu^+_0  (\tilde{ \varphi}^{\rm h})
\end{equation}
where ${\bf D}_0:=\Pi_0+{\bf D}$  on $L^2(\T) $ and $ \Pi_0(\tilde\varphi^{\rm h}):=  ( \tilde\varphi^{\rm h},1)_2$ denotes the projection on the constants.
 
Applying the Cameron-Martin formula together with \eqref{invDNmap}, we obtain, for any  $\tilde{ \varphi}_1^{\rm h}\in H_{\rm even}^{-s}(\T)$,
\[
\E[F((X_{\D^+}^\#  + P_\D \tilde{\varphi}_1^{\rm h})\circ f |_{\T}+Q\omega)]=\E[F(X^\#_{\D^+}\circ f|_{\T})e^{({\bf D}_{\D^+,\mc{C}^+}(p+Q\omega),X^{\#}_{\D^+}\circ f)_2-\frac{1}{2}({\bf D}_{\D^+,\mc{C}^+}(p+Q\omega),p+Q\omega)_2}]
\] 
where $p:=P_{\D}\tilde{\varphi}_1^{\rm h}\circ f$ and the expectation is taken over $X^\#_{\D^+}$. 

Next, we use the identity
\begin{equation}\label{DD+mcB}
{\bf D}_{\D^+,\mc{C}^+}\omega=-k_{g_\mathbb{A}}\circ f+\pi_2({\bf D}_{\mathbb{A}^+_f}(0,\omega))
\end{equation} 
where $\pi_2$ denotes the projection onto the second component.  Indeed, this follows from the relations (with $\nu$ the inward pointing unit vector on $(\pl \D^+_f)_D$)
$$
(-\partial_\nu P_{\D_f}\omega)\circ f=-k_{g_\mathbb{A}}\circ f,\qquad  P_{\D_f}\omega=h,
$$
and the identity 
$$
k_{g_{\mathbb{A}}}\circ f(e^{i\theta})=-\frac{|f(e^{i\theta})| } {|f'(e^{i\theta})| }  \partial_{r}(h\circ f(re^{i\theta}))|_{r=1},
$$
see \cite[Eq. A.4]{BGKR1}.

Setting $X^\#:=X_{\D^+}^\#  + P_\D \tilde{\varphi}_1^{\rm h}$, we finally obtain
\begin{align*}
\E [F(X^\#\circ f |_{\T} +Q\omega)]=& e^{-\frac{Q^2}{2}({\bf D}_{\D^+,\mc{C}^+}\omega,\omega)_2-Q({\bf D}_{\mathbb{A}^+_f}(0,\omega) ,(0, p))_2-\tfrac{1}{2}({\bf D}_{\D^+,\mc{C}^+}p,p)_2}\\
& \times \E \Big[F(X^\#_{\D^+}\circ f|_{\T})e^{-\frac{Q}{2\pi}\int_{\mc{C}^+}k_{g_\mathbb{A}}(X_{\D^+}^\#-P_{\D^+}\tilde\varphi_1^{\rm h})\,\dd \ell_{g_\mathbb{A}} +Q({\bf D}_{\mathbb{A}^+_f}(0,\omega),(0, X_{\D^+}^\#\circ f))_2}
e^{({\bf D}_{\D^+,\mc{C}^+}p,X_{\D^+}^\#\circ f)_2}\Big]
\end{align*}

Using now \eqref{densityDN}, for any $\tilde{ \varphi}_1^{\rm h}\in H_{\rm even}^{-s}(\T)$, we obtain
\begin{align}
\label{computation1}
 &\E  [F(X^\#\circ f|_{\T}+Q\omega)]=  C e^{-\frac{Q^2}{2}({\bf D}_{\D^+,\mc{C}^+}\omega,\omega)_2-Q({\bf D}_{\mathbb{A}^+_f}(0,\omega) ,(0, p))_2}e^{\frac{Q}{2\pi}\int_{\mc{C}^+}k_{g_\mathbb{A}}P_\D\tilde\varphi_1^{\rm h}\,\dd \ell_{g_\mathbb{A}}}\\
 & \times \int F(\tilde{ \varphi}^{\rm h}_2)e^{-\frac{Q}{2\pi}\int_{\mc{C}}k_{g_\mathbb{A}}\tilde{\varphi}_2^{\rm h}\circ f^{-1}\,\dd \ell_{g_\mathbb{A}} +Q({\bf D}_{\mathbb{A}^+_f}(0,\omega) ,(0, \tilde{ \varphi}_2^{\rm h}))_2}  e^{({\bf D}_{\D,\mc{C}}p,\tilde{ \varphi}^{\rm h}_2)_2-\tfrac{1}{2}({\bf D}_{\D^+,\mc{C}^+}p,p)_2} e^{-\frac{1}{2}(\tilde{ \varphi}_2^{\rm h}, ({\bf D} _{\D^+,\mc{C}^+} -2{\bf D})\tilde{ \varphi}_2^{\rm h})_2}\dd\mu_0^+  (\tilde{ \varphi}_2^{\rm h})\nonumber
\end{align}
where  $$C=(2\pi)^{-1/2}\det({\bf D}_{\D^+,{\mc{C}^+}}(2{\bf D}_0)^{-1})^{1/2}.$$

\noindent\textbf{Step 1: Operator identities.}
Let $\mc C=f(\T)=\mc C^+\cup \tau(\mc C^+)$, parametrised by
$f:\T\to f(\T)$, and set
\[
p:=P_{\D}\tilde\varphi_1^{\rm h}\circ f .
\]
We identify functions on $\T^+$ with their even extensions to $\T$.
Likewise, boundary data on the half-annulus $\A_f^+$ are identified with
their even extensions to the doubled annulus $\A_f$. With this convention,
the mixed DN map on $\A_f^+$ is the restriction of the
Dirichlet-to-Neumann map on $\A_f$ to even boundary data.  More precisely, for
$\bs u,\bs v\in H^1_{\rm even}(\T)^2$, and for the pairings used in \eqref{unpairing},
\[
({\bf D}_{\A_f^+}\bs u,\bs v)_2
=
\frac12
({\bf D}_{\A_f}\bs u,\bs v)_{2,\ell}.\]
Let ${\bf D}_{\D,\mc C}$ denote  the symmetric Dirichlet-to-Neumann
operator associated with the interior curve $\mc C=f(\T)$. More precisely,
for $u\in H^1(\T)$, let $P_{\D,\mc C}u$ be the unique function which is
harmonic on the two components of $\D\setminus \mc C$, satisfies
\[
P_{\D,\mc C}u|_{\T}=0,
\qquad
P_{\D,\mc C}u|_{\mc C}\circ f=u,
\]
and has finite Dirichlet energy, characterized  by the
quadratic form
\[
2\pi({\bf D}_{\D,\mc C}u,u)_{2,\ell}
=
\int_{\D}|\nabla P_{\D,\mc C}u|^2\,\dd x.
\]

Applying \cite[Lemma A.1]{BGKR1} to the doubled domains and restricting to
the even sector, we obtain, for $\tilde{\bs\varphi}^{\rm h}
=
(\tilde\varphi_1^{\rm h},\tilde\varphi_2^{\rm h})$,
the identities
\begin{align}
( {\bf D}_{\D^+,\mc C^+}
(\tilde\varphi_2^{\rm h}-p),
\tilde\varphi_2^{\rm h}-p)_2
&=
( {\bf D}_{\A_f^+}\tilde{\bs\varphi}^{\rm h},
\tilde{\bs\varphi}^{\rm h})_2
-
({\bf D}\tilde\varphi_1^{\rm h},
\tilde\varphi_1^{\rm h})_2
+
({\bf D}\tilde\varphi_2^{\rm h},
\tilde\varphi_2^{\rm h})_2,
\label{identity1}
\\
({\bf D}_{\D^+,\mc C^+}\omega,p)_2
&=
-
({\bf D}_{\A_f^+}(0,\omega),
(\tilde\varphi_1^{\rm h},0))_2 .
\label{identity2}
\end{align}

\medskip
\noindent\textbf{Step 2: Simplifying the expectation.} Using  \eqref{identity1} and \eqref{computation1}, for any $\tilde{\varphi}_1^{\rm h}\in H^{-s}_{\rm even}(\T)$ with $s>0$, we obtain 
\[\begin{split}
\E [F(X^\#\circ f|_{\T}+Q\omega)]=& C \int F(\tilde{ \varphi}_2^{\rm h})e^{-\frac{Q}{2\pi}\int_{\mc{C}^+}k_{g_\mathbb{A}}(\tilde{ \varphi}^{\rm h}_2\circ f^{-1}-P_\D\tilde\varphi^{\rm h}_1)\,\dd \ell_{g_\mathbb{A}} +Q({\bf D}_{\mathbb{A}^+_f}(0,\omega) ,(0, \tilde{ \varphi}^{\rm h}_2))_2}\\
&\quad \times  e^{-\frac{Q^2}{2}({\bf D}_{\D^+,\mc{C}^+}\omega,\omega)_2-Q({\bf D}_{\mathbb{A}^+_f}(0,\omega) ,(0, p))_2}  e^{-\frac{1}{2}(\tilde{ \bs{\varphi}}^{\rm h} , ({\bf D} _{\mathbb{A}^+_f} - {\bf D})\tilde{ \bs{\varphi}}^{\rm h})_2}\dd\mu_0^+  (\tilde{ \varphi}^{\rm h}_2) .
\end{split}\]
where $\tilde{ \bs{\varphi}}^{\rm h}=(\tilde{\varphi}_1^{\rm h},\tilde{\varphi}_2^{\rm h})$. Moreover, by \eqref{identity1} and \eqref{DD+mcB},  
\[
-\frac{1}{2\pi}\int_{\mc{C}^+}k_{g_\mathbb{A}}P_\D\tilde\varphi^{\rm h}_1\,\dd \ell_{g_\mathbb{A}} +({\bf D}_{\mathbb{A}^+_f}(0,\omega), (0, p) )_2 =({\bf D}_{\D^+,\mc{C}^+}\omega,p)_2=-({\bf D}_{\mathbb{A}^+_f}(0,\omega), (\tilde{\varphi}^{\rm h}_1,0))_2.
\]
Substituting, we get
\begin{align*}
&\E [F(X^\#\circ f|_{\T}+Q\omega)]\\
&=C e^{-\frac{Q^2}{2}({\bf D}_{\D^+,\mc{C}^+}\omega,\omega)_2}\int F(\tilde{ \varphi}^{\rm h}_2)e^{-\frac{Q}{2\pi}\int_{\mc{C}^+}k_{g_\mathbb{A}}\tilde{ \varphi}_2^{\rm h}\circ f^{-1}\,\dd \ell_{g_\mathbb{A}} +Q({\bf D}_{\mathbb{A}^+_f}(0,\omega),\tilde{ \bs{\varphi}}^{\rm h})_2 -\frac{1}{2}(\tilde{\bs{\varphi}}^{\rm h}, ({\bf D}_{\mathbb{A}^+_f} - {\bf D})\tilde{ \bs{\varphi}}^{\rm h})_2}\dd\mu_0^+  (\tilde{ \varphi}^{\rm h}_2).
\end{align*}

\medskip
\noindent\textbf{Step 3: Conditional law and Markov property.} 
By the Markov property of the GFF, the conditional law of the field $X_{\D^+}$ in $\A^+_f$, given its boundary value   $X_{\D^+}\circ f=\tilde{\varphi}^{\rm h}_2$ on $\T^+$, is that of $X_{\A^+_f}+P_{\A^+_f} (0,\tilde{\varphi}^{\rm h}_2)$. Hence,  for any measurable function $G$ on  
$H^{-s}(\D^+)$ depending only on the field restricted to $\A_f^+$, 
\[ 
\E_{ \tilde{ \bs{\varphi}}^{\rm h}}[G((X_{\D^+}+P_\D \tilde{\varphi}^{\rm h}_1)|_{\A_f^+})]=\E_{ \tilde{ \bs{\varphi}}^{\rm h}}[G(X_{\A^+_f}+(P_{\A_f} \tilde{ \bs{\varphi}}^{\rm h}+P_{\A_f} (0,p))|_{\A^+_f})],
\] 
where $\E_{\tilde{ \bs{\varphi}}^{\rm h}}$ denotes    expectation conditional on $\tilde{ \bs{\varphi}}^{\rm h}=( \tilde{\varphi}_1^{\rm h},\tilde{\varphi}_2^{\rm h})$, and we used the relation $P_\D \tilde{\varphi}_1^{\rm h}=P_{\A_f} ( \tilde{\varphi}_1^{\rm h},p)$.

Applying this with the conditioning $X_{\D^+}\circ f=\tilde{\varphi}_2-p-Q\omega$ on $\T^+$, we find that $X^\#$ is distributed as 
$$ X_{\A^+_f}+P_{\A^+_f} \tilde{\bs{\varphi}}^{\rm h}-P_{\A_f}\bs{\omega}\quad \text{on } \A_f^+,\qquad  \bs{\omega}:=(0,\omega).$$

\medskip
\noindent\textbf{Step 4: Integration and cancellation of anomaly term.} 
Therefore
\begin{align*}
&\E_{\tilde{\varphi}^{\rm h}_1}\Big[F((X^\#|_{\T}\circ f+Q\omega ) e^{-\int_{\A^+_f}  M^{\boldsymbol{\mu}}_\gamma(X^\#,\dd x)}\Big]\\
&=Ce^{-\frac{Q^2}{2}({\bf D}_{\D^+,\mc{C}^+}\omega,\omega)_2}\int F( \tilde{ \varphi}^{\rm h}_2)H( \tilde{ \varphi}^{\rm h}_2) e^{-\frac{Q}{2\pi}\int_{\mc{C}^+}k_{g_\mathbb{A}}( \tilde{ \varphi}^{\rm h}_2\circ  f^{-1} )\,\dd \ell_{g_\mathbb{A}}+Q({\bf D}_{\A^+_f}\bs{\omega} ,  \tilde{\bs{ \varphi}}^{\rm h})_2 -\frac{1}{2}(\tilde{ \bs{\varphi}}^{\rm h} , ({\bf D}_{\A^+_f} - {\bf D})\tilde{ \bs{\varphi}}^{\rm h})_2}\dd\mu^+_0  (\tilde{ \varphi}^{\rm h}_2) \\
&= Ce^{J(\omega)}\int F( \tilde{ \varphi}^{\rm h}_2)H( \tilde{ \varphi}^{\rm h}_2) e^{-\frac{Q}{2\pi}\int_{\mc{C^+}}k_{g_\mathbb{A}}((\tilde{ \varphi}^{\rm h}_2-Q\omega)\circ  f^{-1} )\,\dd \ell_{g_\mathbb{A}} +Q({\bf D}\bs{\omega} ,  \tilde{\bs{ \varphi}}^{\rm h} )_2-\frac{1}{2}(\tilde{ \bs{\varphi}}^{\rm h}-Q\bs{\omega} , ({\bf D} _{\A^+_f} - {\bf D})(\tilde{ \bs{\varphi}}^{\rm h}-Q\bs{\omega}))_2}\dd\mu^+_0  (\tilde{ \varphi}^{\rm h}_2)
\end{align*}
where  $H( \tilde{ \varphi}^{\rm h}_2)=\E_{ \tilde{ \bs{\varphi}}^{\rm h}}[e^{-\int_{\A^+_f} M_{\gamma}^{\bs{\mu}}(\phi_{g_\A}-QP_{\A^+_f}\bs{\omega},\dd x)}]$
and 
\[\begin{split}
J(\omega)= &-\frac{Q^2}{2\pi}\int_{\mc{C}^+}k_{g_\A}\omega\circ f^{-1} d\ell_{g_\A}-\frac{Q^2}{2}(({\bf D}_{\D^+,\mc{C}^+}+{\bf D})\omega,\omega)_2+\frac{Q^2}{2}({\bf D} _{\mathbb{A}^+_f}\bs{\omega},\bs{\omega})_2\\
		& =-\frac{Q^2}{2}({\bf D}\omega,\omega)_2-\frac{Q^2}{4\pi}\int_{\mc{C}^+}k_{g_\A}\omega\circ f^{-1} \dd\ell_{g_\A}=0,
\end{split}\]
where we used \eqref{DD+mcB} and  since (recall $ k_{g_{\mathbb{A}}}\circ f(e^{i\theta})=-\frac{|f(e^{i\theta})| } {|f'(e^{i\theta})| }  \partial_{r}(h\circ f(re^{i\theta}))|_{r=1}$)
\[ \int_{\mc{C}^+}k_{g_\A}\omega\circ f^{-1} d\ell_{g_\A}=-\int_0^{\pi} \pl_r \Big(\log\big|r\frac{f'(re^{i\theta})}{f(re^{i\theta})}\big|\Big) 
\log|\frac{f'(re^{i\theta})}{f(re^{i\theta})}| d\theta =-2\pi ({\bf D}\omega,\omega)_2.\]

\medskip
\noindent\textbf{Step 5: Final identification and determinant computation.} 
By definition \eqref{amplitude},  
\[\mc{A}_{\mathbb{A}^+_f,g_{\mathbb{A}},\bzeta^+_f}(\tilde{\bs{\varphi}}^{\rm h}-Q\bs{\omega})=Z_{\mathbb{A}^+_f,g_{\mathbb{A}},{\rm m}}\int H( \tilde{ \varphi}^{\rm h}_2) e^{-\frac{Q}{2\pi}\int_{\mc{C}^+}k_{g_\mathbb{A}}((\tilde{ \varphi}^{\rm h}_2-Q\omega)\circ  f^{-1} )\,\dd \ell_{g_\mathbb{A}} -\frac{1}{2}(\tilde{ \bs{\varphi}}^{\rm h}-Q\bs{\omega} , ({\bf D} _{\A^+_f} - {\bf D})(\tilde{ \bs{\varphi}}^{\rm h}-Q\bs{\omega}))_2}\dd\mu_0^+(\tilde{\varphi}_2),\]
hence
\[{\bf T}^{+}_fF(\tilde{\varphi}^{\rm h}_1)=\frac{C}{Z_{\mathbb{A}^+_f,g_{\mathbb{A}},{\rm m}}}\mc{A}_{\mathbb{A}^+_f,g_{\mathbb{A}},\bzeta^+_f}(\tilde{\bs{\varphi}}^{\rm h}-Q\bs{\omega}).\]
It remains to compute $C/Z_{\mathbb{A}^+_f,g_{\mathbb{A}},{\rm m}}$.
Consider a metric $g_{\mc{C}^+}=e^{-2\rho}g_0$ (with $g_0=|\dd z|^2$) Neumann extendible with $\rho(z)=\log (|f'\circ f^{-1}(z)|.|f^{-1}(z)|)$ near $\mc{C}^+$, so that $g_{\mc{C}^+}$ is admissible near $\mc{C}^+$.
By \cite[Proposition B.3]{GRW1},
\[ \det(\Delta_{\D^+,{\rm m},g_{\mc{C}^+}})=\det(\Delta_{\A_f^+,{\rm m},g_{\mc{C}^+}})\det(\Delta_{\D_f^+,{\rm m},g_{\mc{C}^+}}){\det}_{\rm Fr}({\bf D}_{\D^+,\mc{C}^+}(2{\bf D}_0)^{-1})(2\pi)^{1/2}.
\]
By \eqref{formula: anomaly corner}, we get 
\[ \begin{split}
 \frac{\det(\Delta_{\D^+,{\rm m},g_{\mc{C}^+}})^{1/2}}{\det(\Delta_{\D_f^+,{\rm m},g_{\mc{C}^+}})^{1/2}\det(\Delta_{\A_f^+,{\rm m},g_{\mc{C}^+}})^{1/2}} =& \frac{\det(\Delta_{\D^+,{\rm m},g_0})^{1/2}}{\det(\Delta_{\D_f^+,{\rm m},g_0})^{1/2}\det(\Delta_{\A_f^+,{\rm m},g_0})^{1/2}}e^{\frac{1}{8}\log |f'(1)f'(-1)|}\\
\end{split}. \]
 By using that $(\D_f^+,g_0)$ is isometric to $(\D^+,f^*g_0)$,   the symmetry of $\bar{f}(\bar{z})=f(z)$, and again \eqref{formula: anomaly corner}, one has
\[\begin{split} 
\frac{\det(\Delta_{\D^+,{\rm m},g_0})^{1/2}}{\det(\Delta_{\D_f^+,{\rm m},g_0})^{1/2}}=& e^{\frac{1}{24\pi} \int_{\D^+}|\nabla \log |f'||^2\dd x
+\frac{1}{12\pi}\int_{\T^+}\log|f'|\dd\theta -\frac{1}{16}\log(|f'(1)|.|f'(-1)|)}\\
=& e^{\frac{1}{48\pi} \int_{\D}|\nabla \log |f'||^2\dd x+\frac{1}{12}\log |f'(0)| 
 -\frac{1}{16}\log(|f'(1)|.|f'(-1)|)}\\
\frac{\det(\Delta_{\A_f^+,{\rm m},g_\A})^{1/2}}{\det(\Delta_{\A_f^+,{\rm m},g_0})^{1/2}}= & e^{\frac{1}{48\pi}\int_{\mathbb{A}_f}|\nabla \log |x||^2 \dd x+\frac{1}{24\pi}\int_{\mc{C}^+}k_{g_\A}\log |x|\dd\ell_{g_\A}-\frac{1}{16}\log(|f(1)|.|f(-1)|)}.
\end{split}\]
By combining these identities, we find that
\[
 \frac{{\det}_{\rm Fr}({\bf D}_{\D^+,\mc{C}^+}(2{\bf D}_0)^{-1})}{(2\pi)^{-1/4}\det(\Delta_{\A_f^+,{\rm m},g_\A})^{1/2}}=e^{\frac{1}{48\pi} \int_{\D}|\nabla \log |f'||^2\dd x+\frac{1}{12}\log |f'(0)| 
 +\frac{1}{16}(\omega(1)+\omega(-1))+\frac{1}{48\pi}\int_{\mathbb{A}_f}|\nabla \log |x||^2 \dd x+\frac{1}{24\pi}\int_{\mc{C}^+}k_{g_\A}\log |x|\dd\ell_{g_\A}}.
 \]
The last few lines of the proof of  \cite[Proposition 3.4]{BGKR1} show that 
\begin{align*}
&\frac{1}{48\pi}\Big( \int_{\D}|\nabla \log |f'||^2\dd x+\int_{\mathbb{A}_f}|\nabla \log |x||^2 \dd x+2\int_{f(\T)}k_{g_\A}\log |x|\dd\ell_{g_\A}\Big)\\
&=  \frac{1}{48\pi}\int_\T {\bf D} \omega(e^{i\theta}). \omega (e^{i\theta})\dd\theta -\frac{1}{24}\log |f'(0)|.\end{align*}
Taking into account the prefactor $|f'(0)|^{Q^2/4}$ in the definition
\eqref{FKv} of ${\bf T}_f^+$, we obtain
\[
C_f^+
=
|f'(0)|^{\frac{Q^2}{4}}
\frac{{\det}_{\rm Fr}({\bf D}_{\D^+,\mc{C}^+}(2{\bf D}_0)^{-1})^{1/2}}
{\sqrt{2\pi}}.
\]
Hence, using $c_{\rm L}=1+6Q^2$,
\[
C_f^+
=
(2\pi)^{-\frac{3}{4}}
\exp\left(
\frac{1}{16}(\omega(1)+\omega(-1))
+
\frac{c_{\rm L}}{24}\log|f'(0)|
+
\frac{1}{48\pi}\int_\T {\bf D}\omega\,\omega\,d\theta
\right).
\]
Equivalently, with our convention
\[
\frac{1}{48\pi}\int_\T {\bf D}\omega\,\omega\,d\theta
=
\frac{1}{12}({\bf D}\omega,\omega)_2,
\]
this is exactly \eqref{formula: Cf+}.
\end{proof}
	
By  combining Proposition \ref{prop:Tf} and  Proposition \ref{Weyl2}, we obtain the following:
\begin{proposition}\label{cor:annulus_propagator}
Let $f\in \mc{S}^+_>$ and $g_f$ be the admissible metric given by \eqref{defg_tadmissible}. Then we have for $c_{\rm L}=1+6Q^2$
		\[ (2\pi)^{3/4} e^{\frac{c_\mathrm{L}}{12}W^+(f,g_f)} {\bf T}^{+}_fF(\tilde{\varphi}^{\rm h}_1)= 
		\int \mc{A}_{\mathbb{A}^+_f,g_f,\bzeta^+_f}(\tilde{\varphi}^{\rm h}_1,\tilde{\varphi}^{\rm h}_2)F(\tilde{\varphi}^{\rm h}_2) \,\dd\mu_0^+(\tilde{\varphi}^{\rm h}_2)
		\]
		\begin{equation}\label{def: W+}
			\textrm{ with }\quad W^+(f,g):= -\frac{1}{2}\log|f'(0)|-\frac{1}{2}\Big({\bf D}\log \frac{|f'|}{|f|}|_{\T},\log \frac{|f'|}{|f|}|_{\T}\Big)_{2,\mathrm{h}}-12S_{\rm L}^0(\mathbb{A}^+_f,g,g_{\mathbb{A}}).
		\end{equation}
	
	\end{proposition}

\begin{remark}
We   remark here the relation $W^+(f,g)=\frac{1}{2}W(f,g)$, where the latter was introduced in~\cite[eq. (3.12)]{BGKR1}.
\end{remark}

\subsection{Differentiability of propagators}
 
In this subsection, we study the differentiability of the map $f\mapsto {\bf T}^{+}_f$ under the assumptions given by \eqref{assumptions_gamma_mu}. Naturally, when $\mu=0$, the relevant case is $\mu^{\rm L}+\mu^{\rm R}>0$; otherwise, the situation reduces to the free field theory.

The main result of this section is the following:
\begin{proposition}\label{th:differentiabiliteTf}
The mapping $f\in \mc{S}_\epsilon^+\mapsto {\bf T}^{+}_f  \in \mc{L}(\mc{D}(\mc{Q}^+),\mc{D}(\mc{Q}^+))$   is differentiable with differential  
$$v\in  {{\rm Hol}^{+,\bullet}_\epsilon(\D)}\mapsto D_{v}{\bf T}^{+}_f \in \mc{L}(\mc{D}(\mc{Q}^+),\mc{D}(\mc{Q}^+))$$
			characterized by 
$$\forall F\in \mc{D} (\mc{Q}^+),G\in \mc{D}'(\mc{Q}^+),\quad \cjg D_{v}{\bf T}^{+}_fF,G\cjd_{\mc{H}^+} : = -\mc{Q}^+_{\rm w}(F,{{\bf T}^{+}_f}^* G) $$
			where $\mc{Q}^+_{\rm w}$ is the closure of the $\mc{Q}^+$-continuous bilinear form defined on $\cC_c(\R\times \Omega_{\T^+})\times\cC_c(\R\times \Omega_{\T^+})$ by
			\begin{equation} 
				\mc{Q}^+_{\rm w}(F,G):=-\sum_na_n\cjg{\bf L}^+_nF,G\cjd_{\mc{H}^+} 
			\end{equation}
with 
$${\rm w}=w(z)\pl_z=\frac{v(z)}{f'(z)}\partial_z,\qquad \frac{v(z)}{f'(z)}=\sum_{n\geq 0} a_nz^{n+1},\quad \text{for }|z|=1.$$  Otherwise stated,  $D_{v}{\bf T}^{+}_f=-{\bf T}^{+}_f{\bf H}^{+}_{\rm w}:\mc{D}(\mc{Q}^+)\to \mc{D}(\mc{Q}^+)$  with the operator ${\bf H}^{+}_{\rm w}:\mc{D}(\mc{Q}^+)\to \mc{D}'(\mc{Q}^+)$ defined as described in \eqref{defHv^+} and \eqref{eq:def_L+}.
		\end{proposition}
Before studying the differentiability, we state a few technical results. First we study the continuity of the map $f\in S_\epsilon^+\mapsto {\bf T}^{+}_f$.

\begin{lemma}\label{lemcont+}
For each $\epsilon>0$, the mapping $f\in  \mc{S}_\epsilon^+ \mapsto \mathbf{T}^{+}_f\in\mathcal{L}(\mathcal{H}^+)$ is continuous.
\end{lemma}

\begin{proof} We adapt (and follow the same strategy as) the proof of \cite[Proposition 3.7]{BGKR1} to the boundary setting. Recall from Lemma \ref{contbyold}   that, for any $f\in \mc{S}^+$,  the operator $ \mathbf{T}^{+ }_f$ extends to a bounded operator on $\mathcal{H}^+$. 	
Proposition \ref{cor:annulus_propagator} shows that the continuity of the map follows from the continuity of the operator with  integral kernel given by the amplitude associated to $ \mathbf{T}^{+}_f$, since  the prefactor $W^+(f,g_f)$   appearing there  is continuous in $f\in\mc{S}_\epsilon^+$ for each $\epsilon>0$. Hence, it suffices to prove the continuity of the operator whose integral kernel is the amplitude
\begin{equation}\label{defTamplitude}
\mc{A}_{\mathbb{A}^+_f,g_f,\bzeta_f^+}(\tilde{\bs{\varphi}}^{\rm h})= C(f) e^{ -\frac{1}{2}(\tilde{ \bs{\varphi}}^{\rm h} , (\mathbf{D} _{\A^+_f} - \mathbf{D})\tilde{ \bs{\varphi}}^{\rm h})_2}  \E_{\tilde{\bs{\varphi}}^{\rm h}} \Big[  e^{-\frac{Q}{4\pi}\int_{\A_f^+}K_{g_f}(X^\#_{\D^+}+P\tilde{ \bs{\varphi}}^{\rm h}){\rm dv}_{g_f}} e^{- M^{g,\bs{\mu}}_\gamma (X_{\D^+}^\#+P\tilde{ \bs{\varphi}}^{\rm h},  \A^+_f)   }  \Big]
\end{equation}
where $C(f):=Z_{\A^+_f,g_f,{\rm m}}$  (recall  \eqref{znormal}), $\tilde{\bs{\varphi}}^{\rm h}= (\tilde \varphi_1^{\rm h}, \tilde \varphi_2^{\rm h})= (c_1+\varphi_1^{\rm h},c_2+\varphi_2^{\rm h})$,    and $X_{\D^+}^\#$ is the GFF with mixed b.c.   on  $ \D^+$.
 
First, we claim that the map $f\in \mc{S}^+_\epsilon\mapsto  C(f)=Z_{\A^+_f,g_f,{\rm m}}$ is continuous. 
This follows from the continuity in $f$ of the heat trace ${\rm Tr}(e^{-t\Delta_{\A^+_f,g_f,{\rm m}}})$ for the Laplacian on $(\A^+_f,g_f)$ with mixed boundary condition, and from the continuity in $f$ of the coefficients in its small-time asymptotic expansion. 

Indeed, the heat kernel on $(\A^+_f,g_f)$ can be expressed via the Neumann double $(\A_f,g_f)$ of $(\A^+_f,g_f)$, obtained by symmetric reflection across $\R$, as 
$$
K(t,z,z')=e^{-t\Delta_{\A_f,g_f,{\rm D}}}(z,z')+e^{-t\Delta_{\A_f,g_f,{\rm D}}}(z,\bar{z}').
$$
Consequently, the coefficients (for $j=0,1,2$) of the asymptotic expansion
$$\int_{\A^+_f}K(t,z,z)\,\dd {\rm v}_{g_f}(z)\sim \sum_{j=0}^\infty a_j(f)t^{-1+j/2},\qquad t\to 0^+$$
are smooth functions of the metric coefficients of $g_f$, hence continuous in $f$, provided  $g_f$ is  admissible. This property is shown in the proof of \cite[Lemma B.1]{GRW1} for the local contributions coming from neighborhoods of the corners, while the regular part of the expansion away from the corners follows from the classical analysis of McKean-Singer \cite{McKean-Singer}. Thus, the partition function $C(f)=Z_{\A^+_f,g_f,{\rm m}}$ depends continuously  on $f\in  \mc{S}_\epsilon^+ $.

We consider a   sequence $(f_n)_n$ converging to $f$ in  $\mc{S}^+_\epsilon$. To simplify notations, set 
$$(\Sigma_n,g_n):=(\A_{f_n }^+,g_{f_n}),\qquad (\Sigma,g):=(\A^+_f,g_f),$$
and denote by  $X_n$ and $X$ the mixed b.c. GFFs on $\Sigma_n$ and $\Sigma$,  respectively. Let $P_n \tilde{\bs{\varphi}}^{\rm h}$ and $P \tilde{\bs{\varphi}}^{\rm h}$ be the harmonic extensions on $\Sigma_{n}$ and $\Sigma$ of  $\tilde{\bs{\varphi}}^{\rm h}$; the measures $ {\rm dv}_{g_{n}}$ and $ {\rm dv}_{g_f}$ (resp. $\dd \ell_{g_{n}}$ and $ \dd \ell_{g}$) are denoted  ${\rm dv}_{n}$ and $ {\rm dv}$ (resp. $\dd\ell_n$ and $ \dd\ell$).  Finally, write   $K_n$, $K$ for the scalar curvatures of $g_{n}$, $g$, and $\bzeta^+_n,\bzeta^+$ for the parametrisations of $\pl \Sigma_n$ and $\pl \Sigma$ induced by $({\rm Id},f_n)$ and $({\rm Id},f)$.

For $\xi>0$ small,  define  
$$\Sigma^\xi=\{x\in \Sigma\,|\, d(x,\pl \Sigma_{\rm D})\geq \xi\},$$ where $d(\cdot,\cdot)$ denotes the Euclidean distance. 
For $n$ large enough, $\Sigma^\xi \subset \Sigma_n$. We will decompose each amplitude into a part supported away from the Dirichlet boundary and a part localised near it, and show that both contributions converge appropriately. 

For each $n$ and $\xi>0$, we write
$$\mc{A}_{\Sigma_n, g_n,\bzeta^+_n}( \tilde{\bs{\varphi}})= A^{(1)}_{n,\xi}+A^{(2)}_{n,\xi},$$  where
\begin{equation*}
 A^{(1)}_{n,\xi}= C(f_n)e^{ -\frac{1}{2}(\tilde{ \bs{\varphi}}^{\rm h}, (\mathbf{D} _{\Sigma_{n}} - \mathbf{D})\tilde{ \bs{\varphi}}^{\rm h})_2}  \E_{ \tilde{\bs{\varphi}}^{\rm h}} \left  [  e^{-\frac{Q}{4\pi}\int_{\Sigma_{n}}K_{n} (X_{n}+P_n\tilde{ \bs{\varphi}}^{\rm h}){\rm dv}_{n} }   e^{- M_{\gamma}^{g_n,\bs{\mu}} (X_n+P_n\tilde{ \bs{\varphi}}^{\rm h}, \Sigma^\xi)     }    \right]
\end{equation*}
and 
\begin{align*}
 A^{(2)}_{n,\xi}=& C(f_n)e^{ -\frac{1}{2}(\tilde{ \bs{\varphi}}^{\rm h}, (\mathbf{D}_{\Sigma_{n}} - \mathbf{D})\tilde{ \bs{\varphi}}^{\rm h})_2}  
 \E_{ \tilde{\bs{\varphi}} } \left  [   e^{-\frac{Q}{4\pi}\int_{\Sigma_{n}}K_{n} (X_{n}+P_n\tilde{ \bs{\varphi}}^{\rm h}){\rm dv}_{n} } \big( e^{-  M_{\gamma}^{g_n,\bs{\mu}} (X_n+P_n\tilde{ \bs{\varphi}}^{\rm h}, \Sigma_n)     } -  e^{- M_{\gamma}^{g_n,\bs{\mu}} (X_n+P_n\tilde{ \bs{\varphi}}^{\rm h}, \Sigma^\xi)    }  \big)  \right].
\end{align*}

Now we claim that 
$$
\iint | A^{(2)}_{n,\xi}|^2 (\dd \mu^+_0)^{\otimes 2}\leq C\xi^{1/p}$$
for some $p>1$ and an irrelevant constant $C>0$, so that the operator with integral kernel $A^{(2)}_{n,\xi}$ has operator norm $\mathcal{H}^+\to\mathcal{H}^+$ smaller than $C\xi^{1/p}$. 

This double integral can be interpreted as the gluing of the amplitude with itself, by the gluing lemma  \cite[Prop. 6.2]{GRW1}. Indeed, let us denote by $\Sigma_n'$  the surface obtained by gluing $(\Sigma_{n},g_n,\bs{\zeta}_n^+)$ 
with the surface $(\Sigma_{n},g_n,\bs{\zeta}_n^+\circ(- o))$ (recall $o(z)=1/z$) in the following way:  the incoming D-marked boundary $f_n(\T^+)$ in $(\Sigma_{n},g_n,\bs{\zeta}_n^+)$ is glued to the outgoing D-marked half-circle $f_n(\T^+)$ of 
$(\Sigma_{n},g_n,\bs{\zeta}_n^+\circ(- o))$, and the outgoing D-marked half-circle $\T^+$ of $(\Sigma_{n},g_n,\bs{\zeta}_n^+)$ is glued to the incoming D-marked half-circle $\T^+$ of $(\Sigma_{n},g_n,\bs{\zeta}_n^+\circ(- o))$. 

We keep the notation  $g_n$ and $K_n$ for the resulting  metric and scalar curvature on $\Sigma_n'$, which is a cylinder with two N-marked  boundaries (note that this is the double of $\Sigma_n$, but not a Neumann double), equipped with an antiholomorphic involution denoted by $\tau$ fixing two embedded half-circles. We write $\Sigma_n$ and $\tau(\Sigma_n)$ for the two copies of $\Sigma_n$ contained  inside $\Sigma_n'$. We still denote by $g_n$ and $K_n$ the induced metric and scalar curvature, and $\dd {\rm v}_n$ the Riemannian measure on $\Sigma_n'$. 
By the gluing lemma \cite[Prop. 6.2]{GRW1} (remark that  we use it in the non standard case when $\mu=0$; yet it still holds), combined with the elementary  inequality  $1-e^{- x} \leq   (x^{\eta}\wedge 1)$ for $\eta\in (0,1)$, we deduce that (see \cite[Proposition 3.7]{BGKR1} for details)
\begin{align*}
\iint & | A^{(2)}_{n,\xi}|^2 (\dd \mu^+_0)^{\otimes 2}\\
\leq &C'(f_n)\int \E\Big[e^{-\frac{Q}{4\pi}\int_{\Sigma'_n}K_n(c+X'_n){\rm dv}_{n}}\big(M_{\gamma}^{g_n,\bs{\mu}} (c+X_n' ,  \Sigma_n \setminus  \Sigma^\xi)^\eta\wedge 1\big)\,
\big(M_{\gamma}^{g_n,\bs{\mu}} (c+X_n'  , \tau( \Sigma_n \setminus \Sigma^{\xi})^\eta\wedge 1\big)\,
\\
&   \qquad \qquad\qquad \qquad  \times e^{-  M_{\gamma}^{g_n,\bs{\mu}} (c+X_n' , \Sigma^{\xi}) -M_{\gamma}^{g_n,\bs{\mu}} (c+X_n' ,\tau  (\Sigma^{\eps}))   }  \Big]  \,\dd c 
   \\
   \leq &C'(f_n)   \int \E\Big[e^{-\frac{Q}{4\pi}\int_{\Sigma'_n}K_n(c+X'_n){\rm dv}_{n}}\big(M_{\gamma}^{g_n,\bs{\mu}} (c+X_n'  , \tau( \Sigma_n \setminus \Sigma^{\xi})\big)^\eta
   e^{  - M_{\gamma}^{g_n,\bs{\mu}} (c+X_n' ,\tau  (\Sigma^{\xi}))    }  \Big]  \,\dd c 
\end{align*}
where $C'(f_n):=\big(\frac{{\rm v}_{g_n}(\Sigma_n')}{{\det}'(\Delta_{\Sigma_n',g_n,{\rm N}})}\big)^\hf$ is a metric dependent constant, and $X'_n$ is the  Neumann GFF on $\Sigma_n'$ in the glued metric $g_n$. 

On the cylinder  $\Sigma_n'$, the Gauss-Bonnet formula implies that $\int_{\Sigma_n'}K_n {\rm dv}_{n}=0$, so that the   contribution of the constant mode $c$ in the integral of the first exponential term vanishes. The curvature term involving the Neumann GFF can be handled using   the Cameron-Martin theorem, which induces a shift of the   GFF law by a Green-type function   $G'(f_n)(x):=\int_{\Sigma_n'} K_n(x')G'_{n,{\rm N}}(x,x')  {\rm dv}_{n}(x')$, where $G'_{n,{\rm N}}$ denotes the Neumann  Green function on $(\Sigma_n',g_n)$. The variance of this Cameron-Martin shift is then given by   $V'(f_n):=\int_{\Sigma_n'}K_{n}  G'(f_n)  {\rm dv}_{n} $.  This produces
\begin{multline*}
\iint  | A^{(2)}_{n,\xi}|^2 (\dd \mu_0^+)^{\otimes 2}  \leq  C'(f_n) e^{\frac{Q^2}{32\pi^2}V'(f_n)}
\\
\int \E\Big[ \big(M_{\gamma}^{g_n,\bs{\mu}} (c+X'_{n}+G'(f_n) ,  \Sigma_n \setminus \Sigma^\xi)\big)^\eta 
   e^{  -M_{\gamma}^{g_n,\bs{\mu}}  (c+X'_{n}+G'(f_n),\tau  (\Sigma^\xi))   }  \Big]  \,\dd c 
\end{multline*}
  Recall   that $M_{\gamma}^{g_n,\bs{\mu}}$  denotes the sum  of three GMC: the bulk GMC and the two boundary GMCs, each  multiplied by their respective cosmological constant $\mu,\mu^{\rm L},\mu^{\rm R}$. Recall also the assumption \eqref{assumptions_gamma_mu} (and the discussion following it), where we assume that either $\mu>0$, or $\mu=0$ with $\mu^{\rm L}+\mu^{\rm R}>0$. In particular, at least one of these constants is strictly positive.
 
  For instance, we may assume $\mu>0$, the other cases can be treated analogously. Under this assumption, the expectation above is bounded by the same expression where $M_{\gamma}^{g_n,\bs{\mu}} $ is replaced by its bulk component in the exponential term.  Using the subadditivity of the map $x\in\R_+\to x^\eta$, the last term in the right-hand side can therefore be estimated  in terms of  three contributions corresponding to the bulk and boundary GMCs, namely:
\[\int \E\Big[ M_\gamma^{g_n,\bs{\mu}} (c+X'_n+G'(f_n) ,  \Sigma_{n} \setminus \Sigma^\xi)^\eta 
   e^{  - M_\gamma ^{g_n,\bs{\mu}} (c+X'_n+G'(f_n),\tau  (\Sigma^\xi))   }  \Big]  \,\dd c \leq T_n +T_n^{\rm L}+T_n^{\rm R},\]
   where
\begin{align*}
   T_n:=&\mu \int \E\Big[   M_{\gamma}^{g_n} (c+X'_n+G'(f_n) ,  \Sigma_{n} \setminus \Sigma^\xi)^\eta 
   e^{  -\mu M_{\gamma }^{g_n} (c+X'_n+G'(f_n),\tau  (\Sigma^\xi))   }  \Big]  \,\dd c,
   \\
   T_n^{\rm L}:=&\mu^{\rm L}\int \E\Big[ M_\gamma^{{\rm b},g_n} (c+X'_n+G'(f_n) ,  (\Sigma_{n} \setminus \Sigma^\xi)_{\rm L})^\eta 
   e^{  -\mu M_{\gamma }^{g_n} (c+X'_n+G'(f_n),\tau  (\Sigma^\xi))   }  \Big]  \,\dd c,
   \\
   T_n^{\rm R}:=&\mu^{\rm R} \int \E\Big[ M_\gamma^{{\rm b},g_n} (c+X'_n+G'(f_n) ,  (\Sigma_{n} \setminus \Sigma^\xi)_{\rm R})^\eta 
   e^{  -\mu M_{\gamma }^{g_n} (c+X'_n+G'(f_n),\tau  (\Sigma^\xi))   }  \Big]  \,\dd c
\end{align*}
where the meaning of  $ (\Sigma_{n} \setminus \Sigma^\xi)_{\rm L}$ and $ (\Sigma_{n} \setminus \Sigma^\xi)_{\rm R}$ is as follows: $\Sigma_{n} $ has two Neumann boundary parts, tagged with ${\rm L}$ or ${\rm R}$, so these two sets denote the intersections of $\Sigma_{n} \setminus \Sigma^\xi$ with each of these parts. All of the three terms above can be treated in a similar way, so we   focus on the bulk contribution $   T_n$. We first perform  the change of variables $y=e^{\gamma c}M_{\gamma }^{g_n} (X'_n+G'(f_n),\tau  (\Sigma^\xi))$, then integrate out the variable  $y$, and finally apply   H\"older's inequality  to obtain
 \begin{align*}
T_n
= & \frac{\mu}{\gamma}\Gamma(\eta)  \E\Big[ M_{\gamma}^{g_n} (X_{n}'+G'(f_n) ,  \Sigma_n \setminus \Sigma^\xi)^\eta 
    M_{\gamma }^{g_n} (X'_n+G'(f_n),\tau  (\Sigma^\xi))^{-\eta   }  \Big]   
    \\
 \leq &   \frac{\mu}{\gamma}\Gamma(\eta)  \E\Big[ M_{\gamma}^{g_n} (X'_n+G'(f_n) ,  \Sigma_n \setminus \Sigma^\xi)^{p \eta}\Big]^{1/p} \E\Big[    M_{\gamma }^{g_n} (X'_n+G'(f_n),  \Sigma^\xi)^{-\eta q  }  \Big]   ^{1/q}
\end{align*}
where $p,q> 1$ are conjugate exponents.  Furthermore, since $f_n\to f$ in $\mc{S}_\eps^+$ as $n\to \infty$, by \cite[Lemma B.1]{BGKR1} there is $C>0$ such that for each $n$
\begin{equation}\label{estgreen}
\sup_n\sup_{x\in\Sigma_{n} }|G'(f_n)(x)|\leq C.
\end{equation}
Then we get 
\begin{align*}
T_n
 \leq    \frac{\mu}{\gamma}\Gamma(\eta)e^{2\gamma C }  \E\Big[ M_{\gamma }^{g_n} (X'_n ,  \Sigma_{n} \setminus \Sigma^\xi)^{p\eta} \Big]^{1/p} \E\Big[    M_\gamma^{g_n} (X'_n,  \Sigma^\xi)^{-\eta q  }  \Big]   ^{1/q}. 
 \end{align*}
We have to evaluate both expectations above. First, we treat the first expectation. We choose $p$ so that $p\eta=1$ in order to compute easily this expectation. Recall  that, for $\mu>0$, we assume $\gamma^2<2$, and this ensure that the expectation is finite when $p\eta=1$. By item (1) of Lemma \ref{lemgreen} below and for some irrelevant constant $C$ (which may change along lines),
 \begin{align*}
 \E\Big[ M_{\gamma}^{g_n} (X'_n ,  \Sigma_{n} \setminus \Sigma^\xi) \Big]   \leq C\int_{\Sigma_{n} \setminus \Sigma^\xi  } e^{\frac{\gamma^2}{2}W_{n,g,{\rm N}}} \dd {\rm v}_n    \leq C \xi,
\end{align*}
where $W_{n,g,{\rm N}}$ is the Robin mass on $\Sigma'_n$, see \cite[eq. (2.13) and below]{GRW1}. 

Second, we focus on the second expectation. Consider a non-empty open set $\mc{O}\subset  \Sigma^\xi$ over which $\chi_{f_n}=0$ for all $n$ (recall \eqref{defg_tadmissible}), then 
\[\E\Big[    M_{\gamma}^{g_n} (X'_n,  \Sigma^\xi)^{-\eta q  }  \Big] \leq \E\Big[    M_{\gamma}^{g_{\A}} (X'_n ,  \mc{O})^{-\eta q  }  \Big] .\]
The last expectation can then be shown to be bounded independently of $n$ using the standard argument of Kahane's convexity equality (see e.g. \cite{rhodes2014_gmcReview}), by using item (2) of Lemma \ref{lemgreen}. 
%
%

 The argument is the same for $T_n^{\rm L}$ and $T_n^{\rm R}$. All in all, and taking into account that $ C'(f_n)$ is bounded uniformly in $n$ since $f\mapsto C'(f)$ is continuous, we get that 
\[ \iint   | A^{(2)}_{n,\xi}|^2 (\dd \mu_0^+)^{\otimes 2} \leq C\xi^{1/p}.\]
 This proves that the operator with integral kernel $A^{(2)}_{n,\xi}$ can be made arbitrarily small in operator norm with $\xi$, uniformly in  $n$.

We can write similarly  $\mc{A}_{\mathbb{A}_f^+,g_f,\bzeta^+}( \tilde{\bs{\varphi}}^{\rm h})=  A^{(1)}_{\xi}+A^{(2)}_{\xi}$  where   $A^{(2)}_{\xi}$ yields an operator on $\mc{H}^+$ whose norm is less or equal to $C \xi^{1/p}$ and  $A^{(1)}_{\xi} $ is given by 
\begin{equation*}
A^{(1)}_{\xi} = C(f)e^{ -\frac{1}{2}(\tilde{ \bs{\varphi}}^{\rm h} , (\mathbf{D}_{\Sigma} - \mathbf{D})\tilde{ \bs{\varphi}}^{\rm h})_2}  \E_{ \tilde{\bs{\varphi}}^{\rm h}} \left  [ e^{-\frac{Q}{4\pi}\int_{\Sigma}K(X+P\tilde{ \bs{\varphi}}^{\rm h}){\rm dv}  }   e^{-  M_\gamma^{g,\bs{\mu}} (X+P\tilde{ \bs{\varphi}}^{\rm h},  \Sigma^{\xi})   }  \right].
\end{equation*}
We will estimate $A^{(1)}_{n,\xi}-A^{(1)}_{\xi}$ and, since $C(f)$ is continuous in $f\in \mc{S}^+_\epsilon$, we can forget this term and 
replace $C(f)$ by $1$ in the definition of $A^{(1)}_{\xi}$. 
We write $A^{(1)}_{n,\xi}-A^{(1)}_{\xi}= B^{(1)}_{n,\xi}+B^{(2)}_{n,\xi} $ where 
 \begin{multline*}
B^{(1)}_{n,\xi}= e^{ -\frac{1}{2}(\tilde{ \bs{\varphi}}^{\rm h} , (\mathbf{D}_{\Sigma} - \mathbf{D})\tilde{ \bs{\varphi}}^{\rm h})_2}  \Big( \E_{ \tilde{\bs{\varphi}}^{\rm h}} \Big[ e^{-\frac{Q}{4\pi}\int_{\Sigma_n}K_{n} (X_n+P_n\tilde{ \bs{\varphi}}^{\rm h}){\rm dv}_{n}-   M_\gamma^{g_n,\bs{\mu}}(X_n+P_n \tilde{ \bs{\varphi}}^{\rm h},  \Sigma^\xi)   }  \Big]
 \\
 -\E_{ \tilde{\bs{\varphi}}^{\rm h}}  \Big[   e^{-\frac{Q}{4\pi}\int_{\Sigma}K (X+P\tilde{ \bs{\varphi}}^{\rm h}){\rm dv} -  M_\gamma^{g,\bs{\mu}}(X+P\tilde{ \bs{\varphi}}^{\rm h},  \Sigma^\xi)   }  \Big]   \Big)
\end{multline*}
and 
\begin{equation*}
B^{(2)}_{n,\xi} = ( e^{ -\frac{1}{2}(\tilde{ \bs{\varphi}}^{\rm h} , (\mathbf{D}_{\Sigma_{n}} - \mathbf{D})\tilde{ \bs{\varphi}}^{\rm h})_2} - e^{ -\frac{1}{2}(\tilde{ \bs{\varphi}}^{\rm h} , (\mathbf{D}_{\Sigma} - \mathbf{D})\tilde{ \bs{\varphi}}^{\rm h})_2} )
 \E_{ \tilde{\bs{\varphi}}^{\rm h}} \Big  [  e^{-\frac{Q}{4\pi}\int_{\Sigma_n}K_{n} (X_n+P_n\tilde{ \bs{\varphi}}^{\rm h}){\rm dv}_{n} }   e^{-  M^{g_n,\bs{\mu}}_\gamma (X_n+P_n \tilde{ \bs{\varphi}}^{\rm h},  \Sigma^\xi)   }  \Big].
\end{equation*}
Concerning $B^{(2)}_{n,\xi}$, we can apply again the Cameron-Martin theorem to get that
\begin{equation*}
B^{(2)}_{n,\xi}= ( e^{ -\frac{1}{2}(\tilde{ \bs{\varphi}}^{\rm h} , (\mathbf{D}_{\Sigma_{n}} - \mathbf{D})\tilde{ \bs{\varphi}}^{\rm h})_2} - e^{ -\frac{1}{2}(\tilde{ \bs{\varphi}}^{\rm h} , (\mathbf{D}_{\Sigma} - \mathbf{D})\tilde{ \bs{\varphi}}^{\rm h})_2} ) e^{\frac{Q^2}{32\pi^2}V(f_n)-\frac{Q}{4\pi}\int_{\Sigma_n}K_{n}P_n\tilde{ \bs{\varphi}}^{\rm h}{\rm dv}_{n} }  \E_{ \tilde{\bs{\varphi}}^{\rm h}} \Big  [    e^{-  M_\gamma^{g_n,\bs{\mu}} (X_n+P_n \tilde{ \bs{\varphi}}^{\rm h}+G(f_n),  \Sigma^\xi)   }  \Big]
\end{equation*}
where we have set 
\begin{equation}\label{def:G(f_n)}
G(f_n)(x):=\int_{\Sigma_n} K_{n}(x')G_n(x,x')  {\rm dv}_{n}(x'), \quad  V(f_n):=\int_{\Sigma_n}K_{n}  G(f_n)  {\rm dv}_{n},
\end{equation}
with $G_n$ the Green function on $\Sigma_n$ with mixed boundary conditions, and similarly 
\begin{equation}\label{def:G(f)}
G(f)(x):= \int_{\Sigma} K(x')G(x,x')  {\rm dv}(x'), \quad V(f):=\int_{\Sigma}K  G(f)  {\rm dv}.
\end{equation}
We split the boundary fields as $\tilde{ \bs{\varphi}}^{\rm h}=(c_1,c_2)+  \bs{\varphi}^{\rm h}$, and we set $y=(0,c_2-c_1)$. Then, we have  $P_n \tilde{ \bs{\varphi}}^{\rm h}=c_1+P_n y+P_n   \bs{\varphi}^{\rm h}$ in such a way that  (see item (5) of Lemma \ref{lemgreen})
\begin{equation}\label{minorcurv}
\int_{\Sigma_n}K_{n}P_n\tilde{ \bs{\varphi}}^{\rm h}{\rm dv}_{n}=\int_{\Sigma_n}K_{n}(c_1+P_ny+P_n \bs{\varphi}^{\rm h}){\rm dv}_{n}\geq -C|y|-C_N\|  \bs{\varphi}\|_{H^{-N}_{\rm even}(\T)}
\end{equation} 
for some $C_N,C>0$. Then, we can use again uniform bounds in $n$ for $V(f_n)$,  and bound the expectation by $1$. The case of $B^{(2)}_{n,\xi}$ then follows from item (6) of Lemma \ref{lemgreen}. 

Next we focus on  $B^{(1)}_{n,\xi}$  and we want to prove it converges to $0$ in $\mc{L}(\mc{H}^+)$ as $n\to \infty$. Applying Cameron-Martin gives
 \begin{multline}\label{B1n}
B^{(1)}_{n,\xi}= e^{ -\frac{1}{2}(\tilde{ \bs{\varphi}}^{\rm h} , (\mathbf{D}_{\Sigma} - \mathbf{D})\tilde{ \bs{\varphi}}^{\rm h})_2}  \Big( e^{\frac{Q^2}{32\pi^2}V^+(f_n)-\frac{Q}{4\pi}\int_{\Sigma_n}K_{n}  P_n\tilde{ \bs{\varphi}}^{\rm h}{\rm dv}_{n} } \E_{ \tilde{\bs{\varphi}}^{\rm h}} \Big[   e^{-   M_\gamma^{g_n,\bs{\mu}}(X_n+P_n \tilde{ \bs{\varphi}}^{\rm h}+G(f_n),  \Sigma^\xi)   }  \Big]
 \\
 -e^{\frac{Q^2}{32\pi^2}V^+(f)-\frac{Q}{4\pi}\int_{\Sigma}K  P\tilde{ \bs{\varphi}}^{\rm h}{\rm dv} } \E_{ \tilde{\bs{\varphi}}^{\rm h}}  \Big[   e^{-  M^{g,\bs{\mu}}_\gamma (X+P\tilde{ \bs{\varphi}}^{\rm h}+G(f),  \Sigma^\xi)   }  \Big]   \Big).
\end{multline}
Arguing as above, we can show that the operator with kernel $$e^{ -\frac{1}{2}(\tilde{ \bs{\varphi}}^{\rm h} , (\mathbf{D}_{\Sigma} - \mathbf{D})\tilde{ \bs{\varphi}}^{\rm h})_2}  e^{ -\frac{Q}{4\pi}\int_{\Sigma_n}K_{n}  P_n\tilde{ \bs{\varphi}}^{\rm h}{\rm dv}_{n} } \E_{ \tilde{\bs{\varphi}}^{\rm h}} \Big[   e^{-   M_\gamma^{g_n,\bs{\mu}}(X_n+P_n \tilde{ \bs{\varphi}}^{\rm h}+G(f_n),  \Sigma^\xi)   }  \Big]$$
is bounded as an operator $\mc{H}^+\to\mc{H}^+$ and, because the map $f\in \mc{S}(\D^+)\mapsto V^+(f)$ is continuous, we can thus replace $V^+(f_n)$ by $V^+(f)$. As a consequence, we can assume $V^+(f)=1$ without loss of generality.  In the same way as explained for the analogous term in the proof of \cite[Proposition 3.7]{BGKR1}, we can replace as well the curvature  term $\int_{\Sigma_n}K_{n}  P_n\tilde{ \bs{\varphi}}^{\rm h}{\rm dv}_{n} $ by $\int_{\Sigma_n}K  P\tilde{ \bs{\varphi}}^{\rm h}{\rm dv} $ using  item (6) of Lemma \ref{lemgreen}. This means that we are left with estimating $B^{(1)}_{n,\eps}$ by the "difference of expectations in \eqref{B1n}" times $ e^{ -\frac{1}{2}(\tilde{ \bs{\varphi}}^{\rm h}, (\mathbf{D}_{\Sigma} - \mathbf{D})\tilde{ \bs{\varphi}}^{\rm h})_2}   e^{C(|y|+\|  \bs{\varphi}^{\rm h}\|_{H^{-N}(\T)})}$, which we call $\tilde{B}^{(1)}_{n,\eps}$, the last factor in this expression coming from the estimation for the curvature term similar to \eqref{minorcurv}. We will show that  $\tilde{B}^{(1)}_{n,\eps}$  tends to $0$ in $L^2(\mu_0^{+\otimes 2})$, which is equivalent to the convergence in Hilbert-Schmidt norm for the operator with this kernel.  

Recall that we use the decomposition  $\tilde{ \bs{\varphi}}^{\rm h}=(c_1,c_2)+  \bs{\varphi}^{\rm h}$, and $y=c_2-c_1$. It suffices to prove that $\iint \mathbf{1}_{|c_1|\leq A} |\tilde{B}^{(1)}_{n,\eps}|^2 \dd \mu_0^{\otimes 2}$ tends to $0$ as $n\to \infty$, for some arbitrary large $A$. Indeed, for $c_1$ large, the potential term in each expectation is of the form $e^{-e^{\gamma c_1}\times {\rm GMC}}$. Estimating the expectation of the GMC uniformly in $n$ is done using Kahane's inequality and items (1) and (2) of Lemma \ref{lemgreen}, so that the integration domain $\{c_1>A\}$ can be shown to have a contribution arbitrarily small for large $A$, uniformly in $n$, as detailed in \cite[Proposition 3.7]{BGKR1}. For $c_1$ negatively large, a similar argument occurs. Indeed each expectation is of the form $ \E_{ \tilde{\bs{\varphi}}^{\rm h}}  \big[e^{-e^{\gamma c_1}\times {\rm GMC}}\big]$, and therefore behaves like  $ \E_{ \tilde{\bs{\varphi}}^{\rm h}}  \big[1- e^{\gamma c_1}\times {\rm GMC}\big]$, for negatively large $c_1$. Evaluating the difference of expectations makes the leading order terms (namely $1$) cancel out, and thus yields a  behavior of the form $ \E_{ \tilde{\bs{\varphi}}^{\rm h}}  \big[ e^{\gamma c_1}\times {\rm GMC}\big]$. So, it provides an arbitrarily small contribution  for large $A$, uniformly in $n$.
 
Finally in the region $|c_1|\leq A$, and after using the estimate  $e^{ -\frac{1}{2}(\tilde{ \bs{\varphi}} , (\mathbf{D}_{\Sigma} - \mathbf{D})\tilde{ \bs{\varphi}})_2} \leq e^{-C|y|^2/2}$, we have to evaluate
\begin{align*}
 \int_{|c_1|\leq A}\int_\R  \iint e^{ -C	|y|^2}   e^{C(|y|+\|  \bs{\varphi}^{\rm h}\|_{H^{-N}(\T)})}  D_n(c_1,y,\bs{\varphi}^{\rm h})\dd c_1 \dd y\P^{\otimes 2}(\dd \bs{\varphi}^{\rm h})
\end{align*}
with 
$$D_n(c_1,y,\bs{\varphi}):=\Big| \E_{ \tilde{\bs{\varphi}}} \big[    e^{- M_\gamma^{g_n,\bs{\mu}} (X_n+P_n \tilde{ \bs{\varphi}} +G(f_n), \Sigma^\xi)   }  \big]
 - \E_{ \tilde{\bs{\varphi}} }  \big[   e^{-   M_\gamma^{g,\bs{\mu}} (X+P\tilde{ \bs{\varphi}} +G(f), \Sigma^\xi)   }  \big] \Big| ^2.$$
We have to show that it goes to $0$ as $n\to\infty$. This follows from the dominated convergence theorem as $D_n$ is obviously bounded and converges almost surely in $(c_1,y,\bs{\varphi})$ towards $0$. Indeed, up to the convergences  of the harmonic extensions and the curvature terms, treated with items (4) and (3)  of Lemma \ref{lemgreen}, this amounts to showing that the expectation  of the exponential of the GMC with the field $X_n$ converges to that with the field $X$. Because the function $x\mapsto \exp(-x)$ is convex,  this follows from  the combination of  Kahane's inequalities with item (2) of Lemma \ref{lemgreen}.
\end{proof}

We next prove the technical lemma that has been used in the proposition above.
\begin{lemma}\label{lemgreen}
 Let $(f_n)_n$ be a sequence    converging in $\mc{S}_\epsilon^+$ towards  $f$ as above. Let $P_n \tilde{\bs{\varphi}}^{\rm h}$ and $P \tilde{\bs{\varphi}}^{\rm h}$   denote the harmonic extensions on $\Sigma_{n}$ and $\Sigma$,  $ {\rm dv}_n$ (resp. $\dd \ell_n$) and $ {\rm dv}$ (resp. $\dd \ell$) the corresponding measures (resp. boundary lengths), $G_n$ and $G$ the corresponding Green functions. Let   $\Sigma^\xi$ and $\Sigma'_n$ be as above, and   $W_{n,g,{\rm N}}$ the Robin mass on $\Sigma'_n$. We claim:
 \begin{enumerate}
\item  for some  constant $C>0$,  ${\rm v}_n( \Sigma_{n} \setminus \Sigma^\xi  ) +\ell_n\bigl((\Sigma_n\setminus\Sigma^\xi)
\cap\partial_{\rm N}\Sigma_n\bigr)\leq C\xi$. Moreover, if $\gamma^2<2$, then
$$ \int_{\Sigma_{n}\setminus \Sigma^\xi}e^{\frac{\gamma^2}{2}W_{n,g,{\rm N}}} \dd {\rm v}_n\leq C\xi.$$
\item there exists a sequence $(a_n)_n$ such that $a_n\to 0$ as $n\to\infty$ and $\sup_{(\Sigma^\xi)^2}|G_n-G|\leq  a_n$,
\item  $\sup_{x\in \Sigma^\xi}|G(f_n)(x)-G(f)(x)|\to 0$ as $n\to\infty$, where $G(f_n)$ and $G(f)$ are defined in \eqref{def:G(f_n)} and \eqref{def:G(f)},
\item the sequence $\sup_{\Sigma^\xi}|P_n \tilde{\bs{\varphi}}^{\rm h}-P \tilde{\bs{\varphi}}^{\rm h}|$ converges to $0$ when $n\to\infty$ for any $\tilde{\bs{\varphi}}^{\rm h}\in  (H^{s}(\T))^2 $,
\item for fixed $N\in\R$, we have the estimates, for some constants $C_N>0$ and $C_{N,n}>0$ such that $C_{N,n}\to 0$ as $n\to\infty$ for fixed $N$,
\begin{align*}
{\rm i)} & & \sup_n|\int_{\Sigma_n}K_{n}P_n\tilde{ \bs{\varphi}}^{\rm h}{\rm dv}_{n}| \leq & C_N \|\tilde{ \bs{\varphi}}^{\rm h}\|_{H^{-N}_{\rm even}(\T)}
\\
{\rm ii)}  & &   |\int_{\Sigma_n}K_{n}P_n\tilde{ \bs{\varphi}}^{\rm h}{\rm dv}_{n}-\int_{\Sigma}KP\tilde{ \bs{\varphi}}^{\rm h}{\rm dv}| \leq & C_{N,n} \|\tilde{ \bs{\varphi}}^{\rm h}\|_{H^{-N}_{\rm even}(\T)}
 \\
{\rm iii)}  & &  \sup_{n}\sup_{x\in \Sigma^{\xi}}|P_n\tilde{ \bs{\varphi}}^{\rm h}|+|P\tilde{ \bs{\varphi}}^{\rm h}|\leq  & C_{N} \|\tilde{ \bs{\varphi}}^{\rm h}\|_{H^{-N}_{\rm even}(\T)}
\end{align*}
\item we have the convergence $(e^{ -\frac{1}{2}(\tilde{ \bs{\varphi}}^{\rm h} , (\mathbf{D} _{\Sigma_{n}} - \mathbf{D})\tilde{ \bs{\varphi}}^{\rm h})_2}- e^{ -\frac{1}{2}(\tilde{ \bs{\varphi}}^{\rm h} , (\mathbf{D} _{\Sigma} - \mathbf{D})\tilde{ \bs{\varphi}}^{\rm h})_2}) 
 e^{C(|y|+ \| \bs{\varphi}^{\rm h}\|_{H^{-N}_{\rm even}(\T)}  )}\to 0$ in $\mc{L}(\mc{H}^+,\mc{H}^+)$.
\end{enumerate}
\end{lemma}
\begin{proof} 
For item (1), set
\[
E_{n,\xi}:=\Sigma_n\setminus\Sigma^\xi.
\]
Since $f_n\to f$ in $\mc S_\epsilon^+$, the domains $\Sigma_n$ converge
smoothly to $\Sigma$, and the metrics $g_n$, after pullback to a fixed
surface, converge smoothly to $g$. In particular, the corresponding
Jacobians are uniformly bounded and the Dirichlet and Neumann boundary
components admit uniform collar coordinates.

For every sufficiently small $\xi>0$, and for all $n$ sufficiently large
depending on $\xi$, the smooth convergence of the Dirichlet boundaries
implies
\[
E_{n,\xi}
\subset
\left\{
z\in\Sigma_n:
d_{g_n}(z,\partial_{\rm D}\Sigma_n)\leq C_0\xi
\right\}.
\]
The uniform tubular-neighbourhood estimates therefore give
\[
{\rm v}_n(E_{n,\xi})
+
\ell_n(E_{n,\xi}\cap\partial_{\rm N}\Sigma_n)
\leq C\xi.
\]

We now prove the estimate involving the Neumann Robin mass. Let
$(\Sigma'_n)^{\#2}$ denote the Neumann double of $\Sigma'_n$, and let
$\tau_n$ be its reflection. With the normalization of the Green function
used here, the Neumann Robin mass satisfies
\[
W_{n,g,{\rm N}}(z)
=
W_{(\Sigma'_n)^{\#2},g_n^{\#2}}(z)
+
G_{(\Sigma'_n)^{\#2},g_n^{\#2}}
\bigl(z,\tau_n(z)\bigr).
\]
After identifying the doubles with a fixed closed surface, the metrics
$g_n^{\#2}$ form a smooth precompact family. The uniform local expansion
of the Green function along the diagonal, together with the estimate, for some $c>0$,
\[
 c d_{g_n}\bigl(z,\partial\Sigma'_n\bigr)\leq d_{g_n^{\#2}}\bigl(z,\tau_n(z)\bigr)\leq c^{-1}
d_{g_n}\bigl(z,\partial\Sigma'_n\bigr),
\]
therefore yields constants $C_1,r_0>0$, independent of $n$, such that
\[
W_{n,g,{\rm N}}(z)
\leq
C_1-\log\left(
d_{g_n}(z,\partial\Sigma'_n)\wedge r_0
\right).
\]
Consequently,
\[
e^{\frac{\gamma^2}{2}W_{n,g,{\rm N}}(z)}
\leq
C\left(
d_{g_n}(z,\partial\Sigma'_n)\wedge r_0
\right)^{-\gamma^2/2}.
\]

Away from fixed neighbourhoods of the Dirichlet-Neumann corners, the
distance to $\partial\Sigma'_n$ is uniformly bounded from below. Since
$E_{n,\xi}$ has width $O(\xi)$, the contribution of this region is bounded
by $C\xi$.

Near each Dirichlet-Neumann  corner, choose uniform coordinates
\[
(r,t)\in[0,r_0)^2
\]
such that $r$ is comparable to the distance to the Neumann boundary,
$t$ is comparable to the distance to the Dirichlet boundary, and
$$
\dd{\rm v}_n\leq C\,\dd r\,\dd t.
$$
In these coordinates, $E_{n,\xi}\subset\{0<r<r_0,\ 0<t<C\xi\}$. It follows that
\[
\begin{aligned}
\int_{E_{n,\xi}}
e^{\frac{\gamma^2}{2}W_{n,g,{\rm N}}(z)}
\,\dd{\rm v}_n(z)
&\leq
C\xi
+
C\int_0^{C\xi}\int_0^{r_0}
r^{-\gamma^2/2}\,\dd r\,\dd t
\leq
C\xi\left(
1+\int_0^{r_0}r^{-\gamma^2/2}\,\dd r
\right).
\end{aligned}
\]
Since $\gamma^2<2$, it follows that
\[
\int_{\Sigma_n\setminus\Sigma^\xi}
e^{\frac{\gamma^2}{2}W_{n,g,{\rm N}}(z)}
\,\dd{\rm v}_n(z)
\leq C'\xi.
\]
(2) We can consider the Neumann double $\Sigma_n^{\#2}$ and $\Sigma^{\#2}$, both equipped with anti-holomorphic involutions $\tau_n$ and $\tau$, and use that $G_n(x,x')=G_n^{\#}(x,x')+G_n^{\#}(x,\tau_n(x'))$ and $G(x,x')=G^{\#}(x,x')+G^{\#}(x,\tau(x'))$ respectively for $x,x'\in \Sigma_n\subset \Sigma_n^{\#2}$ and $x,x'\in \Sigma\subset \Sigma^{\#2}$, where $G_n^{\#}$ and $G^\#$ are the Dirichlet 
Green functions of $\Sigma_n^{\#2}$ and $\Sigma^{\#2}$. Then the result follows from \cite[Lemma 3.12]{BGKR1}.\\ 
(3) We use \cite[Lemma 3.12]{GRW1} where the same result is proved but for the case without Neumann boundary. We consider the Neumann double $(\Sigma_n^{\#2},g_n^{\#})$ and $(\Sigma^{\#2},g^{\#})$ as above. We use again the same notation as in (2). 
Using the symmetry of the curvature $K_n^{\#}$ and $K^{\# }$ of the symmetrized metrics $g_n^\#,g^\#$ under the involutions $\tau_n$ and $\tau$,  we have for $x\in  \Sigma^\xi$, 
\[|G(f_n)(x)-G(f)(x)|= \frac{1}{2} \Big|\int_{\Sigma_n^{\#2}} K_n^{\#}(x')G^{\#}_n(x,x'){\rm dv}_{g^{\#2}_n}(x')-\int_{\Sigma^{\#2}} K^{\#}
(x')G^{\#}(x,x'){\rm dv}_{g^{\#}}(x')\Big|.\]
Then, in \cite[Lemma 3.12]{GRW1} it is proved that the right hand side converges to $0$ as $n\to \infty$.
\\
(4) Again, we can use the Neumann double and notice that  $P_n\tilde{\bs{\varphi}}^{\rm h}=(P^{\#}_n\tilde{\bs{\varphi}}^{\rm h})|_{\Sigma_n}$ 
and $P\tilde{\bs{\varphi}}=(P^{\#}\tilde{\bs{\varphi}})|_{\Sigma}$ where $P_n^\#$ and $P^\#$ are the Poisson operator of the Neumann double $\Sigma_n^{\#2}$ and $\Sigma^{\#2}$. The result then follows from \cite[Lemma 3.12]{BGKR1}.\\
(5) The properties (i), (ii), (iii) follow directly from (5) in  \cite[Lemma 3.12]{BGKR1}, using again the Neumann double and the fact that $K^\#_n$ and $K^\#$ are symmetric by $\tau_n$ and $\tau$, respectively.\\ 
(6) By \cite[Lemma 5.5]{GRW1}, the Dirichlet-to-Neumann maps ${\bf D}_{\Sigma_n}$ and ${\bf D}_{\Sigma}$ are exactly the restrictions of the DN maps ${\bf D}_{\Sigma_n^{\#2}}$ and ${\bf D}_{\Sigma^{\#2}}$ of the corresponding Neumann double to functions (or distributions) $(\tilde{\varphi}_1,\tilde{\varphi}_2)$ that are even (i.e. with even Fourier modes). Then, we apply \cite[Lemma 3.12]{BGKR1} which shows the result.
\end{proof}

Now we study some basic continuity properties of the quadratic forms $\mc{Q}_{\rm v}^+$ defined in  \eqref{defQ+v}.   
 \begin{proposition}\label{propA5}
Let $v(z)=-\sum_{n\geq 0}v_nz^{n+1}\in {{\rm Hol}^{+,\bullet}_\epsilon(\D)}$ and ${\rm v}=v(z)\pl_z$. 
Then  the quadratic form $\mc{Q}_{\rm v}^+$ is $\mc{Q}^+ $-continuous  with bound
			$$  |\mc{Q}^+_{\rm v}(F,G)|\leq C \|v\|_{\epsilon=0,3}{\mc{Q}^+}(F,F)^{1/2}{\mc{Q}^+}(G,G)^{1/2}.$$
In particular,  the mapping  $v\in{\rm Hol}^{+,\bullet}_\epsilon(\D)  \mapsto \mc{Q}^+_{\rm v}\in \mathcal{L}(\mc{D}(\mc{Q}^+ )\times \mc{D}(\mc{Q}^+ ))$  is continuous.
\end{proposition}

\begin{proof}	We have   $\mathcal{Q}^+_{\rm v}(F,G): =\cjg \bH_{\rm v}^{+,0} F , G \cjd_{\mc{H}^+} +   \cjg    {\bf V}_{\rm v}^{+,\bs \mu}  F , G \cjd_{\mc{H}^+}$ with $\bH_{\rm v}^{+,0} $ given by \eqref{eq:def_L+}. By \eqref{normLn^+0}, we get $$|\cjg \bH_{\rm v}^{+,0} F , G \cjd_{\mc{H}^+}|\leq C \|v\|_{\epsilon=0,3}{\mc{Q}^+}(F,F)^{1/2}{\mc{Q}^+}(G,G)^{1/2}.$$
Furthermore
$$  \cjg    {\bf V}_{\rm v}^{+,\bs \mu}  F , G \cjd_{\mc{H}^+}\leq 3S \langle{\bf V}^{+,\bs{\mu}} F,F\rangle_{\mc{H}^+}^{1/2} \langle {\bf V}^{+,\bs{\mu}}G,G\rangle_{\mc{H}^+}^{1/2}$$	
with $S=\sup_{\theta\in[0,\pi]}|{\rm Re}(e^{-i\theta}v(e^{i\theta}))|$ since we can bound each of the three terms in ${\bf V}_{\rm v}^{+,\bs \mu} $ by $S  {\bf V}^{+,\bs{\mu}}$.  Our claim follows.
\end{proof}

\subsubsection{Proof of Proposition \ref{th:differentiabiliteTf}}
We will use \cite[Theorem 4.8]{BGKR1}. For this, we need to check the items 1-11 in \cite[Assumption 4.7]{BGKR1}, which we list here.
\begin{enumerate}
\item The Hilbert space is $\mc{H}^+$, equipped with the quadratic form $\mc{Q}^+$ with domain $\mc{D}(\mc{Q}^+)$ and associated contraction semigroup $(e^{-t \mathbf{H}^+})_{t\geq 0}$ on $\mc{H}^+$, which also maps $\mc{D}'(\mc{Q}^+)$ to $\mc{D}(\mc{Q}^+)$ (see \cite[subsect. 7.2.3 and 7.2.4]{GRW1})
\item   $\mc{S}^+$ is a subset of $\mc{S} $, stable by composition.
\item The subspace  $\mc{S}^+$   is preserved by the flow. More precisely, consider  $v\in   {\rm Hol}^{+,\bullet}_\epsilon(\D)$ Markovian, then the solution $f_t$ to the flow $\partial_tf_t=v(f_t)$ with $f_0\in \mathcal{S}_\epsilon^+$ (resp. $f_0(z)=z$) belongs to  $\mathcal{S}_\epsilon^+$ for all $t\geq 0$ (resp. for all $t>0$). The proof follows from \cite[Lemma 4.5]{BGKR1} and the symmetry with respect to $\tau$ of $f_t$, which is a consequence of the symmetry of the vector field $v$ under $\tau$.
\item If $f\in \mc{S}_\epsilon^+$ then for all $t\geq 0$ and $t'\in\R$, $e^{t'}f(e^t\cdot)\in \mc{S}_{\epsilon'}^+$, with $(1+\epsilon')e^t=1+\epsilon$.
\item The map $f \in\mc{S}^+\to \mathbf{T}^+_f\in \mc{L}(L^\infty(\mu_0^+),L^\infty(\mu_0^+))$  obeys the composition rule $\mathbf{T}^+_{f_1\circ f_2}=\mathbf{T}^+_{f_1}\circ \mathbf{T}^+_{f_2}$ (see Prop \ref{prop:semigroup}).
\item  For $f\in \mc{S}_>^+ $, the operator $\mathbf{T}^+_f$ extends to a bounded operator on $\mc{H}^+$  (see Lemma \ref{contbyold}), i.e. $\mathbf{T}^+_f\in \mc{L}(\mc{H}^+,\mc{H}^+)$, and the mapping $f\in  \mc{S}_\epsilon^+ \mapsto \mathbf{T}^+_f\in \mc{L}(\mc{H}^+,\mc{H}^+)$ is continuous for each $\epsilon>0$ (see Lemma \ref{lemcont+}).
\item there is a  linear map $v\in \bigcup_{\epsilon>0} {\rm Hol}^{+,\bullet}_\epsilon(\D)\mapsto \mc{Q}_{\rm v}^+$ where $\mc{Q}_{\rm v}^+$ is a continuous bilinear form on $\mc{D}(\mc{Q}^+)^2$, equipped with the $\mc{Q}^+$-norm. For all  $\epsilon>0$, the restriction of this  map $v\in {\rm Hol}^{+,\bullet}_\epsilon(\D)\mapsto \mc{Q}_{\rm v}^+$ is  $\|\cdot\|_{\epsilon,k=3}$-continuous   (see Prop \ref{propA5}).
\item\label{itemCoer} Let $v(z)=\sum_{n\geq 0}v_nz^{n+1}\in  {\rm Hol}^{+,\bullet}_\epsilon(\D)$. If $ v_0$ is  negatively large enough, meaning there exists $K>0$  such that, if $v$ obeys (see the condition in Prop. \ref{prop_coercive})
$$ 
D(v):=-v_0- K \sum_{n \geq 1}|v_n|n^2>0,
$$
 then  the bilinear form $\mc{Q}^+_{\rm v}$ generates a contraction semigroup $(e^{-t\mathbf{H}^+_{\rm v}})_{t>0}$ on $\mc{H}^+$, for some bounded operator $\mathbf{H}^+_{\rm v}:\mc{D}(\mc{Q}^+)\to \mc{D}'(\mc{Q}^+)$, see the discussion below  \eqref{defQ+v}. Furthermore, we have in this case
\begin{equation*} 
\forall F\in \mc{D}(\mc{Q}^+),\quad {\rm Re}(\mc{Q}^+_{\rm v}(F,F))\geq D(v)\mc{Q}^+ (F,F).
\end{equation*}
Finally, for $v(z)=-z$, we have $\mc{Q}^+_{\rm v}=\mc{Q}^+$.
\item  If $v\in   {\rm Hol}^{+,\bullet}_\epsilon(\D)$ obeys the coercivity condition of item \eqref{itemCoer}, and $f_t$ denotes its flow with $f_0(z)=z$, then $e^{-t\mathbf{H}^+_{\rm v}}=\mathbf{T}^+_{f_t}$ (see Theorem \ref{th:Tft}).

\item  If $f\in \mc{S}^+_\epsilon$ and $v\in {\rm Hol}^{+,\bullet}_\epsilon(\D)$ such that $v/f'$ obeys the coercivity condition of item \eqref{itemCoer}, we set $f_t(z)=f^{-1}(f(z)+tv(z))$. For $\delta$ small enough and for any fixed $\epsilon'<\epsilon$,  the map $(s,t)\in [0,\delta]^2\mapsto \frac{v\circ f_t^{-1}(e^sz)}{f'(z)}\in{\rm Hol}^\bullet_{\epsilon'}(\D^+)$  is continuous, see \cite[Lemma 4.6]{BGKR1}.

\item\label{itemWeird2} if $f\in \mc{S}^+_\epsilon$ and $v\in {\rm Hol}^{+,\bullet}_\epsilon(\D)$, then $-\omega zf'(z)+v(z)\in  {\rm Hol}^{+,\bullet}_\epsilon(\D)$ for all $\omega\in \R$ large enough.
\end{enumerate}
The proof of Theorem \ref{th:differentiabiliteTf} then follows from \cite[Theorem 4.8]{BGKR1}.
\subsection{Differentiability of amplitudes of general half-annuli}\label{sub:derivamp}
To prove differentiability and compute the differential of the amplitudes associated with general half-annuli, we closely follow the strategy of \cite[Section 4.3]{BGKR1}.

For $\delta<1$ let   ${\rm Hol}^+_\delta(\A)$ denote the set of holomorphic maps $f:\A_{\delta,\delta^{-1}}\to f(\A_{\delta,\delta^{-1}})\subset \C$ that are smooth up to the boundary and satisfy $f(\bar{z})=\bar{f}(z)$,  endowed with the topologogy induced by the seminorms \eqref{seminorm} with $(1+\eps):=\delta^{-1}$.

We also consider the open subset  
\[ 
\mc{S}^+_\delta(\A):=\{ f \in {\rm Hol}^+_\delta(\A) \,|\, f \textrm{ biholomorphism},  f(\T)\subset \D^\circ \}.
\]
For $f\in  \mc{S}_\delta(\A^+)$, denote by $\A^+_f\subset \H$  the half-annulus bounded by $\T^+$ and $f(\T^+)$, thus generalising our earlier definition for  $f\in \mc{S}^+_\eps$.  We assign D-marking on $\T^+$ and $f(\T^+)$, and N-marking on $\A^+_f\cap \R$.

For $v\in {\rm Hol}_\delta(\A)$, write ${\rm v}=v(z)\pl_z$ and expand $v(z)=\sum_{n\in \Z}v_nz^{n+1}$. We then define  the operator   
\[ \bH^+_{\rm v}=-\sum_{n\in \Z} v_n\bL_n^+,\]
which is bounded as a map $\mc{D}(\mc{Q}^+)\to \mc{D}'(\mc{Q}^+)$ by \eqref{boundLn}.  

As before, we   denote by $\bs{\zeta}^+_{f}=({\rm Id},f)$ if $f:\T^+\to f(\T^+)\subset \H$ the   analytic parametrisation of the D-marked boundary when $f:\T^+\to f(\T^+)$ is an analytic embedding of a half-circle in  $\D^+$.   The goal of this section is to show the following:
\begin{proposition}\label{l:differential_annuli}
Let $f_0(z)=rz$ for $0<r<1$. Assume that $g_f$ is a family of admissible metrics on $\A^+_{f}$, for $f$ in a neighborhood of $f_0$, such that the mapping $f\mapsto S_{\rm L}^0(\A^+_{f},g_{f},g_{\A})$ is differentiable at $f_0$.  Then the mapping 
$$
f\in\mc{S}^+_\delta(\A) \mapsto  \mc{A}_{\A^+_{f},g_f,\bs{\zeta}^+_{f}}\in\mc{L}( \mc{D}(\mc{Q}^+), \mc{D}(\mc{Q}^+))
$$
 is differentiable at $f=f_0$, with differential given, for any $F\in \mc{D}(\mc{Q}^+)$, by (recall \eqref{def_SL0})
\[ D_v\mc{A}_{\A^+_{f},g_{f},\bs{\zeta}_{f}}|_{f=f_0}(F)=-c_{\rm L}\Big(\frac{v_0}{24 r}+
D_vS_{\rm L}^0(\A^+_{f},g_{f},g_{\A})|_{f=f_0}\Big)\mc{A}_{\A^+_{f_0},g_{\A},\bs{\zeta}^+_{f_0}}(F)-
\mc{A}_{\A^+_{f_0},g_{\A},\bs{\zeta}^+_{f_0}} \big( {\bf H}^+_{{\rm v}/r} F\big)\]
for $v(z)=\sum_{n\in \Z}v_nz^{n+1}\in {\rm Hol}_\delta(\A^+)$ with ${\rm v}=v(z)\pl_z$.
\end{proposition}
\begin{proof}
The proof will proceed in several steps. First, consider the anti-holomorphic involution 
\[ \iota : \hat{\C}\to \hat{\C}, \quad \iota(z)=1/\bar{z},\]
which preserves the upper half-plane $\H$. For $f\in\mc S_>^+$, define
\[
f^*:=\iota\circ f\circ\iota.
\]
Then $f^*$ is a holomorphic and univalent map on $\D^*=\hat{\C}\setminus \D^\circ$, satisfies
\[
f^*(\D^*)\subset(\D^*)^\circ,
\qquad
f^*(\infty)=\infty,
\]
and is symmetric with respect to the real axis. Moreover, since $f$ extends
analytically across $\T$, so does $f^*$.

For $f_v=f_0+v\in\mc S_\delta^+(\A)$ with $v$ sufficiently small,
\cite[Lemma 4.28]{BGKR1} yields a pair
\[
\bs f_v=(f_{{\rm out},v},f_{{\rm in},v})
\in
{\rm Hol}^{\bullet,+}_\epsilon(\D)^2
\]
for some small $\epsilon>0$, and a biholomorphism
\[
\Phi_{f_v}:U_v\longrightarrow V_v,
\]
defined on a neighbourhood $U_v$ of $\overline{\A_{f_v}}$, which maps
$\A_{f_v}$ biholomorphically onto the annulus $\A_{\bs f_v}$ bounded by
$f_{{\rm out},v}^*(\T)$ and $f_{{\rm in},v}(\T)$. The boundary
parametrisations satisfy
\[
\Phi_{f_v}|_{\T}
=
f_{{\rm out},v}^*|_{\T},
\qquad
(\Phi_{f_v}\circ f_v)|_{\T}
=
f_{{\rm in},v}|_{\T}.
\]
 
Moreover, for a fixed $\la\in (r,1)$, the map 
$\Phi_{f_v}$ is uniquely determined by the normalization condition $(f_{{\rm out},v})'(0)=r/\la<1$. The proof of \cite[Theorem 4.32]{BGKR1}
further shows that if $\delta=(1+\eps)^{-1}$ with $\eps>0$ small, 
then the map $$
v\in {\rm Hol}_\delta^+ (\A) \mapsto \bs{f}_v\in {\rm Hol}^{+,\bullet}_\eps(\D)\times {\rm Hol}^{+,\bullet}_\eps(\D)
$$ is differentiable.  For $v=0$, one has $f_{{\rm in},0}(z)=\la z$ and $f_{{\rm out},0}=rz/\la$. Furthermore, $(d  \bs{f}_w)_{w=0}(v)=(v_{\rm out},v_{\rm in})\in {\rm Hol}^{+,\bullet}_\eps(\D)\times {\rm Hol}^{+,\bullet}_\eps(\D)$ determined by
$$v (z)=\frac{r}{\lambda}v_{\rm in}(z)+r\lambda z^2\overline{v_{\rm out}(1/r\bar z)}.$$
It is also proved that $\Phi_{f_v}(\bar{z})=\bar{\Phi}_{f_v}(z)$. Therefore,  $\A^+_{\bs{f}_v}=\A_{\bs{f}_v}\cap \H$ is a half-annulus. We assign the  marking N to the boundary component $\A^+_{\bs{f}_v}\cap \R$ and the marking D  to the   boundary components $f^*_{{\rm out},v}(\T^+)$ and $f_{{\rm in},v}(\T^+)$, which are equipped  with the  parametrisations   $\bs{\zeta}^+_{\bs{f}_v}=(f_{{\rm out},v}^*|_{\T^+},f_{{\rm in},v}|_{\T^+})$ (left is outgoing, right is incoming). Since $\T^+$ cuts $\A_{\bs{f}_{v=0}}$ into two half-annuli $\A^+_{f_{\rm in},v=0}$ and $\iota(\A^+_{f_{\rm out},v=0})$ glued along $\T^+$, it does so for small $v$ as well:  
$\T^+$ cuts the half-annulus $\A_{\bs{f}_v}$ into two annuli $\A^+_{f_{\rm in},v}$ and $\iota(\A^+_{f_{\rm out},v})$ glued along $\T^+$.
By diffeomorphism invariance of boundary Liouville amplitudes (\cite[Proposition  5.9]{GRW1}), and by the gluing property of amplitudes \cite[Proposition 6.1]{GRW1}, for $g_{f_v}$ admissible metric on $\A^+_{f_v}$,
\[\begin{split}
\mc{A}_{\A^+_{f_v},g_{f_v},\bs{\zeta}^+_{f_v}}= &   \mc{A}_{\A^+_{\bs{f}_v},(f_{{\rm out},v}^*)_*g_{f_v},\bs{\zeta}^+_{\bs{f}_v}}
\\
=& (2\pi)^{-\frac{3}{4}}  ( \mc{A}_{\A^+_{f_{{\rm out},v}},\iota_*(f_{{\rm out},v}^*)_*g_{f_v},\bs{\zeta}^+_{f_{{\rm out},v}}})^* \circ \mc{A}_{\A^+_{f_{{\rm in},v}},(f_{{\rm out},v}^*)_*g_{f_v},\bs{\zeta}^+_{f_{{\rm in},v}}}\\
= & (2\pi)^{\frac{3}{4}} e^{ C(v)}({\bf T}^+_{f_{{\rm out},v}})^* {\bf T}^+_{f_{{\rm in},v}}
\end{split}\] 
where we have set 
$$
C(v):=\frac{c_{\rm L}}{12}\Big(W^+(f_{{\rm in},v},(f_{{\rm out},v}^*)_*g_{f_v})+W^+(f_{{\rm out},v},\iota_*(f_{{\rm out},v}^*)_*g_{f_v})\Big).
$$
 In the third line,  we used that exchanging parametrisations of the two boundary half-circles corresponds to taking the adjoint of the associated amplitude, and in the last line we used  Proposition \ref{cor:annulus_propagator}.
Finally, applying Proposition \ref{th:differentiabiliteTf} to differentiate the maps $v\mapsto {\bf T}^+_{f_{{\rm in},v}}$ and $v\mapsto ({\bf T}^+_{f_{{\rm out},v}})^*$ yields
\begin{align*}
D_v\mc{A}_{\A^+_{f},g_{f},\bs{\zeta}_{f}}|_{f=f_0}= & D_vC(0) \mc{A}_{\A^+_{f_0},g_{f_0},\bs{\zeta}_{f_0}}-  (2\pi)^{\frac{3}{4}} e^{ C(0)}({\bf H}^+_{ \frac{\lambda}{r}v_{\rm out}})^*({\bf T}^+_{f_{{\rm out},0}})^* {\bf T}^+_{f_{{\rm in},0}}
\\
&-  (2\pi)^{\frac{3}{4}} e^{ C(0)}({\bf T}^+_{f_{{\rm out},0}})^* {\bf T}^+_{f_{{\rm in},0}}{\bf H}^+_{\frac{1}{\lambda}v_{\rm in}}
\end{align*}
Using that $({\bf H}^+_{\frac{\lambda}{r}v_{\rm out}})^*({\bf T}^+_{f_{{\rm out},0}})^* {\bf T}^+_{f_{{\rm in},0}}=({\bf H}^+_{\frac{\lambda}{r}v_{\rm out}})^*e^{\log r{\bf H}^+}=e^{\log r{\bf H}^+}({\bf H}^+_{\lambda v_{\rm out}(\cdot/r)})^*=e^{\log r{\bf H}^+}{\bf H}^+_{\lambda z^2 \overline{v_{\rm out}(1/r\bar z)}}$, we deduce that
\begin{align*}
D_v\mc{A}_{\A^+_{f},g_{f},\bs{\zeta}_{f}}|_{f=f_0}= & D_vC(0)\mc{A}_{\A^+_{f_0},g_{f_0},\bs{\zeta}_{f_0}}- \mc{A}_{\A^+_{f_0},g_{\A},\bs{\zeta}^+_{f_0}}({\bf H}^+_{{\rm v}/r} \cdot\,).
\end{align*}
Finally the same computation   as in the end of  the proof of \cite[Theorem 4.32]{BGKR1}, yields 
$$D_vC(0)= -c_{\rm L}\Big(\frac{v_0}{24 r}+
D_vS_{\rm L}^0(\A^+_{f},g_{f},g_{\A})|_{f=f_0}\Big).$$
\end{proof}

\section{Spectral resolution of the Hamiltonians}\label{sec:spectral}

\subsection{Spectral resolution for the Liouville (bulk) Hamiltonian}\label{resolution_bulk}

In the following proposition, we collect the relevant results from \cite{GKRV20_bootstrap} and  \cite{BGKRV}, see in particular \cite[Theorem 4.5]{BGKRV} for the statement reproduced below:
\begin{proposition}[\textbf{Spectral resolution of Liouville Hamiltonian}]\label{prop:resolution_bulk_Hamiltonian}
There exists a holomorphic family $\alpha\in \C\mapsto \Psi_{\alpha}$ of functions, called primary states,  belonging to the weighted space 
$e^{-\beta \rho}\mc{D}(\mc{Q})$ for $|{\rm Re}(\alpha)-Q|<\beta$, such that 
$$
{\bf H}\Psi_\alpha=2\Delta_\alpha \Psi_\alpha.
$$ 
Moreover:\\
i) For $\alpha<Q$, and writing $\tilde{\varphi}^\ell=\varphi^\ell+c$, one has
\begin{equation}\label{psi_alpha_proba} 
\Psi_\alpha(\tilde{\varphi}^{\ell})= e^{(\alpha-Q)c}\E_{\varphi}[e^{-\mu e^{\gamma c}\int_\D |x|^{-\gamma \alpha}M_\gamma(X_\D+P\varphi, \dd x)}].
\end{equation}
ii) For all $\nu,\tilde{\nu}\in \mc{T}$,  if $ {\bf L}_{-\nu}={\bf L}_{-\nu(k)} \dots \, {\bf L}_{-\nu(1)}$ and $ \tilde{{\bf L}}_{-\tilde \nu}=\tilde{{\bf L}}_{-\tilde\nu(j)} \dots\, \tilde{{\bf L}}_{-\tilde\nu(1)}$, then
\begin{equation}\label{descendant_bulk_Hamiltonian}
\Psi_{\alpha,\nu,\tilde{\nu}}:={\bf L}_{-\nu}\tilde{{\bf L}}_{-\tilde\nu} \Psi_\alpha \in e^{-\beta \rho}\mc{D}(\mc{Q})
\end{equation}
satisfies 
$$
{\bf H}\Psi_{\alpha,\nu,\tilde{\nu}}=(2\Delta_\alpha+|\nu|+|\tilde{\nu}|)\Psi_{\alpha,\nu,\tilde{\nu}}.
$$
iii) For each $n\in \Z$, the operators ${\bf L}_{n}$ and $\widetilde{{\bf L}}_{n}$ preserve $\mc{W}_\alpha:={\rm span}\{\Psi_{\alpha,\nu,\tilde{\nu}}\,|\, \nu,\tilde{\nu}\in \mc{T}\}$ and satisfy on this space
\[[{\bf L}_n,{\bf L}_m]= (n-m){\bf L}_{n+m}+\frac{c_{\rm L}}{12}\delta_{n,-m}(n^3-n), \quad 
[\widetilde{{\bf L}}_n,\widetilde{{\bf L}}_m]= (n-m)\widetilde{{\bf L}}_{n+m}+\frac{c_{\rm L}}{12}\delta_{n,-m}(n^3-n).\]
iv) For all $u,u'\in \mc{H}$, one has 
\begin{equation}\label{resolution_bulk_Hamiltonian} 
\cjg u,u'\cjd_{\mc{H}}=\frac{1}{2\pi} \sum_{\nu,\nu',\tilde{\nu},\tilde{\nu}'\in\mc{T}}\int_0^\infty \cjg u,\Psi_{Q+iP,\nu,\tilde{\nu}}\cjd_{\mc{H}} \cjg \Psi_{Q+iP,\nu',\tilde{\nu}'},u'\cjd_{\mc{H}} F_{Q+iP}^{-1}(\nu,\nu')F_{Q+iP}^{-1}(\tilde{\nu},\tilde{\nu}')\, \dd P.
\end{equation}
where $F_{\alpha}(\cdot,\cdot)$ are the Schapovalov forms appearing in Proposition \ref{prop:mainvir0}.
\end{proposition}

We note that, due to domain issues, it is a priori not obvious that the Virasoro operators can be iterated on 
$\Psi_\alpha$. This is, however, part of the result established in the above proposition.

\subsection{Spectral resolution for the boundary Liouville Hamiltonian}
For the boundary Liouville Hamiltonian, the following spectral resolution holds:
\begin{proposition}[\textbf{Spectral resolution of boundary Liouville Hamiltonian}]\label{resolution_bdry_Hamiltonian}
There exists a holomorphic family $\alpha \mapsto \Psi^+_{\alpha}$ of functions defined for 
$$
\alpha\in \C,\qquad   {\rm Re}(\alpha)\leq Q, \qquad (\alpha-Q)^2/4\notin -\N,
$$ 
and continuous on $Q+i\R$, called primary states,   belonging to 
$e^{-\beta \rho}\mc{D}(\mc{Q}^+)$ for $|{\rm Re}(\alpha)-Q|<\beta$, such that 
$$
{\bf H}^+\Psi^+_\alpha=\Delta_\alpha \Psi^+_\alpha.$$
Moreover:\\
(i) for $\alpha<Q$, using the notation of Definition \ref{quantTf}, one has for $\tilde{\varphi}^{\rm h}=\varphi^{\rm h}+c$, 
\begin{equation}\label{psi_alpha^+_proba} 
\Psi^+_\alpha(\tilde{\varphi}^{\rm h})= e^{\frac{(\alpha-Q)c}{2}}\E[e^{-\int_{\D^+}  M^{\boldsymbol{\mu}}_\gamma(X_{\D^+}^\#+P\tilde{\varphi}^{\rm h}-(\alpha-Q)\log|x|,\dd x)}]
\end{equation}
(ii) for all $\nu \in \mc{T}$, we denote by $ {\bf L}^+_{-\nu}={\bf L}^+_{-\nu(k)} \dots \, {\bf L}^+_{-\nu(1)}$.  Then, for $\alpha\in Q+i\R^*$, we define
\begin{equation}\label{descendant_boundary_Hamiltonian}
\Psi_{\alpha,\nu}^+:={\bf L}^+_{-\nu} \Psi_\alpha^+,
\end{equation}
which belongs to $ e^{-\beta \rho}\mc{D}(\mc{Q}^+)$ and satisfies 
\begin{equation}\label{primaryH+}
{\bf H}^+\Psi^+_{\alpha,\nu}=(\Delta_\alpha+|\nu|)\Psi^+_{\alpha,\nu}\qquad {\bf L}^+_{n}\Psi_{\alpha}^+=0,\quad n>0.
\end{equation}
(iii) For each $n\in \Z$, the operators ${\bf L}^+_{n}$ preserve $\mc{V}^+_\alpha:={\rm span}\{\Psi^+_{\alpha,\nu}\,|\, \nu \in \mc{T}\}$ and satisfy on this space
\begin{equation}\label{commuteLn^+}
[{\bf L}^+_n,{\bf L}^+_m]= (n-m){\bf L}^+_{n+m}+\frac{c_{\rm L}}{12}\delta_{n,-m}(n^3-n).
\end{equation}
(iv) For all $u,u'\in e^{\delta c_-}\mc{H}^+$ with $\delta>0$, one has 
\begin{equation}\label{resolution_boundary_Hamiltonian} 
\cjg u,u'\cjd_{\mc{H}^+}=\frac{1}{4\pi} \sum_{\nu, \nu' \in\mc{T}}\int_0^\infty \cjg u,\Psi^+_{Q+iP,\nu}\cjd_{\mc{H}^+} \cjg \Psi^+_{Q+iP,\nu'},u'\cjd_{\mc{H}^+} F_{Q+iP}^{-1}(\nu,\nu')\, \dd P
\end{equation}
where $F_{\alpha}(\cdot,\cdot)$ denotes the Schapovalov forms introduced in Proposition \ref{prop:mainvir0}.
\end{proposition}
The proof is written in Section \ref{proof_spec_resolution} in the Appendix.

\subsection{Reverting orientation and rotations}

First, following \cite[subsection 3.3]{GKRV20_bootstrap}, one can  consider the generator of rotations $\mathbf{\Pi}$, defined as the generator of the strongly continuous unitary group on $\mc{H}$ (for $\vartheta\in\R$)
\begin{align}\label{pi:prob}
 e^{i\vartheta \mathbf{\Pi}}F(\tilde{\varphi})=F(\tilde{\varphi}(\cdot+\vartheta)).
\end{align}
By \cite[Prop. 6.6 and 6.7]{GKRV21_Segal}, one has 
\begin{proposition}\label{propPi}
1) For $\beta\in\R $, the group $( e^{i\vartheta \mathbf{\Pi}})_{\vartheta\in\R}$ maps $e^{-\beta  c_-}\mc{H}$ into itself, and commutes with the semigroup $(e^{-t\mathbf{H} } )_{t>0}$ on $e^{-\beta c_-}\mc{H}$.\\
2) For $\beta>Q-{\rm Re}(\alpha)$ and $\nu,\tilde{\nu}\in \mc{T}$, one has $  e^{i\vartheta \mathbf{\Pi}}\Psi_{\alpha,\nu,\tilde{\nu}}=e^{i (|\nu|-|\tilde{\nu}|) \vartheta}\Psi_{\alpha,\nu,\tilde{\nu}}$ in $e^{-\beta c_-}L^2(\R\times \Omega_\T)$.
\end{proposition}

Second,  following~\cite[Section 6.6]{GKRV21_Segal}, we recall the effect of reversing the orientation on the   boundary circles of the amplitudes. Let $o:\mathbb{T}\to\mathbb{T}$ be the involution defined by $o(e^{i\theta})=e^{-i\theta}$. We introduce the operators
\[\forall F\in e^{\beta c_-}\mathcal{H},\quad\quad \mathbf{O}F(\tilde{\varphi}^{\ell})=F(\tilde{\varphi}^{\rm \ell}\circ o),\quad\quad \mathbf{C}F(\tilde{\varphi}^\mathrm{\ell})=\overline{F}(\tilde{\varphi}^\mathrm{\ell}).\]
\begin{proposition}\cite[Proposition 6.11]{GKRV21_Segal}\label{prop: RO bulk}
	For $P\in \R$ and $\nu,\tilde\nu\in\mathcal{T}$, we have the relation
	$$\mathbf{O}\mathbf{C}\Psi_{Q+iP,\nu,\tilde\nu}=\mathbf{C}\mathbf{O}\Psi_{Q+iP,\nu,\tilde\nu}=\Psi_{Q-iP,\nu,\tilde\nu}.$$
\end{proposition}

Now we   analyze the   effect of reversing orientation on the   boundary half-circles of the amplitudes. We introduce the following operators:
$$\forall F\in e^{-\beta c_-}\mathcal{H}^+,\quad\quad \mathbf{O}^+F(\tilde{\varphi}^\mathrm{h})=F(\tilde{\varphi}^{\rm h}\circ(-o)),\quad\quad \mathbf{C}F(\tilde{\varphi}^\mathrm{h})=\overline{F}(\tilde{\varphi}^\mathrm{h}).$$

In the forthcoming analysis, the dependence of several quantities on the boundary cosmological constants will play an essential role.
For this reason, in what follows, we will use the notations $\Psi_{\alpha,\nu}^{+,\mu^{\rm L},\mu^{\rm R}}$, $\mathbf{L}_n^{+,\mu^{\rm L},\mu^{\rm R}}$, and $\mathbf{H}^{+,\mu^{\rm L},\mu^{\rm R}}$ instead of $\Psi_{\alpha,\nu}^{+}$, $\mathbf{L}_n^{+}$, and $\mathbf{H}^+$, respectively.

Recall the notation in Section~\ref{s:semigroup_half}. We have the following lemma:
\begin{lemma}\label{lem: revert orientation}
	The operators $\mathbf{O}^+$ and $\mathbf{C}$ map $\mc{C}_{\infty}(\R\times \Omega_{\T^+})$ into itself. Moreover,  for all $n\in \Z$,
	$$\mathbf{O}^+\mathbf{L}_n^{+,0}=(-1)^n\mathbf{L}_n^{+,0}\mathbf{O}^+,\quad\quad \mathbf{C}\mathbf{L}_n^{+,0}=\mathbf{L}_n^{+,0}\mathbf{C}.$$
	In addition, the operator   $\mathbf{C}$   commutes  with $e^{-t\mathbf{H}^{+,\mu^{\rm L},\mu^{\rm R}}}$ on $e^{-\beta c_-}\mathcal{H}^+$ for $\beta\geq 0$, while the operator  $\mathbf{O}^+$ commutes  with $e^{-t\mathbf{H}^{+,\mu^{\rm L},\mu^{\rm R}}}$ on $e^{-\beta c_-}\mathcal{H}^+$ up to exchanging the boundary cosmological constants, that is,
	$$  \mathbf{O}^+e^{-t\mathbf{H}^{+,\mu^{\rm L},\mu^{\rm R}}}= e^{-t\mathbf{H}^{+,\mu^{\rm R},\mu^{\rm L}}}\mathbf{O}^+.$$
\end{lemma}
\begin{proof}
First,  a direct computation yields the identities	
	$$\mathbf{O}^+\mathbf{A}_n^+=(-1)^n\mathbf{A}_n^+\mathbf{O}^+,\quad\quad \mathbf{C}\mathbf{A}^+_n=-\mathbf{A}^+_n\mathbf{C}.$$
Using the expression of $\mathbf{L}_n^{+,0}$ in terms of the operators $\mathbf{A}_n^+$, we then obtain as an immediate consequence 
$\mathbf{O}^+\mathbf{L}_n^{+,0}=(-1)^n\mathbf{L}_n^{+,0}\mathbf{O}^+$ and $\mathbf{C}\mathbf{L}_n^{+,0}=\mathbf{L}_n^{+,0}\mathbf{C}$.
 The commutation relation 
$e^{-t\mathbf{H}^+}\mathbf{C}=\mathbf{C}e^{-t\mathbf{H}^+}$ follows from the Feynman-Kac representation~\eqref{propagator_half}.
Similarly, the identity $  \mathbf{O}^+e^{-t\mathbf{H}^{+,\mu^{\rm L},\mu^{\rm R}}}= e^{-t\mathbf{H}^{+,\mu^{\rm R},\mu^{\rm L}}}\mathbf{O}^+$ follows again from the Feynman-Kac formula, together with the two facts that $P_{\D^+}(\varphi^{\rm h}\circ(-o))(z))=P_{\D^+}\varphi^{\rm h}(-\bar z)$ for $z\in\D^+$, and that the mixed b.c. GFF satisfies $X_{\D^+,m}(z)=X_{\D^+,m}(-\bar z)$ in law.
\end{proof}
Now we turn to the boundary analogue of Proposition~\ref{prop: RO bulk}.  
\begin{proposition}\label{prop: RO boundary} 
	For $P\in\R$ and $\nu\in\mathcal{T}$, we have
	$$\mathbf{C}\Psi^{+,\mu^{\rm L},\mu^{\rm R}}_{Q+iP,\nu}=\Psi^{+,\mu^{\rm L},\mu^{\rm R}}_{Q-iP,\nu}, \quad\text{ and }\quad \mathbf{O}^+\Psi^{+,\mu^{\rm L},\mu^{\rm R}}_{Q+iP,\nu}=(-1)^{|\nu|}\Psi^{+,\mu^{\rm R},\mu^{\rm L}}_{Q+iP,\nu}.$$
\end{proposition}
\begin{proof} 	It suffices to  establish the statements for real  $\alpha<Q-N$, for some (possibly large) $N>0$, and then to extend it by analytic continuation. 
	We use   formula~\eqref{Psi+_Psi^+0}, which states that, for $\alpha$ real and sufficiently negative, $\Psi^{+,\mu^{\rm L},\mu^{\rm R}}_{\alpha,\nu}=\lim_{t\to +\infty}e^{t(\Delta_\alpha+|\nu|)}e^{-t{\bf H}^{+,\mu^{\rm L},\mu^{\rm R}}}(\mathbf{L}_\nu^{+,0}\Psi^{+,0}_{\alpha})$. Using this representation, we compute:
\begin{align*}
 \mathbf{O}^+\Psi^{+,\mu^{\rm L},\mu^{\rm R}}_{Q+iP,\nu}=& \mathbf{O}^+\lim_{t\to +\infty}e^{t(\Delta_\alpha+|\nu|)}e^{-t{\bf H}^{+,\mu^{\rm L},\mu^{\rm R}}}    \mathbf{L}_{-\nu}^{+,0}\Psi^{+,0}_{\alpha}=\lim_{t\to +\infty}e^{t(\Delta_\alpha+|\nu|)}e^{-t{\bf H}^{+,\mu^{\rm R},\mu^{\rm L}}} \mathbf{O}^+  \mathbf{L}_{-\nu}^{+,0}\Psi^{+,0}_{\alpha}
 \\
 =&(-1)^{|\nu|}\lim_{t\to +\infty}e^{t(\Delta_\alpha+|\nu|)}e^{-t{\bf H}^{+,\mu^{\rm R},\mu^{\rm L}}}   \mathbf{L}_{-\nu}^{+,0}\mathbf{O}^+ \Psi^{+,0}_{\alpha}=(-1)^{|\nu|}\lim_{t\to +\infty}e^{t(\Delta_\alpha+|\nu|)}e^{-t{\bf H}^{+,\mu^{\rm R},\mu^{\rm L}}}   \mathbf{L}_{-\nu}^{+,0}  \Psi^{+,0}_{\alpha}
 \\
 =&\Psi^{+,\mu^{\rm R},\mu^{\rm L}}_{Q+iP,\nu}.
 \end{align*}	
 The second relation $\mathbf{C}\Psi^{+,\mu^{\rm L},\mu^{\rm R}}_{Q+iP,\nu}=\Psi^{+,\mu^{\rm L},\mu^{\rm R}}_{Q-iP,\nu}$ can be treated similarly.
\end{proof}

\section{Ward  identities and Cardy doubling trick}\label{sec:Ward}

In this section, we compute the matrix elements of the amplitudes corresponding to  the complex building blocks of type (A1), (A2), (B1), (B2), (C1) and (C2) introduced in Section \ref{geometric_decomp}, expressed in the bases given by $(\Psi_{Q+ip,\nu,\tilde{\nu}})_{p\in \R^+,\nu\in \mc{T},\tilde{\nu}\in \mc{T}}$ of $\mc{H}$ and $(\Psi^+_{Q+ip,\nu})_{p\in \R^+,\nu\in \mc{T}}$ of $\mc{H}^+$. We regard amplitudes as Hilbert-Schmidt operators as in~\eqref{amplitude_HS_op}. In particular, when the boundary parametrisations are incoming, the amplitudes  are complex linear maps $\mc{H}^{\otimes b^\ell}\otimes (\mc{H}^+)^{\otimes b^{\rm h}}\to \C$, whereas if all parametrisations are outgoing,  the amplitudes can be viewed as elements   of $\mc{H}^{\otimes b^\ell}\otimes (\mc{H}^+)^{\otimes b^{\rm h}}$.
In fact,   the building blocks of types (B1), (B2), (C1), (C2) can be obtained from  the half pairs of pants (A1) and (A2) by gluing discs or half-discs with a marked point  at their center. At the level of Segal amplitudes, this implies that it suffices to analyze the case  of half pairs of pants.
	
\begin{figure}
		\begin{tikzpicture}
			\node[inner sep=0pt] (pant) at (0,0)
			{\includegraphics[width=0.7\textwidth]{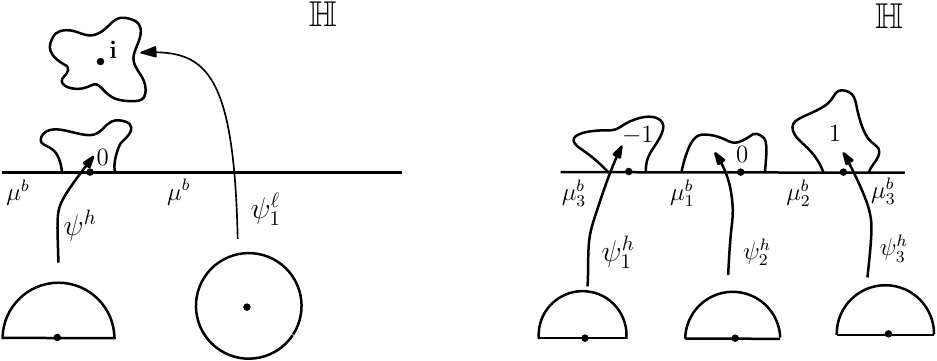}};
		\end{tikzpicture}
		\caption{Left: Case (A1); Right: Case (A2)}\label{fig: type I II}
	\end{figure}

We recall that the half pairs of pants  have the following model forms:
\begin{enumerate}
\item \textbf{Case (A1)}:  let $\bs{\psi}=(\psi_1^\ell,\psi_1^{\rm h})\in {\rm BiHol}_\eps(\D)\times{\rm BiHol}^+_\eps(\D)$ for some $\eps>0$ satisfying
 $$
 (\psi_1^\ell(0),\psi_1^{\rm h}(0))=(i,0),\qquad \psi_1^\ell(\D)\cap \psi_1^{\rm h}(\D^+)=\emptyset.
$$
 We consider $\Sigma_{\bs{\psi}}=\hat{\H}\setminus(\psi_1^\ell(\D)\cup\psi_1^{\rm h}(\D^+))$ with incoming boundary parametrisation $\bs\zeta=(\zeta_1^\ell,\zeta_1^{\rm h})=
(\psi_1^{\ell}|_\T,\psi_1^{\rm h}|_{\T^+})$, where $(\zeta_1^\ell,\zeta^{\rm h})=(\psi_1^\ell(e^{\mathbf{i}\theta}),\psi^{\rm h}(e^{\mathbf{i}\theta}))$. We assign D marking to $\psi_1^\ell(\T)\cup\psi_1^{\rm h}(\T^+)$ and 
N marking to $\R\setminus\psi^{\rm h}(\D)$. The boundary cosmological constant is taken to be a constant $\mu^{\rm b}\geq 0$ on $\psi_1^\ell(\T)\cup\psi_1^{\rm h}(\T^+)$. See Figure~\ref{fig: type I II}.
\item \textbf{Case (A2)}: let $\bs{\psi}=(\psi^{\rm h}_1,\psi^{\rm h}_2,\psi^{\rm h}_3)\in \mathrm{BiHol}^+_\eps(\D)^3$ for some $\eps>0$, satisfying  $$
(\psi_{1}^{\rm h}(0),\psi_{2}^{\rm h}(0),\psi_{3}^{\rm h}(0))=(-1,0,1),\qquad \cap_{k=1}^3\psi^{\rm h}_k(\D)=\emptyset.
$$
We consider $\Sigma_{\bs{\psi}}=\hat{\H}\setminus(\cup_{j=1}^3\psi^{\rm h}_j(\D^+))$  with incoming boundary parameterization $\bs\zeta=(\zeta_1^{\rm h},\zeta_2^{\rm h},\zeta_3^{\rm h})$, where $\zeta_j^{\rm h}=\psi_j^{\rm h}|_{\T^+}$. 
We assign D-marking to $\cup_{j=1}^3\psi_j^{\rm h}(\T^+)$ and N-marking to the intervals in $\Sigma_{\bs{\psi}}\cap \R$. The boundary cosmological constants are taken to be  $\mu^{\rm b}_1\geq 0$ on $\pl\Sigma_{\bs{\psi}}\cap (-1,0)$, 
$\mu^{\rm b}_2\geq 0$ on $\pl\Sigma_{\bs{\psi}}\cap (0,1)$ and $\mu^{\rm b}_3\geq 0$ on 
$\pl\Sigma_{\bs{\psi}}\cap (\R\setminus(-1,1))$. See Figure~\ref{fig: type I II}.
\end{enumerate}

In this section, we restrict to amplitudes with incoming boundaries only, since Proposition~\ref{prop: RO boundary} ensures that the general case follows from this one. We now fix the notation for meromorphic vector fields and their pullbacks.  
For $\psi\in \mathrm{BiHol}_\epsilon(\D)$, we define the pullback of the meromorphic vector field $\mathrm{v}=v(z)\partial_z $ by
\[\psi^*\mathrm{v} =\frac{{v}(\psi(z))}{\psi'(z)}\partial_z=\sum_{k\geq -N} [\psi^*\mathrm{v} ]_k\cdot z^{k+1}\partial_z,\]
where $N$ is some integer depending on $\psi$ and $\mathrm{v}$.

Let us finally introduce some additional notation. Let $P=P(x,\nabla_x)$ be a first-order differential operator on an open set
$\Omega\ \subset \C$, and $Q=Q(s,\nabla_s)$ a first-order differential operator on an open set
$I\ \subset \R$.
Let $F:(\Omega^m\times I^{m^{\rm b}})\to A$ be a function with values in a Banach space $A$, which is differentiable 
 in ${\bf x}=(x_1,\dots,x_m)$ (resp.  in ${\bf s}=(s_1,\dots,s_{m^{\rm b}})$) away from the diagonals
\[\{{\bf x}=(x_1,\dots,x_m,s_1,\dots,s_{m^{\rm b}})\,|\, \exists i,j, x_i=x_j\} \quad (\textrm{resp. }\{{\bf s}=(s_1,\dots,s_{m^{\rm b}})\,|\, \exists i,j, s_i=s_j\}).\] 
We then define the action of $P$ or $Q$ on the $i$-th variable by
\begin{equation}\label{action_ith}
\begin{gathered} 
 P_{\bf x}^{(i)}F({\bf x},{\bf s}):=P(x,\nabla_x)F(x_1,\dots,x_{i-1},x,x_{i+1},\dots, x_m,{\bf s})|_{x=x_i},\\
 Q_{\bf s}^{(i)}F({\bf x},{\bf s}):=Q(s,\nabla_s)F({\bf x},s_1,\dots,s_{i-1},s,s_{i+1},\dots,s_{m^{\rm b}})|_{s=s_i}.
\end{gathered}
\end{equation}
In particular, if ${\rm v}=v(z)\partial_z$ is a holomorphic vector field obeying $\bar v(z)=v(\bar z)$ then ${\rm v}_{\bf s}^{(i)}F({\bf x},{\bf s})=\frac{1}{2}v(s_i)\partial_{s_i}F({\bf x},{\bf s})$.

\subsection{Variation formula for amplitudes of complex building blocks with no corner}\label{sec: variation_no_corner}\label{sub:ward1}

First, we recall the results of \cite{BGKR1,BGKR2,WX25}, which describe the variations of Liouville amplitudes of building blocks of type (A), (B), (C), that is,  in the case where  $\partial\Sigma_N=\emptyset$.

Consider a surface $\Sigma=\hat{\C}\setminus \cup_{k=1}^b\psi_k(\D)$ of type (A), (B), or (C) in model form, where each $\psi_k\in {\rm BiHol}_\eps(\D)$ for some $\eps>0$, with disjoint images $\psi_k(\D)\cap\psi_{k'}(\D)=\emptyset$ for all $k\not=k'$. We assume that  $\psi_k(0)=z^0_k$, with ${\bf z}^0=(z^0_1,z^0_2,z^0_3)=(-1,0,1)$, and denote by  $\zeta_j=\psi_j|_{\T}$ the  parametrisations of the D-marked  boundary components.  Let  ${\bf x}=(x_1,\dots,x_m)$ be  the marked points (with $m+b=3$) and assume 
that $x_k=z^0_{b+k}$. 

We consider an admissible  metric $g$ on $\Sigma$, and extend it smoothly inside each $\psi_k(\D\setminus\{0\})$ by setting $(\psi_k)_*(|\dd z|^2/|z|^2)$.  We denote by $\hat{g}$ the smooth extension of $g$ to $\hat{\Sigma}:=\hat{\C}\setminus \{z^0_1,\dots,z^0_b\}$.

The following Ward identity is established in \cite[Corollary 3.6]{BGKR2} (see also the proof of Proposition \ref{prop: ward H} below, where the details are provided in the more general case with corners):
\begin{proposition}\label{prop:localward move point}
Let ${\rm v}=v(z)\pl_z$ be a holomorphic vector field defined on a neighborhood of $\Sigma$ in $\hat{\C}$. For $j=1,\dots,b$, let ${\rm v}_j=v_j\pl_z:=\psi_j^*{\rm v}$ with $v_j(z)=\sum_{n\in \Z}v_{jn}z^{n+1}$. Then, for 
any $\bs\alpha=(\alpha_1,\cdots,\alpha_m)$ satisfying the Seiberg bounds, one has
\begin{align*}
0= &-\mathcal{A}_{\Sigma,g,\bf{x},\boldsymbol{\alpha},\bs\zeta}\left(\mathbf{H}_{\mathrm{v}_1} \otimes \mathrm{Id} \otimes \cdots \otimes \mathrm{Id}\right)(F) - \cdots 
- \mathcal{A}_{\Sigma,g,\bf{x},\boldsymbol{\alpha},\bs\zeta}\left(\operatorname{Id} \otimes \cdots \otimes \operatorname{Id} \otimes \mathbf{H}_{\mathrm{v}_b}\right)(F)\\
&+  \Big( \sum_{i=1}^m\Delta_{\alpha_i}{\rm div}_g(V)(x_i)+V_{\bf x}^{(i)}\Big) \mathcal{A}_{\Sigma,g,\bf{x},\boldsymbol{\alpha},\bs\zeta}(F)
		-c_{\rm L}\Big(\frac{1}{48\pi}\int_{\Sigma}K_g{\rm div}_{g}(V){\rm dv}_g+ \sum_{j=1}^b \frac{{\rm Re}(v_{j0})}{12}\Big)\mathcal{A}_{\Sigma,g,\bf{x},\boldsymbol{\alpha},\bs\zeta}(F)
	\end{align*}
	for all $F\in \mathcal{D}(\mathcal{Q})^b$, where $V={\rm v}+\bar{{\rm v}}$, ${\rm div}_{g}$ denotes the divergence with respect to $g$ and $V_{\bf x}^{(i)}$ denotes the vector field $V$ acting in the i-th variable of a function of $m$ variables (see \eqref{action_ith}).
\end{proposition}

The following proposition establishes the relation between the variation of the Liouville action and the Schwarzian derivative. This result was proved in~\cite{WX25} and can also be deduced as a corollary of Ward identities, see for instance~\cite[Proposition 11.3]{GKRV21_Segal}.
 \begin{lemma}\label{lem:schwarzian}	
With the same notations as in Proposition \ref{prop:localward move point}, let us define ${\rm div}_{g}({\rm v})=\frac{1}{2}({\rm div}_g(V)-{\rm i}\, {\rm div}_g(W))$ with $W={\rm i}({\rm v}-\bar{{\rm v}})$. Then
\begin{align*}
\frac{1}{24\pi}\int_{\Sigma}K_g{\rm div}_g({\rm v}){\rm dv}_g + \frac{1}{12} \sum_{j=1}^bv_{j0} = \frac{{\rm i}}{12\pi} \sum_{j=1}^b \oint_{\zeta_j(\T)} {v}(z){\rm S}(\psi_j^{-1})(z)\dd z,
\end{align*}
where ${\rm S}(f)(z):=(\frac{f''(z)}{f'(z)})'-\frac{1}{2}(\frac{f''(z)}{f'(z)})^2$ is the Schwarzian derivative of $f$ and the contour integral has the orientation induced by $\zeta_j$.
\end{lemma}

 As a consequence, the multiplicative factor in front of the last term in the statement of Proposition~\ref{prop:localward move point} can be replaced by the real part of the Schwarzian term given above.

\begin{proof} We  briefly reproduce here the computations in \cite{WX25} for completeness.
 Let us consider a conformal and admissible metric  $g=e^{\sigma(z)}|d z|^2$ on $\Sigma$. Then 
\[\frac{1}{24\pi}\int_{\Sigma}K_g{\rm div}_g({\rm v}){\rm dv}_g =\frac{1}{24\pi}\int_{\Sigma} K_{g}\left(\partial_z v+v \partial_z \sigma\right) {\rm dv}_{g}\]
We  recall that 
$$K_{g} {\rm dv}_{g}=-4 \partial_z \partial_{\bar{z}} \sigma \dd x=-2i \partial_z \partial_{\bar{z}} \sigma \dd z\wedge \dd\bar z.$$
Applying Stoke's theorem yields
	\begin{align*}
		\int_{\Sigma} \partial_z \partial_{\bar{z}} \sigma\left(\partial_z v+v \partial_z \sigma\right) \dd z\wedge \dd\bar z& =\int_{\Sigma} v\left(\partial_z \sigma \partial_z \partial_{\bar{z}} \sigma-\partial_z^2 \partial_{\bar{z}} \sigma\right)\dd z\wedge \dd \bar z-\oint_{\partial \Sigma} v \partial_z \partial_{\bar{z}} \sigma \dd \bar{z} \\
		& =\int_{\Sigma} \partial_{\bar{z}} v\big(\partial_z^2 \sigma-\frac{1}{2}\left(\partial_z \sigma\right)^2\big)\dd z\wedge \dd\bar z- \oint_{\partial \Sigma} v\big(\partial_z^2 \sigma-\frac{1}{2}\left(\partial_z \sigma\right)^2\big) \dd z+ v \partial_z \partial_{\bar{z}} \sigma\dd \bar{z}.
	\end{align*}
	where the contour integral is oriented according to the maps  $\zeta_j$ defining    the boundary circles of  $\pl \Sigma$.
The first term vanishes since the vector field v is holomorphic on $\Sigma$. Because $g$ is admissible,  we have $\psi_j^*g=|\dd z|^2 /|z|^2$ near $\T$, and we can write $\sigma \circ \psi_j(z)=-2 \log \left|z \psi_j^{\prime}(z)\right|$ near $\T$. Letting $w=\psi_j(z)$ and  $\kappa=\psi_j^{-1}$ be the inverse map, we obtain $\sigma(w)=-2 \log |\kappa(w)|+2 \log \left|\kappa^{\prime}(w)\right|$. Since $\sigma$ is harmonic near each boundary component $\psi_j(\T)$, we have $\oint_{\partial \Sigma} v \partial_z \partial_{\bar{z}} \sigma \dd \bar{z}=0$. Moreover  $\partial_w^2\left(2 \log \left|\kappa^{\prime}(w)\right|\right)-\frac{1}{2}\left(\partial_w\left(2 \log \left|\kappa^{\prime}(w)\right|\right)\right)^2={\rm S}(\kappa)(w)$, so that
\[\partial_w^2 \sigma-\frac{1}{2}\left(\partial_w \sigma\right)^2={\rm S}(\kappa)(w)+\frac{\kappa^{\prime 2}(w)}{2 \kappa^2(w)}.
\]
	By definition, if we write $\psi_j^* \mathrm{v}=\sum_{n\in \Z} v_{jn} z^{n+1}\partial_z$, then 
	$$v_{j0} =\frac{1}{2\pi {\rm i}}\oint_{\mathbb{T}} \frac{v \circ \psi_j(z)}{\psi_j^{\prime}(z)}   \frac{1}{z^2} \dd z=\frac{1}{2\pi {\rm i}}\oint_{\partial_j \Sigma} v(w)\left(\frac{\kappa^{\prime}(w)}{\kappa(w)}  \right)^2 \dd w.$$
	 The result follows.
\end{proof}

For ${\bs\alpha}=(\alpha_1,\dots, \alpha_m)\in \C^{m}$ and $\bs{\psi}=(\psi_1,\dots,\psi_b)\in {\rm BiHol}_\delta(\A)$, for $\delta<1$ close to $1$,  we introduce the following differential operator of order $1$ acting on functions $F\in C^1(\Sigma^{m},A)$, with values in a Banach space $A$, by
\begin{align}
	\mathbf{D}(\bs\alpha,\bs\psi,\mathrm{v})F:=\sum_{i=1}^m(\Delta_{\alpha_i}{\rm div}_{g}({\rm v})+{\rm v})_{\bf x}^{(i)}F+c_{\rm L}\big(\frac{1}{24\pi {\rm i}}\sum_{j=1}^b\oint_{\zeta_j(\T)}\mathrm{v}S(\psi_j^{-1})\big)F.
\end{align}
We also notice that by the chain rule of Schwarzian derivative, the last term can also be written as $\int_{\zeta_j(\T)}\mathrm{v}{\rm S}(\psi_j^{-1})=-\int_\T\psi_j^*\mathrm v \cdot {\rm S}(\psi_j)$ (viewed as a pairing between holomorphic vector field and quadratic differential).
As a consequence of Proposition \ref{prop:localward move point} and Lemma \ref{lem:schwarzian}, and using 
that 
\[\frac{1}{2}({\bf H}_{\rm v}-{\rm i}{\bf H}_{{\rm i}{\rm v}})={\bf L}_{\rm v}:=-\sum_n v_n{\bf L}_n\] 
with ${\rm v}=(\sum_n v_nz^{n+1})\pl_z$, we have: 
\begin{corollary}\label{final_ward_bulk_pant}
With the same assumptions and notations as in Proposition \ref{prop:localward move point}, the following holds true:
\begin{align*}
\mathbf{D}(\bs\alpha,\bs\psi,\mathrm{v}) \mathcal{A}_{\Sigma,g_0,\bf{x},\boldsymbol{\alpha},\bs\zeta}(F)
= &\mathcal{A}_{\Sigma,g_0,\bf{x},\boldsymbol{\alpha},\bs\zeta}\left(\mathbf{L}_{\mathrm{v}_1} \otimes \mathrm{Id} \otimes \cdots \otimes \mathrm{Id}\right)(F) + \cdots 
+ \mathcal{A}_{\Sigma,g_0,\bf{x},\boldsymbol{\alpha},\bs\zeta}\left(\operatorname{Id} \otimes \cdots \otimes \operatorname{Id} \otimes \mathbf{L}_{\mathrm{v}_b}\right)(F)\end{align*}
	for any $F\in \mathcal{D}(\mathcal{Q})^b$.
	\end{corollary}

\subsection{Variation formula for amplitudes of complex building blocks with corners}\label{sec: variation_corner} 

Consider the surface 
$$\Sigma=\hat{\H}\setminus (\cup_{k=1}^{b^\ell}\psi_k^{\ell}(\D)\cup \cup_{k=1}^{b^{\rm h}}\psi_k^{\rm h}(\D^+))$$
 of type 
(A1),(A2), (B1), (B2), (C1) or (C2), in model form as described in Section \ref{geometric_decomp}, with $\psi_k^{\ell}\in {\rm BiHol}_\eps(\D)$ and $\psi_k^{\rm h}\in {\rm BiHol}_\eps^+(\D)$.  The interior marked points are denoted by ${\bf x}=(x_1,\dots,x_m)$ and boundary marked points ${\bf s}=(s_1,\dots,s_{m^{\rm b}})$. 

We consider an admissible metric $g$ on $\Sigma$, and extend it to
$\psi_k^\ell(\D\setminus\{0\})$ and
$\psi_k^{\rm h}(\D^+\setminus\{0\})$ by setting
\[
g=(\psi_k^\ell)_*
\left(\frac{|\dd z|^2}{|z|^2}\right)
\quad\text{and}\quad
g=(\psi_k^{\rm h})_*
\left(\frac{|\dd z|^2}{|z|^2}\right),
\]
respectively.  We denote by $\hat g$ the resulting extension to
\[
\hat\Sigma
:=
\hat\H\setminus\{z_1^0,\ldots,z_b^0\}.
\]
Let $\tau:\hat\C\to\hat\C$ be the antiholomorphic involution
$\tau(z)=\bar z$. The Neumann double of $\Sigma$ is naturally identified
with
\[
\Sigma^{\#2}
=
\hat\C\setminus
\left[
\bigcup_{k=1}^{b^\ell}
\left(
\psi_k^\ell(\D)
\cup
\tau\bigl(\psi_k^\ell(\D)\bigr)
\right)
\;\cup\;
\bigcup_{k=1}^{b^{\rm h}}
\psi_k^{\rm h}(\D)
\right].
\]
Since $g$ is admissible, it is Neumann extendible, and both $g$ and
$\hat g$ extend smoothly to the lower half-plane by reflection. By abuse
of notation, we keep the same symbols $g$ and $\hat g$ for these
$\tau$-invariant extensions.
We now establish the following Ward identity for surfaces with corners, whose proof parallels that of Proposition \ref{prop:localward move point} (see \cite{BGKR1}).
\begin{proposition}\label{prop: ward H}
Let ${\rm v}=v(z)\pl_z$ be a holomorphic vector field defined  on a neighborhood of $\Sigma^{\#2}$ satisfying $\bar{v}(z)=v(\bar{z})$.
Define 
$${\rm v}_j^\ell= v_j^{\ell}\pl_z:=(\psi^{\ell}_j)^*{\rm v},\quad {\rm v}_j^{\rm h}=v_j^{\rm h}\pl_z:=(\psi^{\rm h}_j)^*{\rm v}, \quad \text{with 
 } v_j^{\ell}(z)=\sum_{n\in \Z}v_{jn}^{\ell}z^{n+1},\quad v_j^{\rm h}(z)=\sum_{n\in \Z}v_{jn}^{\rm h}z^{n+1}.$$
Assume that the weights $(\bs\alpha,\bs\beta)$ associated with ${\bf x},{\bf s}$ satisfy the Seiberg bounds. Then  
\begin{align*}
0= &-\mathcal{A}_{\Sigma,g_0,{\bf x},\boldsymbol{\alpha},{\bf s},\boldsymbol{\beta},\bs\zeta}\left(\mathbf{H}_{{\rm v}^\ell_1} \otimes \mathrm{Id} \otimes \cdots \otimes \mathrm{Id}\right)(F) - \ldots 
- \mathcal{A}_{\Sigma,g,{\bf x},\boldsymbol{\alpha},{\bf s},\boldsymbol{\beta},\bs\zeta}\left(\operatorname{Id} \otimes \cdots \otimes \operatorname{Id} \otimes \mathbf{H}_{{\rm v}^\ell_{b^\ell}}\otimes \cdots \otimes {\rm Id}\right)(F)\\
&-\mathcal{A}_{\Sigma,g,{\bf x},\boldsymbol{\alpha},{\bf s},\boldsymbol{\beta},\bs\zeta}\left( \mathrm{Id} \otimes \cdots \otimes \mathbf{H}^+_{{\rm v}^{\rm h}_1} \otimes\cdots \otimes \mathrm{Id}\right)(F) - \ldots 
- \mathcal{A}_{\Sigma,g,{\bf x},\boldsymbol{\alpha},{\bf s},\boldsymbol{\beta},\bs\zeta}\left(\operatorname{Id} \otimes \cdots \otimes \operatorname{Id} \otimes \mathbf{H}^+_{{\rm v}^{\rm h}_b}\right)(F)\\
&+  \Big( \sum_{i=1}^{m}(\Delta_{\alpha_i}{\rm div}_g(V)+V)_{\bf x}^{(i)}+ \sum_{i=1}^{m^{\rm b}}( \frac{\Delta_{\beta_i}}{2} {\rm div}_g({\rm V})+{\rm V})_{\bf s}^{(i)}\Big) \mathcal{A}_{\Sigma,g,{\bf x},\boldsymbol{\alpha},{\bf s},\boldsymbol{\beta},\bs\zeta}(F)\\
&		- \mathrm{c_L }\Big(\frac{1}{48\pi}\int_{\Sigma}K_g{\rm div}_{g}(V){\rm dv}_g+ \sum_{j=1}^{b^\ell} \frac{{\rm Re}(v^{\ell}_{j0})}{12}+ \sum_{j=1}^{b^{\rm h}} \frac{ v^{\rm h}_{j0}}{24}\Big)\mathcal{A}_{\Sigma,g,{\bf x},\boldsymbol{\alpha},{\bf s},\boldsymbol{\beta},\bs\zeta}(F)
	\end{align*}
for any $F\in \mathcal{D}(\mathcal{Q})^{b^\ell}\times\mathcal{D}(\mathcal{Q}^+)^{b^{\rm h}}$, where 
$V={\rm v}+\bbar{{\rm v}}$.
\end{proposition}

\begin{proof} 
Let $f_t$ denote the flow of the vector field ${\rm v}$ on $\hat\Sigma$, defined for small $t$ in a neighborhood of $\Sigma$ by  $\partial_t f_t=v(f_t)$, with $f_0(z)=z$. Set $\Sigma_t=f_t(\Sigma)$ with boundary parametrisation 
$\boldsymbol{\zeta}_t=(f_t\circ \zeta^{\ell}_{1},\dots,f_t\circ \zeta^{\ell}_{b^\ell},f_t\circ \zeta^{{\rm h}}_{1},\dots,f_t\circ \zeta^{{\rm h}}_{b^{\rm h}})$.

Let $g_t$ be a smooth one-parameter family of  metrics defined on a neighborhood of $ \Sigma$, conformal to $\hat g$, such that   $f_t^*g_t=e^{\omega_t}\hat g$ with $\omega_t=0$ near $\pl\Sigma$, and $g_t=g$ outside a small neighborhood $U$ of  $\pl\Sigma$. We may choose $U$ so that $\Sigma':=\Sigma\setminus U$ has boundary  
$$\partial\Sigma'=\bigcup_j \psi^{\ell}_j((1+\eps)\T)\cup \bigcup_j \psi^{\rm h}_j((1+\eps)\T^+)$$ for small $\epsilon>0$; denote its boundary parametrisation by  $\bs{\zeta}'$.
 
The surface $(\Sigma_t,g_t)$ can be decomposed as the gluing of the $t$-independent surface $(\Sigma',g')$ 
with $(U\cap\Sigma_t,g_t)$, which is diffeomorphic to a disjoint union of $b^\ell$ annuli and $b^{\rm h}$ half-annuli:
\[(U\cap\Sigma_t,g_t) \simeq \sqcup_{j=1}^{b^\ell}(\A_{rf_{j,t}^{\ell}},(\psi_j^\ell\circ r^{-1})^*g_t) \sqcup \sqcup _{j=1}^{b^{\rm h}}(\A^+_{rf_{j,t}^{\rm h}},(\psi_j^{\rm h}\circ r^{-1})^*g_t) \]
for  fixed $r=1/(1+\eps)$  and small $t$,  where $f_{j,t}^{\ell}:=(\psi_j^\ell)^{-1}\circ f_t\circ \psi_j^{\ell}:\T\to \C$, and $f_{j,t}^{{\rm h}}:=(\psi_j^{\rm h})^{-1}\circ f_t\circ \psi_j^{\rm h}:\T^+\to \C$. Here $\psi^\ell\circ r^{-1}(z):=\psi^\ell(r^{-1}z)$ and the same for $\psi^{\rm h}\circ r^{-1}$.

 By diffeomorphism invariance, $$\mathcal{A}_{\Sigma_t,g_t,{\bf x},\boldsymbol{\alpha},{\bf s},\boldsymbol{\beta},\bs{\zeta}_t}=\mathcal{A}_{\Sigma,f_t^*g_t,{\bf x}_t,\boldsymbol{\alpha},{\bf s}_t,\boldsymbol{\beta},\bs{\zeta}},$$ with ${\bf x}_t=f_{-t}({\bf x})$ and ${\bf s}_t=f_{-t}({\bf s})$. The desired identity follows by differentiating this equality at $t=0$.
 
Indeed, using \cite[Proposition 6.1]{GRW1} (see also \eqref{glue_ampl}), the amplitude $\mathcal{A}_{\Sigma_t,g_t,{\bf x},\boldsymbol{\alpha},{\bf s},\boldsymbol{\beta},\bs\zeta_t}$ decomposes as the composition of 
$\mathcal{A}_{\Sigma',g,{\bf x},\boldsymbol{\alpha},\bf{s},\boldsymbol{\beta},\bs\zeta'}$ with the amplitudes of  annuli and half-annuli. Hence, its differential   reduces to that of the annular and half-annular amplitudes.
Applying \cite[Theorem 4.32]{BGKR1} (for annuli) and Proposition~\ref{l:differential_annuli} (for  half-annuli) yields
\begin{align*}
\partial_t\mathcal{A}_{\Sigma_t,g_t,{\bf x},\boldsymbol{\alpha},{\bf s},\boldsymbol{\beta},\bs\zeta_t}|_{t=0}F=&-\mathcal{A}_{\Sigma,g,\bf{x},\boldsymbol{\alpha},\bf{s},\boldsymbol{\beta},\bs\zeta}\left(\mathbf{H}_{\mathrm{v}^\ell_1} \otimes \mathrm{Id} \otimes \cdots \otimes \mathrm{Id}\right)F- \cdots -\mathcal{A}_{\Sigma,g,\bf{x},\boldsymbol{\alpha},\bf{s},\boldsymbol{\beta},\bs\zeta}\left(\mathrm{Id} \otimes \cdots \otimes\mathbf{H}_{\mathrm{v}^\ell_{b^\ell}} \otimes \cdots \mathrm{Id}\right)F\\
&-\mathcal{A}_{\Sigma,g,\bf{x},\boldsymbol{\alpha},\bf{s},\boldsymbol{\beta},\bs\zeta}\left(\operatorname{Id} \otimes\mathbf{H}^{+}_{\mathrm{v}^{\rm h}_{1}}\otimes \cdots \otimes \operatorname{Id}  \right)F-\cdots-\mathcal{A}_{\Sigma,g_0,\bf{x},\boldsymbol{\alpha},\bf{s},\boldsymbol{\beta},\bs\zeta}\big(\operatorname{Id} \otimes \cdots \otimes \operatorname{Id} \otimes \mathbf{H}^{+}_{\mathrm{v}^h_{b_{h}}}\big)F \\
&+c_{\rm L}\big(- \sum_{j=1}^{b^\ell} \frac{{\rm Re}(v_{j0}^\ell)}{12}- \sum_{j=1}^{b^{\rm h}} \frac{ v_{j0}^{\rm h}}{24}+\partial_t S^0_{\rm L}(\Sigma_t,\hat g,g_t) \big)\big)\mathcal{A}_{\Sigma,g,\bf{x},\boldsymbol{\alpha},\bf{s},\boldsymbol{\beta},\bs\zeta}F.
	\end{align*}
On the other hand, by Weyl covariance \eqref{Weyl_amp},  
	\begin{align*}
		\mathcal{A}_{\Sigma,f_t^*g_t,{\bf x}_t,\boldsymbol{\alpha},{\bf s}_t,\boldsymbol{\beta},\bs\zeta}=\exp\Big(c_{\rm L} S_{\rm L}^0(\Sigma, g,f_t^*g_t)-\sum_{i=1}^m\Delta_{\alpha_i}\omega_t(f_{-t}(x_i))-\frac{1}{2}\sum_{i=1}^{m^{\rm b}}\Delta_{\beta_i}\omega_t(f_{-t}(s_i)\Big)\mathcal{A}_{\Sigma,g,{\bf x}_t,\boldsymbol{\alpha},{\bf s}_t,\boldsymbol{\beta},\bs\zeta}.
	\end{align*} 
Next, we have that (note that $S^0_{\rm L}(\Sigma_t,\hat g,g_t) =-S^0_{\rm L}(\Sigma_t,g_t,\hat g)$)
\[S_{\rm L}^0(\Sigma_t,\hat g,g_t)-S_{\rm L}^0(\Sigma,g,f_t^*g_t)=S_{\rm L}^0(\Sigma,f_t^*g,f_t^*g_t)+S_{\rm L}^0(\Sigma,f_t^*g_t,g)=S_{\rm L}^0(\Sigma,f_t^*g,g),\]	 
whose variation is 
\[  \pl_tS_{\rm L}^0(\Sigma,f_t^*g,g)|_{t=0}=-\frac{1}{48\pi}\int_{\Sigma}K_{g}{\rm div}_g(V){\rm dv}_{g}.\]
Moreover, $\partial_t\omega_t(x)|_{t=0}= {\rm div}_g(V)(x)$ and $\partial_t \omega_t(s)|_{t=0}= {\rm div}_g(V)(s)$, while $\partial_t f_{-t}(x_i)|_{t=0}=-v(x_i) $. Combining these relations gives the claimed result.
\end{proof}

\subsection{Matrix coefficients of amplitudes of pairs of pants}\label{sec: pair of pants}\par
Now, we compute the coefficients of the pair-of-pants amplitude in model form, expressed in the basis $\Psi_{Q+ip,\nu,\nu'}$. This is already proved in \cite[Theorem 11.1]{GKRV21_Segal} using the stress tensor, but we provide another proof that is going to be closely related to the case of half pairs of pants, and will later serve to establish the Cardy doubling trick.

\begin{proposition}[\textbf{Type (A)}]\label{thm: ward identity}
Let  $\bs{\psi}^\ell=(\psi_1^\ell,\psi_2^\ell,\psi_3^\ell)\in \mathrm{BiHol}_\eps(\D)^3$ for some $\eps>0$ such that the images $\psi^\ell_j(\D)$ are pairwise disjoint, and assume that
 $$\bs{\psi}(0)={\bf z}^0=(z^0_1,z^0_2,z^0_3)\in\{(-1,0,1),(-i,0,i)\}.$$
Set $\Sigma=\hat{\C}\setminus \cup_{j=1}^3\psi_j^\ell(\D)$, with boundary parametrisation 
 $\bs{\zeta}=(\zeta_1,\zeta_2,\zeta_3):=(\psi^\ell_1|_{\T},\psi^\ell_2|_\T,\psi^\ell_3|_\T)$ and let $g$ be an admissible metric on $\Sigma$.
Then, for $\alpha_j=Q+ip_j$ with $p_j\in \R$, the following identity holds:
\begin{equation}\label{matrix_pant}
\mathcal{A}_{\Sigma,g,\bs\zeta}(\Psi_{\alpha_1,\nu_1,\tilde\nu_1}, \Psi_{\alpha_2,\nu_2,\tilde\nu_2}, \Psi_{\alpha_3,\nu_3,\tilde\nu_3})=\omega(\Sigma,\bs\Delta ,\bs\nu)\overline{\omega(\Sigma,\bs\Delta ,\bs{\tilde\nu})}\mathcal{A}_{\Sigma,g,\bs\zeta}(\Psi_{\alpha_1}, \Psi_{\alpha_2}, \Psi_{\alpha_3})
\end{equation}
where $\bs\Delta =(\Delta_{\alpha_1},\Delta_{\alpha_2},\Delta_{\alpha_3})$, and 
$\omega(\Sigma,\bs\Delta ,\bs\nu)$ is a polynomial in $\bs\Delta $, independent of $g$ and of the cosmological constant $\mu$, satisfying   $\omega(\Sigma,\bs\Delta ,\emptyset)=1$.
Moreover, if $\nu_k=\tilde{\nu}_k=0$ for some $k\in\{1,2,3\}$,  the identity \eqref{matrix_pant} extends holomorphically 
to $\{ {\rm Re}(\alpha_k)\in (0,Q]\}$, and similarly, if $\nu_k=\tilde{\nu}_k=\nu_{k'}=\tilde{\nu}_{k'}$, it extends holomorphically to $$\{ {\rm Re}(\alpha_k)\leq Q, 
{\rm Re}(\alpha_{k'})\leq Q, {\rm Re}(\alpha_k+\alpha_{k'})>Q\}.$$
\end{proposition}	   
   
\begin{proof} 
It suffices to prove the   identity for real $(\alpha_1,\alpha_2,\alpha_3)$, and use an analytic continuation in the $\alpha_j$'s. 
For $n\geq 1$, we consider the holomorphic  vector field  
\begin{equation}\label{vectorv0}
\mathrm{v}_{z^0_1}^{(n)}:=(z-z^0_1)^{-n+1}(z-z_2^0)(z-z^0_3)\partial_z
\end{equation}
on $\hat{\C}\setminus \{z^0_1\}$, with a $(n-1)$-th order pole at $z^0_1$,  and which vanishes at $z_2^0$ and $z_3^0$ (we will consider in the following the vector fields $\mathrm{v}_{z^0_1}^{(n)}$, $\mathrm{v}_{z^0_1}^{(n)}$ as well, which are defined similarly, in an obvious manner). It has the  expansions:
\[ {\rm v}_1=(\psi^\ell_1)^*(\mathrm{v}^{(n)}_{z^0_1})=\sum_{k\geq -n} [(\psi^\ell_1)^*(\mathrm{v}^{(n)}_{z^0_1})]_k\cdot z^{k+1} \partial_z\text{ with } [(\psi^\ell_1)^*(\mathrm{v}^{(n)}_{z^0_1})]_{-n}=\frac{c_1}{(\psi_1^\ell)'(0)^n},\]
\[{\rm v}_2=(\psi^\ell_2)^*(\mathrm{v}_{z^0_1}^{(n)})=\sum_{k\geq 0} [(\psi^\ell_2)^*(\mathrm{v}_{z^0_1}^{(n)})]_k\cdot z^{k+1} \partial_z\text{ and  } {\rm v}_3=(\psi_3^\ell)^*(\mathrm{v}_{z^0_1}^{(n)})=\sum_{k\geq 0} [(\psi_3^\ell)^*(\mathrm{v}_{z^0_1}^{(n)})]_k\cdot z^{k+1}\partial_z\]
where $c_1=(z^0_1-z_2^0)(z^0_1-z_3^0)\not=0$.
We apply Corollary \ref{final_ward_bulk_pant} with ${\rm v}={\rm v}^{(n)}_{z_1^0}$. Since 
\[ {\bf L}_{{\rm v}_1}=\frac{1}{2}({\bf H}_{{\rm v}_1}-{\rm i}{\bf H}_{{\rm i}{\rm v}_1})=-\frac{c_1}{{(\psi_1^\ell)^{'}(0)}^n} {\bf L}_{-n}-\sum_{k > -n} [(\psi_1^\ell)^*(\mathrm{v}^{(n)}_{z^0_1})]_k{\bf L}_k,\] 
 this  gives
\begin{equation}\label{induction_ward}
\begin{split}
-\frac{c_1}{{(\psi_1^\ell)^{'}(0)}^n} \mathcal{A}_{\Sigma,g, \boldsymbol{\zeta}}(\bL_{-n}\otimes \mathrm{Id}\otimes\mathrm{Id})= 
&\sum_{k > -n} [(\psi_1^\ell)^*(\mathrm{v}^{(n)}_{z^0_1})]_k\cdot\mathcal{A}_{\Sigma,g,\boldsymbol{\zeta}}(\bL_{k}\otimes \mathrm{Id}\otimes\mathrm{Id})\\
&+\sum_{k\geq 0}[(\psi_2^\ell)^*(\mathrm{v}_{z^0_1}^{(n)})]_k\cdot \mathcal{A}_{\Sigma,g, \bs\zeta}(\mathrm{Id}\otimes \mathbf{L}_{k}\otimes\mathrm{Id})\\
& +\sum_{k \geq 0} [(\psi_3^\ell)^*(\mathrm{v}_{z^0_1}^{(n)})]_k\cdot\mathcal{A}_{\Sigma,g, \bs\zeta}(\mathrm{Id}\otimes\mathrm{Id}\otimes \mathbf{L}_{k})\\
&+c_{\rm L}\frac{1}{24 \pi {\rm i}} \sum_{j=1}^3 \oint_{\psi_j(\T)} \mathrm{v}_{z^0_1}^{(n)}{\rm S}((\psi_j^\ell)^{-1})\cdot \mathcal{A}_{\Sigma,g, \boldsymbol{\zeta}}(\mathrm{Id}\otimes\mathrm{Id}\otimes \mathrm{Id}).
\end{split}
\end{equation}
Now, using the commutation relations of Virasoro algebra (see \cite[Proposition 3.6]{BGKRV}), we have that 
\begin{equation}\label{LkPsi}
\textrm{ if }k>0, \,\,  {\bf L}_k\Psi_{\alpha,\nu,\tilde{\nu}}= \sum_{j=1}^N C(\Delta_{\alpha},\nu,\nu'_j,k) \Psi_{\alpha,\nu'_j,\tilde{\nu}}, \quad 
{\bf L}_0\Psi_{\alpha,\nu,\tilde{\nu}}=( \Delta_{\alpha}+|\nu|)\Psi_{\alpha,\nu,\tilde{\nu}}
\end{equation}
for some $\nu_j'$ with length $|\nu_j'|=|\nu|-k$ and some constant $C(\Delta_{\alpha},\nu,\nu'_j,k)\in \R$. Using \eqref{induction_ward}, we see that 
\begin{equation}\label{full_coef}
\mathcal{A}_{\Sigma,g,\bs\zeta}(\Psi_{\alpha_1,\nu_1,\tilde\nu_1}, \Psi_{\alpha_2,\nu_2,\tilde\nu_2}, \Psi_{\alpha_3,\nu_3,\tilde\nu_3})
\end{equation}
can be expressed as a linear combination of 
$\mathcal{A}_{\Sigma,g,\bs\zeta}(\Psi_{\alpha_1,\nu'_1,\tilde\nu_1}, \Psi_{\alpha_2,\nu'_2,\tilde\nu_2}, \Psi_{\alpha_3,\nu'_3,\tilde\nu_3})$ for some 
Young diagrams $\nu_1'$ of length $|\nu'_1|<|\nu_1|$ and $\nu_j'$ of length $|\nu'_j|\leq |\nu_j|$ for $j=2,3$, with coefficients depending polynomially on the conformal weights $\Delta_{\alpha_j}$ 
and  the central charge $c_{\rm L}$,  and depending on the maps $\psi_j^\ell$ parametrising the pair of pants in model form. 
Iterating the argument, we see that
\begin{equation}\label{first_iteration} 
\mathcal{A}_{\Sigma,g,\bs\zeta}(\Psi_{\alpha_1,\nu_1,\tilde\nu_1}, \Psi_{\alpha_2,\nu_2,\tilde\nu_2}, \Psi_{\alpha_3,\nu_3,\tilde\nu_3})=\sum_{\nu'_2,\nu'_3}C'(\nu_1,\nu'_2,\nu'_3,\bs\Delta ,\bs{\psi})\mathcal{A}_{\Sigma,g,\bs\zeta}(\Psi_{\alpha_1,\emptyset,\tilde\nu_1}, \Psi_{\alpha_2,\nu'_2,\tilde\nu_2}, \Psi_{\alpha_3,\nu'_3,\tilde\nu_3})\end{equation}
where $\nu'_2$ and $\nu'_3$ run over the set of Young diagrams such that $|\nu'_j|\leq |\nu_j|$, 
and the constants $C'(\nu_1,\nu'_2,\nu'_3,\bs\Delta ,\bs{\psi})$ depend polynomially in $\bs\Delta $. 
Repeating the argument with the vector fields $v^{(n)}_{z_0^2}$, and then with  $v^{(n)}_{z_0^3}$, we see that 
\[ \mathcal{A}_{\Sigma,g,\bs\zeta}(\Psi_{\alpha_1,\nu_1,\tilde\nu_1}, \Psi_{\alpha_2,\nu_2,\tilde\nu_2}, \Psi_{\alpha_3,\nu_3,\tilde\nu_3})=\omega(\Sigma,\bs\Delta ,\bs{\nu})\mathcal{A}_{\Sigma,g,\bs\zeta}(\Psi_{\alpha_1,\emptyset,\tilde\nu_1}, \Psi_{\alpha_2,\emptyset,\tilde\nu_2}, \Psi_{\alpha_3,\emptyset,\tilde\nu_3})\]
and the coefficients $\omega(\Sigma,\bs\Delta ,\bs{\nu})$ depend polynomially on $\bs\Delta $ (with $\bs{\nu}=(\nu_1,\nu_2,\nu_3)$). 

We can then remove the $\tilde{\nu}_j$ terms by using twice Proposition \ref{prop:localward move point} with the same vector fields ${\rm v}={\rm v}^{(n)}_{z_1^0}$ and ${\rm i}{\rm v}$  as before, but then we take a sum instead of the difference. For $\tilde{\nu_1}$ one remarks that 
\begin{equation}\label{tildeLn}
\frac{1}{2}({\bf H}_{{\rm v}_1}+{\rm i}{\bf H}_{{\rm i}{\rm v}_1})=-\frac{\bar{c}_1}{\bbar{{(\psi_1^\ell)^{'}(0)}}^n} \tilde{{\bf L}}_{-n}-\sum_{k > -n} \bbar{[(\psi^\ell_1)^*(\mathrm{v}^{(n)}_{z^0_1})]_k}\tilde{{\bf L}}_k,
\end{equation}
and we observe that the procedure for removing $\tilde{\nu}_1$ reproduces the same algorithm as for removing $\nu_1$, except that the coefficients $[(\psi_j^\ell)^*(\mathrm{v}^{(n)}_{z^0_1})]_k$ are replaced by their complex conjugate. The Schwarzian coefficients, by Lemma \ref{lem:schwarzian}, are also replaced by their complex conjugate, and $\Delta_{\alpha_j}\in \R$ thus 
\[ \mathcal{A}_{\Sigma,g,\bs\zeta}(\Psi_{\alpha_1,\emptyset,\tilde\nu_1}, \Psi_{\alpha_2,\emptyset,\tilde\nu_2}, \Psi_{\alpha_3,\emptyset,\tilde\nu_3})=\sum_{\tilde\nu'_2,\tilde\nu'_3}\bbar{C_1(\tilde{\nu}_1,\tilde{\nu}'_2,\tilde{\nu}'_3,\bs\Delta ,\bs{\psi})}\mathcal{A}_{\Sigma,g,\bs\zeta}(\Psi_{\alpha_1}, \Psi_{\alpha_2,\emptyset,\tilde\nu'_2}, \Psi_{\alpha_3,\emptyset,\tilde\nu'_3})\]
where $\tilde{\nu}'_2$ and $\tilde{\nu}'_3$ run over the set of Young diagrams such that $|\tilde{\nu}'_j|\leq |\tilde{\nu}_j|$, 
and $C(\tilde\nu_1,\tilde\nu'_2,\tilde\nu'_3,\bs\Delta ,\bs{\psi})$ are the constants, polynomial in $\bs\Delta $, defined above.
Repeating the argument with the vector fields $v_{z_2^0}^{(n)}$ we remove  the $\tilde{\nu}'_2$ terms, 
and then  the vector fields $v_{z_3^0}^{(n)}$ allow to remove  the $\tilde{\nu}'_3$ terms. This finally gives
 \[\mathcal{A}_{\Sigma,g,\bs\zeta}(\Psi_{\alpha_1,\nu_1,\tilde\nu_1}, \Psi_{\alpha_2,\nu_2,\tilde\nu_2}, \Psi_{\alpha_3,\nu_3,\tilde\nu_3})=\omega(\Sigma,\bs\Delta ,\bs{\nu})\bbar{\omega(\Sigma,\bs\Delta ,\widetilde{\bs{\nu}})}\mathcal{A}_{\Sigma,g,\bs\zeta}(\Psi_{\alpha_1}, \Psi_{\alpha_2}, \Psi_{\alpha_3}).\]
 This concludes the proof.
\end{proof}
This proposition also allows us to compute the matrix coefficients of the amplitudes associated with an annulus with one marked point and with a disk with two marked points, i.e. the building blocks of type (B) and (C) described in Section \ref{geometric_decomp}.
Indeed, consider an annulus in model form  $\Sigma'=\hat{\C}\setminus \cup_{j=1}^2\psi_j(\D)$ with a marked point $x'=z_3^0\in \{1,-{\rm i}\}$. Observe that $\Sigma$ can be obtained by gluing $\Sigma'_{\rm hole}=\hat{\C}\setminus \cup_{j=1}^3\psi^\ell_j(\D)$ with the disk $\psi^\ell_3(\D)$, where $\psi^\ell_3(z)=z_3^0+rz$ for some sufficiently small $r>0$. The boundary parametrisations are, as before, given by $\zeta_j=\psi_j^\ell|_\T$, and we write  $\bs{\zeta}'=(\zeta_1,\zeta_2)$.  For uniformity of notation, we also set $({\bf x}',\bs{\alpha}')=(x',\alpha_3)$.

The surface $\Sigma'$ can be equipped with an admissible metric $g$ that is also
admissible near the boundary circle $\psi^\ell_3(\T)$ of $\Sigma'_{\rm hole}$, that is 
\begin{equation}\label{metric_condition1}
(\psi^\ell_3)^*g=|\dd z|^2/|z|^2\text{ near }\T, \qquad (\psi_3^\ell)^*g=|\dd z|^2 \text{ near } z=0.
\end{equation} 
By \eqref{amplitude} and Proposition \ref{Weyl2}, the amplitude of the disk is related to $\Psi_{\alpha_3}$ by 
\[\mathcal{A}_{\psi_3^\ell(\D), g, x',\alpha_3,\zeta_3}=e^{c_{\rm L} S_{\rm L}^0(\D,g_\D,(\psi_3^\ell)^*g)}\mathcal{A}_{\psi_3^\ell(\D), (\psi_3^\ell)_*g_\D, x',\alpha_3,\zeta_3}=e^{c_{\rm L} S_{\rm L}^0(\D,g_\D,(\psi_3^\ell)^*g)}Z_{\D,g_\D} \Psi_{\alpha_3}.\]
By the gluing formula for amplitudes \eqref{glue_ampl}
\[\mc{A}_{\Sigma',g,{\bf x}',\bs{\alpha}',\bs{\zeta}'}(\Psi_{\alpha_1,\nu_1,\tilde\nu_1}, \Psi_{\alpha_2,\nu_2,\tilde\nu_2})=\frac{1}{\sqrt{2}\pi}e^{c_{\rm L} S_{\rm L}^0(\D,g_\D,(\psi_3^\ell)^*g)}Z_{\D,g_\D} \mc{A}_{\Sigma'_{\rm hole},g,\bs{\zeta}}(\Psi_{\alpha_1,\nu_1,\tilde\nu_1}, \Psi_{\alpha_2,\nu_2,\tilde\nu_2},\Psi_{\alpha_3}),\]
which allows us  to get the factorisation formula for the annulus amplitude.

Similarly, for a disk in model form $\Sigma''=\hat{\C}\setminus \psi^\ell_1(\D)$ with two marked points ${\bf x}''=(z_2^0,z_3^0)\in \{(0,1),(0,-i)\}$, 
  it corresponds to the gluing of $\Sigma''_{\rm hole}$ with the disks $\psi^\ell_2(\D)$ and $\psi^\ell_3(\D)$, where 
$\psi^\ell_j(z)= z_j^0+rz$ for some sufficiently small $r>0$. Let $\bs{\zeta}''=\zeta''=\psi_1^\ell|_{\T}$. As before, if $g$ is admissible and satisfies
\begin{equation}\label{metric_condition2}
(\psi^\ell_j)^*g=|\dd z|^2/|z|^2 \text{ near }\T,\qquad  (\psi^\ell_j)^*g=|\dd z|^2 \text{ near } 0,\quad j=2,3,
\end{equation}
 we get, for $\bs{\alpha}''=(\alpha_2,\alpha_3)$,
\[\mc{A}_{\Sigma'',g,{\bf x}'',\bs{\alpha}'',\bs{\zeta}''}(\Psi_{\alpha_1,\nu_1,\tilde\nu_1})=e^{c_{\rm L} \sum_{j=2}^3 S_{\rm L}^0(\D,g_\D,(\psi_j^\ell)^*g)}Z_{\D,g_\D}^{2}\mc{A}_{\Sigma''_{\rm hole},g,\bs{\zeta}}(\Psi_{\alpha_1,\nu_1,\tilde\nu_1}, \Psi_{\alpha_2},\Psi_{\alpha_3}).\]
This in turn yields the following expressions for the matrix coefficients of the amplitudes associated with the geometric building blocks of type (B) and (C) in model form:
\begin{corollary}[\textbf{Type (B) and (C)}]\label{cor:case_with_mp}
With the definitions above, in particular the conditions on the metric $g$  \eqref{metric_condition1} in the case of type (B) and \eqref{metric_condition2} in the case of type (C),
\begin{description}
\item[Type (B)] for $\alpha_3<Q$ and $\alpha_1,\alpha_2 \in Q+{\rm i}\R$, one has
\begin{equation}\label{ward_with_1point}
\mathcal{A}_{\Sigma',g,{\bf x}',\bs{\alpha}',\bs{\zeta}'}( \Psi_{\alpha_1,\nu_1,\tilde\nu_1}, \Psi_{\alpha_2,\nu_2,\tilde\nu_2})=\omega(\Sigma'_{\rm hole},\bs\Delta ,\bs\nu)\overline{\omega(\Sigma'_{\rm hole},\bs\Delta ,\bs{\tilde\nu})} \mathcal{A}_{\Sigma',g,{\bf x}',\bs{\alpha}',\bs{\zeta}'}( \Psi_{\alpha_1}, \Psi_{\alpha_2})
\end{equation}
with $\bs\nu=(\nu_1,\nu_2,\emptyset)$, $\tilde{\bs\nu}=(\tilde\nu_1,\tilde\nu_2,\emptyset)$ and $\bs\Delta =(\Delta_{\alpha_1},\Delta_{\alpha_2},\Delta_{\alpha_3})$. 
\item[Type (C)] for  $\bs{\alpha}''=(\alpha_2,\alpha_3)$ with $\alpha_j<Q$ for $j=2,3$, and $\alpha_1 \in Q+{\rm i}\R$, one has 
\begin{equation}
\mathcal{A}_{\Sigma'',g,{\bf x}'',\bs{\alpha}'',\bs\zeta''}(\Psi_{\alpha_1,\nu_1,\tilde\nu_1})=\omega(\Sigma''_{\rm hole},\bs\Delta ,\bs\nu)\overline{\omega(\Sigma''_{\rm hole},\bs\Delta ,\bs{\tilde\nu})} \mathcal{A}_{\Sigma'',g,{\bf x}'',\bs{\alpha}'',\bs\zeta''}( \Psi_{\alpha_1})
\end{equation}
with $\bs\nu=(\nu_1,\emptyset,\emptyset)$, $\tilde{\bs\nu}=(\tilde\nu_1,\emptyset,\emptyset)$ and $\bs\Delta =(\Delta_{\alpha_1},\Delta_{\alpha_2},\Delta_{\alpha_3})$.
\end{description}
\end{corollary}

\subsection{Matrix coefficients of amplitudes of half pairs of pants}\label{sec: boundary ward 1}

We next compute the coefficients of the amplitude of a half pair of pants of type (A2) in model form,  in the basis  $\Psi_{Q+ip,\nu}^+$ of $\mc{H}^+$.

\begin{proposition}[\textbf{Type (A2)}]\label{ward case I}
Let  $\bs{\psi}^{\rm h}=(\psi_1^{\rm h},\psi_2^{\rm h},\psi_3^{\rm h})\in \mathrm{BiHol}^+_\eps(\D)^3$ for some $\eps>0$ such that the images $\psi_j^{\rm h}(\D)$ are pairwise disjoint, and
 $$
 \bs{\psi}^{\rm h}(0)={\bf z}^0=(z^0_1,z^0_2,z^0_3)=(-1,0,1).
 $$
  Let $\Sigma=\hat{\H}\setminus \cup_{j=1}^3\psi^{\rm h}_j(\D^+)$ with boundary parametrisation  $\bs{\zeta}=(\zeta_1^{\rm h},\zeta_2^{\rm h},\zeta_3^{\rm h})=(\psi_1^{\rm h}|_{\T^+},\psi_2^{\rm h}|_{\T^+},\psi_3^{\rm h}|_{\T^+})$, and let $g$ be an admissible metric on $\Sigma$. On the ${\rm N}$-marked boundary $\pl\Sigma_{\rm N}=\pl \Sigma\cap (\R\cup \infty)$, we assign a   function $\mu^{\rm b}\geq 0$, constant over each maximal interval contained in $\pl\Sigma_{\rm N}$.
Then for $\beta_j=Q+{\rm i}p_j$ with $p_j\in \R$, the following holds 
\begin{equation}\label{matrix_half_pant1}			
\mathcal{A}_{\Sigma,g,\bs{\zeta}}(\Psi^{+}_{\beta_1,\nu_1}, \Psi^{+}_{\beta_2,\nu_2}, \Psi^{+}_{\beta_3,\nu_3})=
\omega(\Sigma^{\#2},\bs{\Delta} ,\bs\nu)\mathcal{A}_{\Sigma,g,\bs{\zeta}}(\Psi^{+}_{\beta_1}, \Psi^{+}_{\beta_2}, \Psi^{+}_{\beta_3})
\end{equation}
where $\bs{\Delta} =(\Delta_{\beta_1},\Delta_{\beta_2},\Delta_{\beta_3})$ and $\bs\nu=(\nu_1,\nu_2,\nu_3)$, and $\omega(\Sigma^{\#2},\bs{\Delta} ,\bs\nu)$ is the constant defined in \eqref{matrix_pant}, associated with the Neumann double $\Sigma^{\#2}$ of $\Sigma$. 

Here the states 
$\Psi^{+}_{\beta_j,\nu_j}$ in the $j$-th position are associated to the boundary cosmological constants 
$$
\mu^{\rm L}=\lim_{\eps\to 0^+}\mu^{b}(z^0_j-\eps), \qquad \mu^{\rm R}=\lim_{\eps\to 0^+}\mu^{b}(z^0_j+\eps),$$ corresponding respectively to the left and right intervals of $z_j^0$ along the real boundary.
\end{proposition}
\begin{proof}
We proceed exactly as in the proof of Proposition~\ref{thm: ward identity}, and use the same notations. 
We apply Proposition \ref{prop: ward H} with ${\rm v}={\rm v}_{z^0_1}^{(n)}$ for $n\geq 1$ (recall \eqref{vectorv0}): 
notice that ${\rm v}=v(z)\pl_z$ with $\bar{v}(z)=v(\bar{z})$ and the $z^{k+1}$ coefficients 
$[(\psi_j^{\rm h})^*{\rm v}]_k$ of $(\psi_j^{\rm h})^*{\rm v}$ are real-valued, by assumption on $\psi_j^{\rm h}$ and ${\rm v}$, thus we get (for the same $c_1$ as in \eqref{induction_ward})
\begin{equation}\label{induction_ward_bord}
\begin{split}
-\frac{c_1}{((\psi_1^{\rm h})^{'}(0))^n} \mathcal{A}_{\Sigma,g, \boldsymbol{\zeta}}(\bL^+_{-n}\otimes \mathrm{Id}\otimes\mathrm{Id})= 
&\sum_{k > -n} [(\psi_1^{\rm h})*(\mathrm{v}^{(n)}_{z^0_1})]_k\cdot\mathcal{A}_{\Sigma,g,\boldsymbol{\zeta}^{\rm h}}(\bL^+_{k}\otimes \mathrm{Id}\otimes\mathrm{Id})\\
& +\sum_{k\geq 0}[(\psi_2^{\rm h})^*(\mathrm{v}_{z^0_1}^{(n)})]_k\cdot \mathcal{A}_{\Sigma,g, \bs\zeta}(\mathrm{Id}\otimes \mathbf{L}^+_{k}\otimes\mathrm{Id})\\
& +\sum_{k \geq 0} [(\psi_3^{\rm h})^*(\mathrm{v}_{z^0_1}^{(n)})]_k\cdot\mathcal{A}_{\Sigma,g, \bs\zeta}(\mathrm{Id}\otimes\mathrm{Id}\otimes \mathbf{L}^+_{k})\\
&- \mathrm{c_L }\Big(\frac{1}{48\pi}\int_{\Sigma}K_g{\rm div}_{g}(V){\rm dv}_g+ \sum_{j=1}^{b^{\rm h}} \frac{v^{\rm h}_{j0}}{24}\Big)\mathcal{A}_{\Sigma,g,\bs{\zeta}}({\rm Id}^{\otimes 3})
\end{split}
\end{equation}
with $v_{j0}^{\rm h}:=[(\psi_j^{\rm h})^*(\mathrm{v}_{z^0_1}^{(n)})]_0$.
Note that, since $V=\mathrm{v}_{z^0_1}^{(n)}+\tau^*\mathrm{v}_{z^0_1}^{(n)}$ (recall $\tau(z)=\bar{z}$) and $g$ is admissible, one can extend $g$ to $\Sigma^{\#2}$ in a $\tau$-symmetric way (and smoothly), and we get
\begin{equation}\label{divergence_term}
\frac{1}{48\pi}\int_{\Sigma}K_g{\rm div}_{g}(V){\rm dv}_g=\frac{1}{48\pi}\int_{\Sigma^{\#2}}K_g{\rm div}_{g}(\mathrm{v}_{z^0_1}^{(n)}){\rm dv}_g.
\end{equation}
We apply the same argument as in the proof of Proposition \ref{thm: ward identity}, now using identity \eqref{induction_ward_bord} together with the relation ${\bf L}^+_k\Psi^+_{\alpha,\nu}= \sum_{j=1}^N C(\Delta_\alpha,\nu,\nu'_j,k) \Psi^+_{\alpha,\nu'_j}$, which follows from \eqref{descendant_boundary_Hamiltonian}, \eqref{primaryH+} and \eqref{commuteLn^+}. Here the Young diagrams  $\nu_j'$ all have length $|\nu_j'|=|\nu|-k$ and the same constant $C(\Delta_\alpha,\nu,\nu'_j,k)\in \R$ are the same as in \eqref{LkPsi}. Using this, 
the matrix coefficient $\mathcal{A}_{\Sigma,g,\bs{\zeta}}(\Psi^{+}_{\beta_1,\nu_1}, \Psi^{+}_{\beta_2,\nu_2}, \Psi^{+}_{\beta_3,\nu_3})$ can be expressed in terms of coefficients with only $\nu_1$ removed:
\[
\mathcal{A}_{\Sigma,g,\bs{\zeta}}(\Psi^{+}_{\beta_1,\nu_1}, \Psi^{+}_{\beta_2,\nu_2}, \Psi^{+}_{\beta_3,\nu_3})=\sum_{\nu'_2,\nu'_3}C'(\nu_1,\nu'_2,\nu'_3,\bs{\Delta}_{\bs{\beta}},\bs{\psi}^{\rm h})\mathcal{A}_{\Sigma,g,\bs{\zeta}}(\Psi^{+}_{\beta_1}, \Psi^{+}_{\beta_2,\nu'_2}, \Psi^{+}_{\beta_3,\nu'_3})
\]
where $\nu'_2$, $\nu'_3$ run over Young diagrams with $|\nu'_j|\leq |\nu_j|$. The coefficients $C'(\nu_1,\nu'_2,\nu'_3,\bs{\Delta}_{\bs{\beta}},\bs{\psi}^{\rm h})$ coincide with those in \eqref{first_iteration} except that   $\bs{\psi}^{\rm h}$ replaces
$\bs{\psi}^\ell$ and $\bs{\beta}$ replaces $\bs{\alpha}$. This equality of coefficients is obtained by comparing the factors in \eqref{induction_ward} and \eqref{induction_ward_bord} and using \eqref{divergence_term} together with Lemma \ref{lem:schwarzian}; in particular one has
\[
\frac{1}{48\pi}\int_{\Sigma}K_g{\rm div}_{g}(V){\rm dv}_g+ \sum_{j=1}^{3} \frac{v^{\rm h}_{j0}}{24}=\frac{1}{48\pi}\int_{\Sigma^{\#2}}K_g{\rm div}_{g}(\mathrm{v}_{z^0_1}^{(n)}){\rm dv}_g+ \sum_{j=1}^{3} \frac{v^{\rm h}_{j0}}{24}=\frac{{\rm i}}{24 \pi} \sum_{j=1}^3 \int_{\zeta_j(\T)} \mathrm{v}_{z^0_1}^{(n)}{\rm S}((\psi^{\rm h}_j)^{-1}).
\]
We next iterate the same procedure using Proposition \ref{prop: ward H} with the vector fields ${\rm v}={\rm v}^{(n)}_{z^0_2}$ and  
${\rm i}{\rm v}$ for $n\geq 0$ to remove the elements of the Young diagram $\nu_2$ one by one, and then with ${\rm v}={\rm v}^{(n)}_{z^0_3}$ and   ${\rm i}{\rm v}$ for $n\geq 0$ to remove iteratively the elements of $\nu_3$. This yields
\[\mathcal{A}_{\Sigma,g,\bs{\zeta}}(\Psi^{+}_{\beta_1,\nu_1}, \Psi^{+}_{\beta_2,\nu_2}, \Psi^{+}_{\beta_3,\nu_3})=
\omega(\Sigma,\bs{\Delta} ,\bs\nu)\mathcal{A}_{\Sigma,g,\bs{\zeta}}(\Psi^{+}_{\beta_1}, \Psi^{+}_{\beta_2}, \Psi^{+}_{\beta_3})\]
for some constants $\omega(\Sigma,\bs{\Delta} ,\bs\nu)$. Since the algorithm is identical to that used in Proposition \ref{thm: ward identity} and the coefficients in the cornered case for $\Sigma$ agree with those for the Neumann double  $\Sigma^{\#2}$ (applied with $\bs{\alpha}=\bs{\beta}$), we conclude
 $$
 \omega(\Sigma,\bs{\Delta},\bs\nu)=\omega(\Sigma^{\#2},\bs{\Delta} ,\bs\nu),
 $$
which proves \eqref{matrix_half_pant1}.
\end{proof}
Next, to express the matrix coefficients of the amplitudes for the building blocks of type (B2) and (C2) we apply exactly the same reasoning as in the proof of Corollary \ref{cor:case_with_mp}. We obtain:

\begin{corollary}[\textbf{Type (B2) and (C2)}]\label{cor:case_with_mp_corner} The following factorisation properties hold true:
\begin{description}
\item[Type (B2)]  Let $\psi_j^{\rm h}$ be as in Proposition \ref{ward case I} for  $j=1,2$, and  define 
$$\Sigma'=\hat{\H}\setminus \cup_{j=1}^2\psi^{\rm h}_j(\D)
$$
 with boundary parametrisation $\bs{\zeta}'=(\psi_1^{\rm h},\psi_2^{\rm h})|_{\T^+}$. Let the  marked point and weight be $({\bf s}',\bs{\beta}')=(1,\beta_3)$ with $\beta_3<Q$. Let 
 $$
\Sigma'_{\rm hole}=\hat{\H}\setminus \cup_{j=1}^3\psi^{\rm h}_j(\D^+),\qquad \psi_3^{\rm h}(z)=1+rz
$$
 for some $r>0$ small enough, so that $\psi_3^{\rm h}(\D^+)\cap (\cup_{j=1}^2\psi^{\rm h}_j(\D^+))=\emptyset$,  and let $g$ be an admissible metric on $\Sigma'_{\rm hole}$ such that $(\psi_3^{\rm h})^*g=|dz|^2$ near $z=0$. For  $\beta_1,\beta_2 \in Q+{\rm i}\R$ and $\bs{\Delta} =(\Delta_{\beta_1},\Delta_{\beta_2},\Delta_{\beta_3})$, the following factorisation holds:
\begin{equation}\label{ward_with_1point_corner}
\mathcal{A}_{\Sigma',g,{\bf s}',\bs{\beta}',\bs{\zeta}'}( \Psi^+_{\beta_1,\nu_1}, \Psi^+_{\beta_2,\nu_2})=\omega((\Sigma'_{\rm hole})^{\#2},\bs\Delta,\bs\nu)\mathcal{A}_{\Sigma',g,{\bf s}',\bs{\beta}',\bs{\zeta}'}( \Psi^+_{\beta_1}, \Psi^+_{\beta_2})
\end{equation}
with $\bs\nu=(\nu_1,\nu_2,\emptyset)$. 
\item[Type (C2)]  Let $\psi^{\rm h}_1$ be as in Proposition \ref{ward case I}, and define 
$$\Sigma''=\hat{\H}\setminus \psi^{\rm h}_1(\D)
$$ with boundary parametrisation $\bs{\zeta}''=\psi_1^{\rm h}|_{\T^+}$. Let the  marked points and weights be 
$({\bf s}'';\bs{\beta}'')=(0,1;\beta_2,\beta_3)$ with $\beta_2,\beta_3<Q$. Let
$$
\Sigma''_{\rm hole}=\hat{\H}\setminus \cup_{j=1}^3\psi^{\rm h}_j(\D^+), \qquad \psi_j^{\rm h}(z)=z^0_j+rz, \quad j=2,3,
$$ 
for some $r>0$ small enough so that $\psi_j^{\rm h}(\D^+)\cap \psi_k^{\rm h}(\D^+)=\emptyset$ for all $j\not=k\in \{1,2,3\}$, and
 let $g$ be an admissible metric on $\Sigma''_{\rm hole}$ such that $(\psi_j^{\rm h})^*g=|dz|^2$ near $z=0$, for $j=2,3$. For  $\beta_1 \in Q+{\rm i}\R$ and $\bs{\beta}=(\beta_1,\beta_2,\beta_3)$, the following factorisation holds:
\begin{equation}\label{ward_with_1point_cornerC2}
\mathcal{A}_{\Sigma'',g,{\bf s}'',\bs{\beta}'',\bs{\zeta}''}( \Psi^+_{\beta_1,\nu_1})=\omega((\Sigma''_{\rm hole})^{\#2},\bs\Delta,\bs\nu)\mathcal{A}_{\Sigma'',g,{\bf s}'',\bs{\beta}'',\bs{\zeta}''}( \Psi^+_{\beta_1})
\end{equation}
where $\bs{\Delta}=(\Delta_{\beta_1},\Delta_{\beta},\Delta_{\alpha})$,  and $\bs\nu=(\nu_1,\emptyset,\emptyset)$.
\end{description}
\end{corollary}

Finally we consider the case of half pairs of pants of type (A1). 
\begin{proposition}[\textbf{Type (A1)}]\label{ward case II}
Let  $\bs{\psi}=(\psi^{\ell},\psi^{\rm h})\in \mathrm{BiHol}_\eps(\D)\times \mathrm{BiHol}^+_\eps(\D)$ for some $\eps>0$, such that the images satisfy $\psi^{\rm h}(\D^+)\cap\psi^\ell(\D)=\emptyset$  and 
 $\bs{\psi}(0)=(z^0_1,z^0_2)=({\rm i},0)$. Define 
 $$\Sigma=\hat{\H}\setminus (\psi^\ell(\D)\cup\psi^{\rm h}(\D))$$ 
 with boundary parametrisation  $\bs{\zeta}=(\zeta^{\ell},\zeta^{\rm h})=(\psi^{\ell}|_{\T},\psi^{\rm h})|_{\T^+}$, and let $g$ be an admissible metric on $\Sigma$. 
 
Then for $\alpha=Q+{\rm i}p_1$ and $\beta=Q+{\rm i}p_2$ with $p_j\in \R$, the following factorisation holds:
\begin{equation}\label{matrix_half_pant2}			
\mathcal{A}_{\Sigma,g,\bs{\zeta}}(\Psi_{\alpha,\nu_1,\tilde\nu_1}, \Psi^{+}_{\beta,\nu_2})=
\omega(\Sigma^{\#2},\bs{\Delta},\bs\nu)\mathcal{A}_{\Sigma,g,\bs{\zeta}}(\Psi_{\alpha}, \Psi^{+}_{\beta})
\end{equation}
where $\bs{\Delta}=(\Delta_{\alpha},\Delta_{\beta},\Delta_{\alpha})$, $\bs{\nu}=(\nu_1,\nu_2,\tilde\nu_1)$ and the constant $\omega(\Sigma^{\#2},\bs{\Delta},\bs\nu)$ is the same as in~\eqref{matrix_pant}, hence independent of $\mu^{\rm b}$. The corresponding Neumann double is
 $$\Sigma^{\#2}=\hat{\C}\setminus \cup_{j=1}^3 \psi_j^{\ell}(\D), \quad \psi_1^{\ell}=\psi^{\ell},  \quad \psi_2^{\ell}=\psi^{\rm h},  \quad \psi_3^{\ell}=\tau\circ \psi^\ell\circ \tau,
 $$
 and $(\psi^{\ell}_1,\psi^{\ell}_2,\psi^{\ell}_3) $ induce the parametrisation of $\pl \Sigma^{\#2}$.
\end{proposition}

\begin{proof}
We proceed exactly as in the proof of Proposition~\ref{ward case I}. Let ${\bf z}^0=(z^0_1,z^0_2,z^0_3)=({\rm i},0,-{\rm i})$.
We apply Proposition~\ref{prop: ward H} first with the vector field ${\rm v}={\rm v}_{z^0_2}^{(n)}=z^{-n+1}(z-i)(z+i)\pl_z$. This yields
\begin{equation}\label{induction_ward_bord1}
\begin{split}
-\frac{1}{(\psi^{{\rm h}})^{'}(0)^n} \mathcal{A}_{\Sigma,g, \boldsymbol{\zeta}}(\mathrm{Id}\otimes \bL^+_{-n})= 
&\sum_{k > -n} [(\psi^{{\rm h}})^*{\rm v}]_k\cdot\mathcal{A}_{\Sigma,g,\boldsymbol{\zeta}}( \mathrm{Id}\otimes \bL^+_{k}) +\sum_{k\geq 0}[(\psi^{\ell})^*{\rm v}]_k\cdot \mathcal{A}_{\Sigma,g, \bs\zeta}(\bL_k\otimes \mathrm{Id})\\
&+\sum_{k\geq 0}\bbar{[(\psi^{\ell})^*{\rm v}]}_k\cdot \mathcal{A}_{\Sigma,g, \bs\zeta}(\tilde{\bL}_k\otimes \mathrm{Id})\\
&- \mathrm{c_L }\Big(\frac{1}{48\pi}\int_{\Sigma}K_g{\rm div}_{g}(V){\rm dv}_g+ \frac{{\rm Re}(v^{\ell}_{0})}{12}+\frac{v^{\rm h}_{0}}{24}\Big)\mathcal{A}_{\Sigma,g,\bs{\zeta}}({\rm Id}^{\otimes 3})
\end{split}
\end{equation}
where $v_0^\ell=[(\psi^{\ell})^*(\mathrm{v}_{z^0_2}^{(n)})]_0$ and $v_0^{\rm h}=[(\psi^{\rm h})^*(\mathrm{v}_{z^0_2}^{(n)})]_0$.
As in the proof of Proposition~\ref{thm: ward identity}, an inductive argument allows one to express the matrix coefficient $\mathcal{A}_{\Sigma,g,\bs{\zeta}}(\Psi_{\alpha,\nu_1,\tilde{\nu}_1}, \Psi^{+}_{\beta,\nu_2})$ in the form 
\begin{equation}\label{first_constant_Sigma}
\mathcal{A}_{\Sigma,g,\bs{\zeta}}(\Psi_{\alpha,\nu_1,\tilde{\nu}_1}, \Psi^{+}_{\beta,\nu_2})=\sum_{\nu'_1,\tilde{\nu}_1'}C_1(\nu'_1,\tilde{\nu}_1',\nu_2,\bs{\Delta},\bs{\psi})\mathcal{A}_{\Sigma,g,\bs{\zeta}}(\Psi_{\alpha,\nu_1',\tilde{\nu}'_1}, \Psi^{+}_{\beta})
\end{equation}
where $\nu'_1,\tilde{\nu}'_1$  run over   Young diagrams satisfying $|\nu'_1|\leq |\nu_1|$ and $|\tilde{\nu}_1'|\leq |\tilde{\nu}_1|$, 
and $C_1(\nu'_1,\tilde{\nu}_1',\nu_2,\bs{\Delta},\bs{\psi})$ are constants polynomial in $\bs{\Delta}$, determined by the coefficients in \eqref{induction_ward_bord1}. 

Since $V={\rm v}+\tau^*{\rm v}$ (recall $\tau(z)=\bar{z}$) and $g$ is admissible, one can extend $g$ smoothly to the Neumann double $\Sigma^{\#2}$ in a $\tau$-symmetric manner, yielding \eqref{divergence_term}. 

 We now compare this with the computation of the matrix coefficient $\mathcal{A}_{\Sigma^{\#2},g,\bs{\zeta}^\#}(\Psi_{\alpha,\nu_1,\emptyset}, \Psi_{\beta,\nu_2,\emptyset}, \Psi_{\alpha,\tilde{\nu}_1,\emptyset})$ of the double $\Sigma^{\#2}$, where $\bs{\zeta}^\#=(\psi_1^{\ell}|_\T,\psi_2^{\ell}|_\T,\psi_3^{\ell}|_\T)$ denotes the  boundary parametrisation. Following the same algorithm as in the proof of Proposition~\ref{thm: ward identity}, but starting with the vector field ${\rm v}=\mathrm{v}_{z^0_2}^{(n)}$   to first remove $\nu_2$, and using \eqref{divergence_term} together with $[(\psi_3^{\ell})^*{\rm v}]_k=\bbar{[(\psi^\ell)^*\tau^*{\rm v}]}_k$, we obtain
\begin{equation}\label{induction_ward_double}
\begin{split}
-\frac{1}{(\psi^{\rm h})'(0)^n} \mathcal{A}_{\Sigma^{\#2},g, \boldsymbol{\zeta}^\#}(\mathrm{Id}\otimes \bL_{-n}\otimes {\rm Id})= 
&\sum_{k > -n} [(\psi^{{\rm h}})^*{\rm v}]_k\cdot\mathcal{A}_{\Sigma^{\#2},g, \boldsymbol{\zeta}^\#}( \mathrm{Id}\otimes \bL_{k}\otimes {\rm Id})\\
& +\sum_{k\geq 0}[(\psi_{1}^{\ell})^*{\rm v}]_k\cdot \mathcal{A}_{\Sigma^{\#2},g, \boldsymbol{\zeta}^\#}(\bL_k\otimes \mathrm{Id}\otimes {\rm Id})\\
&+\sum_{k\geq 0}\bbar{[(\psi_{3}^{\ell})^*{\rm v}]}_k\cdot \mathcal{A}_{\Sigma^{\#2},g, \boldsymbol{\zeta}^\#}({\rm Id}\otimes \mathrm{Id}\otimes \bL_k)\\
&- \mathrm{c_L }\Big(\frac{1}{48\pi}\int_{\Sigma^{\#2}}K_g{\rm div}_{g}({\rm v}){\rm dv}_g+ \frac{v^{\ell}_{0}}{24}+ \frac{ \bar v^{\ell}_{0}}{24}+\frac{v^{\rm h}_{0}}{24}\Big)\mathcal{A}_{\Sigma^{\#2},g, \boldsymbol{\zeta}^\#}({\rm Id}^{\otimes 3}).
\end{split}
\end{equation}
Since the coefficients appearing in front of the amplitudes in \eqref{induction_ward_bord1} are equal to the coefficients in front of the amplitudes in \eqref{induction_ward_double}, we deduce that 
\begin{equation}\label{first_reduction_double}
\mathcal{A}_{\Sigma^{\#2},g, \boldsymbol{\zeta}^\#}(\Psi_{\alpha,\nu_1,\emptyset}, \Psi_{\beta,\nu_2,\emptyset},\Psi_{\alpha,\tilde{\nu}_1,\emptyset})=\sum_{\nu'_1,\tilde{\nu}_1'}C_1(\nu'_1,\tilde{\nu}_1',\nu_2,\bs{\Delta},\bs{\psi})\mathcal{A}_{\Sigma^{\#2},g, \boldsymbol{\zeta}^\#}(\Psi_{\alpha,\nu_1',\emptyset}, \Psi_{\beta},\Psi_{\alpha,\tilde{\nu}_1',\emptyset})
\end{equation}
for the same constants $C_1(\nu'_1,\tilde{\nu}_1',\nu_2,\bs{\Delta},\bs{\psi})$ as in \eqref{first_constant_Sigma}, and $\nu'_1,\tilde{\nu}'_1$  run over the set of Young diagrams such that $|\nu'_1|\leq |\nu_1|$ and $|\tilde{\nu}_1'|\leq |\tilde{\nu}_1|$.
To complete the argument, we need to eliminate the Young diagrams $\nu_1$ and $\tilde{\nu}_1$  in the coefficients $\mathcal{A}_{\Sigma,g,\bs{\zeta}}(\Psi_{\alpha,\nu_1',\tilde{\nu}'_1}, \Psi^{+}_{\beta})$ by applying the Ward identity with a suitable vector field. 
However, since this vector field must be $\tau$-symmetric, it cannot possess a single pole off the real line. This restriction becomes problematic when attempting to identify the coefficients $\omega$  with those associated with the Neumann double. To overcome this difficulty, we employ the following trick, which essentially relies on the observation that the Ward identities can be extended to non-symmetric vector fields by means of polarisation identities.
 
 We will use the vector fields 
${\rm v}:={\rm v}_{z_1^0}^{(n)}+ {\rm v}_{z_3^0}^{(n)}$ and ${\rm w}:={\rm i}({\rm v}_{z_1^0}^{(n)}- {\rm v}_{z_3^0}^{(n)})$, which both are of the form 
${\rm v}=v(z)\pl_z$ and ${\rm w}=w(z)\pl_z$ with $v(\bar{z})=\bar{v}(z)$ and  $w(\bar{z})=\bbar{w}(z)$. 
Then one has 
\begin{align*}
& {\rm v}_1:=(\psi_1^{\ell})^*{\rm v}=\sum_{k\geq -n}[(\psi_1^{\ell})^*{\rm v}^{(n)}_{z^0_1}]_kz^{k+1}\pl_z+\sum_{k\geq 0}  [ (\psi_1^{\ell})^* {\rm v}^{(n)}_{z_3^0}]_kz^{k+1}\pl_z,
\\
& {\rm w}_1:=\psi_1^*{\rm w}={\rm i}\sum_{k\geq -n}[(\psi_1^{\ell})^*{\rm v}_{z^0_1}^{(n)}]_kz^{k+1}\pl_z-{\rm i}\sum_{k\geq 0}   [ (\psi_1^{\ell})^* {\rm v}^{(n)}_{z_3^0}]_kz^{k+1}\pl_z.
\end{align*}
Taking the identity of  Proposition \ref{prop: ward H} with the vector field ${\rm v}$ and adding $-{\rm i}$ times the identity of  Proposition \ref{prop: ward H}  applied with the vector field ${\rm w}$, we obtain (for $c_1\not =0$)
\begin{align}\label{induction_ward_bord2}
& -\frac{c_1}{(\psi_1^{\ell})'(0)^n}  \mathcal{A}_{\Sigma,g, \boldsymbol{\zeta}}({\bf L}_{-n}\otimes {\rm Id})= \nonumber\\
&\sum_{k > -n} [(\psi_1^{\ell})^*{\rm v}^{(n)}_{z^0_1}]_k\cdot\mathcal{A}_{\Sigma,g,\boldsymbol{\zeta}}({\bf L}_{k}\otimes {\rm Id}) +\sum_{k\geq 0}
\bbar{ [ (\psi_1^{\ell})^* {\rm v}_{z_3^0}^{(n)}]_k}
\cdot \mathcal{A}_{\Sigma,g, \bs\zeta}(\tilde{\bL}_k\otimes \mathrm{Id})\\
& +\sum_{k\geq 0} [(\psi_2^{\ell})^*{\rm v}^{(n)}_{z^0_1}]_k\cdot \mathcal{A}_{\Sigma,g, \bs\zeta}({\rm Id}\otimes \bL_k^+)
\nonumber\\
& - \mathrm{c_L }\Big(\frac{1}{48\pi}\int_{\Sigma^{\#2}}K_g{\rm div}_{g}({\rm v}_{z^0_1}^{(n)}){\rm dv}_g+ \frac{1}{24}\big([(\psi_1^{\ell})^*{\rm v}_{z_1^0}^{(n)}]_0+\bbar{[(\psi_1^{\ell})^*{\rm v}_{z_3^0}^{(n)}]}_0 +[(\psi_2^{\ell})^*{\rm v}_{z_1^0}^{(n)}]_0\big)\Big)\mathcal{A}_{\Sigma,g,\bs{\zeta}}({\rm Id}^{\otimes 2}). \nonumber
\end{align}
Note that $(\psi_1^{\ell})^*{\rm v}_{z^0_3}=\tau^*(\psi_3^{\ell})^*\bbar{{\rm v}_{z_1^0}}$  (where $\tau(z)=\bar{z}$ as usual), which implies $\bbar{[(\psi_1^{\ell})^*{\rm v}_{z_3^0}^{(n)}]}_0=[(\psi_3^{\ell})^*{\rm v}_{z_1^0}^{(n)}]_0$. 
Applying the same inductive argument as above, we can remove the $\nu_1$ factor in \eqref{first_constant_Sigma}
\begin{equation}\label{nu_1reductionhalfpant}
\mathcal{A}_{\Sigma,g,\bs{\zeta}}(\Psi_{\alpha,\nu_1',\tilde{\nu}'_1}, \Psi^{+}_{\beta})=\sum_{\tilde{\nu}_1''}C_2(\nu'_1,\tilde{\nu}''_1,\bs{\Delta},\bs{\psi})\mathcal{A}_{\Sigma,g,\bs{\zeta}}(\Psi_{\alpha,\emptyset,\tilde{\nu}''_1}, \Psi^{+}_{\beta})
\end{equation}
for some constants $C_2(\nu'_1,\tilde{\nu}''_1,\bs{\Delta},\bs{\psi})$ and $\bs{\psi}=(\psi_1,\psi_2,\psi_3)$, the sum being over $\tilde \nu_1''$ of size $|\tilde \nu_1''|\leq |\tilde \nu_1'|$. On the double $\Sigma^{\#2}$, one applies Proposition 
\ref{prop:localward move point} with the vector fields ${\rm v}$ and ${\rm w}$ above, this gives 
\begin{equation}\label{induction_ward_bord3}
\begin{split}
-\frac{c_1}{(\psi_1^{\ell})'(0)^n}  \mathcal{A}_{\Sigma^{\#2},g, \boldsymbol{\zeta^\#}}({\bf L}_{-n}\otimes {\rm Id}^{\otimes 2})= 
&\sum_{k > -n} [(\psi_1^{\ell})^*{\rm v}_{z^0_1}^{(n)}]_k\cdot\mathcal{A}_{\Sigma^{\#2},g,\boldsymbol{\zeta}^\#}({\bf L}_{k}\otimes {\rm Id}^{\otimes 2}) \\
& +\sum_{k\geq 0}[(\psi_3^{\ell})^*{\rm v}_{z_1^0}^{(n)}]_k\cdot \mathcal{A}_{\Sigma^{\#2},g, \bs\zeta} (\mathrm{Id}^{\otimes 2}\otimes \bL_k)\\
&+\sum_{k\geq 0}[(\psi_2^{\ell})^*{\rm v}^{(n)}_{z^0_1}]_k\cdot \mathcal{A}_{\Sigma^{\#2},g, \bs\zeta^\#}({\rm Id}\otimes \bL_k\otimes {\rm Id})\\
&- \mathrm{c_L }\Big(\frac{1}{48\pi}\int_{\Sigma^{\#2}}K_g{\rm div}_{g}({\rm v}_{z^0_1}^{(n)}){\rm dv}_g+ \sum_{j=1}^3 \frac{[(\psi_j^{\ell})^*{\rm v}_{z_1^0}^{(n)}]_0}{24}\Big)\mathcal{A}_{\Sigma^{\#2},g,\bs{\zeta^\#}}({\rm Id}^{\otimes 3}).
\end{split}
\end{equation}
Using this iteratively and the fact that the coefficients in \eqref{induction_ward_bord3} are the same as those in \eqref{induction_ward_double}, we obtain  
\begin{equation}\label{nu_1reductionpant}
\mathcal{A}_{\Sigma^{\#2},g,\bs{\zeta}^\#}(\Psi_{\alpha,\nu_1',\emptyset}, \Psi_{\beta},\Psi_{\alpha,\tilde{\nu}_1',\emptyset})=\sum_{\tilde{\nu}_1''}C_2(\nu'_1,\tilde{\nu}_1'',\bs{\Delta},\bs{\psi})\mathcal{A}_{\Sigma^{\#2},g,\bs{\zeta}^\#}(\Psi_{\alpha}, \Psi_{\beta},\Psi_{\alpha,\tilde{\nu}_1'',\emptyset})
\end{equation}
where the sum runs over Young diagrams  $\tilde \nu_1''$ of size $|\tilde \nu_1''|\leq |\tilde \nu_1'|$, and $C_2(\nu'_1,\bs{\Delta},\bs{\psi})$ are the same constants as those appearing in \eqref{nu_1reductionhalfpant}.

Finally, by taking the identity of Proposition~\ref{prop: ward H} with the vector field ${\rm v}$ and adding $+{\rm i}$ times the identity of  Proposition \ref{prop: ward H}  applied with the vector field ${\rm w}$, we obtain (for $c_2\not =0$)
\begin{align}\label{induction_ward_bord2bis}
-\frac{c_2}{\bbar{(\psi_1^{\ell})'(0)}^n} & \mathcal{A}_{\Sigma,g, \boldsymbol{\zeta}}(\tilde{{\bf L}}_{-n}\otimes {\rm Id})= 
\nonumber\\
&\sum_{k > -n} \bbar{[(\psi_1^{\ell})^*{\rm v}^{(n)}_{z^0_1}]}_k\cdot\mathcal{A}_{\Sigma,g,\boldsymbol{\zeta}}(\tilde{{\bf L}}_{k}\otimes {\rm Id}) +\sum_{k\geq 0}[(\psi_1^{\ell})^*{\rm v}^{(n)}_{z_3^0}]_k\cdot \mathcal{A}_{\Sigma,g, \bs\zeta}(\bL_k\otimes \mathrm{Id}) 
\nonumber\\
&+\sum_{k\geq 0} [(\psi_2^{\ell})^*{\rm v}_{z_3^0}^{(n)}]_k\cdot \mathcal{A}_{\Sigma,g, \bs\zeta}({\rm Id}\otimes \bL_k^+)
\nonumber\\
&- \mathrm{c_L }\Big(\frac{1}{48\pi}\int_{\Sigma^{\#2}}K_g{\rm div}_{g}(\bbar{{\rm v}}^{(n)}_{z^0_1}){\rm dv}_g+ \frac{1}{24} \sum_{j=1}^3 \bbar{[(\psi_j^{\ell})^*{\rm v}_{z_1^0}^{(n)}]}_0 \Big)\mathcal{A}_{\Sigma,g,\bs{\zeta}}({\rm Id}^{\otimes 2}).
\end{align}
By the same inductive argument as before, there is $C_3(\tilde\nu''_1,\bs{\Delta},\bs{\psi})$ polynomial in $\bs{\Delta}$ such that
\begin{equation}\label{nu_1"reduction}
\mathcal{A}_{\Sigma,g,\bs{\zeta}}(\Psi_{\alpha,\emptyset,\tilde{\nu}''_1}, \Psi^{+}_{\beta})=C_3(\tilde\nu''_1,\bs{\Delta},\bs{\psi})\mathcal{A}_{\Sigma,g,\bs{\zeta}}(\Psi_{\alpha}, \Psi^{+}_{\beta}).
\end{equation}
For the Neumann double, we apply the Ward identity (Prop. \ref{prop:localward move point}) to the vector fields ${\rm v}$, ${\rm w}$, and combine the corresponding relations as $(\text{Rel}({\rm v})+{\rm i} \text{Rel}({\rm w}))/2$ to get
\begin{equation}\label{induction_ward_bord4}
\begin{split}
-\frac{c_1}{(\psi_3^{\ell})'(0)^n}  \mathcal{A}_{\Sigma^{\#2},g, \boldsymbol{\zeta^\#}}({\rm Id}^{\otimes 2}\otimes {\bf L}_{-n})= 
&\sum_{k > -n}  [(\psi_3^{\ell})^*{\rm v}_{z^0_3}]_k\cdot\mathcal{A}_{\Sigma^{\#2},g,\boldsymbol{\zeta}^\#}({\rm Id}^{\otimes 2}\otimes {\bf L}_{k}) \\
& +\sum_{k\geq 0} [(\psi_1^{\ell})^*{\rm v}_{z_3^0}]_k\cdot \mathcal{A}_{\Sigma^{\#2},g, \bs\zeta^\#} ( \bL_k\otimes \mathrm{Id}^{\otimes 2})\\
&+\sum_{k\geq 0}  [(\psi_2^{\ell})^*{\rm v}^{(n)}_{z^0_3}]_k\cdot \mathcal{A}_{\Sigma^{\#2},g, \bs\zeta^\#}({\rm Id}\otimes \bL_k\otimes {\rm Id})\\
&- \mathrm{c_L }\Big( \frac{1}{48\pi}\int_{\Sigma^{\#2}}K_g{\rm div}_{g}({\rm v}_{z^0_3}^{(n)}){\rm dv}_g+\frac{1}{24} \sum_{j=1}^3  [(\psi_j^{\ell})^*{\rm v}_{z_3^0}^{(n)}]_0\Big)\mathcal{A}_{\Sigma^{\#2},g,\bs{\zeta^\#}}({\rm Id}^{\otimes 3}).
\end{split}
\end{equation}
Now we observe that 
\begin{align*}
(\psi_3^{\ell})'(0)=&\bbar{(\psi_1^{\ell})'(0)},\qquad  [(\psi_3^{\ell})^*{\rm v}_{z^0_3}]=\bbar{ [(\psi_1^{\ell})^*{\rm v}_{z^0_1}]},\qquad  [(\psi_2^{\ell})^*{\rm v}_{z^0_3}]=\bbar{ [(\psi_2^{\ell})^*{\rm v}_{z^0_1}]},
\\
& \int_{\Sigma^{\#2}}K_g{\rm div}_{g}({\rm v}_{z^0_3}){\rm dv}_g=\int_{\Sigma^{\#2}}K_g{\rm div}_{g}(\bbar{{\rm v}}_{z^0_1}){\rm dv}_g,
\end{align*} 
which leads to the following reduction
\begin{equation}\label{lastreduction}
\mathcal{A}_{\Sigma^{\#2},g,\bs{\zeta}^\#}(\Psi_{\alpha}, \Psi_{\beta},\Psi_{\alpha,\tilde{\nu}''_1,\emptyset })=C_3(\tilde\nu''_1,\bs{\Delta},\bs{\psi})\mathcal{A}_{\Sigma^{\#2},g,\bs{\zeta}^\#}(\Psi_{\alpha}, \Psi^{+}_{\beta})
\end{equation}
with the same constant $C_3(\tilde\nu''_1,\bs{\Delta},\bs{\psi})$ as in \eqref{nu_1"reduction}. Combining \eqref{first_constant_Sigma}, \eqref{nu_1reductionhalfpant} and \eqref{nu_1"reduction}, and alternatively combining \eqref{first_reduction_double}, 
\eqref{nu_1reductionpant} and \eqref{lastreduction}, we obtain \eqref{matrix_half_pant2}, which completes the proof.
\end{proof}
As in the proof of Corollary \ref{cor:case_with_mp_corner}, we get for building blocks of type (B1) or (C1):
\begin{corollary}[\textbf{Type (B1) or (C1)}]\label{cor:case_with_mp_corner1} The following factorisation properties hold true:
\begin{description}
\item[Type (B1)] Let $\psi^\ell$ be as in Proposition \ref{ward case II},  and define $\Sigma'=\hat{\H}\setminus \psi^{\ell}(\D)$ with boundary parametrisation $\bs{\zeta}'=\psi^{\ell}|_\T$. Let the  marked
 point and weight be $({\bf s}',\bs{\beta}')=(0,\beta)$ for $\beta<Q$. Let $\Sigma'_{\rm hole}=\hat{\H}\setminus (\psi^{\ell}(\D)\cup \psi^{\rm h}(\D^+))$ with $\psi^{\rm h}(z)=rz$ for some $r>0$ small enough, 
so that $\psi^{\rm h}(\D^+)\cap \psi^{\ell}_1(\D)=\emptyset$,   and let $g$ be an admissible metric on $\Sigma'_{\rm hole}$ such that $(\psi^{\rm h})^*g=|dz|^2$ near $z=0$. For  $\alpha\in Q+{\rm i}\R$, 
 the following factorisation holds true
\begin{equation}\label{ward_with_1point_corner_1B1}
\mathcal{A}_{\Sigma',g,{\bf s}',\bs{\beta}',\bs{\zeta}'}(\Psi_{\alpha,\nu,\tilde\nu})=\omega((\Sigma'_{\rm hole})^{\#2},\bs\Delta,\bs\nu)\mathcal{A}_{\Sigma',g,{\bf s}',\bs{\beta}',\bs{\zeta}'}(\Psi_{\alpha})
\end{equation}
with $\bs\nu=(\nu,\emptyset,\tilde\nu)$ and $\bs\Delta=(\Delta_\alpha,\Delta_\beta,\Delta_\alpha)$.
\item[Type (C1)] Let $\psi^{\rm h}$ be as in Proposition \ref{ward case II}, and define $\Sigma''=\hat{\H}\setminus \psi^{\rm h}(\D)$ with boundary parametrisation $\bs{\zeta}''=\psi^{\rm h}|_{\T^+}$. Let the   marked point and weight be
$({\bf x}'',\bs{\alpha}'')=({\rm i},\alpha)$ for $\alpha<Q$. Let $\Sigma''_{\rm hole}=\hat{\H}\setminus \psi^{\rm h}(\D^+)$, with $\psi^\ell(z)={\rm i}+rz$
for some $r>0$ small enough so that $\psi^\ell(\D)\cap \psi^{\rm h}(\D^+)=\emptyset$, and let $g$ be an admissible metric on $\Sigma''_{\rm hole}$ such that $(\psi^\ell)^*g=|dz|^2$ near $z=0$. For  $\beta \in Q+{\rm i}\R$, the following factorisation holds true
\begin{equation}\label{ward_with_1point_cornerC1}
\mathcal{A}_{\Sigma'',g,{\bf x}'',\bs{\alpha}'',\bs{\zeta}''}( \Psi^+_{\beta,\nu})=\omega((\Sigma''_{\rm hole})^{\#2},\bs\Delta,\bs\nu)\mathcal{A}_{\Sigma'',g,{\bf x}'',\bs{\alpha}'',\bs{\zeta}''}( \Psi^+_{\beta})
\end{equation}
with  $\bs\nu=(\emptyset,\nu,\emptyset)$ and $\bs\Delta=(\Delta_\alpha,\Delta_\beta,\Delta_\alpha)$.
  \end{description}
\end{corollary}
{\bf Ward identities for outgoing parametrisations.}
We end this section with the case of outgoing parametrisations.
Consider first a half pair of pants $(\Sigma,\bs{\zeta})$ of type (A2)
under model form, with parametrisations 
$$
\zeta_j=\psi_j^{\rm h}\circ (-o)^{(1-\sigma_{j}^{\rm h})/2}|_{\T^+},\quad j=1,2,3,
$$ 
where $\psi_j^{\rm h}\in {\rm BiHol}^+_\eps(\D)$. Without loss of generality, we assume that the outgoing components are $\zeta_j$ for $1\leq j\leq b^+$.
Let $\bs{\zeta}^{\rm in}$ be given by the collection $\zeta_j=\psi_j^{\rm h}|_{\T^+}$, and denote by $(\Sigma^{\rm in},\bs{\zeta}^{\rm in})$ the associated model form building block. Let  $((\Sigma^{\rm in})^{\#2},(\bs{\zeta}^{\rm in})^{\#})$ be the Neumann double, viewed as a pair of pants building block of type (A). Then, by Proposition \ref{prop: RO boundary}, for $\beta_j\in Q+{\rm i}\R$, we have
\begin{equation}\label{Ward_out_to_in}
\begin{split}
\cjg \mathcal{A}_{\Sigma,g,\bs{\zeta}}(\otimes_{j=b^++1}^3\Psi^{+}_{\beta_j,\nu_j}),\otimes_{j=1}^{b^+}\Psi^{+}_{\beta_{j},\nu_{j}}\cjd_{\otimes^{b^+}\mc{H}^+} =&\mathcal{A}_{\Sigma,g,\bs{\zeta}^{\rm in}}(\otimes_{j=1}^{b^+}{\bf O}^+{\bf C}\Psi^{+}_{\beta_j,\nu_j},\otimes_{j=b^++1}^3\Psi^{+}_{\beta_j,\nu_j})\\
=&(-1)^{\sum_{j=1}^{b^+}|\nu_j|}\mathcal{A}_{\Sigma^{\rm in},g,\bs{\zeta}^{\rm in}}(\otimes_{j=1}^{b^+}\Psi^{+}_{\bar{\beta}_j,\nu_j}, \otimes_{j=b^++1}^3\Psi^{+}_{\beta_j,\nu_j}).
\end{split}\end{equation}
By Proposition \ref{ward case I}, we also have (noting that $\Delta_{\bar{\beta}_j}=\Delta_{\beta_j}$ for $\beta_j\in \R\cup (Q+{\rm i}\R)$)
\begin{equation}\label{out_vers_in}
\mathcal{A}_{\Sigma^{\rm in},g,\bs{\zeta}^{\rm in}}(\otimes_{j=1}^{b^+}\Psi^{+}_{\bar{\beta}_j,\nu_j}, \otimes_{j=b^++1}^3\Psi^{+}_{\beta_j,\nu_j})=
\omega((\Sigma^{\rm in})^{\#2},\bs{\Delta} ,\bs{\nu}) \mathcal{A}_{\Sigma^{\rm in},g,\bs{\zeta}^{\rm in}}(\otimes_{j=1}^{b^+}\Psi^{+}_{\bar{\beta}_j}, \otimes_{j=b^++1}^3\Psi^{+}_{\beta_j}).
\end{equation}
We also have, by Proposition \ref{thm: ward identity},
\[\mathcal{A}_{(\Sigma^{\rm in})^{\#2},g,(\bs{\zeta}^{\rm in})^{\#}}(\otimes_{j=1}^{b^+}\Psi_{\bar{\beta}_j,\nu_j,\emptyset}, \otimes_{j=b^++1}^3\Psi_{\beta_j,\nu_j,\emptyset})=\omega((\Sigma^{\rm in})^{\#2},\bs{\Delta} ,\bs{\nu})\mathcal{A}_{(\Sigma^{\rm in})^{\#2},g,(\bs{\zeta}^{\rm in})^{\#}}(\otimes_{j=1}^{b^+}\Psi_{\bar{\beta}_j}, \otimes_{j=b^++1}^3\Psi_{\beta_j}).\]
Next, we observe from the end of Section \ref{geometric_decomp} that the double  $(\Sigma^{\#2},\bs{\zeta}^\#)$ 
has model form which can be transformed into purely incoming parametrisations $((\Sigma^{\#2})^{\rm in},(\bs{\zeta}^\#)^{\rm in})$ with   boundary parametrisations given by
\[ (\psi_1^{\rm h}\circ e^{{\rm i}\pi},\dots, \psi_{b^+}^{\rm h}\circ e^{{\rm i}\pi}, \psi_{b^++1}^{\rm h},\dots, \psi_3^{\rm h})|_{\T}. \]
This implies, using Lemma \ref{propPi}, that
\begin{align*}
\mathcal{A}_{(\Sigma^{\rm in})^{\#2},g,(\bs{\zeta}^{\rm in})^{\#}}(\otimes_{j=1}^{b^+}\Psi_{\bar{\beta}_j,\nu_j,\emptyset}, \otimes_{j=b^++1}^3\Psi_{\beta_j,\nu_j,\emptyset})
=&
\mathcal{A}_{(\Sigma^{\#2})^{\rm in},g,(\bs{\zeta}^\#)^{\rm in}}(\otimes_{j=1}^{b^+}e^{i\pi \bs{\Pi}}\Psi_{\bar{\beta}_j,\nu_j,\emptyset}, \otimes_{j=b^++1}^3\Psi_{\beta_j,\nu_j,\emptyset})
\\
=&(-1)^{\sum_{j=1}^{b^+}|\nu_j|}\mathcal{A}_{(\Sigma^{\#2})^{\rm in},g,(\bs{\zeta}^\#)^{\rm in}}(\otimes_{j=1}^{b^+}\Psi_{\bar{\beta}_j,\nu_j,\emptyset}, \otimes_{j=b^++1}^3\Psi_{\beta_j,\nu_j,\emptyset}),
\end{align*}
which in turn gives
\[\omega((\Sigma^{\rm in})^{\#2},\bs{\Delta} ,\bs{\nu})=(-1)^{\sum_{j=1}^{b^+}|\nu_j|}\omega((\Sigma^{\#2})^{\rm in},\bs{\Delta} ,\bs{\nu}).\]
Combining this with \eqref{Ward_out_to_in} and \eqref{out_vers_in}, we  obtain:
\[\cjg \mathcal{A}_{\Sigma,g,\bs{\zeta}}(\otimes_{j=b^++1}^3\Psi^{+}_{\beta_j,\nu_j}),\otimes_{j=1}^{b^+}\Psi^{+}_{\beta_{j},\nu_{j}}\cjd_{\otimes^{b^+}\mc{H}^+} =\omega((\Sigma^{\#2})^{\rm in},\bs{\Delta} ,\bs{\nu})
\cjg \mathcal{A}_{\Sigma,g,\bs{\zeta}}(\otimes_{j=b^++1}^3\Psi^{+}_{\beta_j}),\otimes_{j=1}^{b^+}\Psi^{+}_{\beta_j}\cjd_{\otimes^{b^+}\mc{H}^+}.\]
A similar argument holds for the matrix elements of half pairs of pants of type (A2), (B1), (B2), (C1) or (C2). We summarize this in the following proposition.  

\begin{proposition}\label{prop: matrix coefficient outgoing}
In the case of the matrix elements of half pairs of pants amplitudes with possible outgoing boundaries, the statements of   Corollary \ref{ward case I}, Corollary \ref{cor:case_with_mp_corner}, Corollary \ref{ward case II},  Corollary \ref{cor:case_with_mp_corner1} remain valid, provided that the coefficient $\omega(\Sigma^{\#2}_{\rm hole},\bs\Delta,\bs\nu)$ is replaced by $\omega((\Sigma_{\rm hole}^{\#2})^{\rm in},\bs{\Delta},\bs{\nu})$.
\end{proposition}

\begin{remark}
Our notations implicitly imply that  $\Sigma=\Sigma_{\rm hole}$ when $\Sigma$ is a building block of type (A), (A1) or (A2).  
\end{remark}
\begin{figure}
	\begin{tikzpicture}
		\node[inner sep=0pt] (pant) at (0,0)
		{\includegraphics[width=0.7\textwidth]{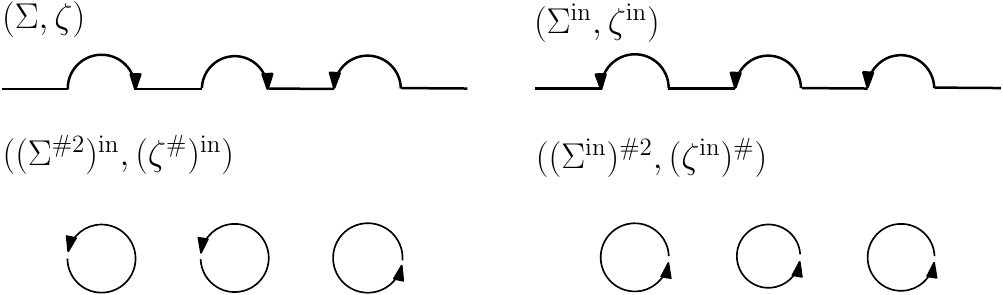}};
	\end{tikzpicture}
	\caption{Four different admissible surfaces in Proposition~\ref{prop: matrix coefficient outgoing}. The   arrow indicates the orientation of the boundary parametrisation and the tip of the arrow head sits on the terminal point of the parametrisation.}
	\label{fig: orientation}
\end{figure}
\section{Proof of the bootstrap formula}
\subsection{Decomposition of correlation functions}\label{sub:graph}
In this section, we apply Segal's gluing formula to express the correlation functions as partial traces of building block amplitudes in a Hilbert space built from tensor products of the spaces $\mc{H}$ and $\mc{H}^+$.

Let $\Sigma$ be a Riemann surface with N-marked boundary $\pl \Sigma$ (and empty Dirichlet boundary), marked points ${\bf x}=(x_1,\dots,x_m)$ in the interior $\Sigma^\circ$ 
and boundary marked points ${\bf s}=(s_1,\dots,s_{m^{\rm b}})$ in $\pl \Sigma$. We consider a maximal non-degenerate system of cutting curves as described in Section~\ref{geometric_decomp}, consisting of parametrised circles $\xi_i^\ell:\T\to \mc{C}_i$, for $i=1,\dots,L$, and half-circles $\xi_i^{\rm h}:\T^+\to \mc{C}_i^+$, for $i=1,\dots,L^+$. Let $g$ be a metric admissible with respect to these cutting curves.

This system of cutting curves partitions $\Sigma$ into geometric building blocks $\mc{P}_j$, which we order 
  so that $\mc{P}_1,\dots,\mc{P}_N$ are of type  (A), (B) or (C) (see Section \ref{geometric_decomp}), while $\mc{P}_{N+1},\dots,\mc{P}_{N+N^+}$  are of type (A1), (A2), (B1), (B2), (C1), or (C2). 
  
 We denote by $\pl_k\mc{P}_j$   the $k$-th  D-marked boundary hypersurface of 
$\mc{P}_j$, corresponding to either a circle $\mc{C}_i$ or a half circle $\mc{C}^+_i$. The boundary parametrisations of $\mc{P}_j$ induced by the maps $\xi_i^\ell$ and $\xi_i^{\rm h}$ are written as
\[\boldsymbol{\zeta}_j=(\zeta_{j1}^\ell,\dots \zeta^\ell_{jb_j^{\ell}},\zeta_{j1}^{\rm h},\dots,\zeta_{jb^{{\rm h}}_j}^{\rm h}) \] 
 where $b_j^\ell$ and $b^{\rm h}_j$ denote respectively the number of D-marked boundary circles and half-circles of $\mc{P}_j$. Similarly, we write ${\bf x}_j=(x_{j1},\dots,x_{jm_j})$ and ${\bf s}_j=(s_{j1},\dots,s_{jm^{\rm b}_j})$ for the interior and boundary marked points of $\Sigma$ contained in $\mc{P}_j$, and we denote by $\bs{\alpha}_j=(\alpha_{j1},\dots,\alpha_{jm_j})$ and $\bs{\beta}_j=(\beta_{j1},\dots,\beta_{jm^{\rm b}_j})$ the corresponding weights.

We  associate an oriented graph $\mc{G}$, whose edges carry two distinct colours, to the non-degenerate cutting as follows: 
\begin{itemize}
\item to each  building block $\mc{P}_j$ we associate a vertex  ${\rm v}_j$, which is a discrete set made up of  $b_j=b_j^\ell+b_j^{\rm h}$ elements in  $\{j\}\times \{1,\dots,b_j\}$;
\item to each circle $\mc{C}_i$, identified with the boundary components $\pl_k\mc{P}_j$ and $\pl_{k'}\mc{P}_{j'}$, we associate a coloured edge ${\rm e}_i=((j,k),(j',k'))$, say blue, oriented so that
$\pl_k\mc{P}_j$ carries the incoming parametrisation and $\pl_{k'}\mc{P}_{j'}$ the outgoing one;
\item to each half-circle $\mc{C}_i^+$, identified with $\pl_k\mc{P}_j$ and $\pl_{k'}\mc{P}_{j'}$, we associate an edge ${\rm e}^+_i=((j,k),(j',k'))$, of a different colour, say red, again oriented from the incoming to the outgoing parametrisation.
\item to each interior marked point $x_i=x_{jk}$ contained in $\mc{P}_j$, we associate a vertex  denoted $x_i$, called phantom vertex, and 
a blue  edge  ${\rm e}_i^*=({\rm v}_j,x_{i})$, called phantom edge. And to each boundary marked point $s_i=s_{jk}$ contained in $(\pl \mc{P}_j)_{\rm N}$, we associate a phantom vertex denoted $s_i$, and a red phantom edge ${\rm e}_i^{+*}=({\rm v}_j,s_{i})$.  
\end{itemize}
The phantom edges and vertices are only used to encode the structure of marked points, but do not play a role in the gluing assignments of building blocks to obtain the surface $\Sigma$.

 \begin{figure}[h] 
\centering
\begin{tikzpicture}[scale=0.8]

\coordinate (Z1) at (-0,3) ;
\coordinate (Z2) at (3,1) ;
\coordinate (Z3) at (3,-1) ;
\coordinate (Z4) at (-0.5,-2) ;
\coordinate (Z5) at (-2.5,-2) ;
\coordinate (Z6) at (-2,3) ;
\draw[draw=black,very thick,fill=lightgray] (Z1) to[out=300,in=180] (Z2) to [out=270,in=90](Z3) to [out=180,in=80](Z4) to [out=180,in=0](Z5) to [out=80,in=280](Z6) to [out=0,in=180](Z1) ;
 \draw[very thick,fill=white,draw=black] (3,0) ellipse (0.5cm and 1cm);
  \draw[very thick,fill=white,draw=black] (-1,3) ellipse (1cm and 0.2cm);
   \draw[very thick,fill=white,draw=black] (-1.5,-2) ellipse (1cm and 0.3cm);
\draw (-1.5,-1.7) node[black]{$\bullet$} ;  
\draw (-1.5,-1.7) node[below,black]{$s_{21}$} ;  
\draw (-0.6,2.75) node[black]{$\bullet$} ;  
\draw (-0.6,2.7) node[above,black]{$s_{51}$} ;  
\draw (-1.2,2.7) node[black]{$\bullet$} ;  
\draw (-1.2,2.7) node[above,black]{$s_{52}$} ;  
\draw (-1,-0.3) node[black]{$\bullet$} ;  
\draw (-1,-0.3) node[below,black]{$x_{11}$} ;
\coordinate (C1) at (-2.3,-1) ; 
\coordinate (C2) at (-0.2,-1.4) ; 
\coordinate (C3) at (0.6,-1) ; 
\coordinate (C4) at (-2,0.6) ; 
\coordinate (C5) at (-1.6,2.8) ; 
\coordinate (C55) at (-0.85,2) ; 
\coordinate (C6) at (-0.3,2.85) ; 
\coordinate (C7) at (-0.1,2.9) ; 
\coordinate (C8) at (2.5,0.3) ; 
\coordinate (C9) at (-1.9,2.9) ; 
\coordinate (C10) at (2.5,-0.2) ; 
\coordinate (C11) at (2.6,-0.7) ; 
\coordinate (C12) at (0,0.5) ; 
\coordinate (C13) at (-1.88,1) ;
\coordinate (C14) at (1.2,1.5) ; 
\coordinate (C15) at (2.7,0.8) ;  
\draw[draw=blue,very thick] (C1) to[out=10,in=130] (C2);
\draw[draw=blue,dashed] (C1) to[out=-20,in=180] (C2);
\draw[draw=blue,very thick] (C3) to[out=110,in=10] (C4);
\draw[draw=blue,dashed] (C3) to[out=160,in=-60] (C4);
\draw[draw=red,very thick] (C5) to[out=270,in=180] (C55) to[out=0, in=270] (C6);
\draw[draw=red,very thick] (C7) to[out=290,in=170] (C8);
\draw[draw=red,very thick] (C9) to[out=290,in=160] (C10);
\draw[draw=red,very thick] (C11) to[out=170,in=-20] (C12) to [out=160,in=0] (C13);
\draw[draw=red,dashed] (C13) to [out=10,in=210](C14);
\draw[draw=red,very thick] (C14) to [out=260,in=170](C15);
 \draw (-1,-1.7) node[above]{$ \mc{P}_2$} ;
  \draw (-1.7,-0.7) node[above]{$ \mc{P}_1$} ; 
    \draw (1.2,-0.8) node[above]{$ \mc{P}_3$} ; 
       \draw (-1.5,1) node[above]{$ \mc{P}_4$} ; 
           \draw (-0.85,2.1) node[above]{$ \mc{P}_5$} ; 
                 \draw (-0.3,1.3) node[above]{$ \mc{P}_6$} ; 
  \end{tikzpicture}
\,\hspace{1cm}
\subfloat[Corresponding oriented multigraph (vertices in gray, phantom vertices in white).]{
\begin{tikzpicture}[xscale=0.75,yscale=0.75] 
\tikzstyle{sommet}=[circle,draw,fill=yellow,scale=0.7] 
\tikzstyle{phantomV}=[circle,draw,fill=cyan,scale=0.6] 
\definecolor{yellow}{gray}{0.8}
\definecolor{cyan}{gray}{1}
\node[red] at (2.7,1.7){$e_1^+$};
\node[red] at (3.3,2.7){$e_2^+$};
\node[red] at (4.7,2.7){$e_3^+$};
\node[red] at (1.3,2.7){$e_2^{+*}$};
\node[red] at (1.3,4.3){$e_3^{+*}$};
\node[red] at (10.3,4.3){$ $};
\node[red] at (-2,4.3){$ $};
\node[sommet] (P1) at (4,0){ $\mathbf{\mathcal{P}_1}$};
\node[sommet] (P3) at (2,1){ $\mathbf{\mathcal{P}_3}$};
\node[sommet] (P2) at (6,1){ $\mathbf{\mathcal{P}_2}$};
\node[sommet] (P4) at (4,2){ $\mathbf{\mathcal{P}_4}$};
\node[sommet] (P5) at (2,3.5){ $\mathbf{\mathcal{P}_5}$};
\node[sommet] (P6) at (4,3.5){ $\mathbf{\mathcal{P}_6}$};
\node[phantomV] (S21) at (8,1){ $s_{21}$};
\node[phantomV] (S51) at (0,4.55){ $s_{51}$};
\node[phantomV] (S52) at (0,2.5){ $s_{52}$};
\node[phantomV] (X11) at (4,-1.5){ $x_{11}$};
\draw[->,>=latex,red] (P2) -- (S21)  node[midway, below]{$e_1^{+*}$};
\draw[->,>=latex,blue] (P1) -- (X11)  node[midway, right]{$e_2^*$};
\draw[->,>=latex,blue] (P1) -- (P2)  node[midway, below]{$e_1$};
\draw[->,>=latex,blue] (P1) -- (P3)  node[midway, below]{$e_2$};
\draw[->,>=latex,red] (P3) -- (P4)  ;
\draw[->,>=latex,red] (P5) -- (P6)  node[midway,above]{$e_4^+$};
\draw[->,>=latex,red] (P5) -- (S51);
\draw[->,>=latex,red] (P5) -- (S52);
\draw[->,>=latex,red] (P4) to[bend left] (P6);
\draw[->,>=latex,red] (P4) to[bend right] (P6) ;
\end{tikzpicture}
}
\caption{Multigraphs associated to marked Riemann surfaces}
\label{fig:partition}
\end{figure}
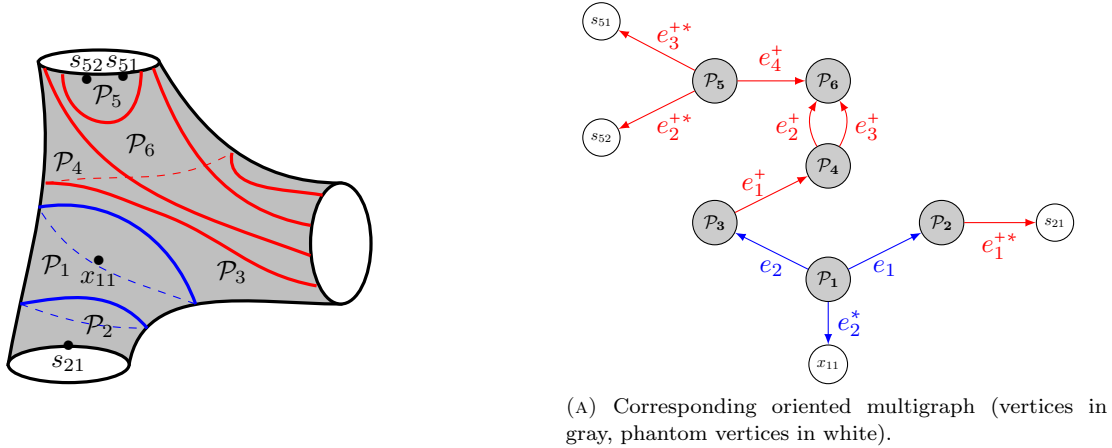

The Neumann double $\Sigma^{\#2}$ has   marked points ${\bf x}^{\#}$   made up of the points ${\bf x}$, with weights $\bs\alpha$, their reflected $\tau_\Sigma({\bf x})$, with weights $\bs\alpha$ too,  and  ${\bf s}$, with weights $\bs\beta$. We collect these weights in $\bs\alpha^\#$. The maximal non-degenerate system of cutting curves   $(\xi_i^\ell)_{i=1,\dots,L}$  and   $(\xi_i^{\rm h})_{i=1,\dots,L^+}$ induces a maximal non-degenerate system of cutting curves for $\Sigma^{\#2}$ made up of $(\xi_i^{\#})_{i=1,\dots,2L+L^+}$ made up of the curves $\xi_i^\ell$ and  $\tau_\Sigma\circ \xi_i^\ell\circ o$ for $i=1,\dots, L$, and the curves defined by $e^{i\theta}\mapsto \zeta_{i}^{\rm h}(e^{i\theta})$ for $\theta\in [0,\pi]$ and $e^{i\theta}\mapsto \tau_\Sigma(\zeta_{i}^{\rm h}(e^{i(2\pi-\theta)}))$ for $\theta\in [\pi,2\pi]$,  for $i=1,\dots,L^+$. 

This family of cutting curves provides a decomposition into building blocks  $(\mc{P}^{\#}_j)_{j\leq 2N+N^+}$ of the closed surface $\Sigma^{\#2}$   made up of the surfaces $(\mc{P}_j)_{j\leq N}$, $(\tau_\Sigma(\mc{P}_j))_{j\leq N}$ and $(\mc{P}_j^{\#2}=\mc{P}_j\# \tau_\Sigma( \mc{P}_j))_{N<j\leq N+N^+}$, with parametrisations given by the curves   $\bs{\zeta}_j^{\#}=(\bs{\zeta}_{jk}^{\#})_{1\leq k\leq 2b_j^\ell+b_j^{\rm h}}$. We denote by $\mc{G}^\#$ the oriented graph associated to the building block decomposition of the Neumann double $\Sigma^{\#2}$.

\subsection{Structure constants associated to the graph}\label{constantpj}

Now we focus on the structure constants attached to this graph. For a building block $\mc{P}_j$ as above, equipped with an  admissible metric $g$, with marked points ${\bf x}_j$, ${\bf s}_j$  and associated weights  $\bs{\alpha}_j$, $\bs{\beta}_j$, and ${\bf p}_j \in \R_*^{b_j}$, with $b_j=b_j^{\ell}+b_j^{\rm h}$,   bulk/boundary  cosmological constant $\mu$ and $\mu^{\rm b}_j$, and orientation signs $\boldsymbol{\sigma}_j$ of the D-marked boundary components of $\partial \mc{P}_j $, we define ${\bf C}^{\mc{P}_j}_{\gamma,\mu,\mu^{\rm b}_j}({\boldsymbol{\alpha}_j},\bs{\beta}_j,{\bf p}_j)$ as follows (recall   \eqref{def_SL0}):
\begin{itemize}
\item if $\mc{P}_j$ is of type (A), (B) or (C) then  we define the structure constant of the block $\mc{P}_j$ to be
\begin{equation}\label{dozzblock}
{\bf C}^{\mc{P}_j}_{\gamma,\mu,\mu^{\rm b}_j}({\boldsymbol{\alpha}_j},\bs{\beta}_j,{\bf p}_j ):={\rm C}^{\rm DOZZ}_{\gamma,\mu}(\alpha_1,\alpha_2,\alpha_3)
\end{equation}
where the DOZZ conbstant is defined in \eqref{theDOZZformula}, and we define the metric dependent constant to be 
\begin{align}\label{metricconstantdozzblockrel}
C_{\mc{P}_j,g}(\boldsymbol{\alpha}_j,\bs{\beta}_j, {\bf p}_j):= C^{{\rm (A)}, b^\ell_j}_{\mc{P}_j,g,{\bf z}^0}( \alpha_1,\alpha_2,\alpha_3).
	\end{align}
Here the function $C^{{\rm (A)}, b^\ell_j}_{\mc{P}_j,g,{\bf z}^0}$ is defined by
\begin{align}\label{metricconstantdozzblock}
C^{{\rm (A)}, b^\ell_j}_{\mc{P}_j,g,{\bf z}^0}( \alpha_1,\alpha_2,\alpha_3):=&\frac{2^{b_j-2}\pi^{b_j-1}}{Z_{\D,g_\D} ^{b_j}}
  \big({\frac{{\rm v}_{g_{\rm dozz}}(\hat\C)}{{\det}'(\Delta_{g_{\rm dozz}})}} \big)^\hf  P({\bf z}^0)\Big(\prod_{k=b_j+1}^{3} g\left(z^0_k\right)^{-\Delta_{\alpha_k}}\Big) \Big(\prod_{k=1}^{b_j} |(\psi_{jk}^\ell)'(0)|^{-\Delta_{\alpha_k}}\Big)   
  \\
 & e^{c_{\rm L}(S^0_{\rm L}( \caP_j,g_{\rm dozz},g)+\sum_{k=1}^{b_j}S^0_{\rm L}(\psi_{jk}^\ell(\D),g_{\rm dozz},(\psi_{jk}^\ell)_*g_\D))}  \nonumber
	\end{align}
where ${\bf z}^0\in\{(-1,0,1);({\rm i},0,-{\rm i})\}$, 
$$ P({\bf z}^0):=  \left|z^0_1-z^0_3\right|^{2\left(\Delta_{\alpha_2}-\Delta_{\alpha_1}-\Delta_{\alpha_3}\right)}\left|z^0_2-z^0_3\right|^{2\left(\Delta_{\alpha_1}-\Delta_{\alpha_2}-\Delta_{\alpha_3}\right)}\left|z^0_1-z^0_2\right|^{2\left(\Delta_{\alpha_3}-\Delta_{\alpha_1}-\Delta_{\alpha_2}\right)},$$
where the parameters in \eqref{dozzblock},   \eqref{metricconstantdozzblockrel} and \eqref{metricconstantdozzblock} are chosen to be
\begin{equation}\label{parameterdozzblock}
(\alpha_1,\alpha_2,\alpha_3):=\left\{
\begin{array}{ll}
(Q+i\sigma_{j1}p_{j1},\alpha_{j1},\alpha_{j2}), & b_j^\ell=1\\[0.1cm]
(Q+i\sigma_{j1}p_{j1},Q+i\sigma_{j2}p_{j2},\alpha_{j1}), & b_j^\ell=2\\[0.1cm]
(Q+i\sigma_{j1}p_{j1},Q+i\sigma_{j2}p_{j2},Q+i\sigma_{j3}p_{j3}), & b_j^\ell=3
\end{array}\right..
\end{equation}
\item if $\mc{P}_j$ is of type (A1), (B1) or (C1) then we define the structure constant of the block $\mc{P}_j$ to be
\begin{equation}\label{Gblock}
{\bf C}^{\mc{P}_j}_{\gamma,\mu,\mu^{\rm b}_j}({\boldsymbol{\alpha}_j},\bs{\beta}_j,{\bf p}_j ):=
\tilde {\rm G}_{\mu,\mu^{\rm b}_j}( \alpha,\beta),  
\end{equation}
where the constant $\tilde{\rm G}_{\mu,\mu^{\rm b}}$ is defined by \eqref{eq:tildeG_zero}, and we define the metric dependent constant to be 
\begin{align}\label{metricconstantGblockrel}
C_{\mc{P}_j,g}(\boldsymbol{\alpha}_j,\bs{\beta}_j, {\bf p}_j):=C^{{\rm (A1)}, b^\ell_j,b^{\rm h}_j}_{\mc{P}_j,g}(\alpha,\beta). 
\end{align}
Here the function $C^{{\rm (A1)}, b^\ell_j,b^{\rm h}_j}_{\mc{P}_j,g}$ is defined by 
\begin{align}\label{metricconstantGblock}
C^{{\rm (A1)}, b^\ell_j,b^{\rm h}_j}_{\mc{P}_j,g}(\alpha,\beta):= &\frac{2^{\frac{b_j^{\ell}}{2}+\frac{3b_j^{\rm h}}{4}-1}\pi^{b_j^\ell-1+\frac{3b_j^{\rm h}}{4}}}{Z_{\D,g_\D} ^{b_j^\ell} Z_{\D^+,g_\D,{\rm m}}^{b_j^{\rm h}} }  |(\psi^\ell_{j1})'(0)|^{2b_j^\ell\Delta_\alpha}  |(\psi^{\rm h}_{j2})'(0)|^{b^{\rm h}_j \Delta_\beta}  g({\rm i})^{-(1-b_j^\ell)\Delta_{\alpha}}g(0)^{-(1-b_j^{\rm h})\frac{\Delta_{\beta}}{2}}2^{-(2\Delta_\alpha-\Delta_\beta)} \nonumber
 \\
 &  \big({\frac{{\rm v}_{g_{\rm dozz}}(\hat\H)}{{\det}'(\Delta_{\hat{\H},g_{\rm dozz},{\rm N}})}} \big)^\hf   e^{c_{\rm L} S_{\rm L}^0(\mc{P}_j,g_{\rm dozz},g)}  e^{c_{\rm L} b_j^\ell S_{\rm L}^0(\psi^\ell_{j1}(\D),g_{\rm dozz},(\psi^\ell_{j1})_*g_\D)+c_{\rm L} b^{\rm h}_jS_{\rm L}^0(\psi^{\rm h}_{j2}(\D^+),g_{\rm dozz},(\psi^{\rm h}_{j2})_*g_\D)} ,
\end{align}
where the parameters in \eqref{Gblock} and \eqref{metricconstantGblockrel} are chosen to be
\begin{equation}\label{parameterGblock}
(\alpha,\beta):=\left\{
\begin{array}{ll}
(Q+i\sigma_{j1}p_{j1},Q+i\sigma_{j2}p_{j2}), & b^{\ell}_j=1,b^{\rm h}_j=1\\[0.1cm]
 (\alpha_{j1},Q+i\sigma_{j2}p_{j2}), & b^{\ell}_j=0,b^{\rm h}_j=1\\[0.1cm]
(Q+i\sigma_{j1}p_{j1},\beta_{j2}), & b^{\ell}_j=1,b^{\rm h}_j=0
\end{array}\right..
\end{equation}
\item if $\mc{P}_j$ is of type (A2), (B2) or (C2) then we define the  structure constant to be  
\begin{equation}\label{Hblock}
{\bf C}^{\mc{P}_j}_{\gamma,\mu,\mu^{\rm b}_j}({\boldsymbol{\alpha}_j},\bs{\beta}_j,{\bf p}_j ):=\widetilde{\rm H}_{\mu^b_{j1},\mu^b_{j2},\mu^b_{j3}}^{\beta_1,\beta_2,\beta_3}
\end{equation}
where $\widetilde{\rm H}_{\mu_{1},\mu_{2},\mu_{3}}^{\beta_1,\beta_2,\beta_3}$ is defined by \eqref{formule_PT} if $\mu>0$ and \eqref{eq:tildeH_zero} if $\mu=0$, and we define the metric dependent constant to be 
\begin{align}\label{metricconstantHblockrel}
C_{\mc{P}_j,g}(\boldsymbol{\alpha}_j,\bs{\beta}_j, {\bf p}_j):=C^{{\rm (A2)},b_j^{\rm h}}_{\mc{P}_j,g}(\beta_1,\beta_2,\beta_3).
\end{align}
Here the function $C^{{\rm (A2)},b_j^{\rm h}}_{\mc{P}_j,g}$ is defined by
\begin{align}\label{metricconstantHblock}
C^{{\rm (A2)},b_j^{\rm h}}_{\mc{P}_j,g}(\beta_1,\beta_2,\beta_3):= &\frac{(2\pi)^{ \frac{3b_j^{\rm h}}{4}-1}}{ Z_{\D^+,g_\D,{\rm m}}^{b_j^{\rm h}} }  
\Big(\prod_{k=b_j^{\rm h}+1}^{3} g\left(z^0_k\right)^{-\frac{\Delta_{\beta_k}}{2}}\Big) 
\Big(\prod_{k=1}^{b_j^{\rm h}} |(\psi_{jk}^{\rm h})'(0)|^{\Delta_{\beta_k}}\Big) 2^{\Delta_{\beta_2}-\Delta_{\beta_1}-\Delta_{\beta_3}}
\nonumber \\
 &  \big({\frac{{\rm v}_{g_{\rm dozz}}(\hat\H)}{{\det}'(\Delta_{\hat{\H},g_{\rm dozz},{\rm N}})}} \big)^\hf      e^{c_{\rm L}(S^0_{\rm L}( \caP_j,g_{\rm dozz},g)+\sum_{k=1}^{b^{\rm h}_j}S^0_{\rm L}(\psi_{jk}^{\rm h}(\D^+),g_{\rm dozz},(\psi_{jk}^{\rm h})_*g_\D))}   ,
\end{align}
where ${\bf z}^0=({-1,0,1})$, and the parameters in \eqref{Hblock} and \eqref{metricconstantHblockrel} are chosen to be
\begin{equation}\label{parameterHblock}
(\beta_1,\beta_2,\beta_3):=\left\{
\begin{array}{ll}
(Q+i\sigma_{j1}p_{j1},\beta_{j1},\beta_{j2}), &  b^{\rm h}_j=1\\[0.1cm]
(Q+i\sigma_{j1}p_{j1},Q+i\sigma_{j2}p_{j2},\beta_{j1}), &  b^{\rm h}_j=2\\[0.1cm]
(Q+i\sigma_{j1}p_{j1},Q+i\sigma_{j2}p_{j2},Q+i\sigma_{j3}p_{j3}), &  b^{\rm h}_j=3
\end{array}\right..
\end{equation}

\end{itemize}

\begin{lemma}\label{structure_amp}
We have\footnote{For building blocks of types (A), (B), (C), we take $({\bf s}_j,\bs{\beta}_j)=(\emptyset,\emptyset)$.}
\begin{align*}
&\big\cjg \mc{A}_{\mc{P}_j,g,{\bf x}_j,\bs{\alpha}_j,{\bf s}_j,\bs{\beta}_j,\bs{\zeta}_j},\bigotimes_{k=1}^{b_j^\ell} 
{\bf C}^{\frac{\sigma_{jk}+1}{2}} \Psi_{Q+ip_{jk}}\bigotimes \bigotimes_{k=1}^{b_j^{\rm h}}{\bf C}^{\frac{\sigma_{jk}+1}{2}}\Psi^+_{Q+ip_{jk}}\big\cjd_{\mc{H}^{\otimes b_j^\ell}\otimes 
(\mc{H}^+)^{\otimes b_j^{\rm h}}}   \\
&= C_{\mc{P}_j,g}(\boldsymbol{\alpha}_j,\bs{\beta}_j, {\bf p}_j)  {\bf C}^{\mc{P}_j}_{\gamma,\mu,\mu^{\rm b}_j}({\boldsymbol{\alpha}_j},\bs{\beta}_j,{\bf p}_j )
\end{align*}
 
\end{lemma}

\begin{proof}  
Because of the diffeomorphism invariance of amplitudes, we may assume that $\mc{P}_j$ is under model form. We treat the case when  $\mc{P}_j$ is a building block of type (A1), the other cases being similar. Furthermore, by Propositions \ref{prop: RO bulk} and \ref{prop: RO boundary}, we may assume as well that the boundary parametrisations are all incoming.  We extend the admissible metric $g$ on $\mc{P}_j$ to the half-plane $\mathbb{H}$.

For $\epsilon$ small enough, let  $\psi^{\ell}_{j1}\in {\rm BiHol}_\eps(\D)$, $\psi^{\rm h}_{j2}\in {\rm BiHol}^+_\eps(\D)$ such that $\mc{P}_j= \hat{\H}\setminus (\psi^{\ell}_{j1}(\D)\cup \psi^{\rm h}_{j2}(\D^+))$,  $(\psi^{\ell}_{j1}(0)  , \psi^{\rm h}_{j2}(0))=( {\rm i} , 0) $, with boundary parametrisations $\bs{\zeta}=(\zeta_{j1}^{\ell},\zeta_{j2}^{\rm h})$ given by
 $$  \psi^{\ell}_{j1}|_{\T^+}=  \zeta_{j1}^{\ell}   ,\qquad \psi^{\rm h}_{j2}|_{\T^+}= \zeta_{j2}^{\rm h}.
 $$

 Let us consider first $\alpha,\beta \in\R$ with $\alpha,\beta<Q$.
By \eqref{amplitude}, \eqref{psi_alpha_proba} and Proposition \ref{Weyl2}, the amplitude of the disk is related to $\Psi_{\alpha}$ by 
\begin{align*}
\mathcal{A}_{\psi^\ell_{j1}(\D), g, {\rm i},\alpha,\zeta_{j1}^{\ell}}=& e^{c_{\rm L} S_{\rm L}^0(\psi^\ell_{j1}(\D),(\psi^\ell_{j1})_*g_\D,g)-\Delta_{\alpha}\omega_{j1}(0)}\mathcal{A}_{\psi^\ell_{j1}(\D), (\psi^\ell_{j1})_*g_\D,  {\rm i},\alpha,\zeta_{j1}^{\ell}}
\\
=& Z_{\D,g_\D} g({\rm i})^{-\Delta_{\alpha}}|(\psi^\ell_{j1})'(0)|^{-2\Delta_\alpha}e^{c_{\rm L} S_{\rm L}^0(\psi^\ell_{j1}(\D),(\psi^\ell_{j1})_*g_\D,g)} \, \Psi_{\alpha }
\end{align*}
where $\omega_{j1}:\D\to \R$ is the smooth map defined by $(\psi^\ell_{j1})^*g=e^{\omega_{j1}}g_{\D}$, and we have used $\caA^0_{\D,g_\D}(\tilde{ \varphi}^\ell)=1$. In the same way,
\begin{align*}
\mathcal{A}_{\psi^{\rm h}_{j2}(\D^+), g, 0,\beta,\zeta_{j2}^{\rm h}}=& e^{c_{\rm L} S_{\rm L}^0(\psi^{\rm h}_{j2}(\D^+),(\psi^{\rm h}_{j2})_*g_\D,g)-\frac{1}{2}\Delta_{\beta }\omega_{j2}(0)}\mathcal{A}_{\psi^{\rm h}_{j2}(\D^+), (\psi^{\rm h}_{j2})_*g_\D, 0,\beta ,\zeta_{j2}^{\rm h}}
\\
=& g(0)^{-\frac{\Delta_\beta}{2}}  |(\psi^{\rm h}_{j2})'(0)|^{- \Delta_\beta}  e^{c_{\rm L} S_{\rm L}^0(\psi^{\rm h}_{j2}(\D^+),(\psi^{\rm h}_{j2})_*g_\D,g) } Z_{\D^+,g_\D,{\rm m}} \Psi^+_{\beta },
\end{align*}
where $\omega_{j2}:\D^+\to \R$ is the smooth map defined by $(\psi^{\rm h}_{j2})^*g=e^{\omega_{j2}}g_{\D}$, and we have used $\caA^0_{\D^+,g_\D}(\tilde{ \varphi}^{\rm h})=1$. Note in particular that the corner terms in Proposition \ref{Weyl2} vanish because $\omega_{j2}|_{\T^+}=0$.

By the gluing formula for amplitudes \eqref{glue_ampl} 
\begin{align*}
\mc{A}_{\hat\H,g,{\rm i},\alpha,0,\beta} =&\frac{1}{2^{1/4}\pi^{3/4}}\big\cjg \mc{A}_{\mc{P}_j,g,\bs{\zeta}_j}, 
{\bf C} \Psi_{\alpha} \otimes  {\bf C} \Psi^+_{\beta}\big\cjd_{\mc{H} \otimes 
 \mc{H}^+}  Z_{\D,g_\D}  Z_{\D^+,g_\D,{\rm m}}  g({\rm i})^{-\Delta_{\alpha}}  g(0)^{-\frac{\Delta_\beta}{2}}
 \\
 &\times  |(\psi^\ell_{j1})'(0)|^{-2\Delta_\alpha}  |(\psi^{\rm h}_{j2})'(0)|^{- \Delta_\beta}  e^{c_{\rm L} S_{\rm L}^0(\psi^\ell_{j1}(\D),(\psi^\ell_{j1})_*g_\D,g)+c_{\rm L} S_{\rm L}^0(\D^+,g_\D,(\psi^{\rm h}_{j2})^*g) } .
 \end{align*}
The amplitude $\mc{A}_{\hat\H,g,{\rm i},\alpha,0,\beta} $ is the bulk-boundary correlation on $\hat\H$ given by \eqref{G2} 
\begin{align*}
	\langle V_{\alpha,g}({\rm i})  V_{\beta,g}(0)  \rangle_{\hat\H,g}=
	g({\rm i})^{-\Delta_{\alpha}}g(0)^{-\frac{\Delta_{\beta}}{2}}\big(\frac{{\rm v}_{g}(\hat\H)}{{\det}'(\Delta_{\hat{\H},g,{\rm N}})}\big)^{\frac{1}{2}} e^{-6Q^2 S_{\rm L}^0(\hat{\H},g,g_{\rm dozz})}  2^{-(2\Delta_\alpha-\Delta_\beta)} \tilde {\rm G}_{\mu,\mu^{\rm b}}(\alpha,\beta) .
\end{align*} 
Combining, we deduce
\begin{align*}
\big\cjg \mc{A}_{\mc{P}_j,g,\bs{\zeta}_j}, 
{\bf C} \Psi_{\alpha} \otimes  {\bf C} \Psi^+_{\beta}\big\cjd_{\mc{H} \otimes 
 \mc{H}^+}   =& 
 \frac{2^{1/4}\pi^{3/4}}{Z_{\D,g_\D}  Z_{\D^+,g_\D,{\rm m}} }  |(\psi^\ell_{j1})'(0)|^{2\Delta_\alpha}  |(\psi^{\rm h}_{j2})'(0)|^{ \Delta_\beta}  \big({\frac{{\rm v}_{g_{\rm dozz}}(\hat\H)}{{\det}'(\Delta_{\hat{\H},g_{\rm dozz},{\rm N}})}} \big)^\hf    2^{-(2\Delta_\alpha-\Delta_\beta)} 
 \\
 &  e^{c_{\rm L} S_{\rm L}^0(\mc{P}_j,g_{\rm dozz},g)}  e^{c_{\rm L} (S_{\rm L}^0(\psi^\ell_{j1}(\D),g_{\rm dozz},(\psi^\ell_{j1})_*g_\D)+S_{\rm L}^0(\psi^{\rm h}_{j2}(\D^+),g_{\rm dozz},(\psi^{\rm h}_{j2})_*g_\D))} \tilde {\rm G}_{\mu,\mu^{\rm b}}(\alpha,\beta),
 \end{align*}
 and the result follows by analytic continuation to $\alpha=Q+{\rm i}p_{j1}$, $\beta=Q+{\rm i}p_{j2}$.
\end{proof}

Combining with Proposition \ref{prop: matrix coefficient outgoing}, we obtain the following holomorphic factorisation
\begin{corollary}\label{corofactor}
Let $\mc{P}_j$ be a building block as above, equipped with an admissible metric $g$. Then
\begin{enumerate}
\item if $\mc{P}_j$ has an empty Neumann boundary, that is of type (A), (B) or (C), then
\begin{align*}
\big\cjg \mc{A}_{\mc{P}_j,g,{\bf x}_j,\bs{\alpha}_j,\bs{\zeta}_j}&,\bigotimes_{k=1}^{b_j^\ell} 
{\bf C}^{\frac{\sigma_{jk}+1}{2}} \Psi_{Q+ip_{jk},\nu_{jk},\tilde{\nu}_{jk} }\big\cjd_{\mc{H}^{\otimes b_j^\ell}}   
\\
= &C_{\mc{P}_j,g}(\boldsymbol{\alpha}_j,\bs{\beta}_j, {\bf p}_j)  {\bf C}^{\mc{P}_j}_{\gamma,\mu,\mu^{\rm b}_j}({\boldsymbol{\alpha}_j},\bs{\beta}_j,{\bf p}_j )\omega((\mc{P}_j)_{\rm hole},\bs\Delta_j ,\bs\nu_j)\overline{\omega((\mc{P}_j)_{\rm hole},\bs\Delta_j ,\bs{\tilde\nu}_j)}
\end{align*}
Note   that there are here no boundary eigenstates or  boundary marked points involved. We keep the notations as such for later compactness of notations.
\item if $\mc{P}_j$ has a non empty Neumann boundary, that is  of  type (A1), (A2), (B1), (B2), (C1) or (C2), then
\begin{align*}
\big\cjg \mc{A}_{\mc{P}_j,g,{\bf x}_j,\bs{\alpha}_j,{\bf s}_j,\bs{\beta}_j,\bs{\zeta}_j}&,\bigotimes_{k=1}^{b_j^\ell} 
{\bf C}^{\frac{\sigma_{jk}+1}{2}} \Psi_{Q+ip_{jk},\nu_{jk},\tilde{\nu}_{jk}}\bigotimes \bigotimes_{k=1}^{b_j^{\rm h}}{\bf C}^{\frac{\sigma_{jk}+1}{2}}\Psi^+_{Q+ip_{jk},\nu_{jk}}\big\cjd_{\mc{H}^{\otimes b_j^\ell}\otimes 
(\mc{H}^+)^{\otimes b_j^{\rm h}}}   
\\
= &C_{\mc{P}_j,g}(\boldsymbol{\alpha}_j,\bs{\beta}_j, {\bf p}_j)  {\bf C}^{\mc{P}_j}_{\gamma,\mu,\mu^{\rm b}_j}({\boldsymbol{\alpha}_j},\bs{\beta}_j,{\bf p}_j )\omega\big(((\mc{P}_j)_{\rm hole}^{\#2})^{\rm in},\bs{\Delta}_j ,\bs{\nu}_j\big). 
\end{align*}
\end{enumerate}
\end{corollary}

\subsection{Traces}

We now fix an ordering operator depending only on the graph $\mc{G}$:
\[ \mc{O}_{\mc{G}}: \otimes_{j=1}^{N+N^+}(\mc{H}^{\otimes b_j^\ell}\otimes (\mc{H}^+)^{\otimes b_j^{\rm h}}) \to \mc{H}^{\otimes \sum_j b_j^\ell}
\otimes (\mc{H}^+)^{\otimes \sum_j b_j^{\rm h}}\]
defined by permuting the tensor components so that all factors of $\mc{H}$ appear first, and, if ${\rm e}_i=((j,k), (j',k'))$, the 
components of $\mc{H}$ corresponding to $\pl_k\mc{P}_j$ and $\pl_{k'}\mc{P}_{j'}$ are mapped  to the $i$-th and the $(i+L)$-th components of the $\mc{H}$-component, respectively. Similarly, if ${\rm e}^+_i=((j,k),(j',k'))$, the 
components of $\mc{H}^+$ associated to $\pl_k\mc{P}_j$ and $\pl_{k'}\mc{P}_{j'}$ are mapped   to the $i$-th and the $(i+L^+)$-th component of the $\mc{H}^+$-component, respectively.

Next, the $\mc{G}$-trace of an element $W\in  \otimes_{j=1}^{N+N^+}(\mc{H}^{\otimes b_j^\ell}\otimes (\mc{H}^+)^{\otimes b_j^{\rm h}}) $ is   defined by the formula
\[ {\rm Tr}^{\mc{G}}_{\mc{H},\mc{H}^+}(W)=\sum_{\substack{(n_1,\dots,n_{L})\in \N^L\\(n^+_1,\dots,n^+_{L^+})\in \N^{L^+}}} (\mc{O}_{\mc{G}}W)(e_{n_1},\dots,e_{n_L},e_{n_1},\dots,e_{n_L},e^+_{n^+_1},\dots,e^+_{n^+_{L^+}},e^+_{n^+_1},\dots,e^+_{n^+_{L^+}})\]
where $(e_{n})_{n\in \N}$ is an orthonormal basis of the dual $\mc{H}^*$ of $\mc{H}$ and $(e^+_n)_{n\in \N}$ is and orthonormal basis of the dual $(\mc{H}^+)^*$ of $\mc{H}^+$. 

By applying \cite[Lemma 7.2]{GKRV21_Segal} as explained in \cite[Section 7.2]{GKRV21_Segal}, this trace is finite in the case when no edge of the form ${\rm e}_i=((j,k),(j,k'))$ or  ${\rm e}^+_i=((j,k),(j,k'))$, for some $j,k,k'$, is involved. Such edges are called loop edges and, in the context of Liouville CFT,  are associated to
gluing of two circles or half-circles in the same geometric building block $\mc{P}_j$ (we call it self-gluing). When there are loop edges, the trace is also finite provided the element 
$W_j\in \mc{H}^{\otimes b_j^\ell}\otimes \mc{H}^{\otimes b_j^{\rm h}}$ in $W$ associated to $\mc{P}_j$, and for which one has a loop edge, has 
the following extra factorisation property:
writing $W_j(\tilde{\bs{\varphi}}^\ell,\tilde{\bs{\varphi}}^{\rm h})$ with $\tilde{\bs{\varphi}}^\ell=(\tilde{\varphi}_1^\ell,\dots,\tilde{\varphi}_{b_j^\ell}^\ell)\in H^{-s}(\T)^{b^\ell_j}$ and $\tilde{\bs{\varphi}}^{\rm h}=(\tilde{\varphi}_{1}^{\rm h},\dots,\tilde{\varphi}_{b_j^{\rm h}}^{\rm h})\in H^{-s}_{\rm even}(\T)^{b^{\rm h}_j}$, and if ${\rm e}_i=((j,k),(j,k'))$ (resp ${\rm e}^+_i=((j,k+b_j^\ell),(j,k'+b_j^\ell))$) is the loop edge associated to a gluing of two circles (resp. half-circles) in $\mc{P}_j$, one assumes that 
\begin{align*}
&W_j(\tilde{\bs{\varphi}}^\ell,\tilde{\bs{\varphi}}^{\rm h})=
\int_{H^{-s}(\T)} W'_j(\tilde{\varphi}_{1}^\ell,\dots,\underbrace{\tilde{\varphi}_{k}^\ell}_{k},\dots,\underbrace{\phi}_{k'},\dots,\tilde{\varphi}_{b_j^\ell}^\ell,\tilde{\bs{\varphi}}^{\rm h})W''_j(\phi,\tilde{\varphi}_{k'}^\ell)d\mu_0(\phi)\\
\Big(\textrm{resp. }\quad &  W_j(\tilde{\bs{\varphi}}^\ell,\tilde{\bs{\varphi}}^{\rm h})=
\int_{H^{-s}_{\rm even}(\T)} W'_j(\tilde{\bs{\varphi}}^\ell,\tilde{\varphi}_{1}^{\rm h},\dots,\underbrace{\tilde{\varphi}_{k}^{\rm h}}_{k},\dots,\underbrace{\phi}_{k'},\dots,\tilde{\varphi}_{b_j^{\rm h}}^{\rm h})W''_j(\phi,\tilde{\varphi}_{k'}^{\rm h})d\mu_0^+(\phi) \Big)
\end{align*}
for some $W'_j\in L^2(\mu_0^{\otimes b_j^{\ell}}\otimes (\mu_0^+)^{\otimes b_j^{\rm h}}),W''_j\in L^2(\mu_0^{\otimes 2})$ 
(resp. $W'_j\in L^2(\mu_0^{\otimes b_j^{\ell}}\otimes (\mu_0^+)^{\otimes b_j^{\rm h}})$, $W''_j\in L^2((\mu^+_0)^{\otimes 2})$). 
The argument for this self-gluing is explained in  \cite[Remark 7.5]{GKRV21_Segal}: the factorisation above reduces to the case with no self-gluing.  

Next, we apply this $\mc{G}$-trace with $W_j:=\mc{A}_{\mc{P}_j,g,{\bf x}_j,\bs{\alpha}_j,{\bf s}_j,\bs{\beta}_j,\bs{\zeta}_j}$,  
the Segal amplitudes associated with $\mc{P}_j$, where the metric $g$ is assumed to be admissible on each building block $\mc{P}_j$ and the weights satisfy the Seiberg bounds so that each $W_j$ is $L^2$ (i.e. Hilbert-Schmidt). 

To handle the self-gluing of circles, we invoke \cite[Lemma 7.6]{GKRV21_Segal}, which establishes the additional factorisation discussed   above. The idea is to decompose $$\mc{P}_j=\mc{P}_j'\# \A_\delta$$ for some annulus $\A_\delta$ ($\delta<1$) bounding the circle $\pl_k\mc{P}_j$ to be glued, and then apply once again  Segal's gluing formula. The amplitude of $\A_\delta$ is not in $L^2(H^{-s}(\T)^2,\mu_0^{\otimes 2})$  due to the $0$-mode, but one can insert an additional weight $e^{\eps c_-}$    (with $\eps>0$)   to make it $L^2$. This modification amounts to   multiplying the amplitude of $\mc{P}_j'$ by $e^{-\eps c_-}$,  without spoiling its Hilbert-Schmidt property if $\eps>0$ is sufficiently small.

 In the case of self-gluing along a half-circle, the same argument applies  after cutting a half-annulus $\A_\delta^+$ to $\mc{P}_j$. The amplitude of $\mc{P}_j'=\mc{P}_j\setminus \A_\delta^+$ then exhibits sufficient  decay in the zero mode 
so that   multiplication by $e^{-\eps c_-}$  preserves the  Hilbert-Schmidt property, by \cite[Proposition 5.12]{GRW1}. We do not elaborate further on this point, as it is essentially identical to the argument in \cite[Lemma 7.6]{GKRV21_Segal}.

Applying the Segal gluing results established in  \cite[Proposition 6.1, 6.2 and 6.3]{GRW1}, we obtain:
\begin{proposition}\label{corel_into_amplitudes}
Let $(\Sigma,g)$ be decomposed into geometric buiding blocks $\mc{P}_j
$ as above in terms of a graph $\mc{G}$, and assume the cutting is non-degenerate and $g$ is an admissible metric on each $\mc{P}_j$. 
Consider some weights $\bs{\alpha}_j=(\alpha_{j1},\dots,\alpha_{jm_j})$, $\bs{\beta}_j=(\beta_{j1},\dots,\beta_{jm_j^{\rm b}})$ associated to the marked points ${\bf x}_j,{\bf s}_j$ as above, and satisfying the Seiberg bounds
\begin{equation}\label{seibergbord}
 \forall j,i, \quad \alpha_{ji}<Q, \quad \beta_{ji}<Q, \quad \sum_{i}\alpha_{ji}+\sum_i\frac{\beta_{ji}}{2}-\frac{Q}{2}\chi(\mc{P}_j^{\#2})>0 
\end{equation}
so that the amplitudes of $(\mc{P}_j,g,{\bf x},\bs{\alpha}_j,{\bf s}_j,\bs{\beta}_j,\bs{\zeta}_j)$ are Hilbert-Schmidt. Then we have 
\begin{equation}\label{corel_TrG} 
\cjg \prod_{i=1}^m V_{\alpha_i}(x_i)\prod_{i=1}^{m^{\rm b}}V_{\beta_i}(s_i)\cjd_{\Sigma,g}=C_{\rm gluing}{\rm Tr}^{\mc{G}}_{\mc{H},\mc{H}^+}\big(\bigotimes_{j=1}^{N+N^+}\mc{A}_{\mc{P}_j,g,{\bf x}_j,\bs{\alpha}_j,{\bf s}_j,\bs{\beta}_j,\bs{\zeta}_j}\big)
\end{equation}
where the gluing constant is 
\begin{equation}\label{Cgluing}
C_{\rm gluing}=2^{-\frac{L}{2}-\frac{3}{4}L^++1}\pi^{-L-\frac{3}{4}L^++1}
\end{equation}
where $L$ is the number of cutting circles of $\Sigma$ and $L^+$ the number of half-circles.
 \end{proposition}

We can define more generally the $\mc{G}$-trace for elements $\hat{W}\in  \otimes_{j=1}^{N+N^+}(\mc{H}_1^{\otimes b_j^\ell}\otimes \mc{H}_2^{\otimes b_j^{\rm h}})$ for $\mc{H}_1,\mc{H}_2$ two separable Hilbert spaces. Because of the spectral resolutions \eqref{resolution_bulk_Hamiltonian} and \eqref{resolution_boundary_Hamiltonian}, it is natural to consider those Hilbert spaces to be given by an $L^2$ space over the space of functions on Young diagrams. More precisely,  the set $\mc{T}$ of Young diagrams  is   countable and can be partitioned as 
\begin{equation}\label{decofT} 
\mc{T}= \bigcup_{n=0}^\infty \mc{T}_n,\quad \mc{T}_n:=\{\nu \in \mc{T}\,| \, |\nu|=n\} \textrm{ if }n>0, \quad \mc{T}_0=\{0\}.
\end{equation}
Set $d_{n}:=|\mc{T}_n|$ its cardinality, we let $\cjg\cdot,\cdot\cjd_{d_n}$ be the canonical Hermitian product on $\C^{d_n}$ and we define 
\begin{equation}\label{defofHT} 
H_\mc{T}:=\bigoplus_{n=0}^\infty \C^{d_n} \textrm{ with Hermitian product }\cjg v,v'\cjd_{\mc{T}}:= \sum_{n=0}^\infty \cjg v_n,v'_n\cjd_{d_n}
\end{equation}
where $v=\sum_n v_n$ with $v_n\in \C^{d_n}$, and $v'=\sum_n v'_n$. 
The Hilbert space 
$(H_\mc{T},\cjg \cdot,\cdot\cjd_{\mc{T}})$ is isomorphic to $\ell^2(\mc{T})$, i.e. the space $L^2(\mc{T},\mu_{\mc{T}})$ where the measure $\mu_{\mc{T}}$ is the counting measure on the set $\mc{T}$, namely
\[  \int_{\mc{T}} f(\nu)\dd\mu_{\mc{T}}= \sum_{\nu\in \mc{T}}f(\nu), \quad    \forall f:\mc{T}\to \R^+.\]

Next, in the trace \eqref{corel_TrG}, we can apply the spectral decomposition \eqref{resolution_bulk_Hamiltonian} of the (bulk) Liouville Hamiltonian $\bH$ for each $\mc{H}$ pairing associated to a glued circle $\mc{C}_i$ (and thus an edge ${\rm e}_i$) and the spectral decomposition \eqref{resolution_boundary_Hamiltonian} of the boundary Liouville Hamiltonian $\bH^+$ for each $\mc{H}^+$ pairing associated to a glued  half-circle $\mc{C}^+_i$ (and thus an edge ${\rm e}^+_i$). This directly produces the formula 
\begin{equation}\label{corel_segal}
\cjg \prod_{i=1}^m V_{\alpha_i}(x_i)\prod_{i=1}^{m^{\rm b}}V_{\beta_i}(s_i)\cjd_{\Sigma,g}=\frac{C_{\rm gluing}}{(2\pi)^L(4\pi)^{L^+}}
\int_{\R_+^L}\int_{\R^{L^+}}{\rm Tr}^{\mc{G}}_{L^2(\mc{T}^2),L^2(\mc{T})} \big(\bigotimes_{j=1}^{N+N^+}\hat{\mc{A}}_{\mc{P}_j,g,{\bf x}_j,\bs{\alpha}_j,{\bf s}_j,\bs{\beta}_j,\bs{\zeta}_j}({\bf p}_j)\big)\dd{\bf p}\dd{\bf p}^+
\end{equation}
where ${\bf p}=(p_1,\dots,p_L)\in\R_+^L$, ${\bf p}^+=(p^+_1,\dots,p^+_{L^+})\in\R_+^{L^+}$  are linked to ${\bf p}_j=(p_{j1},\dots,p_{jb_j})$, with $b_j=b_j^{\ell}+b_j^{\rm h}$, by the relation 
\[p_i=p_{jk}=p_{j'k'}, \textrm{ if }{\rm e}_i=((j,k),(j',k')), \quad p^+_i=p_{jk}=p_{j'k'}, \textrm{ if }{\rm e}^+_i=((j,k),(j',k'))\] 
and, for $\bs{\nu}_j=(\nu_{j1},\dots,\nu_{jb^\ell_j})\in \mc{T}^{b_j^\ell}$,  $\tilde{\bs{\nu}}_j=(\tilde{\nu}_{j1},\dots,\tilde{\nu}_{jb^\ell_j})\in 
\mc{T}^{b_j^\ell}$ and $\bs{\nu}^+_j=(\nu^+_{j1},\dots,\nu^+_{jb^{\rm h}_j})\in \mc{T}^{b_j^{\rm h}}$,
\begin{align*}
&\hat{\mc{A}}_{\mc{P}_j,g,{\bf x}_j,\bs{\alpha}_j,{\bf s}_j,\bs{\beta}_j,\bs{\zeta}_j}({\bf p}_j;\bs{\nu}_j,\tilde{\bs{\nu}}_j,\bs{\nu}_j^+)\\
&:=
\big\cjg \mc{A}_{\mc{P}_j,g,{\bf x}_j,\bs{\alpha}_j,{\bf s}_j,\bs{\beta}_j,\bs{\zeta}_j},\bigotimes_{k=1}^{b_j^\ell} 
{\bf C}^{\frac{\sigma_{jk}+1}{2}} H_{Q+ip_{jk},\nu_{jk},\tilde{\nu}_{jk}}\bigotimes \bigotimes_{k=1}^{b_j^{\rm h}}{\bf C}^{\frac{\sigma_{jk}+1}{2}}H^+_{Q+ip_{jk},\nu^+_{jk}}\big\cjd_{\mc{H}^{\otimes b_j^\ell}\otimes 
(\mc{H}^+)^{\otimes b_j^{\rm h}}}
\end{align*}
where $H_{Q+ip,\nu,\tilde{\nu}}$ and $H^+_{Q+ip,\nu}$ are the following orthonormal basis elements of $\mc{H}$ and $\mc{H}^+$
\[H_{Q+ip,\nu,\tilde{\nu}}: =  \sum_{|\nu'|=|\nu|, |\tilde{\nu}'|= |\tilde{\nu}|}  
F_{Q+ip}^{-1/2}(\nu, \nu')F_{Q+ip}^{-1/2}(\tilde{\nu}, \tilde{\nu}') 
\Psi_{Q+ip,\nu',\tilde{\nu}'}, \quad H^+_{Q+ip,\nu}:=  \sum_{|\nu'|=|\nu|}  
F_{Q+ip}^{-1/2}(\nu, \nu')
\Psi^+_{Q+ip,\nu'}.\]
Here $F_{Q+ip}$ is the Schapovalov quadratic form introduced in \eqref{scapo}.

\subsection{Conformal blocks and bootstrap}\label{CBB}
 For
 ${\bf n}_j=(n_{j1},\dots,n_{jb_j^\#})\in\N^{b_j^\#}$, with $b_j^\#=2b_j^{\ell}+b_j^{\rm h}$, we define ${\bf w}_{{\bf n}_j}^{\mc{P}_j}(\boldsymbol{\alpha}_j,\boldsymbol{\beta}_j,{\bf p}_j)\in \bigotimes_{k=1}^{b_j^\#}\C^{d_{n_{jk}}}$  to be the vector with components 
\begin{equation}\label{defwnj} 
{\bf w}_{{\bf n}_j}^{\mc{P}_j}(\boldsymbol{\alpha}_j,\boldsymbol{\beta}_j,{\bf p}_j):=
\big(\omega(((\mc{P}_j)_{\rm hole}^{\#2})^{\rm in},\bs{\Delta}_j ,\bs{\nu}_j)\big)_{\boldsymbol{\nu}_j\in \mc{T}_{{\bf n}_j}}
\end{equation} 
where $\mc{T}_{{\bf n}_j}:=\mc{T}_{n_{j1}}\times\dots\times \mc{T}_{n_{jb_j^\#}}$,  $\mc{T}_{n}$ is the set of Young diagrams $\nu$ of size $|\nu|=n$,  and  $d_{n}:=|\mc{T}_n|$.
Next, recall from \cite[Section 6.3]{GKRV21_Segal} that for $n\in\N$ and $p\in\R_+$,  
the Schapovalov form ${\bf F}_{Q+ip,n}$ is a positive definite quadratic form on $\C^{d_n}$, with matrix coefficients denoted by ${\bf F}_{Q+ip,n}:=(F_{Q+ip}(\nu,\nu'))_{\nu,\nu'\in\mc{T}_n}$. Denote by $\bs\sigma_j^\#$ the orientations of the boundary parametrisations of the building block $(\mc{P}_j)_{\rm hole}^{\#2}$ and define
\begin{equation}\label{defBtilde}
\tilde{\mc{B}}_{\mc{P}_j}(\boldsymbol{\alpha}_j,\boldsymbol{\beta}_j,{\bf p}_j)=\sum_{{\bf n}_j\in\N^{b_j^\#}} \Big(\bigotimes_{k=1}^{b_j}{\bf C}^{(\sigma_{jk}^\#+1)/2}\bbar{{\bf F}^{-1/2}_{Q+ip_{jk},n_{jk}}}\Big){\bf w}_{{\bf n}_j}^{\mc{P}_j}(\boldsymbol{\alpha}_j,\boldsymbol{\beta}_j,{\bf p}_j) .
\end{equation}

The normalised conformal block of $(\Sigma^{\#^2},g,{\bf x}^\#,\boldsymbol{\alpha}^\#,\boldsymbol{\zeta}^\#)$
is defined for almost all ${\bf p}^\#:=({\bf p},{\bf p}',{\bf p}^+)\in \R_+^{L}\times \R_+^{L}\times \R_+^{L^+}$ as  
\begin{equation}
\tilde{\mc{F}}^{\mc{G}^\#}_{\bf p^\#}(\boldsymbol{\alpha}^\#):= {\rm Tr}^{\mc{G}^\#}_{L^2(\mc{T}^2),L^2(\mc{T})} \big(\bigotimes_{j=1}^{2N+N^+} 
\tilde{\mc{B}}_{\mc{P}_j}(\boldsymbol{\alpha}_j,\boldsymbol{\beta}_j,{\bf p}_j) \big) .
\end{equation}
The almost sure convergence of this series  in ${\bf p}^\#$ is proved in \cite[Section 8]{GKRV21_Segal}. It corresponds to the conformal block of the Neumann double $\Sigma^{\#2}$. By Corollary \ref{corofactor}, the formula \eqref{corel_segal} involves this conformal block evaluated on the diagonal ${\bf p}={\bf p}'$ , while establishing the almost sure convergence of $\tilde{\mc{F}}_{\bf p,\bf p,\bf p^+}(\boldsymbol{\alpha}^\#)$. 

We can now state the main result of this manuscript

\begin{theorem}[\textbf{Conformal bootstrap}]\label{th:fullexpressioncorrel} 
Assume \eqref{seibergbord} and $g$ is admissible with respect to the maximal system of cutting curves.

The correlation function of the   surface  $(\Sigma,g)$ with $m$ 
interior insertions $({\bf x},\boldsymbol{\alpha})$ and $m^{\rm b}$ boundary insertions $({\bf s},\boldsymbol{\beta})$ is given by the expression
\begin{equation}\label{theoremmodifiedconfblock2}
\cjg \prod_{i=1}^m V_{\alpha_i}(x_i)\prod_{i=1}^{m^{\rm b}}V_{\beta_i}(s_i)\cjd_{\Sigma,g} =\frac{C_{\rm gluing}}{(2\pi)^L(4\pi)^{L^+}}\int_{\R_+^L\times \R_+^{L^+}}C_{\Sigma,g}(\boldsymbol{\alpha},\boldsymbol{\beta},{\bf p}) \boldsymbol{\rho}(\boldsymbol{\alpha},\boldsymbol{\beta},{\bf p},{\bf p}^+)
\tilde{\mc{F}}^{\mc{G}^\#}_{\bf p,\bf p,\bf p^+}(\boldsymbol{\alpha}^\#)\dd{\bf p}\dd{\bf p}^+,
\end{equation}
with $C_{\rm gluing}$ given by \eqref{Cgluing}, and
\begin{align*}
 \boldsymbol{\rho}(\boldsymbol{\alpha},\boldsymbol{\beta},{\bf p},{\bf p}^+) :=  \prod_{j=1}^{N+N^+}{\bf C}^{\mc{P}_j}_{\gamma,\mu,\mu^{\rm b}_j}({\boldsymbol{\alpha}_j},\bs{\beta}_j,{\bf p}_j )|_{p_{{\rm v}_1(e_i)}=p_{{\rm v}_2(e_i)}=p_i}, 
 \\
  C_{\Sigma,g}(\boldsymbol{\alpha},\boldsymbol{\beta},{\bf p},{\bf p}^+):=\prod_{j=1}^{N+N^+}
C_{\mc{P}_j,g}(\boldsymbol{\alpha}_j,\bs{\beta}_j, {\bf p}_j)
 |_{p_{{\rm v}_1(e_i)}=p_{{\rm v}_2(e_i)}=p_i}
 \end{align*}
where $ \boldsymbol{\rho}(\boldsymbol{\alpha},\boldsymbol{\beta},{\bf p},{\bf p}^+)$ and $  C_{\Sigma,g}(\boldsymbol{\alpha},\boldsymbol{\beta},{\bf p},{\bf p}^+)$ are respectively the product of structure constants and metric constants of the building blocks associated to the graph $\mc{G}$ defined in Subsection \ref{constantpj}.  The conformal block  $\tilde{\mc{F}}^{\mc{G}^\#}_{\bf p,\bf p,\bf p^+}(\boldsymbol{\alpha}^\#)$ is almost surely  finite in ${\bf p},{\bf p}^+$.
\end{theorem}


 \section{Special cases}\label{special}
In this section we concentrate on two key examples: the disk with four boundary insertions and the annulus with a single boundary insertion. We allow the boundary fields to be arbitrary Virasoro operators, including both primaries and their descendants. The bootstrap relations for Liouville theory then imply that the conformal blocks corresponding to different channel decompositions (or dual graphs) are connected by integral transforms. The corresponding kernels provide a natural continuum analogue of the Verlinde formula in Liouville CFT,  see~\cite{PostTsiares2025} for physics discussion.
  
\subsection{Crossing symmetry for 4-holes disk amplitudes}
 
Consider $\xi_1,\dots,\xi_4:\T\to \C$ which are analytic parametrisations of curves in $\C$ such that $\bbar{\xi_k(e^{i\theta})}=\xi_k(e^{-i\theta})$ and $\xi_k(\T^+)\subset \H$, and assume that $\xi_k$ are positively oriented (in particular $\xi_k(-1)<\xi_k(1)$ in $\R$). Let $\mc{D}_1,\dots,\mc{D}_4\subset \C$ be the compact domains in $\C$ such that $\pl \mc{D}_k=\xi_k(\T)$, and let $\mc{D}_k^+=\mc{D}_k\cap \H$. We assume that the domains $\mc{D}_k$ are pairwise disjoint and ordered so that $\xi_k(1)<\xi_{k+1}(-1)$ for $k=1,2,3$.
Define $\Sigma:=\hat{\H}\setminus \bigcup_{k=1}^4(\mc{D}_k^+)^\circ$, with 4 D-marked boundary components and 4 N-marked boundary components, where the D-marked components being parametrised by $\bs{\xi}=(\xi_1,\dots,\xi_4)$.  It is equipped with the canonical complex strutcure $J_\C$ on $\hat{\C}$.
We call $\Sigma^{\#2}$ its Neumann double, with involution $z\mapsto \bar{z}$.
Let $\xi:\T\to \Sigma^{\#2}$ be another analytic curve such that $\bbar{\xi(e^{i\theta})}=\xi(e^{-i\theta})$ and $\xi(\T^+)\subset \Sigma$, positively oriented, with $\xi(-1),\xi(1)\in \R$ and such that the compact domain $\mc{D}\subset \C$ bounded by $\xi(\T)$ satisfies
 $\mc{D}_1\cup \mc{D}_2\subset \mc{D}^\circ$ and $\mc{D}\cap(\mc{D}_3\cup \mc{D}_4)=\emptyset$.
The curve $\xi(\T^+)$ separates $\Sigma$ into $\mc{P}_1\#\mc{P}_2$ with 
\[\mc{P}_1= \Sigma \setminus \mc{D}^\circ, \quad \mc{P}_2=\Sigma \cap \mc{D}.\]

\begin{figure}[H]
\centering
\begin{tikzpicture}[scale=0.9]
\fill[gray!10] (-6.5,0) rectangle (6.5,5.0);
\draw[very thick] (-6.5,0) -- (6.5,0);
\draw[very thick,red, fill=white] (-4.05,0) arc (180:0:0.65);
\draw[very thick,red, fill=white] (-1.95,0) arc (180:0:0.65);
\draw[very thick,red,fill=white] (1.95,0) arc (180:0:0.65);
\draw[very thick,red,fill=white] (4.05,0) arc (180:0:0.65);
\node[red!70!black] at (-3.50,0.95){$\xi_1(\mathbb T^+)$};
\node[red!70!black] at (-1.4,0.95){$\xi_2(\mathbb T^+)$};
\node[red!70!black] at (2.50,0.95){$\xi_3(\mathbb T^+)$};
\node[red!70!black] at (4.55,0.95){$\xi_4(\mathbb T^+)$};
\node[red!70!black] at (-3.45,0.25){$\mc{D}^+_1$};
\node[red!70!black] at (-1.30,0.25){$\mc{D}^+_2$};
\node[red!70!black] at (2.50,0.25){$\mc{D}^+_3$};
\node[red!70!black] at (4.70,0.25){$\mc{D}^+_4$};
\draw[very thick,green!50!black](-5.05,0) arc (180:0:2.80);
\fill[green!50!black] (-5.05,0) circle (1.4pt);
\fill[green!50!black] (0.55,0) circle (1.4pt);
\node[green!45!black] at (-2.25,3.10){$\xi(\mathbb T^+)$};
\node at (-2.3,2){$\mathcal P_2$};
\node at (3.7,3.75){$\mathcal P_1$};
\node at (-5.55,-0.35){$\mu_1^{\rm b}$};
\node at (-2.30,-0.35){$\mu_2^{\rm b}$};
\node at (1.05,-0.35){$\mu_3^{\rm b}$};
\node at (3.70,-0.35){$\mu_4^{\rm b}$};
\node[below] at (-6.5,0){$-\infty$};
\node[below] at (6.5,0){$+\infty$};
\node at (0,4){$\displaystyle\Sigma=\widehat{\mathbb H}\setminus\bigcup_{k=1}^4\mathcal D_k^+$};
\end{tikzpicture}
\caption{
The surface $\Sigma$ with its four D-marked boundary components
$\xi_k(\mathbb T^+)$. The cutting curve  separates
$\mathcal D_1^+\cup\mathcal D_2^+$ from
$\mathcal D_3^+\cup\mathcal D_4^+$, giving
$\Sigma=\mathcal P_1\#\mathcal P_2$. 
}
\label{pic:geometry-sigma1}
\end{figure}
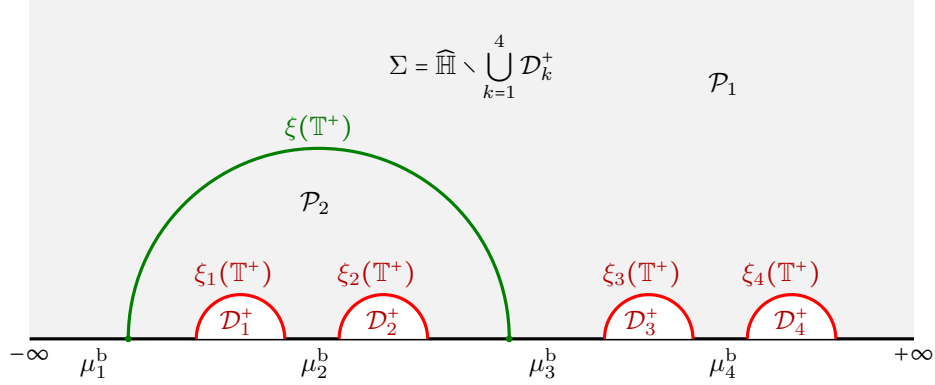
The surface $\mc{P}_1$ has D-marked boundary ($\pl_1\mc{P}_1)_{\rm D}=\xi(\T^+)$, $(\pl_2\mc{P}_1)_{\rm D}=\xi_3(\T^+)$ and $(\pl_3\mc{P}_1)_{\rm D}=\xi_4(\T^+)$, 
parametrised by $\bs{\xi}_1=(\xi_{11},\xi_{12},\xi_{13})=(\xi,\xi_3,\xi_4)$ which are incoming, and 
$\mc{P}_2$ has D-marked boundary $(\pl_1\mc{P}_2)_{\rm D}=\xi(\T^+)$, $(\pl_2\mc{P}_2)_{\rm D}=\xi_1(\T^+)$ and $(\pl_3\mc{P}_2)_{\rm D}=\xi_2(\T^+)$ parametrised by $\bs{\xi}_2=(\xi_{21},\xi_{22},\xi_{23})=(\xi,\xi_1,\xi_2)$ which are incoming except $\xi_{21}$ which is outgoing. All are equipped with the canonical complex structure $J_\C$ induced by $\hat{\C}$. The model form for $(\mc{P}_j,J_\C,\bs{\zeta}_j)$ are denoted 
\[  \bs{\psi}_j=(\psi_{j1},\psi_{j2},\psi_{j3})\]
where $\psi_{jk}\in {\rm BiHol}_\eps^+(\D)$, in the sense that there are biholomorphisms $\Phi_j: \mc{P}_j\to \hat{\H}\setminus \bigcup_{k=1}^3\psi_{jk}((\D^+)^\circ)$ such that $\Phi_j\circ \xi_{jk}=\psi_{jk}|_{\T^+}$ for $(j,k)\not=(2,1)$ and $\Phi_2\circ \xi_{21}=\psi_{21}\circ (-o)|_{\T^+}$.
These $\Phi_j$ extend to the double $\mc{P}_j^{\#2}\to \hat{\C}\setminus \bigcup_{k=1}^3\psi_{jk}(\D^\circ)$ and satisfy $\Phi_j(\bar{z})=\bbar{\Phi_j(z)}$.

 \begin{figure}[H] 
 \centering
\begin{tikzpicture}[scale=0.8]

 \draw [fill=lightgray ,draw=lightgray] (-3,-4) rectangle (3,-1);
 \draw[draw=black,very thick] (-3,-4) -- (3,-4);
    \draw[fill=white,draw=red,very thick] (-2,-4) circle (0.6) ;
    \draw[fill=white,draw=red,very thick] (0,-4) circle (0.5) ;
\draw[fill=white,draw=red,very thick] (2,-4) circle (0.7) ;
 \draw [fill=white,draw=white] (-3,-4.05) rectangle (3,-5);
 \draw (0.8,-2) node[above]{$\hat{\H}\setminus \bigcup_{k=1}^3\psi_{1k}(\D^+)$} ; 
   \draw (-2,-4) node[below]{$-1$} ; 
      \draw (0,-4) node[below]{$0$} ;
         \draw (2,-4) node[below]{$1$} ;  
 \draw (-2,-4) node{$\times$}; 
  \draw (0,-4) node{$\times$}; 
    \draw (2,-4) node{$\times$}; 
 
 \draw [fill=lightgray ,draw=lightgray] (6,-4) rectangle (12,-1);
  \draw[draw=black,very thick] (6,-4) -- (12,-4);
   \draw[fill=white,draw=blue,very thick] (7,-4) circle (0.6) ;
    \draw[fill=white,draw=red,very thick] (9,-4) circle (0.8) ;
\draw[fill=white,draw=red,very thick] (11,-4) circle (0.5) ;
 \draw [fill=white,draw=white] (6,-4.05) rectangle (12,-5);
  \draw (9.8,-2) node[above]{$\hat{\H}\setminus \bigcup_{k=1}^3\psi_{2k}(\D^+)$} ; 
   \draw (7,-4) node[below]{$-1$} ; 
      \draw (9,-4) node[below]{$0$} ;
         \draw (11,-4) node[below]{$1$} ;  
 \draw (7,-4) node{$\times$}; 
  \draw (9,-4) node{$\times$}; 
    \draw (11,-4) node{$\times$}; 
    \end{tikzpicture}
\caption{Model form of building blocks $\mc{P}_1$ and $\mc{P}_2$. Incoming boundary are red, Outgoing boundaries are blue, Neumann boundaries are black.}
\label{pic:disk4pt}
\end{figure}
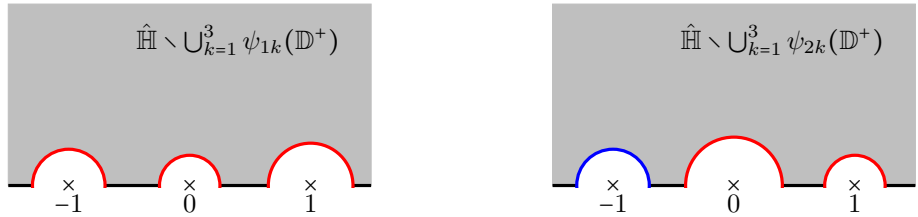  

The model form of the double $(\mathcal P_1^{\#2},J_{\mathbb C},\boldsymbol\xi_1^\#)$ is $(\psi_{11},\psi_{12},\psi_{13})$.
For the second building block, we have $\Phi_2\circ\xi_{21}=\psi_{21}\circ(-o)$.
Thus, after replacing its outgoing parametrisation $\xi_{21}$ by $\xi_{21}\circ o$, the all-incoming double
$(\mathcal P_2^{\#2})^{\rm in}$ has model form $(\psi_{21}\circ R_\pi,\psi_{22},\psi_{23})$, where $R_\pi(z)=-z$.

Assume $\bs\beta=(\beta_{jk})_{j=1,2;k=2,3}$, with   $\beta_{jk}\in (-\infty,Q)\cup (Q+{\rm i}\R)$, obeys
\[{\rm Re}(\beta_{12}+\beta_{13})>Q,\qquad {\rm Re}(\beta_{22}+\beta_{23})>Q,\]
and $\bs\nu=(\nu_{jk})_{j=1,2;k=2,3}$, with $\nu_{jk}=\emptyset$ if $\beta_{jk}\in\R$.
Consider the coefficients 
\[ \omega(\mc{P}_1^{\#2},\bs{\Delta}_1,\bs{\nu}_1) ,\quad   \omega((\mc{P}_2^{\#2})^{\rm in},\bs{\Delta}_2,\bs{\nu}_2)\]
obtained in \eqref{thm: ward identity} where the $(\mc{P}_2^{\#2})^{\rm in}$ has model form given by the biholomorphisms $(\psi_{j1}\circ R_\pi,\psi_{j2},\psi_{j3})$, and where $\bs{\nu}_1=(\nu,\nu_{3},\nu_{4})$, $\bs{\nu}_2=(\nu',\nu_1,\nu_2)$, $\bs\Delta_1=(\Delta_{Q+{\rm i}p},\Delta_{\beta_3},\Delta_{\beta_{4}})$, $\bs\Delta_2=(\Delta_{Q+{\rm i}p},\Delta_{\beta_{1}},\Delta_{\beta_{2}})$.  The normalised conformal block associated to $\Sigma^{\#2}$ cut into $\mc{P}_1^{\#2}\#\mc{P}_2^{\#2}$ (whose gluing graph is denoted $\mc{G}$) is defined by 
\begin{equation}\label{defConfBlock1}
\tilde{\mc{F}}_{ p}^{\mc{G}}(\bs{\beta},\bs{\nu}):= \sum_{n\geq 0}  \sum_{|\nu|=|\nu'|=n}\omega(\mc{P}_1^{\#2},\bs{\Delta}_1,\bs{\nu}_1)\omega((\mc{P}_2^{\#2})^{\rm in},\bs{\Delta}_2 ,\bs{\nu}_2){\bf F}_{Q+{\rm i}p,n}^{-1}(\nu,\nu') 
\end{equation}

On the ${\rm N}$-marked boundary $\pl\Sigma_{\rm N}=\pl \Sigma\cap (\R\cup \infty)$, we assign a  function $\mu^{\rm b}\geq 0$, constant over each maximal interval contained in $\pl\Sigma_{\rm N}$, equal to $\mu^{\rm b}_1$ on $(-\infty,\xi_1(-1))\cup (\xi_4(1),\infty)$, $\mu^{\rm b}_2$ on $(\xi_1(1),\xi_2(-1))$, $\mu^{\rm b}_3$ on $(\xi_2(1),\xi_3(-1)$ and $\mu^{\rm b}_4$ on $(\xi_3(1),\xi_4(-1))$.  In what follows, we will attach parameters to the half-disks determined by $\psi_{jk}^{\rm h}$: these parameters will be indexed similarly with the indices $j,k$ to identify clearly to which half-disk they are assigned to.
 
For each $k$, let $U_k$ be the holomorphy domain in $\alpha$, described in Lemma \ref{lem:boundary_primary_meromorphic}, of the primary
boundary state $\Psi_{\alpha}^+$ of the Hamiltonian ${\bf H}^{+}={\bf H}^{+,0}+{\bf V}^{+,\bs{\mu}}$ with 
 ${\bf \mu}=(\mu=0,\mu^{\rm L}=\mu_k^{\rm b},\mu^{\rm R}=\mu_{k+1}^{\rm b})$ adjacent to
the $k$-th boundary component (with the convention $\mu_5^{\rm b}:=\mu_1^{\rm b}$), and set
\begin{equation}\label{defU}
U:=\bigcap_{k=1}^4 U_k.
\end{equation}

\begin{proposition}[\textbf{First channel bootstrap}]\label{propcut1}
Consider the setting above, $\gamma\in (0,2)$ and assume that the bulk cosmological constant vanishes $\mu=0$, and that $\sum_{k=1}^4\mu^{\rm b}_k>0$. Assume that $g$ is an admissible metric on $\Sigma$, which is admissible with respect to the cut $\xi$. 
Assume $\bs\beta=(\beta_{k})_{k=1,\dots,4}\in U^4$, satisfies 
\begin{equation}\label{Seiberg1}
|{\rm Re}(\beta_1)-Q|+|{\rm Re}(\beta_2)-Q|<Q, \quad |{\rm Re}(\beta_3)-Q|+|{\rm Re}(\beta_4)-Q|<Q
\end{equation}
and set $(\beta_{12},\beta_{13},\beta_{22},\beta_{23})=(\beta_3,\beta_4,\beta_1,\beta_2)$. For $\bs\nu=(\nu_1,\dots,\nu_4)\in \mc{T}^4$, set $(\nu_{12},\nu_{13},\nu_{22},\nu_{23}):=(\nu_3,\nu_4,\nu_1,\nu_2)$ and as before, $\nu_k=\emptyset$ when $\beta_k\in\R$. The amplitudes of  $(\Sigma,g)$  evaluated on the descendant states are given by
\begin{equation}\label{fusionblock1}
\begin{split}
\big\cjg  \mc{A}_{\Sigma,g,\bs{\xi}} , \bigotimes_{k=1}^4 {\bf C}\Psi^+_{\beta_k,\nu_k}
\big\cjd_{(\mc H^+)^{\otimes4}}
=&\frac{(2\pi)^{-\frac{3}{4}}}{ 4\pi} \int_{I} C_{\Sigma,g}(\bs\beta,p)
\widetilde{\rm H}_{\mu^b_{1},\mu^b_{3},\mu^b_{4}}^{Q+{\rm i}p,\beta_{3},\beta_{4}}\widetilde{\rm H}_{\mu^b_{3},\mu^b_{1},\mu^b_{2}}^{Q-{\rm i}p,\beta_{1},\beta_{2}} 
\tilde{\mc{F}}_{p}^{\mc{G}}(\boldsymbol{\beta},\bs\nu) \dd p,\\
=&\frac{(2\pi)^{-\frac{3}{4}}}{ 4\pi} \int_{I} C_{\Sigma,g}(\bs\beta,p)
\widetilde{\rm H}_{\mu^b_{1},\mu^b_{3},\mu^b_{4}}^{Q-{\rm i}p,\beta_{3},\beta_{4}}\widetilde{\rm H}_{\mu^b_{3},\mu^b_{1},\mu^b_{2}}^{Q+{\rm i}p,\beta_{1},\beta_{2}} 
\tilde{\mc{F}}_{p}^{\mc{G}}(\boldsymbol{\beta},\bs\nu) \dd p
\end{split}
\end{equation}
where $I=\R_+$ if $\mu_1^{\rm b}+\mu_3^{\rm b}>0$ and $I=\R$ if $\mu_1^{\rm b}+\mu_3^{\rm b}=0$.
The metric dependent constant is given by 
\[
C_{\Sigma,g}(\bs\beta,p):=C^{{\rm (A2)},3}_{\mc{P}_1,g}( Q+{\rm i}p,\beta_{3},\beta_{4})C^{{\rm (A2)},3}_{\mc{P}_2,g}( Q+{\rm i}p, \beta_{1},\beta_{2}),
 \]
 where $C^{{\rm (A2)},3}_{\mc{P}_j,g}:=C^{{\rm (A2)},3}_{\Phi_j(\mc{P}_j),(\Phi_j)_*g}$ are the metric dependent constants defined in \eqref{metricconstantHblock}, and the conformal block  is given by \eqref{defConfBlock1}. The metric coefficient has the form
 \[C_{\Sigma,g}(\bs\beta,p)=C_0e^{c_{\rm L}S(g)}\big|\frac{1}{4}(\psi_{11})'(0)(\psi_{21})'(0)\big|^{\Delta_{Q+{\rm i}p}}\times \Big(\prod_{j=1}^22^{\Delta_{\beta_{j2}}-\Delta_{\beta_{j3}}}\prod_{k=2}^{3} |(\psi_{jk})'(0)|^{\Delta_{\beta_{jk}}}\Big) \]
where $C_0$ is a universal constant and 
\[\begin{split}
S(g)= & \sum_{j=1}^2S^0_{\rm L}( \caP_j,(\Phi_j)^*g_{\rm dozz},g) +\sum_{j=1}^2\sum_{k=1}^{3}S^0_{\rm L}(\psi_{jk}(\D^+),g_{\rm dozz},(\psi_{jk})_*g_\D))
\end{split} \]
\end{proposition}
\begin{proof} We start with the case $\beta\in Q+{\rm i}\R$. Then the proof proceeds exactly as in Theorem~\ref{th:fullexpressioncorrel}, based on the spectral resolution of Propososition \ref{resolution_bdry_Hamiltonian}, the Ward identity in Proposition \ref{ward case I} and 
the computation of Lemma \ref{structure_amp} in terms of structure constants and metric dependent terms.
The main fact is that  the proof of Theorem~\ref{th:fullexpressioncorrel} shows that the formula \eqref{fusionblock1} remains valid with
\[
\tilde{\mc{F}}_{ p}^{\mc{G}}(\boldsymbol{\beta},\bs\nu):={\rm Tr}^{\mc{G}}_{L^2(\mc{T})} \big(\bigotimes_{j=1}^{2} 
\tilde{\mc{B}}_{\mc{P}_j}( \boldsymbol{\beta}_j,p) \big).
 \]
 We also note that the spectral resolution can be used with both $\Psi^+_{Q+{\rm ip},\nu}$ or with $\Psi^+_{Q-{\rm ip},\nu}$, leading to the two identities of \eqref{fusionblock1}. To extend this result to  $\beta_k$ in $U$, we use the following estimate, which follows from Proposition \ref{resolution_bdry_Hamiltonian}  (see also Proposition \ref{holomorphic_eigenfunctions}):
for each $u, u'\in\mathcal{H}^+$, we have
\begin{equation}\label{bound_decomposition}
\int_{I}\sum_{n\geq 0}|\sum_{|\nu|=|\nu'|=n}\langle \Psi^+_{Q+ip,\nu},u\rangle_{\mc{H}^+}F^{-1}_{Q+ip,n}(\nu,\nu')\langle u',\Psi^+_{Q+ip,\nu'}\rangle_{\mc{H}^+}|\dd p\leq 4\pi ||u||_{\mathcal{H}^+}||u'||_{\mathcal{H}^+}.
\end{equation}
 Fix a family of Young diagrams $\boldsymbol\nu=(\bs{\nu}_1,\bs{\nu}_2)=(\nu_{12},\nu_{13},\nu_{22},\nu_{23})=(\nu_1,\nu_2,\nu_3,\nu_4)$ as above, and assume that $\bs{\beta}\in U^4$ satisfies \eqref{Seiberg1}
 and define the series (with positive coefficients)
 \[\tilde{\mc{F}}_{p}^{\mc{G},{\rm abs}}(\bs{\beta},\bs{\nu}):= \sum_{n\geq 0} \Big| \sum_{|\nu|=|\nu'|=n}\omega(\mc{P}_1^{\#2},\bs{\Delta}_1,\bs{\nu}_1)\omega((\mc{P}_2^{\#2})^{\rm in},\bs{\Delta}_2 ,\bs{\nu}_2)F_{Q+{\rm i}p}^{-1}(\nu,\nu')\Big|.\]
By \cite[Proposition 5.12]{GRW1}, and Theorem~\ref{thm:boundary_descendants_meromorphic}, we have 
\[
\begin{aligned}
u:=\mc A_{\mc P_1,g,\bs\xi_1}
(\,\cdot\,,\Psi^+_{\beta_3,\nu_3},
\Psi^+_{\beta_4,\nu_4})\in\mc H^+, \quad 
u’:=\mc A_{\mc P_2,g,\bs\xi_2}
(\,\cdot\,,\Psi^+_{\beta_1,\nu_1},
\Psi^+_{\beta_2,\nu_2})\in\mc H^+.
\end{aligned}
\]
are $\mathcal{H}^+$-valued holomorphic functions of $\bs{\beta}\in U^4$ in the region \eqref{Seiberg1}. The left hand side  of \eqref{fusionblock1} then extends holomorphically in $\bs{\beta}\in U^4$ in the region \eqref{Seiberg1}, its extension being given by $\cjg u',\bbar{u}\cjd_{\mc{H}^+}$.
The Ward identities for young diagrams $\nu$ of size $|\nu|=n$ gives that 
\begin{align*}
&\sum_{|\nu|=|\nu'|=n}\langle \Psi^+_{Q+ip,\nu},\bar{u}\rangle_{\mc{H}^+}F^{-1}_{Q+ip}(\nu,\nu')
\langle u',\Psi^+_{Q+ip,\nu'}\rangle_{\mc{H}^+}\\
&= C_{\Sigma,g}(\bs\beta,p)\widetilde{\rm H}_{\mu^b_{1},\mu^b_{3},\mu^b_{4}}^{Q+{\rm i}p,\beta_{3},\beta_{4}}\widetilde{\rm H}_{\mu^b_{3},\mu^b_{1},\mu^b_{2}}^{Q-{\rm i}p,\beta_{1},\beta_{2}} \sum_{|\nu|=|\nu'|=n}\omega(\mc{P}_1^{\#2},\bs{\Delta}_1,\bs{\nu}_1)\omega((\mc{P}_2^{\#2})^{\rm in},\bs{\Delta}_2 ,\bs{\nu}_2)F_{Q+{\rm i}p}^{-1}(\nu,\nu')
\end{align*}
Applying \eqref{bound_decomposition}, we deduce that 
\[\int_{I} |C_{\Sigma,g}(\bs\beta,p)
\widetilde{\rm H}_{\mu^b_{1},\mu^b_{3},\mu^b_{4}}^{Q+{\rm i}p,\beta_{3},\beta_{4}}\widetilde{\rm H}_{\mu^b_{3},\mu^b_{1},\mu^b_{2}}^{Q-{\rm i}p,\beta_{1},\beta_{2}}| 
\tilde{\mc{F}}_{p}^{\mc{G},{\rm abs}}(\boldsymbol{\beta},\bs\nu) \dd p\leq 4\pi \|u\|_{\mc{H}^+}\|u'\|_{\mc{H}^+}\]
This shows that the right hand side of \eqref{fusionblock1} is holomorphic in $\bs{\beta}\in U^4$ in the region \eqref{Seiberg1}, and proves the announced result.
\end{proof}

 Consider now another analytic parametrised curve $\xi^*:\T\to \Sigma^{\#2}$ which is symmetric, i.e. $\xi^*(e^{-{\rm i}\theta})=\bbar{\xi^*(e^{{\rm i}\theta})}$, with $\xi^*(\T^+)\subset \H$ and positively oriented (in particular $\xi^*(-1)<\xi^*(1)$ on $\R$), and which is
 not intesersecting $\xi_k(\T)$ for $k=1,\dots,4$. Let $\mc{D}^*$ be the compact domain bounded by $\xi^*(\T)$. Assume that 
 $\mc{D}_2\cup \mc{D}_3\subset (\mc{D}^*)^\circ$ and $\mc{D}^*\cap (\mc{D}_1\cup \mc{D}_4)=\emptyset$.
 The curve $\xi^*(\T^+)$ separate $\Sigma$ into $\mc{P}_1^*\#\mc{P}_2^*$ with 
\[\mc{P}^*_1= \Sigma \setminus (\mc{D}^*)^\circ, \quad \mc{P}^*_2=\Sigma \cap \mc{D}^*.\]
The surface $\mc{P}_1^*$ has D-marked boundary ($\pl_1\mc{P}_1^*)_{\rm D}=\xi^*(\T^+)$, $(\pl_2\mc{P}_1^*)_{\rm D}=\xi_4(\T^+)$ and $(\pl_3\mc{P}_1^*)_{\rm D}=\xi_1(\T^+)$, 
parametrised by $\bs{\xi}^*_1=(\xi^*_{11},\xi^*_{12},\xi^*_{13})=(\xi^*,\xi_4,\xi_1)$ which are incoming, and 
$\mc{P}^*_2$ has N-marked boundary $(\pl_1\mc{P}_2^*)_{\rm D}=\xi^*(\T^+)$, $(\pl_2\mc{P}_2^*)_{\rm D}=\xi_2(\T^+)$ and $(\pl_3\mc{P}^*_2)_{\rm D}=\xi_3(\T^+)$ parametrised by $\bs{\xi}_2^*=(\xi^*_{21},\xi^*_{22},\xi^*_{23})=(\xi^*,\xi_2,\xi_3)$ which are incoming except $\xi^*_{21}$ which is outgoing. All are equipped with the canonical complex structure $J_\C$ induced by $\hat{\C}$.

 The model form for $(\mc{P}^*_j,J_\C,\bs{\zeta}_j^*)$ are denoted 
\[ \bs{\psi}_j^*:= (\psi^*_{j1},\psi_{j2}^*,\psi_{j3}^*)\]
where $\psi_{jk}^*\in {\rm BiHol}_\eps^+(\D)$, in the sense that there are biholomorphisms $\Phi_j^*: \mc{P}_j^*\to \hat{\H}\setminus \cup_{k=1}^3\psi_{jk}^*((\D^+)^\circ)$ such that $\Phi^*_j\circ \xi^*_{jk}=\psi^*_{jk}|_{\T^+}$ for $(j,k)\not=(2,1)$ and $\Phi^*_2\circ \xi^*_{21}=\psi^*_{21}\circ (-o)|_{\T^+}$.
These $\Phi_j^*$ extend to the double $(\mc{P}_j^*)^{\#2}\to \hat{\C}\setminus \cup_{k=1}^3\psi_{jk}^*(\D^\circ)$ and satisfy $\Phi_j^*(\bar{z})=\bbar{\Phi_j^*(z)}$. The normalized conformal block  associated to $\Sigma^{\#2}$ cut into $(\mc{P}_1^*)^{\#2}\#(\mc{P}_2^*)^{\#2}$ (whose gluing graph is denoted $\mc{G}^*$) is defined by 
\begin{equation}\label{defConfBlock2}
\tilde{\mc{F}}_{p}^{\mc{G}^{*}}(\bs{\beta},\bs{\nu}):= \sum_{n\geq 0}  \sum_{|\nu|=|\nu'|=n}\omega((\mc{P}_1^*)^{\#2},\bs{\Delta}_1^*,\bs{\nu}^*_1)\omega(((\mc{P}_2^*)^{\#2})^{\rm in},\bs{\Delta}_2^* ,\bs{\nu}^*_2){\bf F}_{Q+{\rm i}p,n}^{-1}(\nu,\nu') 
\end{equation}
with $\bs{\Delta}_1^*:=(\Delta_{Q+ip},\Delta_{\beta_4},\Delta_{\beta_1})$ and $\bs{\Delta}_2^*:=(\Delta_{Q+ip},\Delta_{\beta_2},\Delta_{\beta_3})$, $\bs{\nu}_1^*=(\nu,\nu_4,\nu_1)$ and $\bs{\nu}_2^*=(\nu',\nu_2,\nu_3)$.
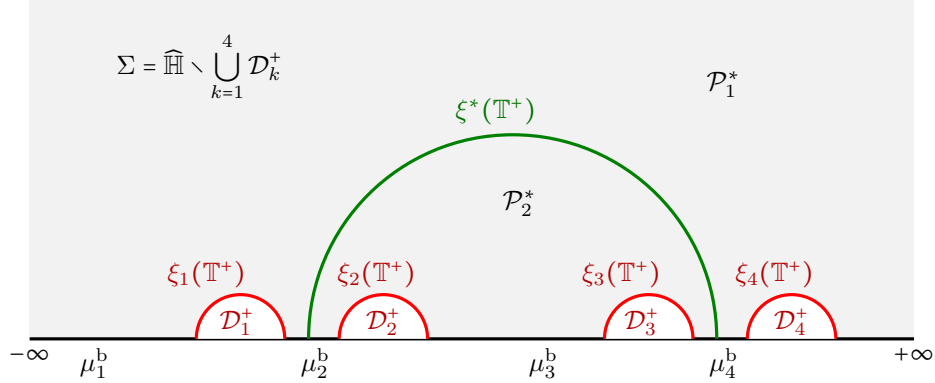
\begin{figure}[H]
\centering
\begin{tikzpicture}[scale=0.9]
\fill[gray!10] (-6.5,0) rectangle (6.5,5.0);
\draw[very thick] (-6.5,0) -- (6.5,0);
\draw[very thick,red, fill=white] (-4.05,0) arc (180:0:0.65);
\draw[very thick,red, fill=white] (-1.95,0) arc (180:0:0.65);
\draw[very thick,red,fill=white] (1.95,0) arc (180:0:0.65);
\draw[very thick,red,fill=white] (4.05,0) arc (180:0:0.65);
\node[red!70!black] at (-3.90,0.95){$\xi_1(\mathbb T^+)$};
\node[red!70!black] at (-1.40,0.95){$\xi_2(\mathbb T^+)$};
\node[red!70!black] at (2.20,0.95){$\xi_3(\mathbb T^+)$};
\node[red!70!black] at (4.45,0.95){$\xi_4(\mathbb T^+)$};
\node[red!70!black] at (-3.45,0.25){$\mc{D}^+_1$};
\node[red!70!black] at (-1.30,0.25){$\mc{D}^+_2$};
\node[red!70!black] at (2.50,0.25){$\mc{D}^+_3$};
\node[red!70!black] at (4.70,0.25){$\mc{D}^+_4$};
\draw[very thick,green!50!black] (-2.40,0) arc (180:0:3.00);
\node[green!45!black] at (0.35,3.35) {$\xi^*(\mathbb T^+)$};
\node at (0.7,2){$\mathcal P_2^*$};
\node at (3.7,3.75){$\mathcal P_1^*$};
\node at (-5.55,-0.35){$\mu_1^{\rm b}$};
\node at (-2.30,-0.35){$\mu_2^{\rm b}$};
\node at (1.05,-0.35){$\mu_3^{\rm b}$};
\node at (3.70,-0.35){$\mu_4^{\rm b}$};
\node[below] at (-6.5,0){$-\infty$};
\node[below] at (6.5,0){$+\infty$};
\node at (-4,4) {$\displaystyle \Sigma=\widehat{\mathbb H}
    \setminus\bigcup_{k=1}^4\mathcal D_k^+$};
\end{tikzpicture}

\caption{
The surface $\Sigma$ with its four D-marked boundary components
$\xi_k(\mathbb T^+)$. The curve $\xi^*(\mathbb T^+)$ surrounds
$\mathcal D_2^+\cup\mathcal D_3^+$.}
\label{pic:geometry-sigma2}
\end{figure}

\begin{proposition}[\textbf{Second channel bootstrap}]\label{propcut2}
Consider the setting above and assume that  $\gamma\in (0,2)$, $\mu=0$ and $\sum_{k=1}^4\mu_k^{\rm b}>0$, with in addition $\mu^{\rm b}_1+\mu^{\rm b}_4>0$.
Assume that $g^*$ is an admissible metric on $\Sigma$, which is admissible with respect to the cut $\xi^*$.
Assume $\bs\beta=(\beta_{k})_{k=1,\dots,4}\in U^4$, satisfies 
\begin{equation}\label{Seiberg2}
|{\rm Re}(\beta_1)-Q|+|{\rm Re}(\beta_4)-Q|<Q, \quad |{\rm Re}(\beta_2)-Q|+|{\rm Re}(\beta_3)-Q|<Q
\end{equation}
and define $(\beta^*_{12},\beta^*_{13},\beta^*_{22},\beta^*_{23})=(\beta_4,\beta_1,\beta_2,\beta_3)$
and $\bs\nu=(\nu_{jk})_{j=1,2;k=2,3}$, with $\nu_{jk}=\emptyset$ if $\beta^*_{jk}\in\R$.  Then, the amplitudes of  $(\Sigma,g^*)$  evaluated on the descendant states are given by
\begin{equation}\label{fusionblock2}
\begin{split}
\big\cjg  \mc{A}_{\Sigma,g^*,\bs{\xi}} ,   \bigotimes_{k=1}^4 {\bf C}\Psi^+_{\beta_k,\nu_k}\big\cjd_{ (\mc{H}^+)^{\otimes4}}   
=&\frac{(2\pi)^{-\frac{3}{4}}}{4\pi} \int_{I^*} C_{\Sigma,g^*}(\bs\beta,p)
\widetilde{\rm H}_{\mu^b_{2},\mu^b_{4},\mu^b_{1}}^{Q+{\rm i}p,\beta_{4},\beta_{1}}\widetilde{\rm H}_{\mu^b_{\otimes4},\mu^b_{2},\mu^b_{3}}^{Q-{\rm i}p,\beta_{2},\beta_{3}} 
\tilde{\mc{F}}_{p}^{\mc{G}^{*}}(\boldsymbol{\beta},\bs\nu) \dd p,\\
=& \frac{(2\pi)^{-\frac{3}{4}}}{ 4\pi} \int_{I^*} C_{\Sigma,g^*}(\bs\beta,p)
\widetilde{\rm H}_{\mu^b_{2},\mu^b_{4},\mu^b_{1}}^{Q-{\rm i}p,\beta_{4},\beta_{1}}\widetilde{\rm H}_{\mu^b_{\otimes4},\mu^b_{2},\mu^b_{3}}^{Q+{\rm i}p,\beta_{2},\beta_{3}} 
\tilde{\mc{F}}_{p}^{\mc{G}^{*}}(\boldsymbol{\beta},\bs\nu) \dd p
\end{split}
\end{equation}
where $I^*=\R$ if $\mu_1^{\rm b}+\mu_4^{\rm b}=0$ and  $I^*=\R_+$ if $\mu_1^{\rm b}+\mu_4^{\rm b}>0$. 
The metric dependent constant is given by 
\[
C_{\Sigma,g^*}(\bs\beta,p):=C^{{\rm (A2)},3}_{\mc{P}_1^*,g^*}( Q+{\rm i}p,\beta_{4},\beta_{1})C^{{\rm (A2)},3}_{\mc{P}_2^*,g^*}( Q+{\rm i}p, \beta_{2},\beta_{3}),
 \]
 where $C^{{\rm (A2)},3}_{\mc{P}^*_j,g^*}$ are the metric dependent constants defined in \eqref{metricconstantHblock}, and the conformal block  is given by \eqref{defConfBlock2}. The metric coefficient has the form
 \[C_{\Sigma,g^*}(\bs\beta,p)=C_0e^{c_{\rm L}S^*(g^*)}\big|\frac{1}{4}(\psi^*_{11})'(0)(\psi^*_{21})'(0)\big|^{\Delta_{Q+{\rm i}p}}\times \Big(\prod_{j=1}^22^{\Delta_{\beta^*_{j2}}-\Delta_{\beta^*_{j3}}}\prod_{k=2}^{3} |(\psi_{jk}^*)'(0)|^{\Delta_{\beta_{jk}^*}}\Big) \]
where $C_0$ is the same  universal constant as in Proposition \ref{propcut1} and 
\[S^*(g^*)=  \sum_{j=1}^2S^0_{\rm L}( \caP_j^*,(\Phi_j^*)^*(g_{\rm dozz}),g^*) +\sum_{j=1}^2\sum_{k=1}^{3}S^0_{\rm L}(\psi^*_{jk}(\D^+),g_{\rm dozz},(\psi_{jk}^*)_*g_\D))\]
\end{proposition}
\begin{proof} The proof is excatly the same as that of Proposition \ref{propcut1}.
 \end{proof}
Using the Weyl anomaly formula \eqref{confan3} and Propositions \ref{propcut1} and \ref{propcut2}, 
we obtain:
\begin{corollary}\label{cor_for_blocks_and_fusion}
Let $\gamma,\mu,\mu^{\rm b}$ and $\Sigma,g,g^*$ be as in Propositions \ref{propcut1} and \ref{propcut2}, and let $\bs{\beta}\in U^4$ satisfy \eqref{Seiberg1} and \eqref{Seiberg2}. Then one has 
\begin{align*}
&\int_{I^*} 
\widetilde{\rm H}_{\mu^b_{2},\mu^b_{4},\mu^b_{1}}^{Q+{\rm i}p,\beta_{4},\beta_{1}}\widetilde{\rm H}_{\mu^b_{4},\mu^b_{2},\mu^b_{3}}^{Q-{\rm i}p,\beta_{2},\beta_{3}} 
|\frac{1}{4}(\psi^*_{11})'(0)(\psi^*_{21})'(0)|^{\Delta_{Q+{\rm i}p}}\tilde{\mc{F}}_{p}^{\mc{G}^*}(\boldsymbol{\beta},\bs\nu) \dd p\\
&=e^{c_{\rm L}(S(g)-S^*(g^*)+S_{\rm L}^0(\Sigma,g,g^*))}C(\bs{\psi}_1,\bs{\psi}_2,\bs{\beta})\int_{I} \widetilde{\rm H}_{\mu^b_{1},\mu^b_{3},\mu^b_{4}}^{Q+{\rm i}p,\beta_{3},\beta_{4}}\widetilde{\rm H}_{\mu^b_{3},\mu^b_{1},\mu^b_{2}}^{Q-{\rm i}p,\beta_{1},\beta_{2}} |\frac{1}{4}(\psi_{11})'(0)(\psi_{21})'(0)|^{\Delta_{Q+{\rm i}p}}
\tilde{\mc{F}}_{p}^{\mc{G}}(\boldsymbol{\beta},\bs\nu) \dd p
\end{align*}
where
\[C(\bs{\psi}_1,\bs{\psi}_2,\bs{\beta}):=\frac{\prod_{j=1}^22^{\Delta_{\beta_{j2}}-\Delta_{\beta_{j3}}}\prod_{k=2}^{3} |(\psi_{jk})'(0)|^{\Delta_{\beta_{jk}}}}{\prod_{j=1}^22^{\Delta_{\beta^*_{j2}}-\Delta_{\beta^*_{j3}}}\prod_{k=2}^{3} |(\psi_{jk}^*)'(0)|^{\Delta_{\beta_{jk}^*}}}.\]
\end{corollary}

Consider now a slightly different case: here $\bs{\xi}$ is as above, except that $\xi_4:\T \to \C$ is such that $\xi_4(1)<\xi_1(-1)$ and $ \xi_4(-1)>\xi_3(1)$, and $\hat{\C}\setminus 
\mc{D}_4^\circ$ is a compact region in $\C$ containing $\mc{D}_1\cup \mc{D}_2\cup \mc{D_3}$ in its interior. We can consider $\Sigma=\mc{D}_4\setminus (\cup_{k=1}^3 \mc{D}_3^\circ)$ with the incoming boundary parametrisations $\bs{\xi}$. The cut $\xi(\T)$ and $\xi^*(\T)$ are before, i.e. $\xi(\T)$ is separating $\mc{D}_1\cup \mc{D_2}$ from $\mc{D}_3\cup \mc{D}_4$ and  $\xi^*(\T)$ is separating $\mc{D}_2\cup \mc{D_3}$ from $\mc{D}_1\cup \mc{D}_4$. Choose a finite real point $z_4$ in the interior of the exterior
hole, and put $M(z)=-1/(z-z_4)$. This orientation-preserving
M\"obius transformation preserves $\H$ and mapping $\mc{D}_4$ to a compact region in $\C$
 it is possible to map $(\Sigma,g,\bs{\xi})$ to the situation considered before.
The results of  Propositions \ref{propcut1} and \ref{propcut2} 
also hold in this case, since
\[\begin{gathered}
\cjg  \mc{A}_{\Sigma,g,\bs{\xi}} ,  \bigotimes_{k=1}^4 {\bf C}\Psi^+_{\beta_k,\nu_k}\cjd_{ (\mc{H}^+)^{\otimes 4}}=\cjg  \mc{A}_{M(\Sigma),M_*g,M\circ \bs{\xi}} ,  \bigotimes_{k=1}^4 {\bf C}\Psi^+_{\beta_k,\nu_k}\cjd_{ (\mc{H}^+)^{\otimes4}}\\
\cjg  \mc{A}_{\Sigma,g^*,\bs{\xi}} ,  \bigotimes_{k=1}^4 {\bf C}\Psi^+_{\beta_k,\nu_k}\cjd_{ (\mc{H}^+)^{\otimes4}}=\cjg  \mc{A}_{M(\Sigma),M_*g^*,M\circ\bs{\xi}} ,  \bigotimes_{k=1}^4 {\bf C}\Psi^+_{\beta_k,\nu_k}\cjd_{ (\mc{H}^+)^{\otimes4}}
\end{gathered}\]
We now consider a particular case:  we let $\xi_1,\dots,\xi_4$ be defined by $\xi_k=\psi_k|_{\T^+}$ and $\xi=\psi|_{\T}$, 
$\xi^* :=\psi^*|_{\T}$
where 
\[\psi_1(z)=r_1 z,\quad \psi_2(z)=s+r_2 z,\quad \psi_3(z)=1+r_3 z,\quad \psi_4(z)=-\frac{R}{z}, \quad \psi (z)=rz,\quad \psi^{*}(z)=1+r^*z.\]
with $\mc{D}_k=\psi_k(\D)$ and $\mc{D}_k^+=\mc{D}_k\cap \hat{\H}$ for $k=1,\dots,4$, $\Sigma=\hat{\H}\setminus \cup_{k=1}^4\mc{D}_k^\circ$ and the parameters satisfy
\[\begin{gathered}
r_i>0, \qquad r_1+r_2<s, \qquad r_2+r_3<1-s,\qquad 1+r_3<R,\\
\max\{r_1, s+r_2\}<r<1-r_3, \qquad \max\{r_3,1-s+r_2\}<r^*<1-r_1,
\qquad 1+r^*<R.
\end{gathered}\]
The model forms for $\mc{P}_1$ are 
\[ (\psi_{11}(z),\psi_{12}(z),\psi_{13}(z))= (\frac{rz-1}{rz+1}, \frac{r_3z}{2+r_3z}, \frac{R+z}{R-z}) ,\quad
(\psi_{21}(z),\psi_{22}(z),\psi_{23}(z))= (\frac{-r}{r+2sz}, \frac{r_1z}{2s-r_1z}, \frac{s+r_2z}{s-r_2z})  \]
and a direct calculation gives 
\[\frac{1}{4}|(\psi_{11})'(0)(\psi_{21})'(0)|=s, \quad \frac{1}{4}|(\psi_{11}^*)'(0)(\psi_{21}^*)'(0)|=1-s,\quad C(\bs{\psi}_1,\bs{\psi}_2,\bs{\beta}):=(1-s)^{\Delta_{\beta_2}+\Delta_{\beta_3}}s^{-\Delta_{\beta_1}-\Delta_{\beta_2}}.\]
We notice that $\Phi_j$ is a M\"obius transformation and is thus defined on $\hat{\C}$, then we have 
\[  S^0_{\rm L}( \hat{\H},g_{\rm dozz},\hat{g}_j)=S^0_{\rm L}( \hat{\H},(\Phi_j)_*g_{\rm dozz},\hat{g}_j)+S^0_{\rm L}( \hat{\H},g_{\rm dozz},(\Phi_j)_*g_{\rm dozz})\]
where $\hat{g}_j$ is the metric equal to $(\Phi_j)_*g$ on $\Phi_j(\mc{P}_j)$ and $(\psi_{jk})_*|\dd z|^2$ on the disk $\psi_{jk}(\D^+)$. A direct calculation using that $\Phi_j$ is M\"obius shows that $S^0_{\rm L}( \hat{\H},g_{\rm dozz},(\Phi_j)_*g_{\rm dozz})=0$.
We can then rewrite $S(g)$ in the form 
\[S(g)=  \sum_{j=1}^2S^0_{\rm L}( \caP_j,g_{\rm dozz},g) +\sum_{j=1}^2\sum_{k=1}^{3}S^0_{\rm L}(\Phi_j^{-1}(\psi_{jk}(\D^+)),g_{\rm dozz},(\Phi_j)^*(\psi_{jk})_*|\dd z|^2))\]
and similarly for $S^*(g^*)$. 
Since $\sum_{j=1}^2S_{\rm L}^0(\mc{P}_j,g_{\rm dozz},g)=S_{\rm L}^0(\Sigma,g_{\rm dozz},g)$ and $g=g^*$ near $\pl \Sigma$, we have 
\[\begin{split}
S(g)-S^*(g^*)=&S_{\rm L}^0(\Sigma,g_{\rm dozz},g)-S_{\rm L}^0(\Sigma,g_{\rm dozz},g^*)+\sum_{j=1}^2\sum_{k=1}^{3}S^0_{\rm L}(\Phi_j^{-1}(\psi_{jk}(\D^+)),g_{\rm dozz},(\Phi_j)^*(\psi_{jk})_*|\dd z|^2))\\
&-\sum_{j=1}^2\sum_{k=1}^{3}S^0_{\rm L}((\Phi_j^*)^{-1}(\psi_{jk}(\D^+)),g_{\rm dozz},(\Phi_j^*)^*(\psi_{jk})_*|\dd z|^2))
\end{split}\]
We notice that $\Phi_j^{-1}\circ \psi_{jk}$ is the uniformising map of $\Phi_j^{-1}\circ \psi_{jk}(\D)$, which is one of the discs $\mc{D}_\ell$. This means that, in the difference above, the contribution for the 4 discs $\mc{D}_1^+,\dots,\mc{D}_3^+$ cancel each other and we are left with the terms 
\begin{align*}
&\sum_{j=1}^2\sum_{k=1}^{3}S^0_{\rm L}(\Phi_j^{-1}(\psi_{jk}(\D^+)),g_{\rm dozz},(\Phi_j)^*(\psi_{jk})_*|\dd z|^2))
-S^0_{\rm L}((\Phi_j^*)^{-1}(\psi_{jk}(\D^+)),g_{\rm dozz},(\Phi_j^*)^*(\psi_{jk})_*|\dd z|^2))\\
&= S^0_{\rm L}(\mc{D}\cap \hat{\H},g_{\rm dozz},\psi_*|\dd z|^2))+S^0_{\rm L}(\hat{\H}\setminus \mc{D},g_{\rm dozz},(\psi\circ (-o))_*|\dd z|^2))\\
& \quad  -S^0_{\rm L}(\mc{D}^*\cap \hat{\H},g_{\rm dozz},(\psi^*)_*|\dd z|^2))-S^0_{\rm L}(\hat{\H}\setminus (\mc{D}^*),g_{\rm dozz},(\psi^*\circ (-o))_*|\dd z|^2))
\end{align*}
Using the explicit expression for $\psi$ and $\psi^*$, we see that the first two terms are equal to $0$. Arguing as above, in the last two terms we can replace $g_{\rm dozz}$ by $(\psi^*)_*g_{\rm dozz}$, and we see that the last two terms are equal to the first two, thus vanish.
We have thus obtained the following:
\begin{corollary}\label{cor_for_blocks_and_fusion2}
Let $\gamma,\mu,\mu^{\rm b}$ and $\Sigma,g,g^*$ be as in Propositions \ref{propcut1} and \ref{propcut2}, and let $\bs{\beta}\in U^4$ satisfy \eqref{Seiberg1} and \eqref{Seiberg2}. Then one has 
\begin{align*}
&(1-s)^{-\Delta_{\beta_2}-\Delta_{\beta_3}}\int_{I^*} 
\widetilde{\rm H}_{\mu^b_{2},\mu^b_{4},\mu^b_{1}}^{Q+{\rm i}p,\beta_{4},\beta_{1}}\widetilde{\rm H}_{\mu^b_{4},\mu^b_{2},\mu^b_{3}}^{Q-{\rm i}p,\beta_{2},\beta_{3}} 
(1-s)^{\Delta_{Q+{\rm i}p}}\tilde{\mc{F}}_{p}^{\mc{G}^*}(\boldsymbol{\beta},\bs\nu) \dd p\\
&=s^{-\Delta_{\beta_1}-\Delta_{\beta_2}}\int_{I} \widetilde{\rm H}_{\mu^b_{1},\mu^b_{3},\mu^b_{4}}^{Q+{\rm i}p,\beta_{3},\beta_{4}}\widetilde{\rm H}_{\mu^b_{3},\mu^b_{1},\mu^b_{2}}^{Q-{\rm i}p,\beta_{1},\beta_{2}} s^{\Delta_{Q+{\rm i}p}}
\tilde{\mc{F}}_{p}^{\mc{G}}(\boldsymbol{\beta},\bs\nu) \dd p
\end{align*}
\end{corollary}
We also notice that, if $I=\R_+$ (and similarly for $I^*$), the integral above can be reduced as an integral on $\R$ with a $1/2$ factor, since the integrand is even with respect to $p$, by what we discussed above.
We also observe that, when $\bs{\nu},\emptyset$,  the conformal block is equal to (as a function of $s$)
\[\begin{split}
\tilde{\mc{F}}^1_{\bs{\beta}}(p,s):=& s^{\Delta_{Q+{\rm i}p}-\Delta_{\beta_1}-\Delta_{\beta_2}}\tilde{\mc{F}}_{p}^{\mc{G}}(\boldsymbol{\beta},\emptyset)\\
=& s^{\Delta_{Q+{\rm i}p}-\Delta_{\beta_1}-\Delta_{\beta_2}}\sum_{\nu\in \mc{T}}\omega_{\D}(\Delta_{\beta_1},\Delta_{\beta_2},\Delta_{Q+{\rm i}p},\nu)\omega_{\D}(\Delta_{\beta_3},\Delta_{\beta_4},\Delta_{Q-{\rm i}p},\nu)s^{|\nu|},
\end{split}\]
where the coefficients $\omega_{\D}(\cdot,\cdot,\cdot,\nu)$ are obtained from the Ward identity and defined in \cite[Proposition 7.1]{GKRV20_bootstrap}; here $\omega_{\D}(\cdot,\cdot,\cdot,0)=1$. Similarly, one has 
\[ \mc{F}^{2}_{\bs{\beta}}(p,s):=(1-s)^{-\Delta_{\beta_2}-\Delta_{\beta_3}+\Delta_{Q+ip}}\tilde{\mc{F}}_{p}^{\mc{G}^*}(\boldsymbol{\beta},\emptyset)=\tilde{\mc{F}}_{p}^{\mc{G}}(\beta_3,\beta_2,\beta_1,\beta_4,\emptyset;1-s).\]
The result of Corollary \eqref{cor_for_blocks_and_fusion2} in that special case $\bs{\nu}=\emptyset$ 
can be rewritten as 
\begin{equation}\label{crossing_no_nu}
\int_{I^*} 
\widetilde{\rm H}_{\mu^b_{2},\mu^b_{4},\mu^b_{1}}^{Q+{\rm i}p,\beta_{4},\beta_{1}}\widetilde{\rm H}_{\mu^b_{4},\mu^b_{2},\mu^b_{3}}^{Q-{\rm i}p,\beta_{2},\beta_{3}} 
\mc{F}_{\boldsymbol{\beta}}^{2}(p,s) \dd p=\int_{I} \widetilde{\rm H}_{\mu^b_{1},\mu^b_{3},\mu^b_{4}}^{Q+{\rm i}p,\beta_{3},\beta_{4}}\widetilde{\rm H}_{\mu^b_{3},\mu^b_{1},\mu^b_{2}}^{Q-{\rm i}p,\beta_{1},\beta_{2}} s^{\Delta_{Q+{\rm i}p}}
\mc{F}_{\boldsymbol{\beta}}^{1}(p,s) \dd p
\end{equation}
\begin{remark}
In a forthcoming work \cite{GRSSbloc}, the authors use the two formulas (with $\bs \nu=\emptyset$) in Corollary \ref{cor_for_blocks_and_fusion} and  \ref{cor_for_blocks_and_fusion2} to get a probabilistic expression for the conformal blocks. They also obtain   Ponsot and Teschner's formula for the fusion kernel
of the four-point spherical blocks, and Teschner's formula for the modular kernel of the one-point torus block  (the latter being based on Proposition \ref{propcuttorus} below). In \cite{BGKRX}, the result of Corollary \ref{cor_for_blocks_and_fusion} is also used, combined with the method of \cite{GRSSbloc}, to extract the existence of the fusion kernel for conformal blocks and construct a unitary projective representation of Moore-Seiberg groupoid and the mapping class group.
\end{remark}

 \subsection{Modular invariance for 1-hole  annulus amplitude} 
  
\begin{figure}[H] 
 \centering
\begin{tikzpicture}[scale=1]
 
 \draw[fill=lightgray,draw=lightgray] (0.01,0) rectangle (5,4);
 \draw[draw=yellow,very thick] (0,3.6) -- (5,3.6); 
  \draw[->,>=stealth,draw=yellow,very thick] (0,3.6) -- (2.5,3.6); 
 \draw[draw=green,very thick] (0,1.5) -- (0,4);
  \draw[draw=green,very thick] (5,1.5) -- (5,4);
    \draw[draw=black,very thick] (1.5,0) -- (3.5,0);
     \draw[draw=black,very thick] (0,4) -- (5,4);
   \draw[fill=white,draw=white,very thick] (0,0) -- (1.5,0)  arc [start angle=0, end angle=90, radius=1.5] ; 
     \draw[fill=white,draw=white,very thick] (3.5,0) -- (5,0) -- (5,1.5) arc [start angle=90, end angle=180, radius=1.5] ; 
      \draw [->,>=stealth,draw=red,very thick] (1.5,0) arc [start angle=0, end angle=90, radius=1.5] ;
         \draw [->,>=stealth,draw=red,very thick] (5,1.5) arc [start angle=90, end angle=180, radius=1.5] ;
\draw (0,0) node{$\times$}; 
\draw (5,0) node{$\times$}; 
\draw (0,0) node[below]{$0$}; 
\draw (5,0) node[below]{$1$}; 
\draw (0,4) node[left]{$\frac{\tau}{2}$};
\draw (5,4) node[right]{$1+\frac{\tau}{2}$}; 
\draw (2.5,0) node[below]{$\mu_1$}; 
\draw (1,3.55) node[above]{$\mc{P}_2$};  
\draw (2.5,4) node[above]{$\mu_2$}; 
\draw (2.5,1) node[above]{$\mc{P}_1$};  
\draw (2.4,3.55) node[below]{\textcolor{yellow}{in}};  
\draw (2.4,3.6) node[above]{\textcolor{yellow}{out}};  
 \draw[fill=lightgray,draw=lightgray] (7.01,0) rectangle (12,4);
 \draw[draw=yellow,very thick] (9.5,0) -- (9.5,4); 
  \draw[->,>=stealth,draw=yellow,very thick] (9.5,0) -- (9.5,2); 
 \draw[draw=green,very thick] (7,1.5) -- (7,4);
  \draw[draw=green,very thick] (12,1.5) -- (12,4);
    \draw[draw=black,very thick] (8.5,0) -- (10.5,0);
     \draw[draw=black,very thick] (7,4) -- (12,4);
   \draw[fill=white,draw=white,very thick] (7,0) -- (8.5,0)  arc [start angle=0, end angle=90, radius=1.5] ; 
     \draw[fill=white,draw=white,very thick] (10.5,0) -- (12,0) -- (12,1.5) arc [start angle=90, end angle=180, radius=1.5] ; 
      \draw [->,>=stealth,draw=red,very thick] (8.5,0) arc [start angle=0, end angle=90, radius=1.5] ;
         \draw [->,>=stealth,draw=red,very thick] (12,1.5) arc [start angle=90, end angle=180, radius=1.5] ;
\draw (7,0) node{$\times$}; 
\draw (12,0) node{$\times$}; 
\draw (7,0) node[below]{$0$}; 
\draw (12,0) node[below]{$1$}; 
\draw (7,4) node[left]{$\frac{\tau}{2}$};
\draw (12,4) node[right]{$1+\frac{\tau}{2}$}; 
\draw (8.5,2.5) node[above]{$\mc{P}^*_1$};  
\draw (9.5,0) node[below]{$\mu_1$}; 
\draw (9.5,4) node[above]{$\mu_2$};  
\draw (9.5,1.5) node[left]{\textcolor{yellow}{out}};  
\draw (9.5,1.5) node[right]{\textcolor{yellow}{in}};  
    \end{tikzpicture}
\caption{The annulus with one half-disk removed (boundary in red) together with the two possible cuts (in yellow): horizontal cut on the left, and vertical cut on the right. Vertical green lines are identified. Black boundary components carry Neumann boundary conditions.  }
\label{pic:newgluing2}
\end{figure}
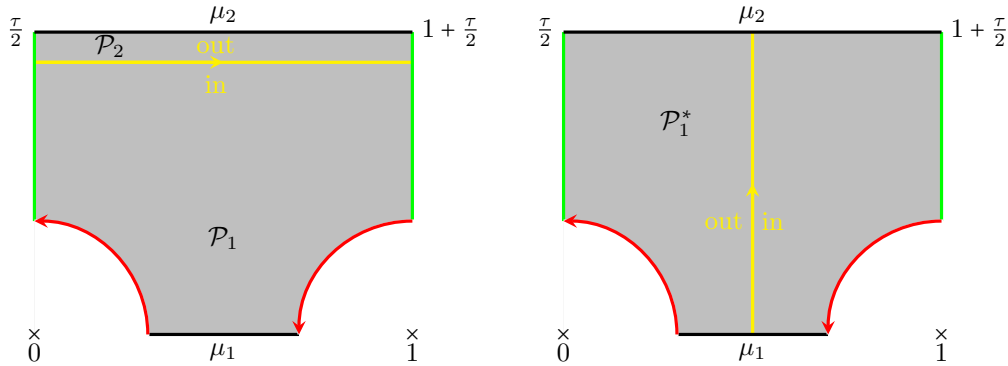

For $T>0$ and $\tau=iT\in i\R_+$, we consider the cylinder (or annulus)
\begin{equation}\label{Ntau}
A(\tau)=\{z\in\C\,|\, 0\leq{\rm Im}(z)\leq T/2\}/\Z,
\end{equation}
and let $\pi_A$ denote the quotient map from this full strip.
For $q_1\in(0,\min\{1/4,T/2\})$, set
\[
\Sigma=A(\tau)\setminus\pi_A((q_1\D^+)^\circ),
\qquad \zeta(e^{i\theta})=\pi_A(q_1e^{i\theta}).
\]
On the rectangle $[0,1]\times[0,T/2]$, the map
$z\mapsto e^{2\pi(z-1)/T}$ is a biholomorphism onto the
half-annulus $\A^+_{e^{-2\pi/T}}$.
 We assign the boundary cosmological constants $\mu_1^{\rm b}$ to $\partial_1A_\tau$ and $\mu_2^{\rm b}$ to $\partial_2A_\tau$. 
We also set  $\psi(z)=q_1z$. 
We will consider two possible cuts for $\Sigma$, and state the two related bootstrap formulae, see Figure \ref{pic:newgluing2}.

The first cut is given by $\xi:\T\to \mc{C}:=\xi(\T)\subset A(\tau)\setminus \psi(\D^\circ)$, the real-analytic curve defined by 
$\xi(e^{{\rm i}\theta})=\theta/(2\pi)+i(T-\eps)/2$ for $\eps>0$ arbitrarily small, which cuts $\Sigma$ into two geometric building blocks $\mc{P}_j$, for $j=1,2$, each of which containing the boundary component with assigned boundary cosmological constant $\mu_j^{\rm b}$. Note that  the cut $\xi$ is degenerate as the building block $\mc{P}_2$ is degenerate.
The  boundary parametrisations of $\mc{P}_1$ of are given by $\bs{\zeta}_1=(\zeta_{11},\zeta_{12})$, with $\zeta_{11}(e^{{\rm i}\theta})=\theta/(2\pi)+i(T-\eps)/2$  mod $1$ being  incoming and  $\zeta_{12}=\zeta$ being incoming. The parametrisation of the annulus $\mc{P}_2$ with one N-marked boundary is
$\bs{\zeta}_2=\zeta_{21}$ with $\zeta_{11}(e^{{\rm i}\theta})=\theta/(2\pi)+i(T-\eps)/2$ outgoing. The Neumann double $\Sigma^{\#2}$ can be represented by $N(\tau)\setminus \psi(\D^\circ)$ with 
\[N(\tau)=  \{z\in \C \,|\, {\rm Im}{z}\in [-T/2,T/2], {\rm Re}(z)\in [0,1]\}/ (\Z+\tau \Z).
\]
Denote by $\pi_{\mc{P}_j}:\mc{P}_j^{\#2}\to \Sigma^{\#2}$  the canonical projection and we keep the same notation for its restriction to $\mc{P}_j$. Note that  the 
anti-holomorphic involution $z\mapsto \bar{z}$ induces an involution $\iota_{\mc{P}_j}:\mc{P}_j^{\#2}\to \mc{P}_j^{\#2}$ satisfying $\pi_{\mc{P}_j}\circ\iota_{\mc{P}_j}=c\circ\pi_{\mc{P}_j}$, where $c([z])=[\bar z]$.
 
We call $\mc{G}$ the oriented graph of the Neumann double $\Sigma^{\#2}$ cut into $\mc{P}_1^{\#2}\# \mc{P}_2^{\#2}$. We also denote 
by $\mc{P}$ the building block associated to the surface $\Sigma^{\#2}=N(\tau)\setminus \psi(\D^\circ)$ cut along the curve 
$e^{i\theta}\mapsto \tau/2+\theta/2\pi$, with boundary parametrisations $\zeta^{\mc{P}}_1(e^{i\theta})=\theta/(2\pi)+iT/2$, 
$\zeta^{\mc{P}}_2(e^{i\theta})=\zeta(e^{i\theta})$ and $\zeta^{\mc{P}}_3(e^{i\theta})=\theta/(2\pi)-iT/2$.
Now consider a model form $(\psi^{\mc{P}}_1,\psi^{\mc{P}}_2,\psi^{\mc{P}}_3)$ for $\mc{P}$, with $(\psi^{\mc{P}}_1(0),\psi^{\mc{P}}_2(0),\psi^{\mc{P}}_3)(0))=(i,0,-i)$, such that there is a biholomorphism
\[\Phi^{\mc{P}}: \mc{P}\to  \hat{\C}\setminus \bigcup_{k=1}^3\psi_k^{\mc{P}}(\D^\circ), \quad \Phi^{\mc{P}}\circ \zeta^{\mc{P}}_k=\psi_k^{\mc{P}}, \textrm{ for } k=1,2,3.\]
Notice also that $\Phi^{\mc{P}}:\mc{P}_1\to \hat{\H}\setminus (\psi_1^{\mc{P}}(e^{-\pi\eps}\D^\circ)\cup \psi_2^{\mc{P}}(\D^\circ))$ is the biholomorphism which put $\mc{P}_1$ under model form, with the model form being $\psi_{11}(z)=\psi_1^{\mc{P}}(e^{-\pi\eps}z)$ and $\psi_{12}=\psi_2^{\mc{P}}(z)$. In our case, it is easy to see that 
\[
\Phi^{\mc{P}}(z)= i\frac{1-e^{2\pi iz}}{e^{2\pi iz}+1}, \qquad 
\psi_{1}^{\mc{P}}(z)={\rm i}\frac{1-e^{-\pi T}z}{1+e^{-\pi T}z}, \qquad \psi_{2}^{\mc{P}}(z):=\Phi^{\mc{P}}(q_1z).
\]
The (incoming boundary) model of the double is obtained by reversing
the outgoing third parametrisation. For the upper building
block cut at height $(T-\varepsilon)/2$, the model maps are
\[
\psi_{11}(z)=\psi_1^{\mc P}(e^{\pi\varepsilon}z),
\qquad \psi_{12}(z)=\psi_2^{\mc P}(z).
\]
For $\nu^+ \in \mc{T}$, we define the normalised conformal block of the Neumann double $\Sigma^{\#2}$ by
\begin{equation}\label{blockchannel1torus}
\tilde{\mc{F}}_{p}^{\mc{G}}(\boldsymbol{\beta},\nu^+):=
 \sum_{n\geq 0}  \sum_{\substack{\nu,\nu'\in \mc{T}\\ |\nu|=|\nu'|=n}}\omega(\mc{P},\bs{\Delta},\bs\nu)  {\bf F}_{Q+ip,n}^{-1}(\nu,\nu') 
 \end{equation}
 where $\bs\Delta=(\Delta_{Q+ip},\Delta_\beta,\Delta_{Q+ip})$, $\bs{\nu}=(\nu,\nu^+, \nu' )$ and $\omega(\mc{P},\bs{\Delta},\bs\nu)$ are the Ward identity coefficients in Proposition \ref{thm: ward identity}.

To remedy the fact that $\mc{P}_2$ is degenerate (thus the amplitude is not Hilbert-Schmidt), for $g$ an admissible metric, Neumann extendible, we differentiate the amplitude $\mc{A}_{\Sigma,g,\zeta}$ with respect to $\mu_2^{\rm b}$, and we focus on this quantity.  We also denote $g$ for the Neumann double of the metric $g$ on $\mc{P}$.

\begin{proposition}[\textbf{First channel bootstrap}]\label{propcuttorus}
Consider the setting above, with $\mu=0$ and $\mu^{\rm b}_j>0$ for $j=1,2$, and assume that the admissible metric satisfies
$g=4\pi^2|\dd z|^2$ on $\mc{P}_2$.
Assume $\beta \in U$  obeys $|{\rm Re}(\beta-Q)|<Q$ with $\beta\not=Q$, and $ \nu^+$ is a Young diagram, with $\nu^+=\emptyset$ if $\beta\in\R$.  
Then, the following holds
\begin{align}\label{fusionblocktorus1}
\big\cjg \partial_{\mu_2}\mc{A}_{\Sigma,g, \zeta} ,  {\bf C} \Psi^+_{\beta,\nu^+}\big\cjd_{ \mc{H}^+}   
=C_{\Sigma,g}(\beta)  \int_{\R} e^{-2\pi T\Delta_{Q+ip}}  
\tilde {\rm G}_{\mu,\mu^{\rm b}_1}(Q+{\rm i}p,\beta)  \partial_{\mu^{\rm b}_2}\widetilde{\rm U}_{\mu,\mu^{\rm b}_2}(Q-{\rm i}p)
\tilde{\mc{F}}_{p}^{\mc{G}}( \bs \beta,\bs \nu^+) \dd p,
\end{align}
The metric dependent constant is given by 
\[\begin{split}
C_{\Sigma,g}(\beta):=&\frac{(2\pi)^{-2}2^{\frac{1}{4}}\pi^{\frac{3}{4}}}{Z_{\D,g_\D} Z_{\D^+,g_\D,{\rm m}}} \big({\frac{{\rm v}_{g_{\rm dozz}}(\hat\H)}{{\det}'(\Delta_{\hat\H,g_{\rm dozz},{\rm N}})}} \big)^\hf   (2\pi q_1)^{\Delta_\beta}    \\
  & \times e^{c_{\rm L}(\frac{1}{2} S_{\rm L}^0(\mc{P},(\Phi^{\mc{P}})^*g_{\rm dozz},g)+S_{\rm L}^0(\psi^{\mc{P}}_{1}(\D),g_{\rm dozz},(\psi^{\mc{P}}_{1})_*g_\D)+S_{\rm L}^0(\psi^{\mc{P}}_2(\D^+),g_{\rm dozz},(\psi^{\mc{P}}_2)_*g_\D))} ,
\end{split} \]
where the conformal block  is given by \eqref{blockchannel1torus}. 
\end{proposition}

\begin{proof} Again, we follow the proof of Theorem~\ref{th:fullexpressioncorrel}. We start with the case where $\Delta_\beta$ is real valued. The amplitude $\partial_{\mu_2}\mc{A}_{\Sigma,g, \zeta} $ is the gluing of the two building block amplitudes $\mc{A}_{\mc{P}_1,g ,\bs{\zeta}_1}$ and $\partial_{\mu_2} \mc{A}_{\mc{P}_2,g ,\bs{\zeta}_2} $, which are both Hilbert-Schmidt. The map $h:z\mapsto e^{2\pi iz}$ maps $A(\tau)$ on the annulus $\A_{e^{-\pi T}}$, where $\mc{P}_2$ is mapped to 
the annulus $h(\mc{P}_2)=\{e^{-\pi T}\leq |z|\leq e^{-\pi(T-\eps)}\}$ and $\mc{P}_1$ is mapped to $\A_{e^{-\pi(T-\eps)}}\setminus h(\psi(\D^+)^\circ)$. 
Note that $(h(\mc{P}_2),g,\bs{\zeta}_2)$ is isometric to $(\A_{e^{-\pi\eps}},g_\A,\zeta_\A)$ with $\zeta_\A(e^{i\theta})=e^{-\eps+i\theta}$. As in (the proof of) Theorem \ref{th:fullexpressioncorrel}, using the spectral resolution \eqref{fcomplete}
\begin{equation}\label{variationdmu_2}
\partial_{\mu_2}\mc{A}_{\Sigma,g, \zeta} (\Psi^+_{\beta,\nu'})
\frac{1}{\sqrt{2}\pi}\frac{1}{2\pi}\int_{\R}\sum_{\nu,\tilde\nu} \mathcal{A}_{\mc{P}_1,g,\bs{\zeta}_1}(H^0_{Q+ip,\nu,\tilde\nu},\Psi^+_{\beta,\nu'})\partial_{\mu_2^b}\mathcal{A}_{\A_{e^{-\pi\eps}},g_\A,\zeta_\A}(H^0_{Q-ip,\nu,\tilde\nu}) \dd p
\end{equation}
where 
\[{H}^0_{Q+ip, \nu, \tilde{\nu}}:=\sum_{|\nu'|=|\nu|, |\tilde{\nu}'|=|\tilde{\nu}|}F_{\Delta_{Q+i p},|\nu|}^{-1 / 2}\left(\nu, \nu'\right)F_{\Delta_{Q+i p},|\tilde\nu|}^{-1/2}\left(\tilde{\nu}, \tilde{\nu}'\right)\Psi^0_{Q+i p, \nu', \tilde{\nu}'}.\]
By \cite[Corollary 6.11]{WuConfBoot}, we have
\begin{equation}\label{propBaojun}
\partial_{\mu_2^b}\mathcal{A}_{\A_{e^{-\pi\eps}},g_\A,\zeta_\A}(H^0_{Q-ip,\nu,\tilde\nu})=\frac{1}{\sqrt{2}}\partial_{\mu_2^{\rm b}}(\widetilde{\mathrm{U}}_{\mu,\mu_2^{\rm b}}(Q-i p))e^{-\pi \eps(-\frac{c_{\rm L}}{12}+2\Delta_{Q+ip})}e^{-\pi \eps 2|\nu|}\mathrm{1}_{\{\nu=\tilde\nu\}}.
\end{equation}

Using \eqref{matrix_half_pant2}	 and \eqref{Gblock}, we get
\begin{equation}\label{AP_1matrix}
\big\cjg \mc{A}_{\mc{P}_1,g ,\bs{\zeta}_1},   {\bf C}\Psi^0_{Q+{\rm i}p,\nu,\tilde\nu } \otimes {\bf C} \Psi^+_{\beta,\nu^+}\big\cjd_{  \mc{H}\otimes\mc{H}^+ } =
C_{\mc{P}_1,g}(Q+ip, {\beta})\omega(\mc{P}_1^{\#2},\bs{\Delta},\bs\nu) \tilde{\rm G}_{\mu,\mu^{\rm b}_1}( Q+ip,\beta),  
\end{equation}
where $\bs{\Delta}=(\Delta_{Q+ip},\Delta_{\beta},\Delta_{Q+ip})$, $\bs{\nu}=(\nu,\nu^+,\tilde\nu )$,   the constant $\omega(\mc{P}_1^{\#2},\bs{\Delta},\bs\nu)$ is the same as in~\eqref{matrix_pant}, and  $ C_{\mc{P}_1,g}(Q+ip, {\beta})$ is  the metric dependent constant defined in \eqref{metricconstantGblock} using the model form of $\mc{P}_1$. Observe now that, for $\bs{\nu}'=(\nu',\nu^+,\tilde{\nu}')$, 
\[\sum_{\nu\in\mc{T}} \sum_{|\nu'|=|\nu|=|\tilde{\nu}'|}\mathbf{F}_{\Delta_{Q+i p},|\nu|}^{-1 / 2}(\nu, \nu')\mathbf{F}_{\Delta_{Q+i p},|\nu|}^{-1 / 2}
(\nu, \tilde{\nu}') \omega(\mc{P}_1^{\#2},\bs{\Delta},\bs\nu')e^{-\pi \eps 2|\nu|}= \sum_{n\geq 0}  \sum_{\substack{\nu,\nu'\in \mc{T}\\ |\nu|=|\nu'|=n}}\omega(\mc{P},\bs{\Delta},\bs\nu)  {\bf F}_{Q+ip,n}^{-1}(\nu,\nu'),\]
which is exactly $\tilde{\mc{F}}_{p}^{\mc{G}}(\boldsymbol{\beta},\nu^+)$.
 This last identity just follows from the Ward identity in Proposition \ref{thm: ward identity} and the fact that gluing an annulus $\A_{e^{-2\pi\eps }}$ to 
 $\mc{P}_1^{\#2}$ at $\pl_1\mc{P}_1^{\#2}$ (which gives the building block $\mc{P}$) amounts to multiply its Ward coefficient $ \omega(\mc{P}_1^{\#2},\bs{\Delta},\bs\nu')$ by $e^{-\pi \eps 2|\nu|}$.
The Liouville anomaly term in $C_{\mc{P}_1,g}$ is
\[ e^{S_1(g,\eps)}:=e^{c_{\rm L}S_{\rm L}^0(\mc{P}_1,(\Phi^{\mc{P}})^*g_{\rm dozz},g)}e^{c_{\rm L} S_{\rm L}^0(\psi_{11}(\D),g_{\rm dozz},(\psi_{11})_*g_\D)+c_{\rm L} S_{\rm L}^0(\psi_{12}(\D^+),g_{\rm dozz},(\psi_{12})_*g_\D)} ,\]
and we observe that 
\[ \lim_{\eps\to 0} e^{S_1(g,\eps)}e^{\pi \eps \frac{c_{\rm L}}{12}}=e^{\frac{c_{\rm L}}{2}S_{\rm L}^0(\mc{P},(\Phi^{\mc{P}})^*g_{\rm dozz},g)}e^{c_{\rm L} S_{\rm L}^0(\psi_1^{\mc{P}}(\D),g_{\rm dozz},(\psi_1^{\mc{P}})_*g_\D)+c_{\rm L} S_{\rm L}^0(\psi_2^{\mc{P}}(\D^+),g_{\rm dozz},(\psi_2^{\mc{P}})_*g_\D)}. \]
Combining this with \eqref{propBaojun}, \eqref{AP_1matrix} and \eqref{variationdmu_2}, we obtain the result with
\[\begin{split}
C_{\Sigma,g}(\beta)e^{-2\pi T\Delta_{Q+ip}}=&\frac{(2\pi)^{-2}2^{\frac{1}{4}}\pi^{\frac{3}{4}}}{Z_{\D,g_\D} Z_{\D^+,g_\D,{\rm m}}}\big({\frac{{\rm v}_{g_{\rm dozz}}(\hat\H)}{{\det}'(\Delta_{\hat\H,g_{\rm dozz},{\rm N}})}} \big)^\hf    |(\psi^{\mc{P}}_{1})'(0)|^{2\Delta_{Q+ip}}  |(\psi^{\mc{P}}_2)'(0)|^{\Delta_\beta}  2^{-(2\Delta_{Q+ip}-\Delta_\beta)}    \\
  & \times e^{\frac{c_{\rm L}}{2} S_{\rm L}^0(\mc{P},(\Phi^{\mc{P}})^*g_{\rm dozz},g)+c_{\rm L} S_{\rm L}^0(\psi^{\mc{P}}_{1}(\D),g_{\rm dozz},(\psi^{\mc{P}}_{1})_*g_\D)+c_{\rm L} S_{\rm L}^0(\psi^{\mc{P}}_2(\D^+),g_{\rm dozz},(\psi^{\mc{P}}_2)_*g_\D)}.
\end{split} \]
 We conclude by computing $ |(\psi^{\mc{P}}_{1})'(0)|=2e^{-\pi T}$ and $|(\psi^{\mc{P}}_2)'(0)|=q_1\pi$.
 
 The extension to the case where $|{\rm Re}(\beta)-Q|<Q$ and $\beta\not=Q$ follows as in the proof of Proposition \ref{propcut1}: the identity \eqref{AP_1matrix} extends holomorphically in $\beta$ in the region just mentionned, using  \cite[Proposition 5.12]{GRW1}, Theorem~\ref{thm:boundary_descendants_meromorphic}. Then the argument of Proposition \ref{propcut1} shows the convergence of the right hand side of \eqref{fusionblocktorus1} in this $\beta$ extended region, and we get the result.
\end{proof}
 
 Next, we analyse the bootstrap in the second channel, which means with the cutting curve $\mc{C}^*$ parametrised by $\zeta^*(e^{i\theta})=\frac{1}{2}+i\frac{\theta T}{2\pi}$, with $\theta\in (0,\pi)$, see Figure \ref{pic:newgluing2}. It  decomposes $\Sigma$ into a single geometric building block $\mc{P}_1^*$ of type (A2).  Let $\bs{\zeta}^*=(\zeta^*_1,\zeta^*_2,\zeta_3^*)$ be the three boundary parametrisations of $\mc{P}_1^*$, where, considering $([-1/2,1/2]\times [0,T/2])\setminus \setminus \psi(\D^\circ)$ as a fundamental domain $F(\tau)$ for the quotient by the map $z\mapsto z+1$, we set  $\zeta^*_1(e^{i\theta})=\zeta^*(e^{i\theta})$ as an outgoing parametrisation, 
 $\zeta^*_2(e^{i\theta})=-1+\zeta^*(e^{i\theta})$ as an incoming parametrisation (which projects to the same curve under the identification $z\sim z+1$, and finally $\zeta^*_3(e^{i\theta})=\psi(e^{i\theta})=q_1e^{i\theta}$.
The biholomorphism that puts $\mc{P}_1^*$ under model form and the model form biholomorphisms are given by 
\begin{equation}\label{model_form_second_channel}
\Phi^*(z)= \frac{e^{\frac{2\pi z}{T}}-1}{e^{\frac{2\pi z}{T}}+1}, \quad \psi_1^*(z)=\frac{e^{-\frac{\pi}{T}}z-1}{e^{-\frac{\pi}{T}}z+1}, \quad \psi_2^*(z)=\Phi^*(q_1z), \quad \psi_3^*(z)=\frac{e^{\frac{\pi}{T}}+z}{e^{\frac{\pi}{T}}-z}.
\end{equation}

We equip  $\Sigma$ with an admissible metric $g^*$ with respect to the cut $\mc{C}^*$, and we call $\mc{G}^*$ the oriented graph of the Neumann double associated to this cut. If $\mc{P}^*$ denotes the Neumann double of $\mc{P}_1^*$, the conformal block  for $\mc{G}^*$ is defined by  
\begin{align*}
&\tilde{\mc{F}}_{ p}^{\mc{G}^*}(\boldsymbol{\beta},\nu^+):=
 \sum_{n\geq 0}  \sum_{|\nu|=|\nu'|=n}\omega((\mc{P}^*)^{\rm in},\bs{\Delta},\bs{\nu})\ F_{Q+ip}^{-1}(\nu,\nu') 
 \end{align*}
 where $\bs{\nu}=(\nu,\nu^+ ,\nu')$ and $\bs{\Delta}=(\Delta_{Q+{\rm i}p},\Delta_{\beta},\Delta_{Q+{\rm i}p})$.
When there is no Young diagram $\nu^+=\emptyset$ (no desendant), we simply denote the block by $\tilde{\mc{F}}_{ p}^{\mc{G}^*}(\boldsymbol{\beta})$.

\begin{proposition}[\textbf{Second channel bootstrap}]\label{propcuttorus2}
Consider the same assumptions as in Proposition \ref{propcuttorus}, with $g^*$ an admissible metric and let $\beta \in U$ satisfy  $|{\rm Re}(\beta-Q)|<Q$ with $\beta\not=Q$, and $ \nu^+$ is a Young diagram, with $\nu^+=\emptyset$ if $\beta\in\R$.  
Then the following holds:
\begin{align}\label{fusionblocktorus2}
\big\cjg   \partial_{\mu^{\rm b}_2}\mc{A}_{\Sigma,g^*, \zeta} ,  {\bf C} \Psi^+_{\beta,\nu^+}\big\cjd_{ \mc{H}^+}   
= C_{\Sigma,g^*}(\beta)\int_{\R_+} e^{-\frac{2\pi}{T}\Delta_{Q+ip}}
 \partial_{\mu^{\rm b}_2}\widetilde{\rm H}_{\mu^b_{2},\mu^b_{1},\mu^b_{1}}^{Q+{\rm i}p^+,\beta, Q-{\rm i}p^+} 
\tilde{\mc{F}}_{p}^{\mc{G}^*}(\boldsymbol{\beta},\nu^+) \dd p,
\end{align}
where the metric dependent constant is given by 
\[C_{\Sigma,g^*}(\beta):= \frac{(2\pi)^{\frac{1}{2}}}{4\pi Z_{\D^+,g_\D,{\rm m}}^{3}}\Big(\frac{{\rm v}_{g_{\rm dozz}}(\hat{\H})}{\det'(\Delta_{\hat\H,g_{\rm dozz},{\rm N}})}\Big)^{\frac{1}{2}}(\frac{2\pi q_1}{T})^{\Delta_\beta}e^{c_{\rm L}(S^0_{\rm L}( \caP_1^*,(\Phi^*)^*g_{\rm dozz},g^*)+\sum_{k=1}^{3}S^0_{\rm L}(\psi_{k}^*(\D^+),g_{\rm dozz},(\psi_{k}^*)_*g_\D))}.
 \]
\end{proposition}
\begin{proof} As in Theorem~\ref{th:fullexpressioncorrel}, using the self-gluing \eqref{self_gluing_formula} formula and the spectral resolution of Proposition \ref{resolution_bdry_Hamiltonian}  we get 
\begin{equation}\label{variationdmu_2}
\partial_{\mu_2}\mc{A}_{\Sigma,g^*, \zeta} (\Psi^+_{\beta,\nu^+})
=\frac{(2\pi)^{-\frac{3}{4}}}{4\pi}\partial_{\mu^{\rm b}_2}\int_{\R_+}\sum_{\nu,\nu'} \mathcal{A}_{\mc{P}^*_1,g^*,\bs{\zeta}^*}(\Psi^+_{Q+ip,\nu},\Psi^+_{\beta,\nu^+},\Psi^+_{Q-ip,\nu'})F_{Q+ip}^{-1}(\nu,\nu')\dd p
\end{equation}
Using now Proposition \ref{ward case I} and Lemma \ref{structure_amp}, we get 
\[\partial_{\mu_2}\mc{A}_{\Sigma,g^*, \zeta} (\Psi^+_{\beta,\nu^+})=\frac{(2\pi)^{-\frac{3}{4}}}{ 4\pi}\int_{\R_+} C_{\mc{P}^*,g^*}(\beta,p)  \pl_{\mu_2^{\rm b}}
\widetilde{\rm H}_{\mu^b_{2},\mu^b_{1},\mu^b_{1}}^{Q+{\rm i}p^+,\beta, Q-{\rm i}p^+} 
\tilde{\mc{F}}_{p}^{\mc{G}^*}(\boldsymbol{\beta},\nu^+) \dd p,\]
with 
\[\begin{split}
C_{\mc{P}^*,g^*}(\beta,p)  =& \frac{(2\pi)^{\frac{5}{4}}}{Z_{\D^+,g_\D,{\rm m}}^{3}}\Big(\frac{{\rm v}_{g_{\rm dozz}}(\hat{\H})}{\det'(\Delta_{\hat{\H},g_{\rm dozz},{\rm N}})}\Big)^{\frac{1}{2}}e^{-\frac{2\pi}{T}\Delta_{Q+ip}}(\tfrac{2\pi q_1}{T})^{\Delta_\beta} \\
& \times e^{c_{\rm L}(S^0_{\rm L}( \caP_1^*,(\Phi^*)^*g_{\rm dozz},g^*)+\sum_{k=1}^{3}S^0_{\rm L}(\psi_{k}^*(\D^+),g_{\rm dozz},(\psi_{k}^*)_*g_\D))}.
\end{split} \]
The convergence issue for $\beta$ complex is proved as in the previous propositions. This completes the proof.
\end{proof}

Using the Weyl anomaly formula \eqref{confan3} and Propositions \ref{propcuttorus} and \ref{propcuttorus2}, 
we obtain:
\begin{align}
&T^{-\Delta_\beta}e^{-c_{\rm L}(S_{\rm L}^0(\Sigma,g,g^*)+S(g)-S(g^*))}\int_{\R_+} e^{-\frac{2\pi}{T}\Delta_{Q+ip}}
 \partial_{\mu^{\rm b}_2}{\rm H}_{\mu^b_{2},\mu^b_{1},\mu^b_{1}}^{Q+ip,\beta, Q-ip} 
\tilde{\mc{F}}_{p}^{\mc{G}^*}(\boldsymbol{\beta},\nu^+) \dd p\label{identity_two_channel_torus}\\
& =\frac{2^{-\frac{1}{4}}\pi^{-\frac{3}{4}}Z^2_{\D^+,g_\D,{\rm m}}}{Z_{\D,g_\D}}\int_{\R} e^{-2\pi T\Delta_{Q+ip}}  
\tilde {\rm G}_{\mu,\mu^{\rm b}_1}(Q+ip,\beta)  \partial_{\mu^{\rm b}_2}{\rm U}_{\mu,\mu^{\rm b}_2}(Q-ip)
\tilde{\mc{F}}_{p}^{\mc{G}}( \bs \beta,\nu^+) \dd p\nonumber
\end{align}
\begin{align*}
\textrm{with } & S(g^*)=S^0_{\rm L}( \caP_1^*,(\Phi^*)^*g_{\rm dozz},g^*)+\sum_{k=1}^{3}S^0_{\rm L}(\psi_{k}^*(\D^+),g_{\rm dozz},(\psi_{k}^*)_*g_\D) \\
& S(g)=S_{\rm L}^0(\mc{P}\cap \H,(\Phi^{\mc{P}})^*g_{\rm dozz},g)+S_{\rm L}^0(\psi^{\mc{P}}_{1}(\D),g_{\rm dozz},(\psi^{\mc{P}}_{1})_*g_\D)+S_{\rm L}^0(\psi^{\mc{P}}_2(\D^+),g_{\rm dozz},(\psi^{\mc{P}}_2)_*g_\D).
\end{align*}
We now have 
\begin{lemma}
The following identity holds: 
\begin{equation}\label{identite_metric_constant}
\frac{2^{-\frac{1}{4}}\pi^{-\frac{3}{4}}Z^2_{\D^+,g_\D,{\rm m}}}{Z_{\D,g_\D}}e^{c_{\rm L}(S_{\rm L}^0(\Sigma,g,g^*)+S(g)-S(g^*))}=\frac{1}{\sqrt{2}\pi} e^{c_{\rm L}\frac{\pi}{12}(T-T^{-1})}.
\end{equation}
\end{lemma}
\begin{proof} Since the left hand side of \eqref{identite_metric_constant} is independent of $\nu^+$, it suffices to consider the case with $\nu^+=\emptyset$, and we will consider the case where $z=0$ is a marked point with real-valued weight $\beta\in (0,Q)$, and the metric are $g_0=|\dd z|^2$ and $g_0^*=(\frac{2\pi}{T})^2|\dd z|^2$. We then compute $\cjg V_{\beta,g_0}(0)M_\gamma^{{\rm b},g_0}(\pl_2A(\tau))\cjd_{A(\tau),g_0}$ using the two channel cuts. The result of \cite[Corollary 6.12]{WuConfBoot} shows that in the first channel
\begin{equation}\label{baojun_paper}
\cjg V_{\beta,g_0}(0)M_\gamma^{{\rm b},g_0}(\pl_2A(\tau))\cjd_{A(\tau),g_0}=-\frac{(2\pi)^{\Delta_\beta}e^{\frac{\pi T}{12}}}{2\sqrt{2}\pi}\int_{\R} e^{-\pi T\frac{p^2}{2}}  
\tilde {\rm G}_{\mu,\mu^{\rm b}_1}(Q+ip,\beta)  \partial_{\mu^{\rm b}_2}\widetilde{\rm U}_{\mu,\mu^{\rm b}_2}(Q-ip)
\tilde{\mc{F}}_{p}^{\mc{G}}( \bs \beta) \dd p.
\end{equation}
Let $\hat{\mc{P}}_1^*=[-1/2,1/2]\times [0,T/2]$, viewed as a building block with $z=0$ as a marked point.
The same argument as for proving \eqref{fusionblocktorus2} gives for $\beta\in (0,Q)$ and the admissible metric 
\begin{equation}\label{identity_2nd_channel}\begin{split}
\cjg V_{\beta,g^*_0}(0)M_\gamma^{{\rm b},g^*_0}(\pl_2A(\tau))\cjd_{A(\tau),g^*_0}=& -\partial_{\mu^{\rm b}_2}\mc{A}_{\Sigma,g_0^*,0,\beta}
\\
= & - C_*\int_{\R_+} e^{-\frac{2\pi}{T}\Delta_{Q+ip}}
 \partial_{\mu^{\rm b}_2}\widetilde{\rm H}_{\mu^b_{2},\mu^b_{1},\mu^b_{1}}^{Q+{\rm i}p^+,\beta, Q-{\rm i}p^+} 
\tilde{\mc{F}}_{p}^{\mc{G}^*}(\boldsymbol{\beta}) \dd p\end{split}
\end{equation}
with 
\[ C_*=
 \frac{(2\pi)^{\frac{3}{4}}}{4\pi Z_{\D^+,g_\D,{\rm m}}^2}\Big(\frac{{\rm v}_{g_{\rm dozz}}(\hat{\H})}{\det' (\Delta_{\hat{\H},g_{\rm dozz},{\rm N}})}\Big)^{\frac{1}{2}} 
e^{c_{\rm L}(S^0_{\rm L}(\hat{\mc{P}}_1^*,(\Phi^*)^*g_{\rm dozz},g_0^*)+\sum_{k=1}^{2}S^0_{\rm L}(\psi_{k}^*(\D^+),g_{\rm dozz},(\psi_{k}^*)_*g_\D))}.
 \]
 Write $\Phi^*=m\circ h$ where $h(z)=e^{2\pi(z-1/2)/T}$ and $m(z)=(z-e^{-\pi/T})/(z+e^{-\pi/T})$ is a M\"obius transformation. We have 
 $h(\hat{\mc{P}}_1^*)=\A^+_{e^{-2\pi/T}}$ and $h_*g_0^*=g_\A=|\dd z|^2/|z|^2$.
 Let $g'=u(z)|\dd z|^2$ be the metric on $\hat{\H}$ defined by 
 \[ u(z)= |z|^{-2}{\bf 1}_{e^{-2\pi/T}\leq |z|\leq 1} + |z|^{-4}{\bf 1}_{\hat{\H}\setminus \D^+}+e^{4\pi/T}{\bf 1}_{e^{-2\pi/T}\D},\]
 so that 
 \[S^0_{\rm L}(\hat{\mc{P}}_1^*,(\Phi^*)^*g_{\rm dozz},g_0^*)+\sum_{k=1}^{2}S^0_{\rm L}(\psi_{k}^*(\D^+),g_{\rm dozz},(\psi_{k}^*)_*g_\D)=S_{\rm L}^0(\hat{\H},g_{\rm dozz},m_*g').\]
Using that $m$ is a M\"obius map preserving $\H$, it is direct to see that $S_{\rm L}^0(\hat{\H},g_{\rm dozz},m_*g_{\rm dozz})=0$, and thus 
\[S_{\rm L}^0(\hat{\H},g_{\rm dozz},m_*g')=S_{\rm L}^0(\hat{\H},g_{\rm dozz},g').\]
Now $g'=e^{\omega}g_{\rm dozz}$ for $\omega(z)=(4\pi)/T$ if $|z|\leq e^{-2\pi/T}$, $\omega(z)=0$ for $|z|\geq 1$ and $\omega(z)=-2\log |z|$ in 
$\A^+_{e^{-2\pi/T}}$. The scalar curvature of $g_{\rm dozz}$ is supported in $\T^+$ and we then see that 
\[S_{\rm L}^0(\hat{\H},g_{\rm dozz},g')=\frac{1}{96\pi}\int_{\A^+_{e^{-2\pi/T}}}|\dd\omega|^2 \dd z=\frac{1}{24}\int_{e^{-2\pi/T}}^1\frac{1}{r} \dd r
=\frac{\pi}{12T}.
\]
We claim that the following identity holds 
\begin{equation}\label{calcul_det}
Z_{\D^+,g_\D,{\rm m}}^{-2}\Big(\frac{{\rm v}_{g_{\rm dozz}}(\hat{\H})}{\det'(\Delta_{\hat{\H},g_{\rm dozz},{\rm N}})}\Big)^{1/2}=(2\pi)^{\frac{1}{4}}
\end{equation}
Assuming this is true, we obtain  
\[C_*=e^{c_{\rm L}\frac{\pi }{12T}} \frac{(2\pi)^{\frac{3}{4}}}{4\pi Z_{\D^+,g_\D,{\rm m}}^2}\Big(\frac{{\rm v}_{g_{\rm dozz}}(\hat{\H})}{\det'(\Delta_{\hat{\H},g_{\rm dozz},{\rm N}})}\Big)^{\frac{1}{2}}=\frac{1}{2}e^{c_{\rm L}\frac{\pi }{12T}}.\]
Now we also have the identity 
\[\cjg V_{\beta,g^*_0}(0)M_\gamma^{{\rm b},g^*_0}(\pl_2A(\tau))\cjd_{A(\tau),g^*_0}=(\frac{2\pi}{T})^{-\Delta_\beta}\cjg V_{\beta,g_0}(0)M_\gamma^{{\rm b},g_0}(\pl_2A(\tau))\cjd_{A(\tau),g_0}>0\] 
thus 
we obtain, using \eqref{identity_2nd_channel}, 
\begin{align*}
& T^{\Delta_\beta}\frac{e^{\frac{\pi T}{12}}}{2\sqrt{2}\pi}\int_{\R} e^{-\pi T\frac{p^2}{2}}  
\tilde {\rm G}_{\mu,\mu^{\rm b}_1}(Q+ip,\beta)  \partial_{\mu^{\rm b}_2}\widetilde{\rm U}_{\mu,\mu^{\rm b}_2}(Q-ip)
\tilde{\mc{F}}_{p}^{\mc{G}}( \bs \beta) \dd p\\
& =\frac{1}{2}e^{\frac{\pi }{12T}} \int_{\R_+} e^{-\frac{\pi}{T}\frac{p^2}{2}}
 \partial_{\mu^{\rm b}_2}\widetilde{\rm H}_{\mu^b_{2},\mu^b_{1},\mu^b_{1}}^{Q+{\rm i}p^+,\beta, Q-{\rm i}p^+} 
\tilde{\mc{F}}_{p}^{\mc{G}^*}(\boldsymbol{\beta}) \dd p
\end{align*}
Consequently, by comparing with \eqref{identity_two_channel_torus} (with $\nu=\emptyset$) we obtain \eqref{identite_metric_constant}.

We finally prove \eqref{calcul_det}. Choose a radial smooth
metric $\hat{g}=e^{\sigma} g_\D$ on $\D^+$, with $\hat{g}=g_\D$ near $0$,
$\hat{g}=|\dd z|^2/|z|^2$ near $\T^+$, and $\sigma|_{\T^+}=0$.
It is admissible at the cutting boundary and Neumann
extendible at the real boundary. We glue $\hat{g}$ to its image under
$z\mapsto-1/z$, and  still denote the resulting admissible metric on $\hat\H$ by $\hat{g}$.
We use the gluing formula in \cite[Proposition B.3]{GRW1}: this gives 
\[ \frac{\det'(\Delta_{\hat{\H},\hat{g},{\rm N}})}{{\rm v}_{\hat{g}}(\hat{\H})}=\sqrt{\frac{2}{\pi}}\det(\Delta_{\D^+,\hat{g},{\rm m}})^2{\det}_{\rm Fr}({\bf D}_{\hat{\H},\T^+,0}(2{\bf D}_0)^{-1})\]
where ${\det}_{\rm Fr}$ denotes the Fredholm determinant, ${\bf D}_0$ is the Fourier multiplier by $\max(1,|n|)$ on Fourier modes  $\cos(n\theta)$ on $L^2(\T^+)$, and ${\bf D}_{\hat{\H},\T^+,0}$ is the sum of the Dirichlet-to-Neumann operator on $\T^+$ associated to the half circle $\T^+$ cutting $\hat{\H}$, with the spectral projector on its kernel. It is equal in this case to (here the factor $2$ comes from the fact $\hat{\H}$ contains two copies $\D^+$ and $\hat{\H}\setminus \D^+$ of the disk) 
\[   {\bf D}_{\hat{\H},\T^+,0}\cos(n\theta)=\max(1,2|n|)\cos(n\theta).\]
Then we get $\det_{\rm Fr}({\bf D}_{\hat{\H},\T^+,0}(2{\bf D}_0)^{-1})=1/2$. 
We thus have $Z_{\hat\H,\hat{g}}=(2\pi)^{1/4}Z_{\D^+,k,{\rm m}}^2$.
The conformal change from $g_{\rm dozz}$ to $\hat{g}$ is $\sigma$, which vanishes on $\T^+$. Additivity and conformal invariance of the Dirichlet energy give
\[Z_{\hat\H,\hat{g}}=e^{S_{\rm L}^0(\hat\H,g_{\rm dozz},\hat{g})}Z_{\hat\H,g_{\rm dozz}}, \quad Z^2_{\D^+,\hat{g},{\rm m}}=e^{2S_{\rm L}^0(\D^+,g_{\rm dozz},\hat{g})} Z^2_{\D^+,g_\D,{\rm m}}.\]
The two anomaly factors cancel, proving  \eqref{calcul_det}
\end{proof}
This shows the final result of this section:
\begin{corollary}\label{modular_inv_S}
The following holds for $\beta\in U$ with $|{\rm Re}(\beta)-Q|<Q$ and $\beta\not=Q$
\begin{align*}
&e^{\frac{\pi}{12T}}\int_{\R_+} e^{-\frac{\pi}{2T}p^2}
 \partial_{\mu^{\rm b}_2}\widetilde{\rm H}_{\mu^b_{2},\mu^b_{1},\mu^b_{1}}^{Q+ip,\beta, Q-ip} 
\tilde{\mc{F}}_{p}^{\mc{G}^*}(\boldsymbol{\beta},\nu^+) \dd p\\
&= \frac{T^{\Delta_\beta}}{\sqrt{2}\pi}e^{\frac{\pi T}{12}} \int_{\R} e^{-\pi T\frac{p^2}{2}}  
\tilde {\rm G}_{\mu,\mu^{\rm b}_1}(Q+ip,\beta)  \partial_{\mu^{\rm b}_2}\widetilde{\rm H}_{\mu,\mu^{\rm b}_2}(Q-ip)
\tilde{\mc{F}}_{p}^{\mc{G}}( \bs \beta,\bs \nu^+) \dd p
\end{align*}
\end{corollary}

\appendix
\appendix
\section{Special functions for structure constants}

\subsection{Barnes' double Gamma functions }
Here we introduce the special functions involved in the expressions for the various structure constants.

We set $l(z)=\frac{\Gamma (z)}{\Gamma (1-z)}$ where $\Gamma$ denotes the standard Gamma function. We introduce Zamolodchikov's special holomorphic function $\Upsilon_{\frac{\gamma}{2}}(z)$ by the following expression for $0<{\rm Re}(z)< Q$
\begin{equation}\label{def:upsilon}
	\log \Upsilon_{\frac{\gamma}{2}} (z)  = \int_{0}^\infty  \left ( \Big (\frac{Q}{2}-z \Big )^2  e^{-t}-  \frac{( \sinh( (\frac{Q}{2}-z )\frac{t}{2}  )   )^2}{\sinh (\frac{t \gamma}{4}) \sinh( \frac{t}{\gamma} )}    \right ) \frac{dt}{t}.
\end{equation}
The function $\Upsilon_{\frac{\gamma}{2}}$  is then defined on all $\C$ by analytic continuation of the expression \eqref{def:upsilon} as expression \eqref{def:upsilon}  satisfies the following remarkable functional relations: 
\begin{equation}\label{shiftUpsilon}
	\Upsilon_{\frac{\gamma}{2}} (z+\frac{\gamma}{2}) = l( \frac{\gamma}{2}z) (\frac{\gamma}{2})^{1-\gamma z}\Upsilon_{\frac{\gamma}{2}} (z), \quad
	\Upsilon_{\frac{\gamma}{2}} (z+\frac{2}{\gamma}) = l(\frac{2}{\gamma}z) (\frac{\gamma}{2})^{\frac{4}{\gamma} z-1} \Upsilon_{\frac{\gamma}{2}} (z).
\end{equation}
The function $\Upsilon_{\frac{\gamma}{2}}$ has no poles in $\C$ and the zeros of $\Upsilon_{\frac{\gamma}{2}}$ are simple (if $\gamma^2 \not \in \Q$) and given by the discrete set $(-\frac{\gamma}{2} \N-\frac{2}{\gamma} \N) \cup (Q+\frac{\gamma}{2} \N+\frac{2}{\gamma} \N )$. We use the convention $\N=\N_{\geq 0}$.

Now we introduce some variants, namely the functions $\Gamma_{\frac{\gamma}{2}}(x)$ and $S_{\frac{\gamma}{2}}(x)$. For all $\gamma \in (0,2) $ and for ${\rm Re}(x) >0$, $\Gamma_{\frac{\gamma}{2}}(x)$ is defined by the integral formula
\begin{equation}
	\log \Gamma_{\frac{\gamma}{2}}(x) = \int_0^{\infty} \frac{dt}{t} \left[ \frac{ e^{-xt} -e^{- \frac{Qt}{2}}   }{(1 - e^{- \frac{\gamma t}{2}})(1 - e^{- \frac{2t}{\gamma}})} - \frac{( \frac{Q}{2} -x)^2 }{2}e^{-t} + \frac{ x -\frac{Q}{2}  }{t} \right].
\end{equation} 
Since the function $\Gamma_{\frac{\gamma}{2}}(x)$ is continuous it is completely determined by the following two shift equations 
\begin{align}\label{eq:shift_G1}
	\frac{\Gamma_{\frac{\gamma}{2}}(x)}{\Gamma_{\frac{\gamma}{2}}(x + \frac{\gamma}{2}) }&= \frac{1}{\sqrt{2 \pi}}
	\Gamma(\frac{\gamma x}{2}) ( \frac{\gamma}{2} )^{ -\frac{\gamma x}{2} + \frac{1}{2}
	},  &
	\frac{\Gamma_{\frac{\gamma}{2}}(x)}{\Gamma_{\frac{\gamma}{2}}(x + \frac{2}{\gamma}) } = \frac{1}{\sqrt{2 \pi}} \Gamma(\frac{2
		x}{\gamma}) ( \frac{\gamma}{2} )^{ \frac{2 x}{\gamma} - \frac{1}{2} },
\end{align}
and by its value at $\frac{Q}{2}$, $\Gamma_{\frac{\gamma}{2}}(\frac{Q}{2} ) =1$. Furthermore, $x \mapsto \Gamma_{\frac{\gamma}{2}}(x)$ admits a meromorphic extension to all of $\mathbb{C}$ with single poles at $x = -n\frac{\gamma}{2}-m\frac{2}{\gamma}$ for $n,m\in\mathbb N$; the order of a pole is the number of lattice pairs $(n,m)$ representing that point (and hence the poles are simple when $\gamma^2\notin\mathbb Q$) and $\Gamma_{\frac{\gamma}{2}}(x)$ is never equal to $0$.  We also need the double sine function defined by:
\begin{equation} 
	S_{\frac{\gamma}{2}}(x) = \frac{\Gamma_{\frac{\gamma}{2}}(x)}{\Gamma_{\frac{\gamma}{2}}(Q -x)}.
\end{equation}
It obeys the following two shift equations:
\begin{align}\label{eq:shift_S1}
	\frac{S_{\frac{\gamma}{2}}(x+\frac{\gamma}{2})}{S_{\frac{\gamma}{2}}(x)} = 2\sin(\frac{\gamma\pi}{2}x), \quad\textrm{and}\quad \frac{S_{\frac{\gamma}{2}}(x+\frac{2}{\gamma})}{S_{\frac{\gamma}{2}}(x)} = 2\sin(\frac{2\pi}{\gamma}x).
\end{align}
The double sine function admits a meromorphic extension to $\mathbb{C}$ with poles at $x = -n\frac{\gamma}{2}-m\frac{2}{\gamma}$ and with zeros at $x = Q+n\frac{\gamma}{2}+m\frac{2}{\gamma}$ for any $n,m \in \mathbb{N}$.   Coincident lattice points are counted with their full multiplicities; in particular all these zeros and poles are simple when $\gamma^2\notin\mathbb Q$.  The above $\Upsilon_{\frac{\gamma}{2}}(x)=\frac{1}{\Gamma_{\frac{\gamma}{2}}(x)\Gamma_{\frac{\gamma}{2}}(Q-x)}$.


\subsection{DOZZ formula}\label{app:DOZZ}

With the previous notations, the DOZZ formula is defined for $\alpha_1,\alpha_2,\alpha_3 \in \C$ by the following formula where we set $\bar{\alpha}=\alpha_1+\alpha_2+\alpha_3$ 
\begin{equation}\label{theDOZZformula}
	C_{\gamma,\mu}^{{\rm DOZZ}} (\alpha_1,\alpha_2,\alpha_3 )  = (\pi \: \mu \:  l(\frac{\gamma^2}{4})  \: (\frac{\gamma}{2})^{2 -\gamma^2/2} )^{\frac{2 Q -\bar{\alpha}}{\gamma}}   \frac{\Upsilon_{\frac{\gamma}{2}}'(0)\Upsilon_{\frac{\gamma}{2}}(\alpha_1) \Upsilon_{\frac{\gamma}{2}}(\alpha_2) \Upsilon_{\frac{\gamma}{2}}(\alpha_3)}{\Upsilon_{\frac{\gamma}{2}}(\frac{\bar{\alpha}}{2}-Q) 
		\Upsilon_{\frac{\gamma}{2}}(\frac{\bar{\alpha}}{2}-\alpha_1) \Upsilon_{\frac{\gamma}{2}}(\frac{\bar{\alpha}}{2}-\alpha_2) \Upsilon_{\frac{\gamma}{2}}(\frac{\bar{\alpha}}{2} -\alpha_3)   }.
\end{equation}      
The DOZZ formula is meromorphic with poles corresponding to the zeroes of the denominator of the expression \eqref{theDOZZformula}.

\subsection{Bulk-boundary structure constants}\label{app:defG}
\label{app:physical_constants}
 The boundary
cosmological constants below are strictly positive; a vanishing
boundary constant is treated by a one-sided limit, whenever finite.
The first argument of
$\widetilde{\mathrm{G}}_{\mu,\mu^{\rm b}}(\alpha,\beta)$ is the
bulk insertion weight, and the second is the boundary insertion weight. Initially, the weights satisfy
$\alpha,\beta<Q$ and $\alpha+\beta/2>Q$; the formulas then extend
meromorphically in $(\alpha,\beta)$.

\paragraph{The case $\mu>0$.}
The boundary parameter $\sigma$ is related to $\mu^{\rm b}$ by
\begin{equation}\label{eq:mu_to_sigma}
 \mu^{\rm b}=\mu_{\rm b}(\sigma)
 =\sqrt{\frac{\mu}{\sin(\pi\gamma^2/4)}}
     \cos\bigl(\pi\gamma(\sigma-Q/2)\bigr).
\end{equation}
For $\mu^{\rm b}>0$, we choose the branch
\[
 \sigma\in
 \begin{cases}
 [Q/2,Q/2+1/(2\gamma)),
    &0<\mu^{\rm b}\leq\sqrt{\mu/\sin(\pi\gamma^2/4)},\\[1mm]
 Q/2+i[0,\infty),
    &\mu^{\rm b}\geq\sqrt{\mu/\sin(\pi\gamma^2/4)}.
 \end{cases}
\]
The two branches meet at $\sigma=Q/2$. For complex boundary cosmological constant
with ${\rm Re}(\mu^{\rm b})>0$, the corresponding choice of strip is
\[
       Q/2-1/(2\gamma)<{\rm Re}(\sigma)<Q/2+1/(2\gamma).
\]
In this parametrization the Hosomichi formula
\cite{Hosomichi,ang2024derivationstructureconstantsboundary},
in the normalization of \eqref{G2}, reads
\begin{align}\label{eq:tildeG_positive}
 \widetilde{\mathrm{G}}_{\mu,\mu^{\rm b}}(\alpha,\beta)
 =&2\pi\left(\pi\mu(\tfrac\gamma2)^{2-\gamma^2/2}
                    l(\tfrac{\gamma^2}{4})\right)^{\frac{Q-\alpha-\beta/2}{\gamma}}
 \frac{\Gamma_{\frac\gamma2}(2Q-\beta/2-\alpha)
       \Gamma_{\frac\gamma2}(\alpha-\beta/2)
       \Gamma_{\frac\gamma2}(Q-\beta/2)^3}
      {\Gamma_{\frac\gamma2}(Q-\alpha)\Gamma_{\frac\gamma2}(Q-\beta)
       \Gamma_{\frac\gamma2}(\alpha)\Gamma_{\frac\gamma2}(Q)
       \Gamma_{\frac\gamma2}(\beta/2)}\\
&\times \int_{\mathcal C}
 e^{2\pi i(Q-2\sigma) r}
 \frac{S_{\frac\gamma2}\big(\frac{\alpha+\beta/2-Q}{2}\pm r\big)}
      {S_{\frac\gamma2}\big(\frac{\alpha-\beta/2+Q}{2}\pm r\big)}
 \frac{\dd r}{i}.\nonumber
\end{align}
The factor
$2^{\Delta_\beta-2\Delta_\alpha}$ occurring in the fixed-point
correlation at $(i,0)$ is not part of the positional coefficient
$\widetilde{\mathrm{G}}$ in \eqref{G2}.

The contour $\mathcal C$ is oriented from $-i\infty$ to $i\infty$.
It separates the left pole families
\[
 -\frac{\alpha+\beta/2-Q}{2}-n\frac\gamma2-m\frac2\gamma,
 \qquad
 \frac{\alpha-\beta/2-Q}{2}-n\frac\gamma2-m\frac2\gamma
 \quad(n,m\in\mathbb N)
\]
from the right pole families
\[
 \frac{\alpha+\beta/2-Q}{2}+n\frac\gamma2+m\frac2\gamma,
 \qquad
 \frac{Q-\alpha+\beta/2}{2}+n\frac\gamma2+m\frac2\gamma.
\]
Here $\mathbb N$ includes zero. Sufficient strict conditions for
convergence at the two ends are
\[
 {\rm Re}(\beta)<2Q,\qquad
          {\rm Re}(\beta)/4<{\rm Re}(\sigma)<Q-{\rm Re}(\beta)/4.
\]

\paragraph{The case $\mu=0$.}
For $\mu^{\rm b}>0$, the  formula is
\cite[Definition 1.5 and Theorem 1.7]{RemyZhu20}
\begin{equation}\begin{split}\label{eq:tildeG_zero}
 \widetilde{\mathrm{G}}_{0,\mu^{\rm b}}(\alpha,\beta)
 =& \frac2\gamma\,
 \Gamma\Big(\frac{2\alpha+\beta-2Q}{\gamma}\Big)
 \Big(\frac{2\pi\mu^{\rm b}}{\Gamma(1-\gamma^2/4)}
       \Big)^{\frac{2Q-2\alpha-\beta}{\gamma}}
 \Gamma\Big(\frac{\gamma\alpha}{2}+\frac{\gamma\beta}{4}
                                      -\frac{\gamma^2}{4}\Big)\\
    & \times  \frac{\Gamma_{\frac\gamma2}(\alpha-\beta/2)
       \Gamma_{\frac\gamma2}(\alpha+\beta/2)
       \Gamma_{\frac\gamma2}(Q-\beta/2)^2}
      {\Gamma_{\frac\gamma2}(Q-\beta)
       \Gamma_{\frac\gamma2}(\alpha)^2\Gamma_{\frac\gamma2}(Q)}.
\end{split}\end{equation}
Here the boundary cosmological constant is expressed through $\sigma$,
we use the exponential parametrization
\begin{equation}\label{eq:G_zero_sigma}
 \mu^{\rm b}=e^{i\pi\gamma(\sigma-\frac{Q}{2})},\qquad
 \sigma=\frac{Q}{2}+\frac{\log\mu^{\rm b}}{i\pi\gamma}
                 \in \frac{Q}{2}+i\mathbb R.
\end{equation}
For complex $\mu^{\rm b}$ with ${\rm Re}(\mu^{\rm b})>0$, the principal
logarithm gives
$Q/2-1/(2\gamma)<{\rm Re}(\sigma)<Q/2+1/(2\gamma)$.
This is not the cosine parametrization at $\mu=0$.
Formula \eqref{eq:tildeG_zero} includes the zero-mode factor and is
not defined by a pointwise $\mu\downarrow0$ limit on $Q+i\mathbb R$.

\subsection{Boundary three-point structure constant}\label{app:defH}
The coefficient
$\widetilde{\mathrm{H}}_{\mu_1,\mu_2,\mu_3}^{(\beta_1,\beta_2,\beta_3)}$
is the full coefficient in \eqref{G3}; the  bulk cosmological constant
$\mu$ is suppressed in the subscript. With the marked points at
$0,1,\infty$, the boundary constants $\mu_1,\mu_2,\mu_3$ are those
on $(-\infty,0),(0,1),(1,\infty)$, respectively. The initial Seiberg
range is $\beta_j<Q$ and $\beta_1+\beta_2+\beta_3>2Q$.
The notation $f(a\pm b)$ and $f(a\pm b\pm c)$ denotes the product
convention of Section~A.1.

\paragraph{The case $\mu>0$.}
For each $j=1,2,3$, set
\begin{equation}\label{eq:mu-sigma}
 \mu_j=\mu_{\rm b}(\sigma_j)
 =\sqrt{\frac{\mu}{\sin(\pi\gamma^2/4)}}
      \cos\bigl(\pi\gamma(\sigma_j-Q/2)\bigr).
\end{equation}
For positive real $\mu_j$, the branch is chosen componentwise as
\[
 \sigma_j\in
 \begin{cases}
 [Q/2,Q/2+1/(2\gamma)),
    &0<\mu_j\leq\sqrt{\mu/\sin(\pi\gamma^2/4)},\\[1mm]
 Q/2+i[0,\infty),
    &\mu_j\geq\sqrt{\mu/\sin(\pi\gamma^2/4)}.
 \end{cases}
\]
For complex $\mu_j$ with ${\rm Re}(\mu_j)>0$, we use
$Q/2-1/(2\gamma)<{\rm Re}(\sigma_j)<Q/2+1/(2\gamma)$.
The Ponsot--Teschner formula
\cite{PT_boundary_3pt,ang2024derivationstructureconstantsboundary}
is
\begin{equation}\label{formule_PT}
\begin{split}
 \widetilde{\mathrm{H}}_{\mu_1,\mu_2,\mu_3}^{(\beta_1,\beta_2,\beta_3)}
 =&2\pi\left(\pi\mu(\tfrac\gamma2)^{2-\gamma^2/2}
                    l(\tfrac{\gamma^2}{4})\right)^{\frac{2Q-\beta_1-\beta_2-\beta_3}{2\gamma}}
\\
 &\quad\times
 \frac{\Gamma_{\frac\gamma2}
       \left(Q-\frac{\beta_2}{2}\pm\frac{Q-\beta_1}{2}\pm\frac{Q-\beta_3}{2}\right)}
      {\Gamma_{\frac\gamma2}(Q)\prod_{j=1}^3\Gamma_{\frac\gamma2}(Q-\beta_j)}
 \frac1{S_{\frac\gamma2}\left(\frac{\beta_3}{2}-\sigma_1+\frac Q2
                                      \pm(\frac Q2-\sigma_3)\right)
        S_{\frac\gamma2}\left(\frac{\beta_1}{2}+\sigma_1-\frac Q2
                                      \pm(\frac Q2-\sigma_2)\right)}\\
 &\quad\times\int_{\mathcal C}
 \frac{S_{\frac\gamma2}\left(\frac{Q-\beta_2}{2}+\sigma_3
                                      \pm(\frac Q2-\sigma_2)+r\right)
       S_{\frac\gamma2}\left(\frac Q2\pm\frac{Q-\beta_3}{2}
                                      +\sigma_3-\sigma_1+r\right)}
      {S_{\frac\gamma2}\left(\frac{3Q}{2}\pm\frac{Q-\beta_1}{2}
                         -\frac{\beta_2}{2}+\sigma_3-\sigma_1+r\right)
       S_{\frac\gamma2}(2\sigma_3+r)S_{\frac\gamma2}(Q+r)}
       \frac{\dd r}{i}.
       \end{split}
\end{equation}

The contour $\mathcal C$ goes from $-i\infty$ to $i\infty$ and
separates the poles of the numerator double sines from the poles
arising from zeros of the denominator double sines. More precisely,
a numerator factor $S_{\gamma/2}(a+r)$ contributes left poles
$-a-n\gamma/2-2m/\gamma$, and a denominator factor
$S_{\gamma/2}(a+r)$ contributes right poles
$Q-a+n\gamma/2+2m/\gamma$, for $n,m\in\mathbb N$.
Here $a$ denotes 
the printed argument of the relevant factor.
The contour passes to the right of the left poles and to the left
of the right poles, and is continued by deformation when the
parameters vary; see
\cite[Section 2.1]{ang2024derivationstructureconstantsboundary}.

\paragraph{The case $\mu=0$.}
For $\mu_j>0$, the parameters are now given by
\begin{equation}\label{eq:tildeH_zero_parameters}
 \mu_j=e^{i\pi\gamma(\sigma_j-Q/2)},\qquad
 \sigma_j=\frac{Q}{2}+\frac{\log\mu_j}{i\pi\gamma}
                     \in \frac{Q}{2}+i\mathbb R,\quad j=1,2,3.
\end{equation}
In particular, ${\rm Re}(\sigma_j)=Q/2$ and
${\rm Im}(\sigma_j)=-\log\mu_j/(\pi\gamma)$ can have either sign.
For complex $\mu_j$ with ${\rm Re}(\mu_j)>0$, the principal logarithm gives
\[
                   Q/2-1/(2\gamma)<{\rm Re}(\sigma_j)<Q/2+1/(2\gamma).
\]
With these parameters, the zero-bulk formula is
\begin{equation}\begin{split}
\label{eq:tildeH_zero}
 \widetilde{\mathrm{H}}_{\mu_1,\mu_2,\mu_3}^{(\beta_1,\beta_2,\beta_3)}
 =&\Big(\frac{2\pi(\gamma/2)^{1-\frac{\gamma^2}{4}}}
                   {\Gamma(1-\frac{\gamma^2}{4})}
      \Big)^{\frac{2Q-\beta_1-\beta_2-\beta_3}{\gamma}} 
 \frac{ 2\pi\Gamma_{\frac\gamma2}
       \left(Q-\frac{\beta_2}{2}\pm\frac{Q-\beta_1}{2}\pm\frac{Q-\beta_3}{2}\right)}
      {\Gamma_{\frac\gamma2}(Q)\prod_{j=1}^3\Gamma_{\frac\gamma2}(Q-\beta_j)
       S_{\frac\gamma2}(\frac{\beta_1}{2}+\sigma_1-\sigma_2)
       S_{\frac\gamma2}(\frac{\beta_3}{2}+\sigma_3-\sigma_1)}
\\
 &\quad\times e^{\frac{i\pi}{2}
 \big[-(2Q-\frac{\beta_1}{2}-\sigma_1-\sigma_2)
             (Q-\frac{\beta_1}{2}-\sigma_1-\sigma_2)+(Q+\frac{\beta_2}{2}-\sigma_2-\sigma_3)
             (\frac{\beta_2}{2}-\sigma_2-\sigma_3)+(Q+\frac{\beta_3}{2}-\sigma_1-\sigma_3)
             (\frac{\beta_3}{2}-\sigma_1-\sigma_3)
       -2\sigma_3(2\sigma_3-Q)\big]}\\
 &\quad\times\int_{\mathcal C}e^{i\pi(-\beta_2/2+\sigma_2-\sigma_3)r}
 \frac{S_{\frac\gamma2}(Q-\frac{\beta_2}{2}+\sigma_3-\sigma_2+r)
       S_{\frac\gamma2}\left(\frac Q2\pm\frac{Q-\beta_3}{2}
                                      +\sigma_3-\sigma_1+r\right)}
      {S_{\frac\gamma2}\left(\frac{3Q}{2}\pm\frac{Q-\beta_1}{2}
                         -\frac{\beta_2}{2}+\sigma_3-\sigma_1+r\right)
       S_{\frac\gamma2}(Q+r)}\frac{\dd r}{i}.
       \end{split}
\end{equation}
This is \cite[Equation (1.26)]{RemyZhu20} with the zero-mode factor
$\frac2\gamma\Gamma((\beta_1+\beta_2+\beta_3-2Q)/\gamma)$
from Definition 1.5 restored. The Gamma factors cancel, leaving the
prefactor displayed in \eqref{eq:tildeH_zero}.

The contour obeys the same pole-separation rule as above, now for
the three numerator and three denominator factors appearing in
\eqref{eq:tildeH_zero}. An initial sufficient tail condition is
\[
               {\rm Re}(Q-\sigma_3+\sigma_2-\beta_2/2)>0;
\]
it holds for positive real couplings and real $\beta_2<Q$.
See \cite[Section 5.4.3]{RemyZhu20} for continuation of this integral.

\subsection{Bulk one-point formula}\label{app:FZZ}
The coefficient $\widetilde{\mathrm{U}}_{\mu,\mu^{\rm b}}(\alpha)$
is defined by \eqref{probfzz}, with positional denominator
$|x-\bar x|^{2\Delta_\alpha}$. The renormalized probabilistic
formula is initially used for $2/\gamma<\alpha<Q$; the expressions
below specify its meromorphic continuation in $\alpha$.

\paragraph{The case $\mu>0$.}
As in \eqref{eq:mu_to_sigma}, the boundary coupling $ \mu^{\rm b}$ and $\sigma$ are the same as Equation~\ref{eq:mu_to_sigma} and take the same physical branch.
For ${\rm Re}(\mu^{\rm b})>0$, the complex-parameter strip is
$Q/2-1/(2\gamma)<{\rm Re}(\sigma)<Q/2+1/(2\gamma)$.
Writing $s=2\sigma-Q$ as in the FZZ formula, this is equivalently
\begin{equation}\label{eq:def_s}
 \cos\frac{\pi\gamma s}{2}
  =\frac{\mu^{\rm b}}{\sqrt\mu}\sqrt{\sin\frac{\pi\gamma^2}{4}},
 \qquad
 s\in\begin{cases}
 [0,1/\gamma),& (\mu^{\rm b})^2\sin(\pi\gamma^2/4)\leq\mu,\\
 i[0,\infty),& (\mu^{\rm b})^2\sin(\pi\gamma^2/4)\geq\mu.
 \end{cases}
\end{equation}
At the common endpoint one has $\sigma=Q/2$ and $s=0$.
The FZZ formula \cite{FZZ,AngRemySun21_FZZ} in the normalization
of \eqref{probfzz} is
\begin{align}\label{eq:tildeU_positive}
 \widetilde{\mathrm{U}}_{\mu,\mu^{\rm b}}(\alpha)
 ={}&\frac4\gamma
 \left(\pi\mu\,l(\tfrac{\gamma^2}{4})\right)^{\frac{Q-\alpha}{\gamma}}
 \Gamma\left(\frac{\gamma\alpha}{2}-\frac{\gamma^2}{4}\right)
 \Gamma\left(\frac{2(\alpha-Q)}\gamma\right)
 \cos\bigl((\alpha-Q)\pi s\bigr).
\end{align}
At $s=0$, coupling derivatives of the full expression are interpreted
by their removable limits. 

\paragraph{The case $\mu=0$.}
For $\mu^{\rm b}>0$, the zero-bulk formula is
\cite[Definition 1.5 and Theorem 1.6]{RemyZhu20}
\begin{align}\label{eq:tildeU_zero}
 \widetilde{\mathrm{U}}_{0,\mu^{\rm b}}(\alpha)
 ={}&\frac2\gamma
 \Gamma\left(\frac{2(\alpha-Q)}\gamma\right)
 \Gamma\left(\frac{\gamma\alpha}{2}-\frac{\gamma^2}{4}\right)
 \left(\frac{2\pi\mu^{\rm b}}{\Gamma(1-\gamma^2/4)}
       \right)^{\frac{2(Q-\alpha)}\gamma}.
\end{align}
The boundary coupling can also be parametrized by
\[
 \mu^{\rm b}=e^{i\pi\gamma(\sigma-Q/2)},\qquad
 \sigma=Q/2+\frac{\log\mu^{\rm b}}{i\pi\gamma}\in Q/2+i\mathbb R.
\]
For ${\rm Re}(\mu^{\rm b})>0$, using the principal logarithm gives the
same open strip
$Q/2-1/(2\gamma)<{\rm Re}(\sigma)<Q/2+1/(2\gamma)$.
The closed formula \eqref{eq:tildeU_zero} uses $\mu^{\rm b}$ directly;
it does not use the cosine parameter $s$ and is not defined by a
pointwise $\mu\downarrow0$ limit on $Q+i\mathbb R$.

In the normalization of Section~\ref{sub:structure}, both regimes satisfy
\begin{equation}\label{eq:tilde_screening}
 \partial_{\mu^{\rm b}}\widetilde{\mathrm{U}}_{\mu,\mu^{\rm b}}(\alpha)
 =-2\pi\widetilde{\mathrm{G}}_{\mu,\mu^{\rm b}}(\alpha,\gamma).
\end{equation}
Indeed, differentiation inserts the boundary length; then
$\Delta_\gamma=1$ and
$\int_{\mathbb R}|x-r|^{-2}\dd r=2\pi/|x-\bar x|$ give the identity
from \eqref{G2} and \eqref{probfzz}. It extends meromorphically in
$\alpha$.

\section{Proof of the spectral resolution of the Liouville Hamiltonians} \label{sec:hamilton}

In this section, we will use the method of \cite[Section 6]{GKRV20_bootstrap} to prove the spectral resolution of the Hamiltonian of the boundary Liouville theory. In that reference, the proof is quite general and used mostly functional analytic arguments which turn out to apply verbatim to our operator ${\bf H}^{+}$.  For this reason, we write a proof along the lines of \cite[Section 6]{GKRV20_bootstrap} but we decided to deal with a rather general and abstract setting, that contains the case of both the boundary and the bulk Hamiltonian, but that could also be useful for other applications. It can be viewed as an extension of that paper, and 
as there are many aspects of the proof which are exactly the same as in \cite[Section 6]{GKRV20_bootstrap}, we only focus on the new features and refer to that paper for other technical details. We also notice that our proof here simplifies the proof of \cite[Section 6]{GKRV20_bootstrap} for the treatment of the $\{c\geq 0\}$ region.

\subsection{Assumption on the Hamiltonian.}\label{s:assumptions} 
Let $(\Omega,\mathbb{P})$ be a probability space with the space $L^2(\Omega,\mathbb{P})$ separable  and let
$\hat{\mc{H}}=L^2(\R \times \Omega;dc\otimes\mathbb{P})$. We will consider a Hamiltonian of the form 
\begin{equation}\label{H+} 
	\hat{{\bf H}}= \hat{{\bf H}}^{0}+ \sum_{j=1}^J e^{\gamma_j c}V_j
\end{equation}
where $\gamma_j>0$, $\hat{{\bf H}}^{0}$ is an unbounded self-adjoint operator on $\mc{H}^+$ having the form
\[ \hat{{\bf H}}^0=-\pl_c^2+\frac{Q^2}{4}+\hat{{\bf P}}\]
for some $Q>0$,  $\hat{{\bf P}}$ is a self-adjoint non-negative operator with discrete spectrum on $L^2(\Omega,\P)$ 
denoted by $\sigma(\hat{{\bf P}})=\{\la_j\geq 0 \,|\, \la_j<\la_{j+1}\}$ 
and associated to a closed non-negative quadratic form $\mc{Q}_{\hat{{\bf P}}}$ on $L^2(\Omega)$, 
$V_j$ are operators defined by non-negative quadratic forms $\mc{Q}_{V_j}$ defined on a dense subset of $L^2(\Omega)$. Moreover we assume that $\hat{{\bf H}}^{0}$ and $\hat{{\bf H}}$ are the self-adjoint extension (see \cite[Theorem VIII.15]{Reed-Simon}) associated to some closed positive quadratic forms given respectively by 
\begin{align*}
	& \hat{\mc{Q}}^{0}(u)=\int_{\R} \E[|\pl_cu|^2]+\frac{Q^2}{4} \E[u^2]+\mc{Q}_{\hat{{\bf P}}}(u) dc,\\
	& \hat{\mc{Q}}(u)=\hat{\mc{Q}}^{0}(u)+\sum_{j=1}^J \int_\R e^{\gamma_j c}\mc{Q}_{V_j}(u)dc.
\end{align*} 
We denote by $\mc{D}(\mc{Q}_{\hat{{\bf P}}}),\mc{D}(\mc{Q}_{V_j})\subset L^2(\Omega)$ and $\mc{D}(\hat{\mc{Q}}^{0}),\mc{D}(\hat{\mc{Q}})\subset \hat{\mc{H}}$ 
the domains of these forms and let $\mc{D}'(\hat{\mc{Q}}^{0})$ and $\mc{D}'(\hat{\mc{Q}})$ be the respective duals of $\mc{D}(\hat{\mc{Q}}^0)$ and $\mc{D}(\hat{\mc{Q}})$.
Let 
\begin{equation}
	E_k:=\{F\in L^2(\Omega)\,|\, 1_{[0,\la_k]}(\hat{{\bf P}})F=F\}=\bigoplus_{j=0}^{k}\ker(\hat{{\bf P}}-\la_j)
\end{equation}
be the sum of eigenspaces with eigenvalues of $\hat{{\bf P}}$ less or equal to $\la_k$ and $\Pi_{k}: L^2(\Omega)\to E_k$ the orthogonal projection. For convenience (and since this is our case for the boundary Liouville Hamiltonian), we will also assume that $\ker \hat{{\bf P}}=\R$ and $\la_0=0$. 
The spectrum of $\hat{{\bf H}}$ is contained in $[Q^2/4,\infty)$ and 
we consider for $\alpha\in \C$ the spectral parameter 
\begin{equation}\label{Deltaalpha}
	\Delta_\alpha=\frac{\alpha}{2}(Q-\frac{\alpha}{2}).
\end{equation}
For ${\rm Re}(\alpha)<Q$, $\Delta_{\alpha}\in \C\setminus [Q^2/4,\infty)$  we define the resolvent 
\[ \hat{{\bf R}}(\alpha):=(\hat{{\bf H}}-\Delta_\alpha)^{-1} : \hat{\mc{H}}\to \mc{D}(\hat{{\bf H}})\]
is thus bounded and analytic with respect to $\alpha$ if $\mc{D}(\hat{{\bf H}})\subset \mc{D}(\hat{\mc{Q}})$ denotes the domain of $\hat{{\bf H}}$. The resolvent also maps $\mc{D}'(\hat{\mc{Q}})\to \mc{D}(\hat{\mc{Q}})$ continuously. 

For $j=1,2$, let $\chi_j\in C^\infty(\R)$ with ${\rm supp}(\chi)_j\in (-\infty,0)$ and equal to $1$ near $(-\infty,1)$, 
we  make the following assumptions:
\begin{align}
	&  (H_1) \qquad  E_k \subset L^p(\Omega) , \quad \forall p<\infty,   \label{LpEk} \\  
	& (H_2)\qquad  E_k\subset \mc{D}(\mc{Q}_{V_j}), \quad i.e.\,  \exists C_k>0, \forall u\in E_k, \,  \mc{Q}_{V_j}(u)\leq C_k\|u\|_{L^2(\Omega)}^2. \label{domainE_k}\\
	& (H_3) \qquad (1-\chi_1)\hat{{\bf R}}(\alpha_0)(1-\chi_2)  \textrm{ is compact as a map }\hat{\mc{H}}\to \hat{\mc{H}}.
\end{align}   
for some $\alpha_0\in (-\infty,Q)$ such that $|\alpha_0-Q|/2< \min_j \gamma_j/2$

We note that Lemma 6.1 of \cite{GKRV20_bootstrap} applied to the operator $e^{\gamma_jc}V_{j}$  gives the following:
for $\chi\in L^\infty(\R)\cap C^\infty(\R)$ with support in $(-\infty,A)$ for some $A\in\R$ and $\chi'\in L^\infty(\R)$, 
then, for all $\beta\geq -\min_{j}\gamma_j/2$ and $\beta'\in\R$, the following operators are bounded
\begin{equation}\label{boundedness_family}
	\begin{gathered}
		\chi \Pi_k: \mc{D}(\hat{\mc{Q}})\to \mc{D}(\hat{\mc{Q}}) , \quad \chi \Pi_k: \mc{D}'(\hat{\mc{Q}})\to \mc{D}'(\hat{\mc{Q}})\\
		\chi e^{\gamma_j c}V_j\Pi_k : e^{\beta'c}\hat{\mc{H}}\to e^{(\beta'-\beta) c}\mc{D}'(\hat{\mc{Q}}),\\
		\chi e^{\gamma_j c}\Pi_k V_j: e^{(\beta-\beta') c}\mc{D}(\hat{\mc{Q}})\to e^{-\beta' c}L^2(\R;E_k),\\
		\chi \Pi_k e^{\gamma_j c}V_j\Pi_k: e^{\beta'c}\hat{\mc{H}}\to e^{(\beta'-\beta) c}L^2(\R;E_k).
	\end{gathered}
\end{equation}
If $\beta> -\gamma_j/2$ then one also has
\begin{equation}\label{boundedness_anotherone}
	\chi e^{\gamma_j c}V_j\Pi_k : e^{\beta'c}L^\infty(\R\times \Omega)\to e^{(\beta'-\beta) c}\mc{D}'(\hat{\mc{Q}}).
\end{equation}  

\subsection{The boundary Hamiltonian satisfies the assumptions}

In \cite[Section 7.2.1.]{GRW1}, it is shown that the boundary free Hamiltonian $\hat{{\bf H}}^0=\bH^{+,0}$ satisfies the desired properties of 
Section \ref{s:assumptions} with $(\Omega,\P)=(\Omega_{\T^+},\P^+)$, $\la_k=k\in \N_0$, $E_k$ is a finite dimensional space of Hermite polynomials which belong to $L^p(\Omega)$. 
In \cite[Section 7.2.2.]{GRW1} it is shown that  under the assumption 
\[ (\mu=0, \gamma\in (0,2)) \textrm{ or } (\mu>0, \gamma\in (0,\sqrt{2})),\]
the boundary Hamiltonian ${\bf H}^{+}$ is the self-adjoint extension of a quadratic form $\mc{Q}^+=\mc{Q}^{+,0}+\sum_{j=1}^3e^{\gamma_j c}\mc{Q}_{V_j}$ with 
\[\gamma_1=\gamma, \quad \gamma_2=\gamma_3=\gamma/2, \quad Q=\gamma/2+2/\gamma.\]
The assumption $(H_2)$ is shown to be satisfied for the boundary Hamiltonian $\hat{{\bf H}}={\bf H}^{+}$ in \cite[Lemma 7.8]{GRW1}.

We now prove the assumption $(H_3)$ also holds for the boundary Hamiltonian.
\begin{lemma}\label{R_+(0)}
	Let ${\bf H}^{+}$ be the boundary Liouville Hamiltonian of Section \ref{s:Hamiltonian_LCFT_bdry} and ${\bf R}^+(\alpha)=({\bf H}^+-\Delta_\alpha)^{-1}$. Then for ${\rm Re}(\alpha_0)<Q$, its resolvent satisfies that  $(1-\chi_1){\bf R}^+(\alpha_0)(1-\chi_2)$ is compact as a map $\mc{H}^+\to \mc{H}^+$.
\end{lemma}
\begin{proof} We will proceed similarly to \cite[Corollary 3.8]{GGV}. We write 
	\[\begin{split}
		(1-\chi_1){\bf R}^+(\alpha_0)(1-\chi_2)=& \int_0^\infty (1-\chi_1)e^{-t{\bf H}^{+}}(1-\chi_2)e^{-t\Delta_{\alpha_0}}dt\\
		& =\lim_{\eps\to 0}\int_\eps^{\eps^{-1}} (1-\chi_1)e^{-t{\bf H}^{+}}(1-\chi_2)e^{-t\Delta_{\alpha_0}}dt.
	\end{split}\]
	where the limit is in the sense of bounded operators $\mc{H}^+\to \mc{H}^+$ (this follows from the spectral theorem).
	To prove that $(1-\chi_1){\bf R}^+(\alpha_0)(1-\chi_2)$ is compact, it suffices to show that $(1-\chi_1)e^{-t{\bf H}^{+}}(1-\chi_2)$ is Hilbert-Schmidt with Hilbert-Schmidt 
	norm uniformly bounded by some constant $C(\eps)$ for $t\in [\eps,\eps^{-1}]$. The proof of this fact follows the argument of \cite[Lemma 3.7]{GGV}. Indeed, for $q_t:=e^{-t}$ we write the amplitude of $e^{-t{\bf H}^{+}}$ as 
	\[\mc{A}_{\A^+_{q_t}}(c,\varphi^{\rm h},c',{\varphi^{\rm h}}')=\mc{A}^0_{\A^+_{q_t}}(c,\varphi^{\rm h},c',{\varphi^{\rm h}}')\E_{\varphi,\varphi'}[e^{-\mu\int_{\A^+_{q_t}}|x|^{-\gamma Q}M_\gamma(\phi_{\A^+_{q_t}},dx) -\mu^{\rm L}W_{\rm L}(\phi_{\A^+_{q_t}})-\mu^{\rm R} W_{\rm R}(\phi_{\A^+_{q_t}})} ]\]
	where 
	\[W_{\rm L}(\phi_{\A^+_{q_t}}):=\int_{\mathbb{I}_t^{\rm L}}|x|^{-\gamma Q}M_{\gamma,\pl}(\phi_{\A^+_{q_t}},dx), \quad W_{\rm R}(\phi_{\A^+_{q_t}}):=\int_{\mathbb{I}_t^{\rm R}}|x|^{-\gamma Q}M_{\gamma,\pl}(\phi_{\A^+_{q_t}},dx).\]
	Let us first deal with the case $\mu>0$. One can write 
	\begin{equation}\label{int_kernel_prop}
		|\mc{A}_{\A^+_{q_t}}(c,\varphi,c',\varphi')|\leq \mc{A}^0_{\A^+_{q_t}}(c,\varphi,c',\varphi')\E_{\varphi,\varphi'}[e^{-\mu\int_{\A^+_{q_t}}|x|^{-\gamma Q}M_\gamma(\phi_{\A^+_{q_t}},dx) }].
	\end{equation}
	Here $\phi_{\A^+_{q_t}}=X_{\A^+_{q_t},m}+P(c+\varphi^{\rm h},c'+{\varphi^{\rm h}}')$ where $X_{\A^+_{q_t},m}$ is the GFF on $\A^+_{q_t}$ with mixed condition (Neumann on $\mathbb{I}_t^{{\rm R}/{\rm L}}$ and Dirichlet on $\T^+$) and $P_{\A_{q_t}}(c+\varphi^{\rm h},c'+{\varphi^{\rm h}}')$ is the harmonic function in $\A^+_{q_t}$ with boundary values $c+\varphi^{\rm h}$ on $\T^+$ and $c'+{\varphi'}^{\rm h}$ on $q_t\T^+$, with Neumann condition on $\mathbb{I}_t^{{\rm R}/{\rm L}}$.
	The proof of \cite[Lemma 3.7]{GGV} shows that for $p\in [2,1+e^{t})$ there is $C(t)>0$ independent of $c,c'$ such that 
	\begin{equation}\label{boundA^0}
		\|\mc{A}^0_{\A^+_{q_t}}(c,\cdot,c',\cdot)\|_{L^p(\Omega_{\T^+}^2)} \leq C(t)e^{-\frac{(c-c')^2}{4t}} 
	\end{equation}
	We can now bound the potential part in \eqref{int_kernel_prop}. By the maximum principle, if $\bar{c}=(c+c')/2$, 
	$|P_{\A_{q_t}}(c,c')-\bar{c}|\leq |c-c'|$. Using that $e^{-x}\leq C_\theta x^\theta $ for each $\theta>0$, we bound 
	\[ \E_{\varphi,\varphi'}[e^{-\mu\int_{\A^+_{q_t}}|x|^{-\gamma Q}M_\gamma(\phi_{\A^+_{q_t}},\dd x) }]\leq C_\theta \mu^{-\theta}e^{\gamma \theta \bar{c}} \E_{\varphi,\varphi'}\Big[\Big(\int_{\A^+_{q_t}}|x|^{-\gamma Q}M_\gamma(\phi_{\A^+_{q_t}},\dd x)\Big)^{-
		\theta}\Big]\]
	If $B\subset \A^+_{q_t}$ is a ball not intersecting $\pl\A^+_{q_t}$, the random variable $\inf_{x\in B}P_{\A^+_{q_t}}(\varphi^{\rm h},{\varphi^{h}}')$ has finite moments of any order, and by \cite[Prop. 3.6.]{Robert_Vargas} we have for some constant $C_{\theta,t}$ uniformly bounded for $t$ in compact sets of $(0,\infty)$
	\[\E_{\varphi,\varphi'}\Big[\Big(\int_{B}|x|^{-\gamma Q}M_\gamma(X_{\A^+_{q_t},D},dx)\Big)^{-
		\theta}\Big]\leq C_{\theta,t}\]
	For any $q\in [1,\infty)$, one has a constant $C_{\theta,t,q}>0$ uniformly bounded for $t$ in compact sets of $(0,\infty)$
	such that 
	\[ \E\Big[\Big(\E_{\varphi,\varphi'}[e^{-\mu\int_{\A^+_{q_t}}|x|^{-\gamma Q}M_\gamma(\phi_{\A^+_{q_t}},dx) }]\Big)^q \Big]^{1/q} \leq 
	C_{\theta,t,q}e^{-\gamma \theta \bar{c}}e^{\gamma \theta |c-c'|}\]
	which gives, by taking $1/q+1/p=1$ with $p\in (2,1+e^{t})$ and using H\"older inequality and \eqref{boundA^0}
	\[  \|\mc{A}_{\A^+_{q_t}}(c,\cdot,c',\cdot)\|_{L^2(\Omega_{\T^+}^2)}\leq C'_{\theta,t,q} e^{-\gamma \theta \bar{c}+\gamma \theta |c-c'|-\frac{(c-c')^2}{4t}}\]
	which is in $L^2((-1,\infty)^2,dc\otimes dc')$ with norm uniformly bounded for $t$ in compact sets of $(0,\infty)$. 
	Thus $(1-\chi_1)(c)(1-\chi_2)(c')\mc{A}_{\A^+_{q_t}}(c,c',\varphi^{\rm h},{\varphi^{\rm h}}')$ is in $L^2((\R \times \Omega_{\T^+})^2)$ with uniformly bounded for $t$ in compact sets of $(0,\infty)$, 
	means that $(1-\chi_1)e^{-t{\bf H}^{+}}(1-\chi_2)$ is Hilbert-Schmidt with Hilbert-Schmidt norm uniformly bounded by some constant $C(\eps)$ for $t\in [\eps,\eps^{-1}]$.
	
	When $\mu=0$,  but $\mu^{\rm L}+\mu^{\rm R}>0$ we proceed similarly: we take now an open interval 
	$B\subset \mathbb{I}_t^{{\rm R}/{\rm L}}$ not intersecting the boundary of $\mathbb{I}_t^{{\rm R}/{\rm L}}$ and get (by \cite[Prop. 3.6.]{Robert_Vargas}) that the random variable $\inf_{x\in B}P_{\A^+_{q_t}}(\varphi^{\rm h},{\varphi^{{\rm h}}}')$ has finite moments of any order and 
	\[\E_{\varphi,\varphi'}\Big[\Big(\int_{B}|x|^{-\gamma Q}M_\gamma(X_{\A^+_{q_t},D},dx)\Big)^{-\theta}\Big]\leq C_{\theta,t}\]
	with $C_{\theta,t}$ uniformly bounded for $t$ in compact sets of $(0,\infty)$. Then the reste of the proof is as above.
\end{proof}

\subsection{The bulk Hamiltonian satisfies the assumptions}
The bulk Liouville Hamiltonian defined in \eqref{Hamiltonian_bulk} is such that  
\[ 2{\bf H}=-\pl_c^2+ 2{\bf P}+Q^2+2\mu e^{\gamma c}V\]
thus, up to changing $Q$ into $Q/2$, we are in the general setting where the operator of \eqref{H+} is given by 
$\hat{{\bf H}}=2{\bf H}$ and by \cite{GKRV20_bootstrap} the assumptions $(H_1)$, $(H_2)$ are satisfied with $J=1$ and $\gamma_1=\gamma$. The compactness assumption $(H_3)$ is proved exactly as in Lemma \ref{R_+(0)} (up to comestic change, it essentially corresponds to the case $\mu^{\rm L}=\mu^{\rm R}=0$ in that Lemma).

\subsection{Absence of $L^2$-eigenvalues}
The first observation is that the Hamiltonian $\hat{{\bf H}}$ has no $L^2$ eigenvalues 
\begin{lemma}\label{embeddedeig}
	The operator $\hat{{\bf H}}$ does not have non-zero eigenfunction $u\in \mc{D}(\hat{{\bf H}})$.
\end{lemma}
\begin{proof} The proof follows the argument of Lemma 6.2 in \cite{GKRV20_bootstrap}, using a positive commutator argument. We briefly recall the argument. First, the same exact proof as \cite[Lemma 6.2]{GKRV20_bootstrap} gives that if $f\in \hat{\mc{H}}$ and $\pl_cf\in \hat{\mc{H}}$, and $u\in \mc{D}(\hat{\mc{Q}})$ satisfies 
	$\hat{\mc{Q}}(u,v)=\cjg f,v\cjd_{\hat{\mc{H}}}$ for all $v\in \hat{\mc{H}}$, then we claim that $\pl_cu\in \mc{D}(\hat{\mc{Q}})$ and for each $v\in \mc{D}(\hat{\mc{Q}})$
	\begin{equation}\label{ident_commut} 
		\hat{\mc{Q}}(\pl_c u,v)=\cjg \pl_cf,v\cjd_{\hat{\mc{H}}}-\sum_{j=1}^J \gamma_j\int_\R e^{\gamma_j c}\mc{Q}_{V_j}(u,v)dc.
	\end{equation}
	For $u\in \mc{D}(\hat{\mc{H}})$ and $\la\in \R$ such that $(\hat{{\bf H}}-\la)u=0$, we take $f=\la u$ and $v=\pl_c u$ in \eqref{ident_commut}, and we get $\hat{\mc{Q}}(u,\pl_cu)=\la \cjg u,\pl_cu\cjd_{\hat{\mc{H}}}=0$. Combining these identities with 
	\eqref{ident_commut} applied with $v=u$ then gives (using that $\mc{Q}_{V_j}$ are non-negative) that $\int_{\R}e^{\gamma_j}\mc{Q}_{V_j}(u,u)dc=0$ for each $j$, and thus  $\hat{\mc{Q}}^{0}(u,u)=0$. This implies that $u=0$.
\end{proof}

\subsection{Resolvent of ${\bf H}^{+}$ and Poisson operator in physical half-plane}
For ${\rm Re}(\alpha)<Q$, $\Delta_{\alpha}\in \C\setminus [Q^2/4,\infty)$  and the resolvent 
\[ \hat{{\bf R}}(\alpha):=(\hat{{\bf H}}-\Delta_\alpha)^{-1} : \hat{\mc{H}}\to \mc{D}(\hat{{\bf H}})\subset \hat{\mc{H}}\]
is bounded and analytic with respect to $\alpha$, and also maps $\mc{D}'(\mc{Q}^+)\to \mc{D}(\mc{Q}^+)$ continuously. 
We will use also the parameter $p\in \C$ related to $\alpha$ by 
\[\alpha=Q+2ip\]
so that $\Delta_\alpha=Q^2/4+p^2$.
We will denote by $\sqrt{\cdot}$ the square root defined so that ${\rm Im}(\sqrt{z})>0$ if $z\in \C\setminus \R^+$, i.e. $\sqrt{re^{i\theta}}=\sqrt{r}e^{i\theta/2}$ for $\theta\in (0,2\pi)$ and $r>0$. In particular, one has $\sqrt{z^2}=z$ for ${\rm Im}(z)>0$ .\\

In \cite[Lemma 6.5]{GKRV20_bootstrap}, it is shown that the resolvent  $({\bf H}-2\Delta_{\alpha})^{-1}$ 
and the propagator $e^{-t{\bf H}}$ for the Liouville Hamiltonian ${\bf H}$ are bounded on some weighted spaces $e^{\beta \rho}\mc{H}$ (for some $\beta \in\R$ depending on $\alpha$ in the resolvent case)
if $\rho\in C^\infty(\R)$ is a function of the $c$ variable satisfying for some $a\in\R$
\[ \rho(c)=0 \textrm{ if }c\geq 0, \,\, \rho(c)=c-a \textrm{ if }c\leq -1, \quad \rho'(c)\in [0,1]\]
The proof of  \cite[Lemma 6.5]{GKRV20_bootstrap} applies verbatim to $\hat{{\bf H}}$ using the explicit computation for $\beta\in\R$
\[ {\rm Re} \cjg e^{-\beta\rho}\hat{{\bf H}}(e^{\beta \rho}u),u\cjd_{\hat{\mc{H}}}=\hat{\mc{Q}}(u)-\beta^2\|\rho' u\|^2_{\hat{\mc{H}}}\geq \hat{\mc{Q}}(u)-\beta^2\|u\|^2_{\hat{\mc{H}}}.\]
This gives the following properties: for $\beta\in\R$, 
${\rm Re}(\frac{(\alpha-Q)^2}{4})>\beta^2$ and ${\rm Re}(\alpha)<Q$, then
\begin{equation}\label{Ralphaphysical} 
	\hat{{\bf R}}(\alpha): e^{-\beta \rho}\hat{\mc{H}}\to e^{-\beta \rho}\mc{D}(\hat{{\bf H}}), \quad \hat{{\bf R}}(\alpha): e^{-\beta \rho}\mc{D}'(\hat{\mc{Q}})\to e^{-\beta \rho}\mc{D}(\hat{\mc{Q}})
\end{equation}
is bounded and analytic in $\alpha$, it invert the operator $\hat{{\bf H}}-\Delta_\alpha: e^{-\beta \rho}\hat{\mc{H}}\to e^{-\beta \rho}\hat{\mc{H}}$ which is closed with domain $e^{-\beta \rho}\mc{D}(\hat{{\bf H}})$. It also shows that the propagator $e^{-t\hat{{\bf H}}}$ is a  bounded semi-group on $e^{-\beta\rho}\hat{\mc{H}}$ with norm, for each $\eps>0$,
\begin{equation}\label{normeweightedprop}
	\|e^{-t\hat{{\bf H}}}\|_{\mc{L}(e^{-\beta\rho}\hat{\mc{H}})}\leq e^{-t(\frac{Q^2}{4}-\beta^2)}, \quad \|e^{-t\hat{{\bf H}}}\|_{\mc{L}(e^{-\beta \rho}\mc{H}^+,e^{-\beta \rho}\mc{D}(\hat{\mc{Q}}))}\leq 
	C_\epsilon t^{-1/2}e^{-t(\frac{Q^2}{4}-\beta^2-\eps)}
\end{equation}
for some $C_\eps>0$ and one has 
\begin{equation}\label{resolvvspropagator} 
	\hat{{\bf R}}(\alpha)=\int_0^{\infty} e^{-t\hat{{\bf H}}+t \Delta_{\alpha}}\dd t ,\quad  
	\|\hat{{\bf R}}(\alpha)\|_{\mc{L}(e^{-\beta\rho}\hat{\mc{H}})}\leq \Big(\frac{Q^2}{4}-\beta^2-{\rm Re}(\Delta_\alpha)\Big)^{-1}. 
\end{equation}
Note that or all $t>0$ and all $\beta \in \R$
\[
e^{-t\hat{{\bf H}}}: e^{-\beta \rho}\hat{\mc{H}} \to e^{-\beta \rho}\mc{D}(\hat{\mc{Q}}), \quad e^{-t\hat{{\bf H}}}: e^{-\beta \rho}\mc{D}'(\hat{\mc{Q}})\to e^{-\beta \rho}\hat{\mc{H}}
\]
are bounded, thus using $e^{-t\hat{{\bf H}}}=e^{-\frac{t}{2}\hat{{\bf H}}}e^{-\frac{t}{2}\hat{{\bf H}}}$, we obtain the boundedness of 
\begin{equation}\label{boundednesspropag}
	e^{-t\hat{{\bf H}}}: e^{-\beta \rho}\mc{D}'(\hat{\mc{Q}})\to e^{-\beta \rho}\mc{D}(\hat{\mc{Q}}).
\end{equation}

With these properties, we are in position to define the Poisson operator for $\hat{{\bf H}}$. 
We fix $\chi\in C^\infty(\R)$ equal to $1$ in $(-\infty,a-1)$ and  $0$ in $(a,+\infty)$, then for $\alpha=Q+2ip$ with ${\rm Im}(p)>0$ we choose  
\begin{equation}\label{betaell}
	\beta_\ell>{\rm Im}\sqrt{p^2-\la_\ell}-\min_{j}\gamma_j/2, \,\quad \textrm{ and }\beta_\ell\geq 0.
\end{equation}
Then for ${\rm Re}((\alpha-Q)^2)>\beta_\ell^2$ we define the Poisson operator of degree $\ell\in \N_0$ by 
\begin{equation}\label{def_P_ell}\begin{gathered}
		\hat{\mc{P}}_\ell(\alpha):\left\{ \begin{array}{lll}
			E_\ell& \to & e^{-(\beta_\ell+\min_j\gamma_j/2) \rho}\mc{D}(\hat{\mc{Q}})\\
			F= \sum_{j=0}^{\ell} F_j &\mapsto & \chi F_-(\alpha) -\hat{{\bf R}}(\alpha)(\hat{{\bf H}}-\Delta_{\alpha})(\chi F_-(\alpha)),
		\end{array}\right.\\
		\textrm{ with }F_-(\alpha):=\sum_{j=0}^{\ell}F_je^{ic\sqrt{p^2-\la_j}}.
\end{gathered}\end{equation}
Here the function $F_-(\alpha)$ is an eigenfunction of the free Hamiltonian $\hat{{\bf H}}^0$, and  the term $\chi F_-(\la)$ is an ansatz to the equation $(\hat{{\bf H}}-\Delta_\alpha)u=0$ in the region $c\leq 0$, which is corrected to an exact solution by $\hat{{\bf R}}(\alpha)(\hat{{\bf H}}-\Delta_{\alpha})(\chi F_-(\alpha))\in e^{-\beta_\ell \rho}\mc{D}(\hat{\mc{Q}})$:
\[ (\hat{{\bf H}}-\Delta_\alpha)\hat{\mc{P}}_\ell(\alpha)F=0.\]
The fact that $\hat{\mc{P}}_\ell(\alpha)$ is well-defined and has the claimed mapping property follows from the proof of \cite[Lemma 6.7]{GKRV20_bootstrap} applied to $\hat{{\bf H}}$.

The proof of \cite[Proposition 6.9]{GKRV20_bootstrap} also applies to $\hat{{\bf H}}$ and this gives 
\begin{equation}\label{poisson_intertwine}
	\mc{P}_\ell(\alpha)F= \lim_{t\to +\infty}e^{t \Delta_\alpha}e^{-t\hat{{\bf H}}}\Big(e^{ic\sqrt{p^2-\la_j}}\chi(c)F\Big)
\end{equation}
in $e^{-(\beta+\min_j\gamma_j/2) \rho}\mc{H}^+$ provided
${\rm Re}((\alpha-Q)^2/4)>\beta^2$ with $\beta>{\rm Im}(\sqrt{p^2-\la_j})-\min_j\gamma_j/2$ and $\alpha=Q+2ip$.\\

To analyse the resolvent $\hat{{\bf R}}(\alpha)$ as an analytic function of $\alpha$, 
we will need to work on a ramified covering of $\C$. 
We denote by $Z$ the Riemann surface on which the functions $\alpha \mapsto \sqrt{-(\alpha-Q)^2/4-\la_j}$ are single valued  for all $j$ (for $j=0$, $\sqrt{-(\alpha-Q)^2/4}=-i(\alpha-Q)/2$) and so that for $\alpha\in Z$, ${\rm Im}(\sqrt{-(\alpha-Q)^2/4-\la_j})>0$ if $j$ is large enough. This is a  ramified covering of $\C$ with ramification points $\{\alpha=Q\pm 2i\sqrt{\la_j}\, |\, j\in\N\}$, the ramifications being of order $2$,  the covering map is denoted by $\pi: Z\to \C$.  The half-plane ${\rm Re}(\alpha)<Q$ is  embedded in $Z$ and is called the \textbf{physical half-plane}: it is the region where the resolvent $\hat{{\bf R}}(\alpha)$ is analytic and bounded on $\hat{\mc{H}}$. 

\subsection{Resolvent near the spectrum}
The main part of the spectral resolution is to show the following Proposition, similar to \cite[Prop 6.10]{GKRV20_bootstrap}:
\begin{proposition}\label{extensionresolvent}
	Assume that $0\leq \beta<\min \gamma_j/2$. Then the following properties hold true:\\
	1) The resolvent $\hat{{\bf R}}(\alpha):=(\hat{{\bf H}}-\Delta_\alpha )^{-1}$ is bounded as a map 
	$\hat{\mc{H}}\to \mc{D}(\hat{{\bf H}})$ for ${\rm Re}(\alpha)<Q$ and the operator
	$(\hat{{\bf H}}-\Delta_{\pi(\alpha)})^{-1}$ admits a meromorphic continuation to the region 
	\[ \bigcup_{k\in \N}\{\alpha\in Z\,|\, |\pi(\alpha)-Q|^2\leq \la_k^{\frac{1}{4}}, \forall j\leq k, {\rm Im}\sqrt{-(\alpha-Q)^2/4-\la_j}>-\min(\beta,\min\gamma_j/2-\beta)
	\}\] 
	as a map 
	\begin{align*}
		\hat{{\bf R}}(\alpha): e^{\beta \rho}\hat{\mc{H}}\to e^{-\beta \rho}\mc{D}(\hat{{\bf H}}),
	\end{align*}
	and the residue at each pole is a finite rank operator. For each $k\in \N$ and $\psi\in C^\infty(\R)\cap L^\infty(\R)$ satisfying $\psi'\in L^\infty$ and ${\rm supp}(\psi)\subset (-\infty,A)$ for some $A\in \R$, 
	\[\hat{{\bf R}}(\alpha)(1-\Pi_k)\psi: \mc{D}'(\hat{\mc{Q}})\to e^{-\beta \rho}\mc{D}(\hat{\mc{Q}}).\] 
	2) If $f\in e^{\beta \rho}\mc{D}'(\mc{Q}^+)$ is supported in $c\in (-\infty,A)$ for some $A>0$ and is such that $\Pi_kf\in e^{\beta \rho}L^2(\R;E_k)$ 
	and $(1-\Pi_k)f\in \mc{D}'(\hat{\mc{Q}})$,  then for $\alpha$ as above and not a pole, one has in the region $\{c\leq 0\}$
	\begin{equation}\label{asymptoticRf}
		\hat{{\bf R}}(\alpha)f=\sum_{j=0}^k a_j(\alpha,f)e^{-ic\sqrt{p^2-\la_j}}+G(\alpha,f),
	\end{equation}
	with $a_j(\alpha,f)\in \ker (\hat{{\bf P}}-\la_j)$ and $G(\alpha,f)\in \mc{D}(\hat{\mc{Q}})$, all depending meromorphically on $\alpha$ in the region they are defined. Moreover $G(\alpha,f),\pl_cG(\alpha,f)\in e^{\beta\rho}L^2(\R^-; E_k)+L^2(\R^-; E_k^\perp)$.\\
	3) There is no pole of $\hat{{\bf R}}(\alpha)$ in $\{\alpha \in Z\,|\, {\rm Re}(\alpha)\leq Q\}\setminus \cup_{j=0}^\infty \{Q\pm 2i\sqrt{\la_j}\}$ and $Q\pm 2i\sqrt{\la_j}$ can be at most a pole of order $1$.
\end{proposition}

The proof of this proposition needs several parametrices described below:
\begin{enumerate}
	\item first, the region $L^2(\R^+\times \Omega)$ where roughly speaking the resolvent localized in that region is compact and then behaves like the resolvent of an operator with discrete spectrum, 
	\item then, the region $L^2(\R^-)\otimes E_k^\perp$ where the Hamiltonian is invertible modulo a compact operator in a ball of size of the order of $\la_k$ due to the lower bound $\hat{{\bf H}}\geq \hat{{\bf P}}\geq \la_k$ on this space,
	\item finally, the region $L^2(\R^-)\otimes E_k$ where the Hamiltonian behaves roughly as $\hat{{\bf H}}^{0}$ modulo a relatively compact perturbation; this is the scattering regime where the generalized eigenfunctions of $\hat{{\bf H}}$ can be described in  terms of those of $\hat{{\bf H}}^{0}$.
\end{enumerate}
\noindent \textbf{The region $\{c\geq 0\}$.} We will show the following:

\begin{lemma}\label{l:c>--1}
	Let $\chi \in C^\infty(\R;[0,1])$ with support in $(-\infty,0)$, equal to $1$ near $-\infty$ and let $\alpha_0<Q$. 
	There is an analytic family of operators ${\bf Q}(\alpha): \hat{\mc{H}}\to \mc{D}(\hat{\mc{Q}})$, depending on $\alpha_0$, such that for $\alpha\in \C$
	\[(\hat{{\bf H}}-\Delta_{\alpha}){\bf Q}(\alpha)= 1-\chi +{\bf K}_{>0}(\alpha)\]
	with ${\bf K}_{>0}(\alpha):\hat{\mc{H}}\to \mc{D}(\hat{\mc{Q}})$ a compact family of operators, analytic in $\alpha\in \C$, which is also bounded as a map 		
	\begin{equation}\label{K+D'toD} 
		{\bf K}_{>0}(\alpha): \mc{D}'(\hat{\mc{Q}}) \to e^{\beta \rho}\mc{D}(\hat{\mc{Q}})
	\end{equation} 
	for $\beta<|\alpha_0-Q|/2$.
\end{lemma}
\begin{proof}
	Let $\chi_1\in C^\infty(\R;[0,1])$ with support in $(-\infty,0)$ such that $(1-\chi_1)=1$ on the support of $(1-\chi)$.
	First notice that 
	\[ (\hat{{\bf H}}-\Delta_{\alpha})(1-\chi_1)\hat{{\bf R}}(\alpha_0)(1-\chi)=1-\chi+ [\pl_c^2,\chi_1]\hat{{\bf R}}(\alpha_0)(1-\chi)-(\Delta_\alpha-\Delta_{\alpha_0})(1-\chi_1)\hat{{\bf R}}(\alpha_0)(1-\chi).\]
	Here $ [\pl_c^2,\chi_1]=\chi_1''+\chi_1'\pl_c$ is a first order differential  operator in $c$ with compact support.
	Then set 
	\[ {\bf Q}(\alpha):=(1-\chi_1)\hat{{\bf R}}(\alpha_0)(1-\chi)-\hat{{\bf R}}(\alpha_0)([\pl_c^2,\chi_1]\hat{{\bf R}}(\alpha_0)(1-\chi)-(\Delta_\alpha-\Delta_{\alpha_0}) (1-\chi_1)\hat{{\bf R}}(\alpha_0)(1-\chi))\]
	which satisfies 
	\[ \begin{split}
		(\hat{{\bf H}}-\Delta_{\alpha}){\bf Q}(\alpha)=& 1-\chi +(\Delta_\alpha-\Delta_{\alpha_0})\hat{{\bf R}}(\alpha_0)([\pl_c^2,\chi_1]\hat{{\bf R}}(\alpha_0)(1-\chi)-(\Delta_\alpha-\Delta_{\alpha_0}) (1-\chi_1)\hat{{\bf R}}(\alpha_0)(1-\chi))\\
		=:& 1-\chi +{\bf K}_{>0}(\alpha).
	\end{split}\]
	We have for $\chi_2\in C^\infty(\R)$ supported in $(-2,\infty)$, equal to $1$ on ${\rm supp}(\chi_1')\cup {\rm supp}(1-\chi_1)$, that
	\begin{equation}\label{term1} 
		\begin{gathered}
			\hat{{\bf R}}(\alpha_0)[\pl_c^2,\chi_1]\hat{{\bf R}}(\alpha_0)(1-\chi) =\hat{{\bf R}}(\alpha_0)[\pl_c^2,\chi_1] \chi_2\hat{{\bf R}}(\alpha_0)(1-\chi)\\
			\hat{{\bf R}}(\alpha_0)(1-\chi_1)\hat{{\bf R}}(\alpha_0)(1-\chi)=\hat{{\bf R}}(\alpha_0) (1-\chi_1)\chi_2\hat{{\bf R}}(\alpha_0)(1-\chi)
	\end{gathered}\end{equation}
	with $\chi_2\hat{{\bf R}}(\alpha_0)(1-\chi)$ compact on $\hat{\mc{H}}$ by assumption (H3)  and 
	\[\hat{{\bf R}}(\alpha_0)[\pl_c^2,\chi_1] : \hat{\mc{H}}\to e^{\beta \rho}\mc{D}(\hat{\mc{Q}}), \quad \hat{{\bf R}}(\alpha_0)(1-\chi_1)
	: \hat{\mc{H}}\to e^{\beta \rho}\mc{D}(\hat{\mc{Q}})\]
	bounded by  \eqref{Ralphaphysical}. 
	This shows that both operators in \eqref{term1} are compact as map $\hat{\mc{H}}\to \mc{D}(\hat{\mc{Q}})$.  Since $\hat{{\bf R}}(\alpha_0)(1-\chi):\mc{D}'(\hat{\mc{Q}})\to \mc{D}(\hat{\mc{Q}})$, we also obtain the boundedness \eqref{K+D'toD} of ${\bf K}_{>0}(\alpha)$.
\end{proof}

\noindent \textbf{The region $\{c\leq 0\}$ on  $E_k^\perp:=(1-\Pi_k)\hat{\mc{H}}$.}
Let $\chi,\tilde{\chi}\in C^\infty(\R;[0,1])$ with support in $(-\infty,0)$, $\tilde{\tilde{\chi}}=1$ on the support of $\chi$ and $\chi=1$ near $(-\infty,1]$.
The proof of \cite[Lemma 6.11]{GKRV20_bootstrap} can be applied verbatim to our case: it 
uses the quadratic form 
\[ \hat{\mc{Q}}_{k}^{\perp}:=\hat{\mc{Q}}|_{\mc{D}(\hat{\mc{Q}})\cap\ker \Pi_k \cap \ker r_{\R^+}}\]
(with $r_{\R^+}:L^2(\R\times \Omega)\to L^2(\R^+\times \Omega)$ the restriction map), which satisfies for each $\eps\geq 0$
\[ \hat{\mc{Q}}_{k}^\perp(u)\geq \|\pl_cu \|_{\hat{\mc{H}}}^2+\frac{Q^2}{4}\|u\|^2_{\mc{H}^+}+(\la_k-\eps) \|u\|_{\hat{\mc{H}}}^2+\eps \|\hat{{\bf P}}^{1/2}u\|_{\hat{\mc{H}}}\]
and the Lax-Milgram theorem  $\hat{\mc{Q}}_{k}^\perp(u)-\Delta_\alpha \|u\|^2_{\hat{\mc{H}}}$. This gives:
\begin{lemma}\label{l:estimateE_kperp}
	1) Let $\beta\in (0,\min_j \gamma_j/2)$. For all $k\in \N$ large enough, there is an analytic family of operators in $\{{\rm Re}(\alpha)\leq Q \}\cup \{ |\alpha-Q|^2\leq 2\la_k\}$
	\[ \hat{{\bf R}}_k^{\perp}(\alpha): L^2(\R;E_k^\perp)\to L^2(\R^-;E_k^\perp)\]
	such that $\tilde{\chi}\, \hat{{\bf R}}_{k}^{\perp}(\alpha)\chi: \mc{D}'(\hat{\mc{Q}})\to \mc{D}(\hat{\mc{Q}})\cap L^2(\R^-;E_k^\perp)$ and
	\[ (\hat{{\bf H}}-\Delta_{\alpha})\tilde{\chi}\, \hat{{\bf R}}_{k}^{\perp}(\alpha)\chi=(1-\Pi_k)\chi+{\bf J}_k^\perp(\alpha)+{\bf K}_k^\perp(\alpha)\]
	with ${\bf J}_k^\perp(\alpha):\mc{D}'(\hat{\mc{Q}})\to L^2(\R^-;E_k^\perp)$ and ${\bf K}_k^\perp:\mc{D}'(\hat{\mc{Q}})\to e^{\beta \rho}L^2(\R;E_k)$ bounded and holomorphic for $\alpha$ as above, and given by the expression 
	\begin{align*}
		{\bf J}_k^\perp(\alpha):=-[\pl_c^2,\tilde{\chi}]\hat{{\bf R}}_{k}^{\perp}(\alpha)(1-\Pi_k)\chi , \quad {\bf K}_k^\perp(\alpha):=\sum_{j=1}^J e^{\gamma_j c}\tilde{\chi}\Pi_k V_j\hat{{\bf R}}_{k}^{\perp}(\alpha)\chi (1-\Pi_k).
	\end{align*}
	Moreover, in this $\alpha$-region, ${\bf K}_k^\perp(\alpha):\hat{\mc{H}}\to L^2(\R;E_k)$ is compact and there is $C>0$ independent of $\alpha$ and $k$ such that
	\[ \| {\bf J}_k^\perp(\alpha)\|_{\mc{L}(\hat{\mc{H}})}\leq C\la_k^{-1/2}.\]
	2) For $\beta\in\R$, in the region ${\rm Re}((\alpha-Q)^2/4)>\beta^2-\la_k+1$, the operator 
	\[\hat{{\bf R}}_k^\perp(\alpha):e^{-\beta \rho} \mc{D}'(\hat{\mc{Q}})\to e^{-\beta \rho}(L^2(\R^-;E_k^\perp)\cap \mc{D}(\hat{\mc{Q}}))\] 
	is a bounded holomorphic family, 
	${\bf K}_k^\perp(\alpha): e^{-\beta \rho}L^2(\R\times \Omega_\T)\to e^{-\beta \rho}L^2(\R^-;E_k)$ 
	is a compact holomorphic family, 
	${\bf J}_k^\perp(\alpha):e^{-\beta \rho}L^2(\R\times \Omega_\T)\to e^{-\beta \rho}L^2(\R^-;E_k^\perp)$ is bounded analytic with norm
	\[\| {\bf J}_k^\perp(\alpha)\|_{\mc{L}(e^{-\beta \rho}\hat{\mc{H}})}\leq \frac{C(1+|\beta|)}{\sqrt{{\rm Re}((\alpha-Q)^2/4)+\la_k-\beta^2}}\]
	for some $C>0$ independent of $k,\alpha$
\end{lemma}
As in  \cite[Lemma 6.11]{GKRV20_bootstrap}, the operator $\hat{{\bf R}}_k^\perp(\alpha)$ is obtained as the inverse $(\hat{{\bf H}}_{k}^\perp-\Delta_\alpha)^{-1}$ where 
$\hat{{\bf H}}_{k}^\perp$ is the self-adjoint unbounded 
operator $\hat{{\bf H}}_{k}^\perp$ associated to the quadratic form $\hat{\mc{Q}}_{k}^\perp$ bounded below by 
\[ \hat{\mc{Q}}_{k}^\perp(u)\geq (\la_k+Q^2/4)\|u\|_{L^2(\R^-;E_k^\perp)}^2.\]
The restriction to $E_k^\perp$ valued functions implies that this operator $\hat{{\bf H}}_{k}^\perp$ has spectrum in $[\la_k+Q^2/4,\infty)$, which is why $(\hat{{\bf H}}_{k}^\perp-\Delta_\alpha)^{-1}$ exists for $\Delta_\alpha$ in a ball of radius of the order of $\la_k$. To deal with weighted spaces $e^{-\beta \rho}\hat{\mc{H}}$, one computes the quadratic form associated to $e^{\beta \rho }(\hat{{\bf H}}_{k}^\perp-\Delta_\alpha) e^{-\beta \rho}$, whose real part is bounded below by  
\[ \| \pl_c u\|^2_{\hat{\mc{H}}}+(\tfrac{Q^2}{4}-\beta^2+\la_k-{\rm Re}(\Delta_\alpha))\|u\|_{\hat{\mc{H}}}^2+\sum_{j=1}^J \int_{\R} e^{\gamma_jc}\mc{Q}_{V_j}(u,u)\dd c\geq \Big(\frac{{\rm Re}(\alpha-Q)^2}{4}-\beta^2+\la_k\Big)\|u\|_{\hat{\mc{H}}}^2\]

\noindent \textbf{The region $\{c\leq 0\}$ on  $E_k:=(1-\Pi_k)\hat{\mc{H}}$.}
In the region $c\leq 0$ and on $E_k$ (small eigenmodes of $\hat{{\bf P}}$), the operator $\hat{{\bf H}}$ can be seen as a perturbation of 
$\hat{{\bf H}}^{0}$ and we can use as a parametrix for its resolvent in $\{c\leq 0\}$ the resolvent 
of $\hat{{\bf H}}^{0}$ restricted to $E_k$ and with Dirichlet condition at $c=0$:
for ${\rm Re}(\alpha)<Q$ (so that $\Delta_\alpha \in \C\setminus [Q^2/4,\infty)$) we denote this resolvent by 
\[ \hat{{\bf R}}_{k}^{0}(\alpha):=(\hat{{\bf H}}^{0}-\Delta_\alpha)^{-1}\Pi_k : L^2(\R^-;E_k) \to L^2(\R^-;E_k).\]
With $\alpha=Q+2ip$ and ${\rm Im}(p)>0$, using that $\hat{{\bf H}}^{0}$ is diagonal with respect to the eigendecomposition of $\hat{{\bf P}}$
\[ (\hat{{\bf H}}^{0} -\Delta_\alpha)|_{\ker (\hat{{\bf P}}-\la_j)}= (-\pl_c^2-p^2+\la_j)|_{\ker (\hat{{\bf P}}-\la_j)}  \]
we obtain from Sturm-Liouville theory that 
for $j\leq k$ and $f\in L^2(\R^-)$ and $\phi_j\in \ker (\hat{{\bf P}}-\la_j)$
\[\begin{split} 
	(\hat{{\bf H}}^{0}-\tfrac{Q^2}{4}+p^2)^{-1}f(c)\phi_j =\frac{1}{\sqrt{p^2-\la_j}}& \phi_j\Big(\int_{-\infty}^c \sin(c\sqrt{p^2-\la_j})e^{-ic'\sqrt{p^2-\la_j}}f(c')\dd c'\\
	& \quad +\int_{c}^0 e^{-ic\sqrt{p^2-\la_j}}\sin(c'\sqrt{p^2-\la_j})f(c')\dd c'\Big)
\end{split}\]
where we recall that $\sqrt{z}$ is defined with the cut on $\R^+$, so that $\sqrt{p^2}=p$
if ${\rm Im}(p)>0$. From this explicit expression, one infers exactly as in the proof of \cite[Lemma 6.16]{GKRV20_bootstrap} that these operators extend analytically for $\alpha$ on the Riemann surface $Z$ and are bounded on weighted 
$e^{\beta |c|}L^2(\R^-;E_k)$ spaces. Applying the same proof of \cite[Lemma 6.16]{GKRV20_bootstrap} gives: 
\begin{lemma}\label{modelres}
	1) For $k$ and $0<\beta<\min_{j}\gamma_j/2$, the operator  
	\begin{equation}
		\hat{{\bf R}}_{k}^{0}(\alpha): e^{\beta\rho}L^2(\R^{-}\times \Omega)\to e^{-\beta \rho}L^2(\R^-;E_k)
	\end{equation}
	is bounded for ${\rm Im}(p)>0$ and can be holomorphically extended to the region 
	\begin{equation}\label{region_holo_R_k}\{\alpha=Q+2ip \in Z \,|\, \forall j\leq k, {\rm Im}\sqrt{p^2-\la_j}>-\beta\}.
	\end{equation} 
	Moreover, if $\tilde{\chi},\chi \in C^\infty(\R)$ are equal to $1$ in $\{c\leq -1\}$, have support in $\R^-$ and $\tilde{\chi}=1$ on the support of $\chi$, the operator 
	\[\tilde{\chi}\, \hat{{\bf R}}_{k}^{0}(\alpha)\chi:e^{\beta \rho}L^2(\R\times\Omega)\to e^{-\beta \rho}(L^2(\R^-;E_k)\cap \mc{D}(\hat{\mc{Q}})),\]
	satisfies 
	\[ (\hat{\bf H}-\Delta_\alpha)\tilde{\chi}\, \hat{{\bf R}}_{k}^{0}(\alpha)\chi =\Pi_k \chi+{\bf K}_{k,1}(\alpha)+{\bf K}_{k,2}(\alpha)\]
	where ${\bf K}_{k,1}(\alpha)$, ${\bf K}_{k,2}(\alpha)$ are such that for ${\rm Im}\sqrt{p^2-\la_j}>-\min(\beta,\min_{j}\gamma_j/2-\beta)$
	\begin{align*}
		& {\bf K}_{k,1}(\alpha):e^{\beta \rho}\hat{\mc{H}} \to e^{\beta \rho}\hat{\mc{H}}
		\\
		& {\bf K}_{k,2}(\alpha):e^{\beta \rho}\hat{\mc{H}}\to e^{\beta \rho}\mc{D}'(\hat{\mc{Q}})
	\end{align*}
	are holomorphic families of compact operators in the region \eqref{region_holo_R_k}, and  ${\bf K}_{k,i}(\alpha)(1-\Pi_k)=0$ for $i=1,2$, $(1-\Pi_k){\bf K}_{k,1}(\alpha)=0$ and $\Pi_k{\bf K}_{k,2}(\alpha)=0$.\\
	2) There is $C_k>0$ depending on $k$ such that for $f\in e^{\beta \rho}\hat{\mc{H}}$, there are 
	$a_j(\alpha,f)\in \ker (\hat{{\bf P}}-\la_j)$ and  $G(\alpha,f)\in H^1(\R;E_k)\cap \mc{D}(\hat{\mc{Q}})$ depending linearly on $f$ and holomorphic in the region
	\[\{\alpha=Q+2ip \in Z\,|\,  \forall j\leq k, {\rm Im}\sqrt{p^2-\la_j}>-\min(\beta,\min_j \gamma_j/2-\beta)\}\] 
	such that in the region $\{c\leq 0\}$
	\begin{equation}\label{asympR_-}
		(\hat{{\bf R}}_{k}^{0}(\alpha)f)=\sum_{j=0}^k a_j(\alpha,f)
		e^{-ic\sqrt{p^2-\la_j}}+G(\alpha,f),
	\end{equation}
	\[ \|G(\alpha,f)(c)\|_{L^2(\Omega)}+\|\pl_cG(\alpha,f)(c)\|_{L^2(\Omega)}\leq C_k e^{\beta \rho}
	\|e^{-\beta \rho}\Pi_kf\|_{\hat{\mc{H}}}.\]
	3) For each $\beta\in \R$, the operator $\hat{{\bf R}}_{k}^{0}(\alpha)$ extends as a bounded analytic family 
	\[ \hat{{\bf R}}_{k}^{0}(\alpha): e^{-\beta \rho}L^2(\R^-\times \Omega)\to e^{-\beta \rho}(L^2(\R^-;E_k)\cap \mc{D}(\hat{\mc{Q}}))\]
	in the region $\{{\rm Im}(p)>|\beta|\}$ and the following operators are compact analytic families in that region 
	\[{\bf K}_{k,1}(\alpha): e^{-\beta \rho}\hat{\mc{H}}\to e^{-\beta \rho}\hat{\mc{H}}, \qquad {\bf K}_{k,2}(\alpha): e^{-\beta \rho}\hat{\mc{H}}\to e^{-\beta \rho}\mc{D}'(\hat{\mc{Q}}).\] 
\end{lemma}
As in the proof of \cite[Lemma 6.16]{GKRV20_bootstrap}, the operators ${\bf K}_{k,1}(\alpha),{\bf K}_{k,2}(\alpha)$ in that Lemma are given by 
\begin{align*}
	& {\bf K}_{k,1}(\alpha)= [\pl_c^2,\tilde{\chi}]\Pi_k \hat{{\bf R}}_{k}^{0}(\alpha)\chi+
	\sum_{j=1}^J\Pi_ke^{\gamma_j c} V_j\Pi_k\tilde{\chi}\,\hat{{\bf R}}_{k}^{0}(\alpha)\chi, \\
	&  {\bf K}_{k,2}(\alpha)=\sum_{j=1}^J(1-\Pi_k)e^{\gamma_j c}V_j\Pi_k\tilde{\chi}\, \hat{{\bf R}}_{k}^{0}(\alpha)\chi
\end{align*}

We are now in position to prove Proposition \ref{extensionresolvent}:
\begin{proof}[Proof of Proposition \ref{extensionresolvent}]
	Again, being functional analytic in nature, the proof of \cite[Proposition 6.10]{GKRV20_bootstrap} is quite general and applies as well here with only cosmetic changes. We only summarize the argument rather than repeating it and we refer to that paper for details. 
	For $\beta\in \R$, we define the Hilbert space and associated scalar product (for fixed $k$)
	\[\hat{\mc{H}}_{k,\beta}:=e^{\beta \rho}L^2(\R;E_k)\oplus L^2(\R;E_k^\perp);\]
	\[ \cjg u_1,u_2\cjd_{\hat{\mc{H}}_{k,\beta}}:= \int_{\R}e^{-2\beta \rho(c)}\cjg \Pi_k u_1,\Pi_ku_2\cjd_{L^2(\Omega)}\dd c +\cjg(1-\Pi_k)u_1,(1-\Pi_k)u_2 \cjd_{\hat{\mc{H}}}.\] 
	For $\beta\in [0,\min_j \gamma_j/2)$, for $\chi,\tilde{\chi}$ as in Lemma \ref{l:estimateE_kperp}, and $\tilde{\tilde{\chi}}\in C^\infty(\R;[0,1])$ with support in $\R^-$ and equal to $1$ on ${\rm supp}(\tilde{\chi})$, let us
	define the operator 
	\[ \til{{\bf R}}(\alpha):=\tilde{\chi}\, \hat{{\bf R}}_{k}^{\perp}(\pi(\alpha))\chi+{\bf Q}(\alpha)+\tilde{\chi}\, \hat{{\bf R}}_{k}^{0}(\alpha)\chi-\tilde{\tilde{\chi}}\, \hat{{\bf R}}_{k}^{\perp}(\pi(\alpha)){\bf K}_{k,2}(\alpha)\]
	where the operators above are defined in Lemma  \ref{l:c>--1},  \ref{l:estimateE_kperp}, and \ref{modelres}. 
	From these Lemmas, $\til{{\bf R}}(\alpha)$ is bounded and holomorphic for $\alpha$ in 
	\begin{equation}\label{region_of_alpha}
		\{\alpha=Q+2ip \in Z \,|\, |\pi(\alpha)-Q|^2/4<\la_k^{1/4}, \forall j\leq k, {\rm Im}\sqrt{p^2-\la_j}>-\beta\}
	\end{equation}
	as a map $\til{{\bf R}}(\alpha): \hat{\mc{H}}_{k,\beta}\to \hat{\mc{H}}_{k,-\beta}\cap e^{-\beta \rho}\mc{D}(\hat{\mc{Q}})$. It is a parametrix for $(\hat{{\bf H}}-\Delta_{\pi(\alpha)})$ in the sense that 
	\[ \begin{split}
		(\hat{{\bf H}}-\Delta_{\pi(\alpha)})\til{{\bf R}}(\alpha)=& 1+{\bf J}_k^\perp(\pi(\alpha))+{\bf K}_k^\perp(\pi(\alpha))+{\bf K}_{+}(\pi(\alpha))+{\bf K}_{k,1}(\alpha)\\
		& -(\tilde{{\bf J}}_k^\perp(\pi(\alpha))+\tilde{{\bf K}}_k^\perp(\pi(\alpha))){\bf K}_{k,2}(\alpha)\\
		=:& 1+{\bf J}_k^\perp(\pi(\alpha))+ \tilde{{\bf K}}_k(\alpha)
	\end{split}\]
	where ${\bf J}_k^\perp(\pi(\alpha))$ and ${\bf K}_k^\perp(\pi(\alpha))$ are the operators of Lemma  \ref{l:estimateE_kperp}
	and $\tilde{{\bf J}}_k^\perp(\pi(\alpha))$ and $\tilde{{\bf K}}_k^\perp(\pi(\alpha))$ are the operators of Lemma  \ref{l:estimateE_kperp} with $\tilde{\chi}$ (resp. $\chi$) replaced by $\tilde{\tilde{\chi}}$ (resp. $\tilde{\chi}$). By  Lemma  \ref{l:c>--1}, \ref{l:estimateE_kperp}, and \ref{modelres}, $\tilde{{\bf K}}_k(\alpha):\hat{\mc{H}}_{k,\beta}\to \hat{\mc{H}}_{k,\beta}$ is compact  and holomorphic in $\alpha$ in the region \eqref{region_of_alpha}. Note also that 
	\[\tilde{{\bf K}}_k(\alpha)(1-\Pi_k): \mc{D}'(\hat{\mc{Q}})\to \hat{\mc{H}}_{k,\beta}\]
	is bounded as a consequence of these Lemmas. Using that $\Pi_k{\bf J}_k^\perp(\pi(\alpha))={\bf J}_k^\perp(\pi(\alpha))\Pi_k=0$, we obtain from Lemma \ref{l:estimateE_kperp} that for $\alpha$ in \eqref{region_of_alpha}, the operator ${\bf J}_k^\perp(\pi(\alpha))$ has norm 
	\[ \| {\bf J}_k^\perp(\pi(\alpha))\|_{\mc{L}(\hat{\mc{H}}_{k,\beta})}=\| {\bf J}_k^\perp(\pi(\alpha))\|_{\mc{L}(\hat{\mc{H}})}\leq  C\la_k^{-1/2}\]
	for some $C$ independent of $k$. This implies that
	\[({\bf H}^{+}-\Delta_{\pi(\alpha)})\til{{\bf R}}(\alpha)( 1+{\bf J}_k^\perp(\pi(\alpha))^{-1}=1+\tilde{{\bf K}}_k(\alpha)( 1+{\bf J}_k^\perp(\pi(\alpha))^{-1}=: 1+{\bf K}_k(\alpha)\]
	with ${\bf K}_k(\alpha):\hat{\mc{H}}_{k,\beta}\to \hat{\mc{H}}_{k,\beta}$ compact and analytic in $\alpha$ in \eqref{region_of_alpha}. 
	If ${\bf K}_k(\alpha)$ is invertible at one point $\alpha_1$ of \eqref{region_of_alpha}, we can apply the analytic Fredholm theorem to deduce that $(1+{\bf K}_k(\alpha))^{-1}$ exists as a meromorphic function of $\alpha$ in \eqref{region_of_alpha} and then 
	\[\til{{\bf R}}(\alpha)( 1+{\bf J}_k^\perp(\pi(\alpha))^{-1}(1+{\bf K}_k(\alpha))^{-1}\]
	produces a meromorphic extension of ${\bf R}(\alpha)=({\bf H}^{+}-\Delta_{\pi(\alpha)})^{-1}$ from ${\rm Re}(\alpha)<Q$ to the region  \eqref{region_of_alpha}, which is bounded on $\mc{H}_{k,\beta}$. The arguments of the proof of \cite[Proposition 6.10]{GKRV20_bootstrap} can again be applied without any change to prove that there is a finite rank operator ${\bf W}:\hat{\mc{H}}_{k,\beta}\to \hat{\mc{H}}_{k,\beta}$ such that
	\[(\hat{{\bf H}}-\Delta_{\pi(\alpha)})(\til{{\bf R}}(\alpha)( 1+{\bf J}_k^\perp(\pi(\alpha))^{-1}+{\bf W})=1+{\bf K}(\alpha)\]
	for some ${\bf K}(\alpha)$ satisfying the same properties as ${\bf K}_k(\alpha)$ but with $1+{\bf K}(\alpha_0)$ invertible on $\hat{\mc{H}}_{k,\beta}$, and we can thus apply the reasoning above to obtain the meromorphic continuation of ${\bf R}(\alpha)$ in  \eqref{region_of_alpha} 
	as a bounded operator on $\mc{H}_{k,\beta}$. The rest of the proof, in particular point 2) and 3), is exactly the same as the proof of \cite[Proposition 6.10]{GKRV20_bootstrap} and we do not repeat it. 
\end{proof}

\subsection{Poisson operator and generalised eigenfunctions}
Now that the resolvent has been meromorphically extended and its fine mapping properties have been obtained in Proposition \ref{extensionresolvent}, we can apply the same exact proof as \cite[Propositions 6.19 and 6.21]{GKRV20_bootstrap} to our case. This gives:

\begin{proposition}\label{poisson_proposition}
	1) For $0<\beta<\min_{j=1}\gamma_j/2$ and $\ell\in\N$,
	there is an analytic family of operators $\hat{\mc{P}}_\ell(\alpha)$  
	\[\hat{\mc{P}}_\ell(\alpha): E_\ell\to e^{-\beta\rho}\mc{D}(\hat{\mc{Q}})\]
	in the region 
	\begin{equation}\label{region_valide}
		\Big\{\alpha=Q+2ip\in \C\,\Big|\, {\rm Re}(\alpha)< Q, {\rm Im}\sqrt{p^2-\la_\ell}<\beta\Big\}\cup \Big\{Q+2ip\in Q+i\R\, \Big| \, |p|\in \bigcup_{j\geq \ell} (\sqrt{\la_j},\sqrt{\la_{j+1}})\Big\},
	\end{equation}
	it is continuous at each $Q\pm 2i\sqrt{\la_j}$ for $j\geq \ell$, satisfies $(\hat{{\bf H}}-\Delta_\alpha)\hat{\mc{P}}_\ell(\alpha)F=0$ and has asymptotic expansion at $c\to -\infty$ 
	\begin{equation}\label{expansionpoisson}
		\hat{\mc{P}}_\ell(\alpha)F=\sum_{j\leq \ell}\Big(F^-_je^{ic\sqrt{p^2-\la_j}}+F_j^+(\alpha)
		e^{-ic\sqrt{p^2-\la_j}}\Big)+G_\ell(\alpha,F)
	\end{equation}
	with $F^-_j=\Pi_{\ker(\hat{{\bf P}}-\la_j)}F$, $F_j^+(\alpha)\in \ker (\hat{{\bf P}}-\la_j)$, and 
	$G_\ell(\alpha,F),\partial_c G_\ell(\alpha,F)\in e^{\frac{\beta}{2}\rho(c)}\hat{\mc{H}}+L^2(\R;  E_\ell^{\perp})$, $G_\ell(\alpha,F)\in \mc{D}(\hat{\mc{Q}})$. In particular, one has $\hat{\mc{P}}_\ell(\alpha)\in e^{-({\rm Im}(\sqrt{p^2-\la_\ell})+\eps)\rho}\mc{D}(\hat{\mc{Q}})$ for all $\eps>0$.
	For each cutoff function  $\theta\in C_c^\infty(\R)$, one has $\theta \hat{\mc{P}}_\ell(\alpha)F\in \mc{D}(\hat{{\bf H}})$. 
	A solution $u\in e^{-\beta\rho}\hat{\mc{H}}$ to the equation $(\hat{{\bf H}}-\tfrac{Q^2}{4}-p^2)u=0$ with the asymptotic expansion \eqref{expansionpoisson} is unique.
	
	2) The operator $\hat{\mc{P}}_\ell(\alpha)$ admits a meromorphic extension  from the region \eqref{region_valide} to the region 
	\begin{equation}\label{regiondextension}
		\Big\{\alpha=Q+2ip \in Z\, |\,  \forall j\leq \ell, {\rm Im}\sqrt{p^2-\la_j}\in (\beta/2-\min_j\gamma_j,\beta/2)\Big\}\end{equation}
	and $\hat{\mc{P}}_\ell(\alpha)F$ satisfies \eqref{expansionpoisson} with the functions $F_j^+(\alpha)$ depending meromorphically on $\alpha$  in that region.
	
	3) The operator $\hat{\mc{P}}_\ell(\alpha)$ admits an analytic extension  from the region \eqref{region_valide} to the region 
	\begin{equation}\label{region_etendue} 
		\Big\{\alpha=Q+2ip\in \C\,\, |\, \,{\rm Re}(\alpha)<Q\, ,\, {\rm Im}(p)>{\rm Im}(\sqrt{p^2-\la_\ell})-\min_j\gamma_j/2\Big\}
	\end{equation}
	as an function in $e^{-({\rm Im}(\sqrt{p^2-\la_\ell})+\eps)\rho}\mc{D}(\hat{\mc{Q}})$ for all $\eps>0$
\end{proposition}

\subsection{Proof of Proposition \ref{resolution_bdry_Hamiltonian}: spectral resolution, Plancherel formula and generalised eigenstates}\label{proof_spec_resolution}

The proof of \cite[Theorem 6.25]{GKRV20_bootstrap} applied to our case now gives 
\begin{proposition}\label{spectralmeasure}
	For each $j\in \N$, let $(h_{jk})_{k=1,\dots,k(j)}$ be an orthonormal basis of 
	$\ker_{L^2(\Omega)}(\hat{{\bf P}}-\la_j)$. The spectral resolution holds   for all $u,u'\in  e^{\beta\rho}\hat{\mc{H}}$ 
	with $\beta>0$ 
	\begin{equation}\label{spectraldecomposH}
		\cjg u,u'\cjd_{\hat{\mc{H}}} = \frac{1}{2\pi}\sum_{j=0}^{\infty} \sum_{k=1}^{k(j)}\int_{0}^\infty
		\Big\cjg u,\hat{\mc{P}}_j\Big(Q+2i\sqrt{p^2+\la_j}\Big)h_{jk}\Big\cjd_{\hat{\mc{H}}} 
		\Big\cjg \hat{\mc{P}}_j\Big(Q+2i\sqrt{p^2+\la_j}\Big)h_{jk},u'\Big\cjd_{\hat{\mc{H}}} {\rm d}p.
	\end{equation}
	In particular, the spectrum of $\hat{{\bf H}}$ is absolutely continuous.
\end{proposition}

We next apply this to the particular case of the boundary Liouville Hamiltonian, i.e. with $\hat{{\bf H}}={\bf H}^{+}$, 
$\hat{\mc{H}}=\mc{H}^+$ and $\hat{{\bf P}}={\bf P}^+$ in the case  $\gamma\in (0,\sqrt{2})$ if $\mu>0$ and $\gamma\in (0,2)$ if $\mu=0$: one has in that case
\[ \min_{j=1,2,3} \gamma_j=\gamma/2,\]
the eigenfunctions  $h_{jk}$ of ${\bf P}^+$ are given by the family 
$(\psi_{\bf k}^+)_{\bf k}\in \bigcap_{p<\infty} L^p(\Omega_{\T^+})$ with 
\[ \psi_{\bf k}^+=\frac{\hat{\psi}^+_{\bf k}}{\|\hat{\psi}^+_{\bf k}\|_{L^2(\Omega_{\T^+})}}, \quad \hat{\psi}^+_{\bf k}=\prod_{k=1}{\rm He}_{k_n}(x_n)\]
and ${\bf k}=(k_1,\dots,k_n,\dots)$ is sequence of non-negative integers with only finitely many of them being non zero (this set of sequences is denoted $\mc{N}$). The corresponding eigenvalue is $|{\bf k}|:=\sum_{n=1}^\infty nk_n\in \N$ thus $(\la_j)_{j=1}^\infty=\N$.
In this case, we will denote $\mc{P}_\ell^+(\alpha):=\hat{\mc{P}}_\ell(\alpha)$ the Poisson operator for each $\ell\in \N$ and ${\bf R}^+(\alpha):=\hat{\bf R}(\alpha)$ the resolvent constructed in Proposition \ref{extensionresolvent} with $\hat{{\bf H}}={\bf H}^+$.
We set 
\[ \Psi^+_{\alpha,{\bf k}}:=\mc{P}^+_{|{\bf k}|}|(Q-2\sqrt{(Q-\alpha)^2/4-|\k|})\psi_{{\bf k}}^+,\qquad \Delta_{Q-2\sqrt{(Q-\alpha)^2/4-|\k|}}=\Delta_{\alpha}+|\bf k|\] 
where  $\sqrt{z}$ is used here with the convention that the cut is on $\R^-$ (ie. $\sqrt{Re^{i\theta}}=\sqrt{R}e^{i\theta/2}$ for $\theta\in (-\pi,\pi)$). The functions $\Psi^+_{\alpha,{\bf k}}$ are eigenfunctions of ${\bf H}^{+}$ with eigenvalues $\Delta_\alpha+|\k|$. A consequece of Proposition \ref{spectralmeasure} and the properties of $\mc{P}^+_{|{\bf k}|}(\cdot)$ in Proposition \ref{poisson_proposition} is the following:
\begin{proposition}[\textbf{Spectral resolution for the boundary Liouville Hamiltonian}]\label{holomorphic_eigenfunctions}
	Let $\gamma\in (0,\sqrt{2})$. For each  $\k \in \mc{N}$ and  $\beta>(Q-{\rm Re}(\alpha))/2$, the function $\Psi^+_{\alpha,{\bf k}}\in e^{- \beta \rho}\mc{D}(\mc{Q}^+)$ is an eigenfunction of ${\bf H}^{+}$ with eigenvalue $\Delta_\alpha+|{\bf k}|$ and it is holomorphic in the region
	\begin{equation}\label{defWell}
		W_{|{\bf k}|}:= \Big\{ \alpha \in \C \setminus \mc{D}_{|{\bf k}|}\, |\,  
		{\rm Re}(\alpha)\leq Q, {\rm Re}\sqrt{(Q-\alpha)^2/4-|{\bf k}|}>{\rm Re}(Q-\alpha)/2-\gamma/4\Big\}
	\end{equation}
	where $\mc{D}_{|{\bf k}|}=\bigcup_{j\geq |{\bf k}|}\{Q\pm 2i\sqrt{(j-|{\bf k}|)}\}$ is a discrete set where  
	$\Psi^+_{\alpha,{\bf k}}$ is continuous in $\alpha$ with square root singularities.
	The set $W_{|{\bf k}|}$ is a connected subset of the half-plane $\{{\rm Re}(\alpha)\leq Q\}$, containing $(Q+i\R)\setminus \mc{D}_{|{\bf k}|}$ and the real half-line $(-\infty,Q-\frac{4|{\bf k}|}{\gamma}-\frac{\gamma}{4})$. 
	For $u,u'\in \mc{H}^+$, the following holds  
	\eqref{spectraldecomposH} as 
	\begin{equation}\label{spectral_decomposition_H_+} 
		\cjg u,u'\cjd_{\mc{H}^+} = \frac{1}{2\pi}\sum_{{\bf k}\in \mc{N}} \int_{0}^\infty
		\cjg u , \Psi^+_{Q+2iP,{\bf k}}\cjd_{\mc{H}^+} 
		\cjg \Psi^+_{Q+2iP,{\bf k}},u'\cjd_{\mc{H}^+} \dd P.
	\end{equation}
\end{proposition}

Next, we discuss the properties of the descendant states for the Virasoro algebra. For $\nu \in \mc{T}$ a Young diagram, we define the function 
\begin{equation}\label{descendant_boundary}
	\Psi^+_{\alpha,\nu}:=\mc{P}_{|\nu|}^+(Q-2\sqrt{(Q-\alpha)^2/4-|\nu|})\mc{Q}^+_{\alpha,\nu} \qquad \textbf{(boundary Liouville descendant states)}
\end{equation}
in the region  $W_{|\nu|}$ defined by \eqref{defWell} (with $\la_\ell=\ell=|\nu|$) where $\Psi^{+,0}_{\alpha,\nu}=e^{\frac{\alpha-Q}{2}c}\mc{Q}^+_{\alpha,\nu}$ is the free descendant state defined in \eqref{Psi^+0alphanu}. Notice that $\bH^+ \Psi^+_{\alpha,\nu}=(\Delta_\alpha+|\nu|) \Psi^+_{\alpha,\nu}$ and that the region of analyticity of $\Psi^+_{\alpha,\nu}$ contains $(Q+i\R)\setminus \cup_{j\in \N}(Q+2i\sqrt{j})$, and that
$\Psi^+_{\alpha,\nu}$ is continuous on $Q+i\R$.
When $\alpha<Q-N$ for some $N$ large enough (depending on $|\nu|$), recall that
\begin{equation}\label{Psi+_Psi^+0}
	\Psi^+_{\alpha,\nu}=\lim_{t\to +\infty}e^{t(\Delta_\alpha+|\nu|)}e^{-t{\bf H}^+}\Psi^{+,0}_{\alpha,\nu}
\end{equation}
and if $\nu=0$ we have a probabilistic expression for $\Psi^+_\alpha$. For this, we write 
\[ c+\varphi^{\rm h}_t(e^{i\theta})+B_{2t}=X^\#(e^{-t+i\theta}), \quad B_{2t}=\frac{1}{2\pi}\int_0^{2\pi}(X^\#(e^{-t+i\theta})-c)\dd\theta\]
where $X^{\#}=c+X^{\#}_{\D^+}+P\varphi^{\rm h}$, and $B_t$ is a standard Brownian motion with covariance $\E[B_tB_{t'}]=\min(t,t')$
independent of the lateral noise process $\varphi^{\rm h}_t$. Then using that $S^{+,\bs{\mu}}(t)=e^{-t{\bf H}^+}$ with $S^{+,\bs{\mu}}(t)$ the operator \eqref{propagator_half}, we have 
\begin{equation}\label{primary_state}
	\begin{split}
		\Psi^+_{\alpha}(c,\varphi^{\rm h})=&\lim_{t\to +\infty}e^{t\Delta_\alpha}e^{-t{\bf H}^+}(e^{\frac{(\alpha-Q)c}{2}})\\
		=&\lim_{t\to +\infty}e^{t(\Delta_\alpha-\frac{Q^2}{4})} \E_{\varphi^{\rm h}}[e^{\frac{(\alpha-Q)}{2}(c+B_{2t})}e^{-\int_{\A^+_{e^{-t}}}  M^{\boldsymbol{\mu}}_\gamma(X^\#,\dd x)} ]\\
		\Psi^+_{\alpha}(c,\varphi^{\rm h})=& e^{\frac{(\alpha-Q)c}{2}}\E[e^{-\int_{\D^+}  M^{\boldsymbol{\mu}}_\gamma(X^\#-(\alpha-Q)\log|x|,\dd x)} ]
	\end{split}
\end{equation}
where we used Girsanov transform to absorb the term $e^{\frac{(\alpha-Q)}{2}B_{2t}}$ and 
$\E[ X^\#(x)B_{2t}]=-2\log|x|$ for $|x|\geq e^{-t}$.

Let us now  prove that the descendant states \eqref{descendant_boundary} can be expressed in terms of the 
Virasoro operators for the boundary Liouville theory:
\begin{lemma}\label{descendent_states_with_vir}
	For each $\nu=(\nu(1),\dots, \nu(k)) \in \mc{T}$ and $\beta>0$, for $\alpha<Q$ sufficiently negative,   the function $\Psi^+_{\alpha,\nu}\in e^{\beta \rho}\mc{D}(\mc{Q}^+)$  is such that for each $n\in \Z$, we have ${\bf L}^+_{n} \Psi^+_{\alpha,\nu}\in e^{\beta \rho}\mc{D}(\mc{Q}^+)$ 
	and 
	\begin{equation}\label{Ln^+_preserve}
		{\bf L}^+_{n} \Psi^+_{\alpha,\nu}\in {\rm span}\{\Psi^+_{\alpha,\nu'}\,|\, \nu'\in \mc{T}, |\nu'|=|\nu|-n \}
	\end{equation}
	with ${\bf L}^+_{n} \Psi^+_{\alpha,\nu}=0$ if $n>|\nu|$. This property extends analytically to   $W_{|\nu|}$ defined by \eqref{defWell}, 
	containing $(Q+i\R)\setminus \cup_{j\in \N} Q\pm 2i\sqrt{j}$, and 
	the descendant state $\Psi^+_{\alpha,\nu}$ can be written under the form 
	\[ \Psi^+_{\alpha,\nu}={\bf L}^+_{-\nu(k)}\dots {\bf L}^+_{-\nu(1)}\Psi^+_{\alpha}.\]
\end{lemma}

\begin{proof}
	Using \eqref{Psi+_Psi^+0}, it suffices to show that  for all $n\in\Z$, and $\alpha <Q-N$ for some $N$ large enough,
	\begin{align}
		\bL^+_n\Psi_{\alpha,\nu}^+=\underset{t\to\infty}{\lim}\, e^{t(\Delta_\alpha+|\nu|-n)} e^{-t\bH^+}\bL_n^{+,0}\Psi_{\alpha,\nu}^{+,0}
		\label{LnPsi}
	\end{align}
	since $\bL_n^{+,0}\Psi_{\alpha,\nu}^{+,0}$ is a linear combination of elements $\Psi^{+,0}_{\alpha,\nu'}$ for some $\nu'\in \mc{T}$ of length $|\nu'|=|\nu|-n$. 
	First, for $n\geq 0$, consider the Markovian vector field ${\rm v}=v(z)\pl_z$ with  $v(z)=-\omega z-z^{n+1}$ for $\omega>0$ large enough and denote $v_t(z):=e^tv(e^{-t}z)$ and ${\rm v}_t:=v_t\pl_z$ (recall that $z\mapsto e^{-t}z$ is the flow at time $t$ of $-z\pl_z$). By \eqref{Hv_Virasoro}, 
	note that $\bH^+_{\bf v}=\omega {\bf H}^++{\bf L}^+_n$.
	By definition, we have $\bH^+_{\rm v} e^{-t\bH^+}=-\pl_\eps 
	(e^{-\eps\bH^+_{\rm v}}e^{-t\bH^+})|_{\eps=0}$ acting on $\mc{D}(\mc{Q}^+)$. 
	By Proposition \ref{prop:semigroup} and the identity $e^{\eps {\rm v}}e^{-tz\pl_z}=e^{-tz\pl_z}e^{\eps {\rm v}_t}$, we see that 
	$e^{-\eps \bH^+_{\rm v}}e^{-t\bH^+}= e^{-t\bH^+}e^{-\eps \bH^+_{{\rm v}_t}}$  and differentiating at $\eps=0$ we obtain 
	\[\bH^+_{\rm v}e^{-t\bH^+}=e^{-t\bH^+}\bH^+_{{\rm v}_t}\]
	as continuous operators $e^{-\beta \rho}\mc{D}(\mc{Q}^+)\to e^{-\beta \rho}\mc{D}'(\mc{Q}^+)$ for all $\beta\in \R$.  
	By continuity of $\bH_{\rm v}^+:e^{-\beta \rho}\mc{D}(\mc{Q}^+)\to e^{-\beta \rho}\mc{D}'(\mc{Q}^+)$, the following holds 
	in $e^{-\beta \rho} \mc{D}'(\mc{Q}^+)$ as $t\to\infty$:
	\begin{equation}\label{first_intertwinning}
		\bH_{\rm v}^+\Psi^+_{\alpha,\nu}
		=\underset{t\to\infty}{\lim}\,e^{t(\Delta_\alpha+|\nu|)}\bH^+_{\rm v} e^{-t\bH^+}\Psi_{\alpha,\nu}^{+,0}
		=\underset{t\to\infty}{\lim}\,e^{t(\Delta_\alpha+|\nu|)}e^{-t\bH^+}\bH^+_{{\rm v}_t}\Psi_{\alpha,\nu}^{+,0}.
	\end{equation}
	Let us then write $\bH^+_{{\rm v}_t}=\bH_{{\rm v}_t}^{+,0} +V^{+,\bs{\mu}}_{{\rm v}_t}$ and we will show that 
	\[\underset{t\to\infty}{\lim}\,e^{t(\Delta_\alpha+|\nu|)}e^{-t\bH}V^{+,\bs{\mu}}_{{\rm v}_t}\Psi^{+,0}_{\alpha,\nu}=0\]
	in $e^{-\beta \rho}\mc{D}'(\mc{Q}^+)$. Note that this and \eqref{first_intertwinning} imply \eqref{LnPsi} as $\bH^{+,0}_{{\rm v}_t}\Psi_{\alpha,\nu}^{+,0}=\omega (\Delta_{\alpha}+|\nu|)\Psi_{\alpha,\nu}^{+,0}+e^{-nt}{\bf L}^{+,0}_n\Psi_{\alpha,\nu}^{+,0}$.
	We will use Definition \ref{quantTf} to express $e^{-t\bH}V^{+,\bs{\mu}}_{{\rm v}_t}\Psi^{+,0}_{\alpha,\nu}$, and for this we write 
	\[\varphi^{\rm h}_t(e^{i\theta})+B_{2t}=X^\#(e^{-t+i\theta}), \quad B_{2t}=\frac{1}{2\pi}\int_0^{2\pi}X^\#(e^{-t+i\theta})\dd\theta\]
	where $X^{\#}=X^{\#}_{\D^+}+P\varphi^{\rm h}$, and $B_t$ is a standard Brownian motion (vanishing at $t=0$) 
	independent of the lateral noise process $\varphi^{\rm h}_t$.
	Using that $|v_t|\leq C$ for some uniform $C>0$, by Definition \ref{quantTf} and \eqref{defVk} we have for all $c\in\R$ and $t>0$ 
	\begin{align*}
		&\big|\E_{\varphi^{\rm h}} \big[e^{-t\bH^+}( e^{ \gamma  c}V^{+,\cap}_{{\rm v}_t}\Psi^{+,0}_{\alpha,\nu})\big]\big|\\
		&\leq C e^{-\frac{Q^2}{4}t} e^{\frac{(\alpha-Q+2\gamma)c}{2}}\E_{\varphi^{\rm h}}\Big[e^{\frac{(\alpha-Q+2\gamma)}{2}B_{2t}}|\mc{Q}^+_{\alpha,\nu}(\varphi^{\rm h}_t)|
		\int_{\T^+}:e^{\gamma  \varphi^{\rm h}_t(\theta)}\!: |\sin(\theta)|^{-\frac{\gamma^2}{2}}\dd \theta 
		e^{-  \int_{\A^+_{e^{-t}}}|x|^{-\gamma Q}
			M_{\gamma}^{\bs{\mu}}(X^\#,\dd x)} \Big]\\
		&\leq C e^{\frac{t}{4}((\alpha-Q+2\gamma)^2-Q^2)}e^{\frac{(2\gamma+\alpha-Q)c}{2}}\int_0^{\pi}\frac{e^{\gamma P\varphi^{\rm h} (e^{-t+i \theta})}}{|\sin(\theta)|^{\frac{\gamma^2}{2}}}  \E_{\varphi^{\rm h}}   \Big[|\mc{Q}^+_{\alpha,\nu}(\varphi_t^{\rm h}+\gamma c_{t,\theta})|e^{-   \int_{\A^+_{e^{-t}}}\frac{e^{\gamma^2 d_{t,\theta}}}{|x|^{\gamma \alpha}}M_{\gamma}^{\bs{\mu}}(X^\#,\dd x) }\Big] \dd\theta\\
		&\leq C  e^{-t \Delta_\alpha +\gamma t(\gamma +\alpha-Q)}e^{\frac{(2\gamma+\alpha-Q)c}{2}}\ e^{\gamma \sup_{\theta} P\varphi^{\rm h} (e^{-t+i \theta})}  \sup_{\theta} \E_{\varphi^{\rm h}} [|\mc{Q}^+_{\alpha,\nu}(\varphi^{\rm h}_t+\gamma c_{t,\theta})|]
	\end{align*}
	where we used the Girsanov transform in the third line to remove the terms $e^{\frac{(\alpha-Q)}{2}B_{2t}}$ and $e^{\gamma X_{\D^+}^\#(e^{-t+i\theta})}$, 
	and we used the following covariance formulas for $z\in \A^+_{e^{-t}}$
	\[\begin{gathered}
		\E[ X^\#(z)B_{2t}]=-2\log|z| , \quad c_{t,\theta}(\theta'):=\E[ \varphi_t^{\rm h}(\theta')(X^{\#}_{\D^+}(e^{-t+i\theta)}-B_{2t})]= \log\frac{|1-e^{-2t+i(\theta+\theta')}|.|1-e^{-2t+i(\theta-\theta')}|}{|e^{i\theta}-e^{i\theta'}|.|e^{i\theta}-e^{-i\theta'}|}\\
		d_{t,\theta}(z):=\E[X^\#(z)X^\#_{\D^+}(e^{-t+i\theta})]=\log\frac{|1-ze^{-t-i\theta}|.|1-ze^{-t+i\theta}|}{|z-e^{-t+i\theta}|.|z-e^{-t-i\theta}|}.
	\end{gathered}\]
	Each Fourier coefficient of $c_{t,\theta}$ is uniformly bounded in $t>1,\theta\in \T$ thus 
	\begin{equation*}
		\sup_{\theta} \E_{\varphi^{\rm h}} [|\mc{Q}_{\alpha,\nu}(\varphi_t^{\rm h}+c_{t,\theta})|] \leq \E_{\varphi^{\rm h}} [   G(\varphi^{\rm h}_t)   ]
	\end{equation*}
	where $G$ is some positive polynomial  in finitely many Fourier coefficients of $\varphi^{\rm h}_t$. Using that $\P_{\T^+}$ is stationary for the Ornstein-Uhlenbeck processes driving $\varphi^{\rm h}_t$, we deduce that $\E[G(\varphi^{\rm h}_t)^4]=\E[G(\varphi^{\rm h})^4]<\infty$ 
	uniformly in $t\geq 1$. Hence for $t \geq 1$ and by using H\"older inequality 
	\begin{align*}
		\E[   | e^{\gamma \sup_{\theta} P\varphi^{\rm h} (e^{-t+i \theta})}  \E_{\varphi^{\rm h}} [G(\varphi^{\rm h}_t)]  |^2] & \leq \E[   e^{2 \gamma \sup_{\theta} P\varphi^{\rm h} (e^{-t+i \theta})}]^{1/2} \E[ \E_\varphi [G(\varphi_t)] ^4] ^{1/2} \leq C  .
	\end{align*}
	Since  $\int_\R e^{(2\gamma+\alpha-Q)c}e^{2\beta\rho(c)}\dd c<\infty$ when $\beta>Q-\alpha$ and $\alpha-Q+2\gamma<0$, there is $C>0$ such that
	\[e^{t(2\Delta_\alpha+|\nu|)}\|e^{-t\bH^+}(e^{\gamma c}V^{+,\cap}_{{\rm v}_t}\Psi_{\alpha,\nu}^0)\|_{e^{-\beta\rho}\mc{H}^+}\leq  Ce^{t(\gamma(\gamma+\alpha-Q)+|\nu|)} \]
	which converges to 0 for $\alpha <  Q-\gamma-\frac{|\nu|}{\gamma}$. Next, we deal with the 
	$V^{+,{\rm L}}_{{\rm v}_t}$ and  $V^{+,{\rm R}}_{{\rm v}_t}$ remainder terms, and since they are similar, only write the 
	$V^{+,{\rm L}}_{{\rm v}_t}$ case.
	Using \eqref{defVk2} and the Girsanov transform as above,
	\begin{align*}
		& \big|\E_{\varphi^{\rm h}} \big[e^{-t\bH^+}( e^{\frac{\gamma}{2} c}V^{+,{\rm L}}_{{\rm v}_t}\Psi^{+,0}_{\alpha,\nu})\big]\big|\\
		& \leq C e^{-\frac{Q^2}{4}t} e^{\frac{(\alpha-Q+\gamma)c}{2}}\E_{\varphi^{\rm h}}\Big[e^{\frac{(\alpha-Q+\gamma)}{2}B_{2t}}|\mc{Q}^+_{\alpha,\nu}(\varphi^{\rm h}_t)| :e^{\frac{\gamma}{2}\varphi^{\rm h}_t(\pi)}: 
		e^{- \int_{\A^+_{e^{-t}}}|x|^{-\gamma Q}
			M_{\gamma}^{\bs{\mu}}(X^\#,\dd x)} \Big]\\
		& \leq C e^{\frac{t}{4}((\alpha-Q+\gamma)^2-Q^2)}e^{\frac{(\gamma+\alpha-Q)c}{2}}  \E_{\varphi^{\rm h}}   \Big[|\mc{Q}^+_{\alpha,\nu}(\varphi_t^{\rm h}+\gamma c_{t,\pi})|e^{- \int_{\A^+_{e^{-t}}}\frac{e^{\gamma^2 d_{t,\theta}}}{|x|^{\gamma \alpha}}M_{\gamma}^{\bs{\mu}}(X^\#,\dd x) }\Big] \\
		& \leq  C  e^{-t \Delta_\alpha +\frac{\gamma}{4} t(\gamma +2(\alpha-Q))}e^{\frac{(\gamma+\alpha-Q)c}{2}}\ e^{\gamma P\varphi^{\rm h} (-e^{-t})}   \E_{\varphi^{\rm h}} [|\mc{Q}^+_{\alpha,\nu}(\varphi^{\rm h}_t+\gamma c_{t,\pi})|].
		V\end{align*}
	The same argument as above shows that \[e^{t(\Delta_\alpha+|\nu|)}\|e^{-t\bH^+}(e^{\frac{\gamma}{2} c}V^{+,{\rm L}}_{{\rm v}_t}\Psi_{\alpha,\nu}^{+,0})\|_{e^{-\beta\rho}\mc{H}^+}\leq  Ce^{\frac{\gamma}{4} t(\gamma +2(\alpha-Q))+|\nu|t} \]
	which goes to $0$ as $t\to \infty$ for $\alpha< Q-\frac{\gamma}{2}-\frac{2|\nu|}{\gamma}$.
	We deduce that in $e^{-\beta \rho}\mc{D}'(\mc{Q}^+)$
	\[\bH^+_{\rm v}\Psi^+_{\alpha,\nu}=\underset{t\to\infty}{\lim}\, e^{t(\Delta_\alpha+|\nu|)} e^{-t\bH^+}\bH^{+,0}_{{\rm v}_t}\Psi_{\alpha,\nu}^{+,0}.\]
	Next, since by \eqref{boundednesspropag} (applied with $\hat{\bH}=\bH^+$)  $e^{-t\bH^+}:\mc{D}'(\mc{Q}^+)\to\mc{D}(\mc{Q}^+)$ is continuous, then for all $F,G\in\mc{D}(\mc{Q}^+)$
	\[
	\langle(\bH^+_{{\rm v}_t})^*e^{-t\bH^+}F,G\rangle_{\mc{H}^+}
	=\langle F,e^{-t\bH^+}\bH^+_{{\rm v}_t}G\rangle_{\mc{H}^+}
	=\langle F,\bH^+_{\rm v} e^{-t\bH^+}G\rangle_{\mc{H}^+}
	=\langle e^{-t\bH^+}(\bH^+_{\rm v})^*F,G\rangle_{\mc{H}^+},
	\]
	thus, as bounded operators $\mc{D}(\mc{Q}^+)\to\mc{D}'(\mc{Q}^+)$, we obtain $(\bH^+_{{\rm v}_t})^*e^{-t\bH^+}=e^{-t\bH^+}(\bH^+_{\rm v})^*$.
	Replacing $v$ with $v_{-t}$ in this identity, this gives $(\bH^+_{{\rm v}})^*e^{-t\bH^+}=e^{-t\bH^+}(\bH^+_{{\rm v}_{-t}})^*$. Note that 
	$(\bH^+_{{\rm v}})^*=\omega {\bH}^++{\bf L}^+_{-n}$ and as in \eqref{first_intertwinning}
	\[
	(\omega {\bH}^++{\bf L}^+_{-n})\Psi^+_{\alpha,\nu}
	=\underset{t\to\infty}{\lim}\,e^{t(\Delta_\alpha+|\nu|)}e^{-t\bH^+}(\bH^+_{{\rm v}_{-t}})^*\Psi_{\alpha,\nu}^{+,0}=
	\underset{t\to\infty}{\lim}\,e^{t(\Delta_\alpha+|\nu|)}e^{-t\bH^+}(\omega \bH^++e^{tn}{\bf L}_{-n}^+)\Psi_{\alpha,\nu}^{+,0}.
	\] 
	Using $\bH^++e^{tn}{\bf L}_{-n}^+=\bH^{+,0}+e^{tn}{\bf L}_{-n}^{+,0}+(V^{+,\bs{\mu}}_{{\rm v}_{-t}})^*$ we can deduce that
	\[\bL^+_{-n}\Psi_{\alpha,\nu}^+=\underset{t\to\infty}{\lim}\, e^{t(\Delta_\alpha+|\nu|+n)} e^{-t\bH^+}\bL_{-n}^{+,0}\Psi_{\alpha,\nu}^{+,0}\]
	if we can show that 
	\[\underset{t\to\infty}{\lim}\,e^{t(\Delta_\alpha+|\nu|)}e^{-t\bH^+}(V^{+,\bs{\mu}}_{{\rm v}_{-t}})^*\Psi_{\alpha,\nu}^{+,0}=0.
	\]
	But this limit holds by the same exact argument as above if $\alpha$ is negative enough, since 
	$(V^{+,\bs{\mu}}_{{\rm v}_{-t}})^*$ has the same form as $V^{+,\bs{\mu}}_{{\rm v}_{-t}}$ but with an additional $e^{nt}$ growth due to the fact that $v_{-t}=-\omega z-e^{nt}z^{n+1}$.
\end{proof}

\begin{proof}[Proof of Proposition  \ref{resolution_bdry_Hamiltonian}] The eigenstates $\Psi^+_{Q+2iP,{\bf k}}$ for ${\bf k}\in \mc{N}$ and $P\in \R^*$
	can be rewritten as linear combinations of $\Psi^+_{Q+2iP,\nu}$ for $|\nu|=|{\bf k}|$, and the proof of \cite[Lemma 8.1]{GKRV20_bootstrap} applied  with the spectral resolution of Proposition \ref{holomorphic_eigenfunctions} gives directly \eqref{resolution_boundary_Hamiltonian}. Point (i) in Proposition  \ref{resolution_bdry_Hamiltonian} follows from Lemma \ref{descendent_states_with_vir}. The statement (ii) follows from the exact same argument as for the Virasoro generators ${\bf L}_n$, see the proof of \cite[Theorem 4.5]{BGKRV}.
\end{proof}

Next, we refine the asymptotic behaviour described in Proposition \ref{poisson_proposition} of the eigenstates $\Psi^+_{Q+2ip,\nu}(c,\varphi^{\rm h})$ as $c\to -\infty$ for $p\in \R^*$. We will in particular show that there are only two oscillating terms in the asymptotic expansion. In  scattering-theoretic terms, this means that the scattering operator is diagonal (we do not define here the scattering operator, but this can be done exactly as in \cite[Section 6.7]{GKRV20_bootstrap}).

As a consequence of Proposition \ref{poisson_proposition}, there exists some coefficient $R^+(\alpha)\in \C$ such that the eigenstate $\Psi_{Q+2ip}^+$ for $p\in \R$ has the asymptotic behavior 
\begin{equation}\label{asympopt_primary}
	\Psi^+_{Q+2ip}(c,\varphi^{\rm h})=e^{ipc} + R_{\mu}^+(Q+2ip)e^{-ipc}+G_0(Q+2ip), \quad G_0(p)\in e^{\frac{\gamma}{8}\rho}\mc{H}^+\cap \mc{D}(\mc{Q}^+).
\end{equation}
and, the uniqueness of the solution of $({\bf H}^+-\Delta_{Q+2ip})u=0$ in $e^{-\eps \rho}\mc{D}(\mc{Q}^+)$ (for $\eps>0$ small) with asymptotic expansion of the form \eqref{asympopt_primary} shows that for $p>0$ 
\[\Psi^+_{Q-2ip}=R_{\mu}^+(Q-2ip)\Psi^+_{Q+2ip}.\]

\begin{lemma}
	The descendant states $\Psi_{Q+2ip,\nu}^+$ satisfy the asymptotic expansion for $p>0$
	\begin{equation}\label{descendant_asympt}
		\Psi^+_{Q+2ip,\nu}(c,\varphi^{\rm h})=e^{ipc}\mc{Q}^+_{Q+2ip,\nu}(\varphi^{\rm h}) + R_{\mu}^+(Q+2ip)e^{-ipc}\mc{Q}^+_{Q-2ip,\nu}(\varphi^{\rm h})+G_{Q+2ip,\nu}(c,\varphi^{\rm h})
	\end{equation}
	with $G_{Q+2ip,\nu}\in e^{\frac{\gamma}{8}\rho}\mc{H}^+\cap \mc{D}(\mc{Q}^+)$.
\end{lemma}
\begin{proof} The proof is essentially the same as \cite[Theorem 4.2]{BGKRV}. We provide the argument for completeness.
	We proceed by induction. For $\ell\in \N$, let $\mc{I}(\ell)$  be the following property: 
	for all $\nu,\tilde{\nu}\in \mc{T}$ with $|\nu|+|\nu'|\leq \ell$ and $P\in \R_+\setminus \cup_{j=0}^\infty \{\sqrt{j}\}$,
	\begin{equation}\label{assumptionpsi}
		\Psi^+_{Q+2ip,\nu}(c,\varphi^{\rm h})=e^{iPc}\mc{Q}^+_{Q+2ip,\nu}(\varphi^{\rm h})+R_{\mu}^+(Q+2ip)e^{-ipc}\mc{Q}^+_{Q-2ip,\nu}(\varphi^{\rm h})+G_{Q+2ip,\nu}(c,\varphi^{\rm h})
	\end{equation}
	for some $G_{Q+2ip,\nu}\in \mc{D}(\mc{Q}^+)$. First, $\mc{I}(0)$ is satisfied by  \eqref{asympopt_primary}.
	Assume that $\mc{I}(\ell-1)$ holds for $\ell\geq1$. For  $\nu \in \mc{T}$ with $|\nu|=\ell$, we write $\nu=(\nu',n)$, 
	with $|\nu'|+n=|\nu|$ for some $\nu'\in \mc{T}$, 
	For $F\in \mc{C}_{\rm exp}(\R\times \Omega_{\T^+})$, 
	\begin{equation}\label{splitting}
		\begin{split}
			\cjg \Psi^+_{Q+2ip,\nu'},{\bf L}^+_n F\cjd_{\mc{H}^+} =  &\cjg  e^{ipc}\mc{Q}^+_{Q+2ip,\nu'}+R_\mu^+(Q+2ip)e^{-ipc}\mc{Q}^+_{Q-2ip,\nu'}, 
			{\bf L}^{+,0}_nF\cjd_{\mc{H}^+}  +\cjg G_{Q+2ip,\nu'},{\bf L}^+_nF\cjd_{\mc{H}^+}  \\
			& +  \mu\cjg  e^{iPc}\mc{Q}^+_{Q+2ip,\nu',\tilde{\nu}}+R_\mu^+(Q+2ip)e^{-ipc}\mc{Q}^+_{Q-2ip,\nu',\tilde{\nu}}, e^{\gamma c} V^{+,\cap}_n F\cjd_{\mc{H}^+}\\
			&+  \mu^{\rm L}\cjg  e^{ipc}\mc{Q}^+_{Q+2ip,\nu',\tilde{\nu}}+R_\mu^+(Q+2ip)e^{-ipc}\mc{Q}^+_{Q-2ip,\nu',\tilde{\nu}}, e^{\frac{\gamma}{2} c} V^{+,{\rm L}}_{n} F\cjd_{\mc{H}^+}\\
			& + \mu^{\rm R}\cjg  e^{ipc}\mc{Q}^+_{Q+2ip,\nu',\tilde{\nu}}+R_\mu^+(Q+2ip)e^{-ipc}\mc{Q}^+_{Q-2ip,\nu',\tilde{\nu}}, e^{\frac{\gamma}{2} c} V^{+,{\rm R}}_n F\cjd_{\mc{H}^+}
	\end{split}\end{equation}
	by \eqref{assumptionpsi} and \eqref{formulaLn}, where $V^{+,*}_n$ are the potential terms in \eqref{formulaLn} for $*\in \{\cap,{\rm R},{\rm L}\}$. 
	We choose some particular $F$ depending on a parameter $T>0$, namely
	\[ F_T(c,\varphi)=\chi_T(c) e^{ip'c}h(\varphi) , \quad \textrm{ with }h\in E_{\ell}, \quad \chi_T(c)=\chi(c+T)\]
	for $p'\in \R$, and  $\chi \in C_0^\infty(\R;[0,1])$ satisfying $\chi=1$ near $0$, and  let  $\tilde{\chi}$ be a cutoff function with the same property as $\chi$ but with $\tilde{\chi}\chi=\chi$, and let $\tilde{\chi}_T(c)=\tilde{\chi}(c+T)$.
	First, observe that 
	${\bf L}_n^{+,0}(e^{ip'c}h(\varphi))= e^{ip'c}h'(\varphi)$ for some $h'\in E_{\ell-n}$.
	We now consider the asymptotic behaviour, as $T\to \infty$, of \eqref{splitting}. The last four terms in the RHS will go to $0$, and the first term in the RHS is explicit.
	
	First, for all $p\in \R^*$, ${\bf L}^{+,0}_{-n}(e^{\pm ipc}\mc{Q}^+_{Q\pm 2ip,\nu'})= e^{\pm ipc}\mc{Q}^+_{Q\pm 2ip,(\nu',n)}$
	so that for $p,p'\in \R^*$
	\[\begin{split} 
		\cjg  e^{\pm ipc}\mc{Q}^+_{Q\pm 2ip,\nu'}, {\bf L}^{+,0}_n (\chi_Te^{ ip'c}h)\cjd_2=& \int \chi(c+T)e^{ic(\pm p-p')}\dd c 
		\cjg\mc{Q}^+_{Q\pm 2ip,(\nu',n),\tilde{\nu}} , h\cjd_{L^2(\Omega_{\T^+})}\\
		=&e^{iT(p'\mp p)} \hat{\chi}(p'\mp p)\cjg\mc{Q}^+_{Q\pm 2ip,(\nu',n),\tilde{\nu}} , h \cjd_{L^2(\Omega_{\T^+})}
	\end{split}\]
	where $\hat{\chi}$ denotes the Fourier transform.
	The last three lines of \eqref{splitting} go to $0$ as $T\to \infty$: for $*\in \{\cap,{\rm R},{\rm L}\}$
	\[
	|\cjg e^{\pm ipc}\mc{Q}^+_{Q\pm 2ip,\nu'}, e^{\gamma_* c}V^{+,*}_n F_T\cjd_{\mc{H}^+}|\leq C \Big| \int_{-T-1}^{-T+1}e^{\gamma_* c}\dd c\Big|\times 
	|\cjg V_n^{+,*} h,\mc{Q}^+_{Q\pm 2ip,\nu'}\cjd_{L^2(\Omega_{\T^+})}| \to 0
	\]
	where $\gamma_*=\gamma$ if $*=\cap$ and $\gamma_*=\gamma/2$ if $*\in\{{\rm R},{\rm L}\}$, and  
	$\cjg V_n^{+,*} h,\mc{Q}^+_{Q+2ip,\nu'}\cjd_{L^2(\Omega_{\T^+})}$ makes sense since $\mc{Q}^+_{Q+2ip,\nu'}h\in \mc{S}(\Omega_{\T^+})$. 
	We will show below  that $\cjg G_{Q+2ip,\nu'},{\bf L}^+_nF\cjd_{\mc{H}^+}\to 0$ as $T\to \infty$, but let us assume so for now. Then, 
	choosing $p'=-p$ and $\chi$ so that $\int \chi=1$, as $T\to +\infty$
	\begin{equation}\label{asymptoticPsiLnF}
		\begin{split}
			\cjg \Psi^+_{Q+2ip,\nu'},{\bf L}^+_n F_T\cjd_{\mc{H}^+} = & e^{-2iTp}\hat{\chi}(-2p)\cjg\mc{Q}^+_{Q+2ip,(\nu',n)} , h\cjd_{L^2(\Omega_{\T^+})}\\
			& +R^+(Q+2ip)\cjg\mc{Q}^+_{Q- 2ip,(\nu',n)} , h\cjd_{L^2(\Omega_{\T^+})}+o(1).
		\end{split}
	\end{equation}
	Applying the same reasoning to the asymptotic expansion  \eqref{expansionpoisson} with $F_j^-=0$ for $j<\ell$ and $F_\ell^-:=\mc{Q}_{Q+2ip,\nu}^+$, we have as $T\to \infty$ (with $F_j^+$ defined by \eqref{expansionpoisson})
	\[\begin{split} 
		\cjg \Psi_{Q+ip,(\nu',n)},  F_{T}\cjd_{\mc{H}^+}=& e^{-2iTp}\hat{\chi}(-2p)\cjg\mc{Q}^+_{Q+2ip,(\nu',n)} , h\cjd_{L^2(\Omega_{\T^+})}+ \cjg 
		F^+_{\ell}, h\cjd_{L^2(\Omega_{\T^+})}
		\\
		&+\sum_{j<\ell} e^{iT(p-\sqrt{p^2+\ell-j})}\hat{\chi}(p-\sqrt{p^2+\ell-j}) 
		\cjg F_j^+, h\cjd_{L^2(\Omega_{\T^+})}+o(1).
	\end{split}\]
	and this shows that $F_\ell^+=R^+(Q+2ip)\mc{Q}^+_{Q- 2ip,(\nu',n)}$ and $F_{j}^-=0$ for $j<\ell$ since $h$ is arbitrary, which proves \eqref{descendant_asympt}.
	
	To prove that $\cjg G_{Q+2ip,\nu'},{\bf L}^+_nF_T\cjd_{L^2(\R^-\times \Omega_\T)}\to 0$, we decompose ${\bf L}^+_n={\bf L}_n^{+,0}+{\bf V}_n^{+,\bs{\mu}}$ where ${\bf V}_n^{+,\bs{\mu}}$ is the potential term in \eqref{formulaLn}. There is $C>0$ such that as $T\to \infty$
	\begin{equation}\label{firstboundG}
		|\cjg G_{Q+2ip,\nu'},\chi_T {\bf V}_n^{+,\bs{\mu}}e^{ip'c}h\cjd_{\mc{H}^+}|\leq C\|\tilde{\chi}_TG_{Q+2ip,\nu'}\|_{\mc{D}(\mc{Q}^+)} \|\chi_Te^{ip'c}h\|_{\mc{D}(\mc{Q}^+)}
	\end{equation}
	since ${\bs V}_n^{+,\bs(\mu)}:\mc{D}(\mc{Q}^+)\to \mc{D}'(\mc{Q}^+)$ is bounded. It is easy to see that $\|\tilde{\chi}_TG_{Q+2ip,\nu'}\|_{\mc{D}(\mc{Q}^+)}\to 0$ as $T\to \infty$ since $G_{Q+2ip,\nu'}\in \mc{D}(\mc{Q}^+)$  and that 
	$ \|\chi_Te^{ip'c}h\|_{\mc{D}(\mc{Q}^+)}$ is uniformly bounded (see proof of \cite[Theorem 4.2]{BGKRV} for details), thus 
	$\cjg G_{Q+2ip,\nu'}, {\bf V}_n^{+,\bs{\mu}}F_T\cjd_{\mc{H}^+} \to 0$ as $T\to \infty$. Next, 
	\[  |\cjg G_{Q+2ip,\nu'},{\bf L}_n^{+,0}F_T\cjd_{\mc{H}^+} \leq \|\tilde{\chi}_TG_{Q+2ip,\nu'}\|_{\mc{H}^+}\|{\bf L}_n^{+,0}F_T\|_{\mc{H}^+}\]
	and the proof of \cite[Theorem 4.2]{BGKRV} (using the expression \eqref{Ln^+0}) shows that $\|{\bf L}_n^{+,0}F_T\|_{\mc{H}^+}$ is uniformly bounded while $\|\tilde{\chi}_TG_{Q+2ip,\nu'}\|_{\mc{H}^+}$ clearly goes to $0$ as $T\to \infty$.
	This ends the proof.
\end{proof}

It is proved in \cite[Proposition 3.6]{CH26} that the reflection coefficient $\tilde{R}_\mu^+(\alpha)$ is given by the following expression:

\noindent i) Case $\mu>0$:
\begin{equation}\label{eq: mu>0}
\widetilde{R}_{\mu>0}^+(\alpha)=\big(\frac{\pi \mu \Gamma(\frac{\gamma^2}{4})(\frac{\gamma}{2})^{2-\frac{\gamma^2}{2}}}{\Gamma(1-\frac{\gamma^2}{4})}\big)^{\frac{Q-\alpha}{\gamma}}\frac{\Gamma_{\frac{\gamma}{2}}(\alpha-Q)S_{\frac{\gamma}{2}}(2Q-\sigma^{\rm R}-\sigma^{\rm L}-\frac{\alpha}{2})S_{\frac{\gamma}{2}}(\sigma^{\rm R}+\sigma^{\rm L}-\frac{\alpha}{2})}{\Gamma_{\frac{\gamma}{2}}(Q-\alpha)S_{\frac{\gamma}{2}}(\frac{\alpha}{2}+ \sigma^{\rm L}-\sigma^{\rm R})S_{\frac{\gamma}{2}}(\frac{\alpha}{2}- \sigma^{\rm L}+\sigma^{\rm R})}.
\end{equation}
ii) Case $\mu=0$ and $\mu^{\rm L}+\mu^{\rm R}>0$: 
if $\mu^{\rm L},\mu^{\rm R}>0$, choose $\sigma^{\rm L},\sigma^{\rm R}$ by
	\begin{equation}\label{eq:FZZT_parameters_mu_zero_checked}
		\mu^{\rm L}=e^{i\pi\gamma(\sigma^{\rm L}-Q/2)},\qquad
		\mu^{\rm R}=e^{i\pi\gamma(\sigma^{\rm R}-Q/2)},\qquad
		{\rm Re}(\sigma^{\rm L})={\rm Re}(\sigma^{\rm R})=Q/2.
	\end{equation}
	If exactly one of $\mu^{\rm L},\mu^{\rm R}$ vanishes, the formula below is understood in the limiting sense.
\begin{equation}\label{eq: mu=0}
	\widetilde{R}_{\mu=0}^+(\alpha)=-\Gamma\!\Big(1-\frac{2(Q-\alpha)}{\gamma}\Big)\frac{(2\pi)^{\frac{2}{\gamma}(Q-\alpha)-\frac12}
		(2/\gamma)^{\frac{\gamma}{2}(Q-\alpha)-\frac12}
		\Gamma_{\frac{\gamma}{2}}(\alpha-b)e^{i\pi(\sigma^{\rm L}+\sigma^{\rm R}-Q)(Q-\alpha)}}
	{(Q-\alpha)\,\Gamma(1-\gamma^2/4)^{\frac{2}{\gamma}(Q-\alpha)}
		\Gamma_{\frac{\gamma}{2}}(Q-\alpha)
		S_{\frac{\gamma}{2}}(\alpha/2+\sigma^{\rm R}-\sigma^{\rm L})
		S_{\frac{\gamma}{2}}(\alpha/2+\sigma^{\rm L}-\sigma^{\rm R})}.
\end{equation}
where $\Gamma_{\gamma/2}(x)$ is the Barnes double Gamma function, $S_{\gamma/2}(x)=\frac{\Gamma_{\gamma/2}(x)}{\Gamma_{\gamma/2}(Q-x)}$ is the double sine function, and 
\[ \frac{\mu^{\rm L}}{\sqrt{\mu}}= (\sin(\pi \frac{\gamma^2}{4}))^{-\frac{1}{2}}\cos(\pi \gamma(\sigma^{\rm L}-\frac{Q}{2}) ),\quad \frac{\mu^{\rm R}}{\sqrt{\mu}}=(\sin(\pi \frac{\gamma^2}{4}))^{-1/2}\cos(\pi \gamma(\sigma^{\rm R}-\frac{Q}{2}) )\]
A simple computation shows  
\begin{equation}\label{eq:Rplus_at_Q}
	\widetilde{R}_\mu^+(\alpha)\widetilde{R}_\mu^+(2Q-\alpha)=1,\qquad \widetilde{R}_{\mu}^+(Q)=-1.
\end{equation}
\noindent\textbf{An exceptional set.}
Put $\lambda_{m,n}:=m\frac{\gamma}{2}+n\frac{2}{\gamma}$ for $m,n\in\mathbb N_0$, and set
\[
\mathcal E_\mu:=
\begin{cases}
\{\pm(\sigma^{\rm L}+\sigma^{\rm R}-Q),\,
  \pm(\sigma^{\rm L}-\sigma^{\rm R})\},
&\mu>0,\\[1mm]
\{\pm(\sigma^{\rm L}-\sigma^{\rm R})\},
&\mu=0,\quad\mu^{\rm L},\mu^{\rm R}>0,\\[1mm]
\varnothing,
&\mu=0,\quad\mu^{\rm L}\mu^{\rm R}=0.
\end{cases}
\]
We use the discrete set
\[
\mathcal K:=
\bigcup_{m,n\geq0}\{Q-\lambda_{m,n},\,Q+\lambda_{m,n}\}
\;\cup\!
\bigcup_{\substack{m,n\geq0\\d\in\mathcal E_\mu}}
\{2d-2\lambda_{m,n},\,2Q-2d+2\lambda_{m,n}\}.
\]
For fixed cosmological constants, this set is  invariant under
$\alpha\mapsto2Q-\alpha$, and independent of the descendant level.
The first lattice also contains all Kac parameters. 

We recall function $\rho:\R\to \R_-$ satisfies $\rho(c)=0$ for $c\geq 0$, and $\rho(c)=c-a$ for $c\leq -1$, and $\rho'(c)\in [0,1]$ everywhere. In particular, $\rho\leq 0$.
\begin{lemma}\label{lem:boundary_primary_meromorphic}
	Assume $\mu,\mu^{\rm L},\mu^{\rm R}\geq0$ and $\mu+\mu^{\rm L}+\mu^{\rm R}>0$.  The primary family $\alpha\mapsto\Psi^+_\alpha$, initially defined in the physical half-plane ${\rm Re}(\alpha)<Q$, extends holomorphically to an open neighborhood of $\{{\rm Re}(\alpha)\leq Q\}$ and meromorphically to $\mathbb C$. Away from $\mathcal{K}$, we have $\Psi_\alpha^+\in e^{-\beta\rho}\mathcal{D}(\mathcal{Q}^+)$ for $\beta>\frac{1}{2}|\mathrm{Re}(\alpha)-Q|$.
	It satisfies,
	\begin{equation}\label{eq:boundary_primary_reflection_global}
		\Psi^+_{Q-2ip}=\widetilde{R}_\mu^+(Q-2ip)\Psi^+_{Q+2ip}\quad(p\in\mathbb R\setminus\{0\}),
		\qquad
		\Psi^+_{2Q-\alpha}=\widetilde{R}_\mu^+(2Q-\alpha)\Psi^+_\alpha.
	\end{equation}
\end{lemma}

\begin{proof}
	We follow \cite[Lemma 4.3]{BGKRV}, keeping track of the ramification points.  Fix $\beta>0$ and choose $0<\beta_0<\min\{\beta,\gamma/4\}$.  By Propositions \ref{extensionresolvent} and \ref{poisson_proposition} with $\ell=0$, the family $\Psi^+_\alpha=\mc P^+_0(\alpha)1$ has an $e^{-\beta_0\rho}\mc D(\mc Q^+)$-valued meromorphic continuation to a neighborhood in the spectral surface $Z$ of the physical line.  Let $\tau_j$ be a local deck involution above $Q\pm2i\sqrt j$, $j\geq1$.  The function $r_0(\alpha)=-\frac i2(\alpha-Q)$ and the coefficient $R^+(\alpha)$ functions on $\C$ (invariant by the deck transformations).  Hence, at every point $\alpha=Q+2ip$, which is not a ramification point, in a small sepctral interval contained in the continuation domain, the expansion \eqref{asympopt_primary} gives
	\[
	\tau_j^*\Psi^+_\alpha-\Psi^+_\alpha\in\mc D(\mc Q^+).
	\]
	This difference is an eigenfunction of $\bH^+$ with eigenvalue $\Delta_{Q+2ip}$.  This eigenfunction is then in  $\mc D(\bH^+)$, and by Lemma \ref{embeddedeig} it is $0$.  we thus get $\tau_j^*\Psi^+=\Psi^+$ in the common holomorphic continuation domain, so the family descends from $Z$ to the $\alpha$-plane.
	
	We now show there is no square-root singularity.  Near $\alpha_j=Q\pm2i\sqrt j$, write $\alpha-\alpha_j=w^2$.  The meromorphic family on $Z$ has a finite Laurent expansion
	\[
	\Psi^+(w)=\sum_{k=-N}^{\infty}A_k w^k.
	\]
The invariance by the deck transformations on $Z$ reads $\Psi^+(w)=\Psi^+(-w)$, hence $A_k=0$ for every odd $k$.. Using  3) in Proposition \ref{extensionresolvent} we deduce that  $\Psi^+_\alpha$ is holomorphic on an open neighborhood of $\{{\rm Re}(\alpha)\leq Q\}$.
For $p\in \R^*$, by \eqref{asympopt_primary} and the absence of $L^2$-eigenvalue, we deduce that
\[\Psi^+_{Q-2ip}=\widetilde{R}^+(Q-2ip)\Psi^+_{Q+2ip}.\]
This identity produces the mermorphic extension  of $\Psi_\alpha^+$ by setting, for ${\rm Re}(\alpha)\geq Q$,
\[\Psi^+_\alpha:=\widetilde{R}^+(\alpha)\Psi^+_{2Q-\alpha}.\]

	Finally, if ${\rm Re}(\alpha)<Q$, point~3 of Proposition \ref{poisson_proposition} gives, for every $\varepsilon>0$,
	\[
	\Psi^+_\alpha\in e^{-((Q-{\rm Re}(\alpha))/2+\varepsilon)\rho}\mc D(\mc Q^+).
	\]
	For $\alpha\in\mathcal S_\beta:=\{\alpha\in\mathbb C\,|\, |{\rm Re}(\alpha)-Q|<2\beta\}$ choose $\varepsilon$ so that $(Q-{\rm Re}(\alpha))/2+\varepsilon<\beta$; the case ${\rm Re}(\alpha)>Q$ follows from the reflection definition, and the physical line by continuity.  Hence the continuation is $e^{-\beta\rho}\mc D(\mc Q^+)$-valued meromorphic on $\mathcal S_\beta$.  
\end{proof}
\begin{theorem}\label{thm:boundary_descendants_meromorphic}
	Under the assumptions of Lemma \ref{lem:boundary_primary_meromorphic}, for every Young diagram $\nu\in\mc T$,  $\Psi^+_{\alpha,\nu}$ admits a meromorphic continuation to $\alpha \in \C$.  It is holomorphic on an open neighborhood of $\{{\rm Re}(\alpha)\leq Q\}$. Avoiding from $\mathcal{K}$, $\alpha\to \Psi_{\alpha,\nu}^+$  is an $e^{-\beta\rho}\mc D(\mc Q^+)$-valued analytic function for $\beta>\frac{1}{2}|\mathrm{Re}(\alpha)-Q|$.  The following identities hold meromorphically
	\begin{align}
		\bH^+\Psi^+_{\alpha,\nu}&=(\Delta_\alpha+|\nu|)\Psi^+_{\alpha,\nu},\label{eq:boundary_descendant_eigen_global}\\
		\Psi^+_{\alpha,\nu}&=\bL^+_{-\nu(k)}\cdots\bL^+_{-\nu(1)}\Psi^+_\alpha,
		\qquad \nu=(\nu(1),\ldots,\nu(k)),\label{eq:boundary_descendant_vir_global}\\
		\Psi^+_{2Q-\alpha,\nu}&=\widetilde{R}^+(2Q-\alpha)\Psi^+_{\alpha,\nu}.
		\label{eq:boundary_descendant_reflection_global}
	\end{align}
\end{theorem}

\begin{proof}
	We argue as in \cite[Theorem 4.4]{BGKRV}.  Fix $\beta>0$ throughout the proof. The same argument as in  \cite[Theorem 4.4]{BGKRV} proves that the operators
$\bL_n^+:e^{-\beta\rho}\mc D(\mc Q^+)\to e^{-\beta\rho}\mc D'(\mc Q^+)$ and $\bH^+:e^{-\beta\rho}\mc D(\mc Q^+)\to e^{-\beta\rho}\mc D'(\mc Q^+)$ are bounded, while \eqref{boundednesspropag} gives
\[e^{-t\bH^+}:	e^{-\beta\rho}\mc D'(\mc Q^+)\longrightarrow e^{-\beta\rho}\mc D(\mc Q^+),\qquad t>0.\]
We use an induction on $|\nu|$.   Lemma \ref{lem:boundary_primary_meromorphic} deals with $|\nu|=0$.  
Write $\nu=(\nu',n)$ and $\ell=|\nu|$.  By the induction hypothesis,
	\[
	\Phi_{\alpha,\nu}:=\bL^+_{-n}\Psi^+_{\alpha,\nu'}
	\]
	is a $e^{-\beta\rho}\mc D'(\mc Q^+)$-valued meromorphic family on $\mathcal S_\beta:=\{\alpha\in\mathbb{C}\,|\, |{\rm Re}(\alpha)-Q|<2\beta\}$.  On a non-empty open subset of $\mathcal S_\beta\cap W_\ell\cap W_{|\nu'|}\cap\{{\rm Re}(\alpha)<Q\}$, by Lemma \ref{descendent_states_with_vir} 
	\[
	(\bH^+-\Delta_\alpha-\ell)\Phi_{\alpha,\nu}=0,
	\qquad
	e^{-t\bH^+}\Phi_{\alpha,\nu}
	=e^{-t(\Delta_\alpha+\ell)}\Phi_{\alpha,\nu}.
	\]
	The second identity follows from the first by differentiating in $t$ in the weighted spaces.  Both sides are $e^{-\beta\rho}\mc D'(\mc Q^+)$-valued meromorphic on the connected strip $\mathcal S_\beta$, so the identity extends to the whole strip.  Hence, for any fixed $t>0$,
	\[
	\Phi_{\alpha,\nu}
	=e^{t(\Delta_\alpha+\ell)}e^{-t\bH^+}\Phi_{\alpha,\nu},
	\]
	and the right-hand side is in $e^{-\beta\rho}\mc D(\mc Q^+)$.  We define $\Psi^+_{\alpha,\nu}:=\Phi_{\alpha,\nu}\in e^{-\beta\rho}\mc D(\mc Q^+)$. By Lemma~\ref{descendent_states_with_vir} for $\alpha$ very negative,  $\Phi_{\alpha,\nu}$ is the analytic continuation of $\Psi^+_{\alpha,\nu}$ and we get \eqref{eq:boundary_descendant_vir_global}. 
	
	The same formula shows inductively that the family $\Psi^+_{\alpha,\nu}$ is holomorphic with values in $e^{-\beta\rho}\mc{D}(\mc{Q}^+)$, provided the $\Psi^+_{\alpha,\nu'}$ with $|\nu'|<|\nu|$ satisfy this property. By unique continuation from ${\rm Re}(\alpha)$ very negative, we obtain \eqref{eq:boundary_descendant_eigen_global}.  
Finally, applying $\bL^+_{-n}$ to the identity $\Psi^+_{2Q-\alpha,\nu'}=\widetilde{R}^+(2Q-\alpha)\Psi^+_{\alpha,\nu'}$  
proves \eqref{eq:boundary_descendant_reflection_global} for $\nu=(\nu',n)$ and we obtain the desired result by induction.
\end{proof}

		\bibliographystyle{alpha}
		\bibliography{references}  
	\end{document}